\documentclass[11pt,twoside]{book} 
\usepackage{asp2008n}
\usepackage{times}
\usepackage{lscape}
\usepackage{epsf}

\usepackage{graphicx}
\usepackage{amssymb}
\usepackage{longtable}
\usepackage[figuresright]{rotating}
\usepackage{multirow}

\newcommand{\simless}{\mathbin{\lower 3pt\hbox {$\rlap{\raise 5pt\hbox{$\char'074$}}\mathchar"7218$}}}

\mainmatter

\newlength{\deftabcolsep}
\begin{document}



\title{Star Forming Regions in Cepheus}
\author{M\'aria Kun}
\affil{Konkoly Observatory, H-1525 Budapest, P.O. Box 67, Hungary}
\author{Zolt\'an T. Kiss}
\affil{Baja Astronomical Observatory, P.O. Box 766, H-6500 Baja, Hungary}
\author{Zolt\'an Balog\footnotemark}
\affil{Steward Observatory, University of Arizona, 933 N. Cherry Av., Tucson AZ 85721, USA}
\footnotetext{on leave from Dept. of Optics and Quantum Electronics, University
 of Szeged, D\'om t\'er 9, Szeged, H-6720, Hungary}

\begin{abstract}
The northern Milky Way in the constellation of Cepheus ($100\deg \le l \le 120\deg; 0\deg \le b \le 20\deg$)
contains several star forming regions. The molecular clouds of the
Cepheus Flare region at $b > 10\deg$, are sites of low and intermediate
mass star formation located between 200 and 450~pc from the Sun. Three nearby
OB~associations, Cep~OB2, Cep~OB3, Cep~OB4, located at 600--800 pc, are each
involved in forming stars, like the well known high mass star forming region
S\,140 at 900~pc. The reflection nebula NGC\,7129 around 1~kpc harbors young,
compact clusters of low and intermediate mass stars. The giant star forming
complex NGC\,7538 and the young open cluster NGC\,7380, associated with the
Perseus arm, are located at $d > 2$\,kpc.
\end{abstract}

\section{Overview}

\begin{figure}[!ht]
\centerline{
\includegraphics[draft=False,width=9cm]{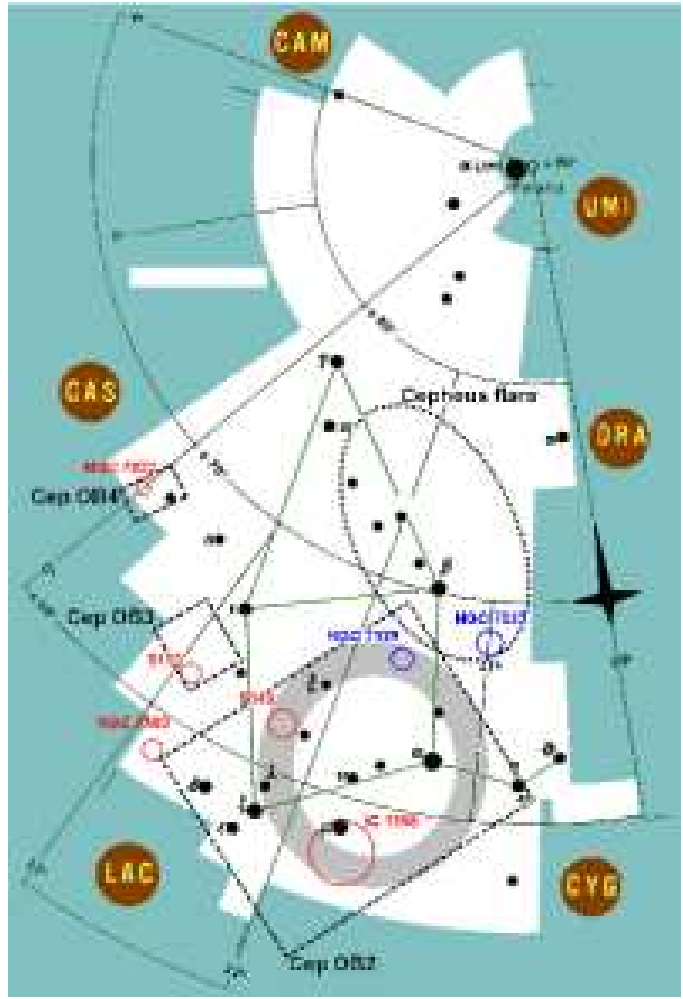}
}
\caption{Positions of the major star forming regions of Cepheus, overplotted on a schematic
drawing of the constellation.}
\label{cepheus:Fig1}
\end{figure}

In this chapter we describe the star forming regions of the constellation of
Cepheus. A large scale map of the constellation, with the boundaries defined
by IAU overlaid, and the most prominent star forming regions indicated,
is shown in Fig.~\ref{cepheus:Fig1}. This huge area of the sky, stretching
between the Galactic latitudes of about 0\deg \ and +30\deg, contains
several giant star forming molecular cloud complexes located at various distances
from the Sun. According to their distance they can be ranged into three large
groups: \\
(1) Clouds nearer than 500\,pc located mainly at $b \ge 10\deg$, in the Cepheus Flare. \\
(2) Three OB~associations, Cep~OB\,2, Cep~OB\,3 and Cep~OB\,4 between 600--900\,pc. \\
(3) Star forming regions associated with the Perseus spiral arm at 2--3\,kpc. \\
In the following we discuss the first two of these groups.

\begin{figure}[htbp!]
\centerline{
\includegraphics[draft=False,width=5.25in]{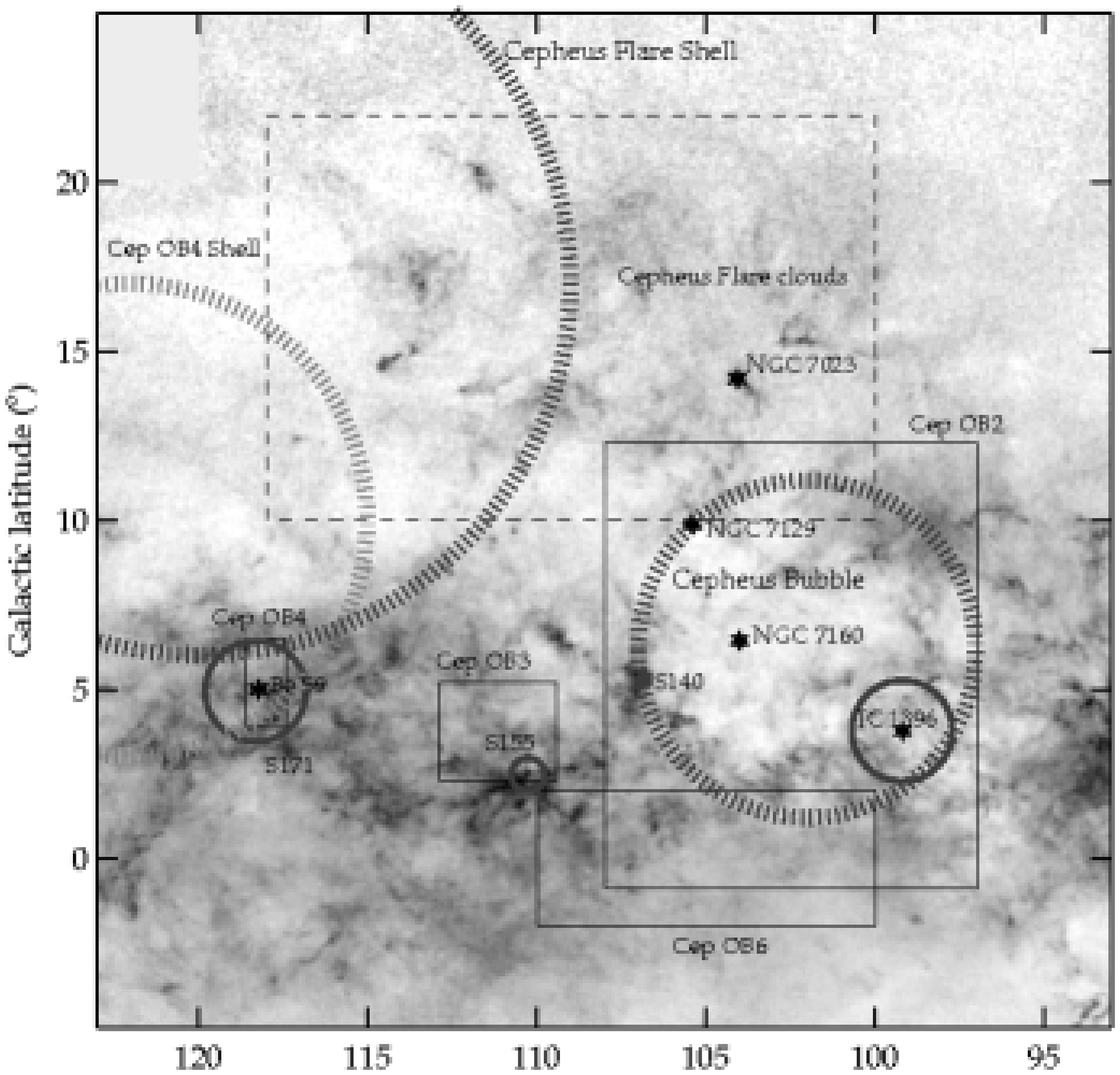}
}
\caption{Distribution of the visual extinction  in the Cepheus region in the [l,b] plane \citep{DUK}
with outlines of the major star forming regions, discussed in this chapter, overplotted.
Solid rectangles  indicate the nominal boundaries of the OB associations
\citep{Humphreys,deZeeuw}, and the dashed rectangle shows the Cepheus Flare cloud complex;
giant HII regions are  marked by solid grey circles, and star symbols indicate
young open clusters.  Three large circles, drawn by radial dashes, show giant shell-like
structures in the interstellar medium. The Cepheus Flare Shell \citep{Olano} belongs to
the nearby Cepheus Flare complex, and the Cepheus Bubble \citep{KBT} is associated with the
association Cep~OB2, and the Cepheus~OB4 Shell \citep{Olano} is associated with Cep~OB4.}
\label{Fig_DUK}
\end{figure}

Fig.~\ref{Fig_DUK} shows the distribution of dark clouds perpendicular to the Galactic
plane \citep{DUK}, with the outlines of the major star forming complexes overplotted.

The large-scale $^{13}$CO observations performed by \citet{YDMOF}
led to a refinement of division of the clouds into complexes. The groups
listed in Table~\ref{Tab_groups}, adopted from  \citet{YDMOF},
were defined on the basis of their positions, radial velocities, and
distances, where distance data were available.

Table~\ref{Tab_cloud} lists the dark clouds
identified in the Cepheus region from \citet{Barnard} to the Tokyo Gakugei University (TGU) Survey
\citep{DUK}, and the molecular clouds, mostly revealed by the millimeter
emission by various isotopes of the carbon monoxide \citep{DBYF,YDMOF}. The cloud name in the
first column is the LDN \citep{LyndsD} name where it exists, otherwise the first appearance of the
cloud in the literature.
Equatorial (J2000) and Galactic coordinates are listed in columns (2)--(5), and the
area of the cloud in square degrees in column~(6). Column (7) shows the radial velocity of the cloud
with respect to the Local Standard of Rest. The number of associated young stellar objects
is given in column (8), and the alternative names, following the system of
designations by {\it SIMBAD\/}, are listed in column (9). We note that the LDN coordinates,
derived from visual examination of the POSS plates, may be uncertain in several cases.

\begin{figure*}[htbp!]
\centering
\includegraphics[draft=False,height=\textwidth,angle=90]{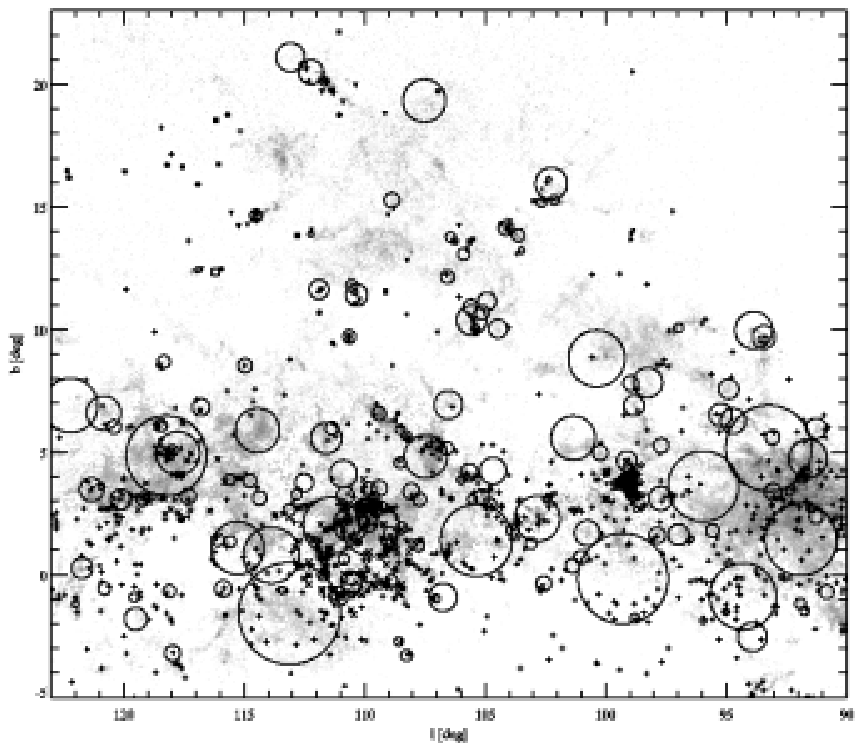}
\caption{Pre-main sequence stars and candidates in the Cepheus region overlaid on the map of visual
extinction obtained from 2MASS data based on interstellar reddening using the NICER method \citep{Lombardi}.
Large circles denote those clouds from Table~\ref{Tab_cloud}
which have been associated with young stars. The meaning of different symbols are as follows:
Filled triangles - T~Tauri stars;
Filled squares - Herbig Ae/Be stars; Filled circles - Weak-line T~Tauri stars;
Diamonds - Tr\,37 ROSAT X-ray sources; Open triangles - PMS members of Tr\,37;
Open squares - Candidate and possible PMS members of Cep OB3b; X - H$\alpha$ emission stars;
 + - T~Tauri candidates selected from a 2MASS color-color diagram.}
\label{fig_yso}
\end{figure*}

Figure~\ref{fig_yso} shows the distribution of the pre-main sequence stars and candidates
over the whole Cepheus region.

\footnotesize
\begin{landscape}
\begin{center}
\begin{longtable}{l@{\hskip2mm}c@{\hskip2mm}c@{\hskip2mm}c@{\hskip2mm}c@{\hskip2mm}c@{\hskip2mm}c@{\hskip2mm}c@{\hskip2mm}l}

\caption{List of clouds catalogued in Cepheus.}\\
\noalign{\smallskip}
\tableline
\noalign{\smallskip}
Cloud name           & RA(2000)   & Dec(2000)  & l      & b      & Area   & v$_\mathrm{LSR}$~~ &  $n_\mathrm{star}$ & Alternative names \\
     & (h m)   & ($\deg$ $\arcmin$)  & ($\deg$)   & ($\deg$)   & (sq.deg.) & (km\,s$^{-1}$)  &  &  \\
\noalign{\smallskip}
\tableline
\noalign{\smallskip}
\endfirsthead

\caption{Continued.}\\
\noalign{\smallskip}
\tableline
\noalign{\smallskip}
Cloud name       & RA(2000)   & Dec(2000)  & l      & b      & Area   & v$_\mathrm{LSR}$~~  &  $n_\mathrm{star}$ & Alternative names \\
     & (h m)   & ($\deg$ $\arcmin$)  & ($\deg$)   & ($\deg$)   & (sq.deg.) & (km\,s$^{-1}$)  &  &  \\
\noalign{\smallskip}
\tableline
\noalign{\smallskip}
\endhead

\noalign{\smallskip}
\endfoot

\noalign{\smallskip}
\tableline
\noalign{\smallskip}
\multicolumn{9}{l}{\parbox{18cm}{\footnotesize 
References:  
$[$ADM95$]$~--~\citet{ADM95};
Barnard~--~\citet{Barnard};
$[$B77$]$~--~\citet{B77};
$[$BKP2003$]$~--~\citet{BKP03};
$[$BM89$]$~--~\citet{BM89};
BFS~--~\citet{BFS};
CB~--~\citet{CB88}; 
Ced~--~\citet{Ced}
DBY~--~\citet{DBYF}; 
$[$DE95$]$~--~\citet{DE95}; 
DG ~--~\citet{DG};
DSH~--~\citet{KTA2006};
$[$ETM94$]$~--~\citet{Eiroa94}; 
$[$FMS2001$]$~--~\citet{FMS01}; 
FSE~--~\citet{FSEM};
$[$G82a$]$~--~\citet{Gy2}
$[$G85$]$~--~\citet{GRS}; 
$[$GA90$]$~--~\citet{GA0}; 
GAL~--~\citet{KLW96};  
GF~--~\citet{GF};
GM ~--~\citet{Magakian03};
GN~--~\citet{Magakian03}; 
GSH~--~\citet{EP2005};
$[$GSL2002$]$~--~\citet{GSL}
HCL ~--~\citet{Heiles}; 
$[$HW84$]$~--~\citet{HW84}; 
$[$IS94$]$~--~\citet{IS94}; 
JWT Core~--~\citet{JWT00}; 
$[$KC97c$]$~--~\citet{KC97}; 
KR~--~\citet{KR80};
$[$KSH92$]$~--~\citet{Kawabe}; 
$[$KTK2006$]$~--~\citet{KTK2006};
LDN~--~\citet{LyndsD}; 
$[$LM99$]$~--~\citet{LM99}; 
$[$M73$]$~--~\citet{Martin73}; 
MBM~--~\citet{MBM}; 
Min ~--~\citet{Min}; 
$[$MPR2003$]$~--~\citet{MPR}; 
$[$NJH2003$]$~--~\citet{NJH03}; 
$[$P85b$]$~--~\citet{Petros85}
$[$PGH98b$]$\,Cloud~--~\citet{Patel98}; 
$[$PGS95$]$~--~\citet{Patel95}; 
PP~--~\citet{PP79}; 
RNO~--~\citet{Cohen}
SFO ~--~\citet{SFO}; 
Sh\,2- ~--~\citet{Sharpless}; 
$[$SMN94$]$~--~\citet{SFN94}; 
TDS~--~\citet{TDS}; 
TGU~--~\citet{DUK}; 
$[$THR85$]$~--~\citet{THR85};
$[$TOH95$]$~--~\citet{TOH95}; 
$[$TW96$]$~--~\citet{TW}; 
$[$WAM82$]$~--~\citet{WAM82};
$[$WC89$]$ ~--~\citet{WC89}; 
WWC~--~\citet{Weikard};
YMD\,CO~--~\citet{YDMOF}; 
$[$YF92$]$~--~\citet{YF92}; 
$[$YNF96$]$~--~\citet{YNF96}. }}
\endlastfoot

LDN 1122         & 20 32.7 & +65 20.3 & ~99.97 & +14.81 &  0.041 &   ~~~4.8~~ &   0 & YMD\,CO\,1, TGU\,597 \\
LDN 1089         & 20 32.8 & +63 30.3 & ~98.40 & +13.78 &  0.244 & ~$-$2.4~~ &   0 & TDS\,366, TGU\,586 \\
LDN 1094         & 20 32.8 & +64 00.3 & ~98.83 & +14.06 &  0.002 &   ~~~0.1~~ &   1 & CB 222 \\
TGU 667          & 20 33.9 & +73 51.6 & 107.53 & +19.33 &  2.460 &  &   1 & $[$KTK2006$]$ G107.1+19.3, [KTK2006]  \\
   &         &          &        &        &        &       &     & G107.5+19.8, $[$KTK2006$]$ G107.9+18.9 \\
LDN 1152         & 20 34.4 & +68 00.4 & 102.37 & +16.14 &  0.025 & ~~~2.9~~   &   0 & YMD\,CO\,8 \\
CB 223           & 20 34.7 & +64 10.7 & 99.10  & +13.98 &  0.003 & ~$-$2.7~~ & 0 \\
LDN 1100         & 20 34.8 & +64 00.4 & ~98.96 & +13.88 &  0.007 & ~$-$2.7~~  &   1 & CB\,224 \\
LDN 1033         & 20 37.2 & +57 10.5 & ~93.44 & ~+9.69 &  0.714 & ~$-$2.4~~ &   0 & TDS\,343, TDS\,345, TDS\,346, \\ 
                 &         &          &        &        &        &       &     & TGU\,549, DBY\,093.1+09.6,  \\
                 &         &          &        &        &        &       &     &  DBY\,093.5+09.4 \\
LDN 1036         & 20 38.2 & +57 10.6 & ~93.52 & ~+9.58 &  0.088 & ~$-$2.1~~ &   1 & DBY\,093.5+09.4 \\
LDN 1041         & 20 38.2 & +57 30.6 & ~93.79 & ~+9.78 &  0.022 & ~$-$2.1~~ &   0 & DBY\,093.5+09.4 \\
LDN 1157         & 20 39.5 & +68 00.7 & 102.65 & +15.75 &  0.005 & ~~~2.9~~ &  1 & $[$DE95$]$\,LDN 1157\,A1, YMD\,CO\,8, TGU\,619 \\
LDN 1147         & 20 40.5 & +67 20.8 & 102.14 & +15.29 &  0.127 & ~~~2.9~~ &   0 & YMD\,CO\,8, TDS\,379, TGU\,619, \\
                 &         &          &        &        &        &       &     &  JWT Core 36, $[$LM99$]$\,340, \\
                 &         &          &        &        &        &       &     &   $[$BM89$]$\,1-99, $[$LM99$]$\,344 \\
LDN 1148         & 20 40.5 & +67 20.8 & 102.14 & +15.29 &  0.015 & ~~~2.9~~ &   1 & YMD\,CO\,8, TGU\,619 \\
LDN 1044         & 20 41.2 & +57 20.8 & ~93.90 & ~+9.35 &  0.006 & ~$-$2.1~~ &   0 & DBY\,093.5+09.4  \\
LDN 1049         & 20 42.2 & +57 30.8 & ~94.12 & ~+9.35 &  0.005 & ~~~0.6~~ &   0 & DBY\,094.1+09.4 \\
LDN 1039         & 20 42.3 & +56 50.8 & ~93.58 & ~+8.94 &  0.079 & ~$-$2.1~~ &   0 & DBY\,093.5+09.4  \\
LDN 1051         & 20 42.7 & +57 30.9 & ~94.16 & ~+9.29 &  0.010 &  ~~~0.6~~  &   0 & DBY\,094.1+09.4 \\
LDN 1155         & 20 43.5 & +67 40.9 & 102.60 & +15.25 &  0.006 &  ~~~2.9~~ &   0 & YMD\,CO\,8, TGU\,619 \\
LDN 1158         & 20 44.5 & +67 41.0 & 102.66 & +15.17 &  0.111 &  ~~~0.9~~ &   0 & TDS\,388, $[$BM89$]$\,1-100, $[$LM99$]$\,346, TGU\,619\\ 
                 &         &          &        &        &        &       &     &   $[$BM89$]$\,1-102, $[$BM89$]$\,1-103, $[$LM99$]$\,348 \\
LDN 1038         & 20 46.3 & +56 21.0 & ~93.53 & ~+8.20 &  0.660 & ~$-$2.1~~ &   0 & $[$BM89$]$\,1-104, DBY\,093.5+09.4 \\
LDN 1076         & 20 49.3 & +59 51.2 & ~96.57 & +10.04 &  0.008 & ~$-$2.2~~ &   0 & DBY\,096.8+10.2 \\
LDN 1082         & 20 51.1 & +60 11.3 & ~96.98 & +10.07 &  0.111 & ~$-$2.6~~ &  8 & Barnard 150, GF\,9, TDS\,362, $[$LM99$]$\,350,  \\
                 &         &          &        &        &        &       &    & DBY\,097.1+10.1, $[$BM89$]$\,1-105, $[$LM99$]$\,351 \\
LDN 1171         & 20 53.5 & +68 19.5 & 103.71 & +14.88 &  0.002 &  ~~~3.4~~ &   0 & CB\,229 \\

LDN 1037         & 20 54.4 & +55 26.5 & ~93.53 & ~+6.75 &  0.591 &       &   0 & TGU\,551 \\
LDN 1168         & 20 56.6 & +67 36.6 & 103.31 & +14.21 &  0.003 &       &   0 &  \\
LDN 1061         & 20 58.3 & +57 21.7 & ~95.36 & ~+7.56 &  0.618 & ~$-$2.1~~ &   0 & DBY\,095.2+07.4, DBY\,095.2+07.4 \\
LDN 1071         & 20 58.3 & +58 11.7 & ~96.00 & ~+8.10 &  0.092 & ~$-$0.4~~ &   0 & Barnard 354, DBY\,096.0+08.1, TGU\,569  \\
LDN 1056         & 20 58.4 & +56 01.7 & ~94.35 & ~+6.69 &  0.036 &       &   0 &  \\
TGU\,730         & 20 58.8 & +78 11.8 & 112.23 & +20.50 &  0.803 & ~$-$8.0~~ & 3 & $[$BM89$]$\,1-108, $[$BM89$]$\,1-109,\\
	         &         &          &        &        &        &       &    & $[$B77$]$\,48, $[$BM89$]$\,1-112, GN 21.00.4, RNO\,129 \\
LDN 1228         & 20 59.0 & +77 31.8 & 111.67 & +20.10 &  0.086 & ~$-$7.6~~ &   4 & YMD\,CO\,66, MBM 162, TGU\,718 \\
LDN 1170         & 21 01.7 & +67 36.9 & 103.63 & +13.84 &  0.215 & ~~~2.8~~ &   0 & TDS\,392, TGU\,629, GSH 093+07+9 \\
	         &         &          &        &        &        &       &     & $[$LM99$]$\,359, $[$LM99$]$\,361 \\
LDN 1058         & 21 02.4 & +56 01.9 & ~94.72 & ~+6.27 &  0.869 & ~$-$1.1~~ &   1 & TDS\,354, TGU\,558 \\
LDN 1174         & 21 02.6 & +68 11.9 & 104.15 & +14.14 &  0.294 & ~~~2.8~~ & $\ga15$~~ & TDS\,393, $[$LM99$]$\,358, $[$BM89$]$1-110, \\  
	         &         &          &        &        &        &       &     & $[$B77$]$\,39, $[$BM89$]$\,1-113, $[$LM99$]$\,360, \\
	         &         &          &        &        &        &       &     & $[$BM89$]$\,1-114, $[$BM89$]$\,3B- 9, \\
	         &         &          &        &        &        &       &     & $[$BM89$]$\,1-111, PP~100, $[$BM89$]$\,1-115 \\		 		 
	         &         &          &        &        &        &       &     & Ced\,187, GN 21.01.0,  YMD\,CO\,14 \\
LDN 1172         & 21 02.7 & +67 41.9 & 103.76 & +13.82 &  0.010 & ~~~2.7~~ & $\ga4$~~ &  YMD\,CO\,14, TGU\,629 \\
LDN 1167         & 21 03.7 & +67 02.0 & 103.30 & +13.32 &  0.062 & ~~~2.8~~  &   0 & YMD\,CO\,11  \\
LDN 1173         & 21 04.7 & +67 42.0 & 103.88 & +13.67 &  0.006 & ~~~2.7~~  &   0 & YMD\,CO\,14 \\
LDN 1068         & 21 06.3 & +57 07.1 & ~95.89 & ~+6.59 &  0.027 & ~~~0.2~~  &   0 & Barnard 359, DBY\,095.9+06.6, TGU\,570 \\
LDN 1063         & 21 07.4 & +56 22.2 & ~95.44 & ~+5.97 &  0.025 & ~~~0.3~~  &   0 & Barnard 151, Barnard 360, \\
	         &         &          &        &        &        &       &     &  DBY\,095.5+06.5, TGU\,561 \\
LDN 1065         & 21 07.4 & +56 32.2 & ~95.56 & ~+6.09 &  0.023 & ~~~0.3~~   &   0 & DBY\,095.5+06.5, TGU\,568 \\
LDN 1069         & 21 08.4 & +56 52.2 & ~95.90 & ~+6.21 &  0.018 & ~~~0.3~~   &   0 & DBY\,095.5+06.5 \\
LDN 1067         & 21 09.4 & +56 42.3 & ~95.87 & ~+6.00 &  0.012 & ~~~0.3~~   &   0 & DBY\,095.5+06.5,  TGU\,568 \\
Sh\,2-129        & 21 11.3 & +59 42.3 & ~98.26 & ~+7.84 &  1.190 & ~~~3.7~~ &   2 & DBY\,097.3+08.5, BFS 9 \\
LDN 1060         & 21 11.5 & +55 17.4 & ~95.02 & ~+4.82 &  0.031 & ~$-$0.9~~ &  0 & DBY\,090.5+02.4 \\
LDN 1119         & 21 13.2 & +61 42.4 & ~99.91 & ~+9.03 &  0.010 & ~~~2.5~~ &   0 & DBY\,100.1+09.3, TGU\,598 \\
LDN 1062         & 21 13.5 & +55 32.4 & ~95.40 & ~+4.79 &  0.003 & ~$-$0.9~~  &   0 & DBY\,090.5+02.4 \\
LDN 1064         & 21 13.5 & +55 37.4 & ~95.46 & ~+4.85 &  0.001 & ~$-$0.9~~ &   0 & DBY\,090.5+02.4 \\
LDN 1125         & 21 14.7 & +61 42.5 & 100.04 & ~+8.90 &  0.010 &  ~~~2.5~~ &   1 & Barnard 152, TDS\,376, $[$LM99$]$\,365, \\ 
                 &         &          &        &        &        &           &     & DBY\,100.1+09.3, TGU\,598 \\

LDN 1072         & 21 16.5 & +56 12.6 & ~96.18 & ~+4.95 &  1.510 & ~$-$0.6~~ &  0 & Barnard 153, TDS\,358, TGU\,574, \\ 
	         &         &          &        &        &        &       &     & DBY\,096.3+05.2 \\
LDN 1177         & 21 18.8 & +68 15.7 & 105.22 & +13.06 &  0.005 & ~~~2.9~~  & 2 & CB\,230, TGU\,641 \\
TGU 589          & 21 19.7 & +59 27.7 & ~98.83 & ~+6.90 &  0.560 &       &   2 &  \\
LDN 1162         & 21 20.0 & +65 02.8 & 102.92 & +10.76 &  0.004 &       &   0 &  \\
LDN 1080         & 21 21.5 & +56 32.8 & ~96.91 & ~+4.69 &  0.001 &       &   0 & Barnard 154 \\
YMD CO\,23       & 21 22.2 & +69 22.6 & 106.27 & +13.60 &  0.056 &~ $-$9.4~~ &   2 & TGU\,656 \\
LDN 1108         & 21 26.4 & +59 33.1 & ~99.49 & ~+6.37 &  0.046 &       &   0 &  \\
DBY\,098.4+05.2  & 21 26.8 & +57 55.9 & ~98.40 & ~+5.17 &  0.107 & ~~~1.0~~   &  0 &  \\
LDN 1086         & 21 28.5 & +57 33.2 & ~98.30 & ~+4.74 &  0.095 & ~$-$4.6~~ & 18 & DBY\,098.8+04.2, TGU\,584, $[$PGS95$]$\,2, \\
 	         &         &          &        &        &        &       &     & IC\,1396\,W, FSE\,1\\
LDN 1176         & 21 31.0 & +66 43.3 & 104.93 & +11.15 &  0.446 & $-$10.4~~ & 0 & YMD\,CO\,17, TGU\,634, \\
 	         &         &          &        &        &        &       &     & JWT Core 44, GAL 104.9+11.2 \\
LDN 1145         & 21 31.3 & +62 43.3 & 102.14 & ~+8.24 &  0.183 &       &   0 & TGU\,622, $[$PGH98b$]$\,Cloud 11 \\
LDN 1146         & 21 31.3 & +62 43.3 & 102.14 & ~+8.24 &  0.183 &       &   0 &  \\
LDN 1096         & 21 31.5 & +58 03.3 & ~98.93 & ~+4.83 &  0.001 &       &   0 & TGU\,590 \\
LDN 1102         & 21 32.5 & +58 03.3 & ~99.03 & ~+4.74 &  0.008 &       &   0 &  \\
LDN 1085         & 21 33.3 & +56 44.6 & ~98.23 & ~+3.69 &  0.010 & ~~~3.4~~   &   0 & WWC\,156 \\
LDN 1093         & 21 33.5 & +57 38.4 & ~98.85 & ~+4.34 &  0.034 & ~$-$4.6~~ &   20 & DBY\,098.8+04.2, TGU\,587, FSE\,3 \\
                     &         &          &        &        &        &       &     & $[$PGS95$]$\,8 \\
LDN 1083         & 21 33.6 & +55 58.0 & ~97.72 & ~+3.11 &  0.761 & ~~~8.7~~ &  & WCC\,178--179 \\
LDN 1098         & 21 34.5 & +57 38.0 & ~98.95 & ~+4.25 & 0.025 & ~$-$4.6~~ &  20 & DBY\,098.8+04.2, FSE\,3 \\
Barnard\,365     & 21 34.9 & +56 43.0 & ~98.36 & ~+3.53 & 0.010 & ~~~7.6~~  & &  WWC\,140 \\
TGU 639          & 21 35.5 & +66 32.0 & 105.13 & +10.70 &  0.280 &       &   1 & $[$KTK2006$]$ G105.0+10.7 \\
LDN 1087         & 21 35.6 & +56 33.5 & ~98.32 & ~+3.35 &  0.013 & ~~~7.4~~ &   2 & TDS\,364 \\
LDN 1090         & 21 35.6 & +56 43.5 & ~98.43 & ~+3.48 &  0.028 &       &   0 &  Barnard 365 \\
LDN 1199         & 21 35.9 & +68 33.5 & 106.57 & +12.15 &  0.235 & $-$11.5~~ &   1 & TDS\,398, TGU\,653 \\
                 &         &          &        &        &        &       &     & $[$KTK2006$]$ G106.4+12.0, \\
                 &         &          &        &        &        &       &     & $[$KTK2006$]$ G106.7+12.3 \\
LDN 1099         & 21 36.5 & +57 23.5 & ~98.98 & ~+3.88 &  0.008 & ~$-$8.0~~ &   22 &  DBY\,099.1+04.0, FSE\,5 \\
LDN 1116         & 21 36.5 & +58 33.5 & ~99.76 & ~+4.75 &  0.034 &       &   3 & $[$IS94$]$\,7, $[$PGS95$]$\,16, $[$IS94$]$\,14, SFO\,35, \\
                 &         &          &        &        &        &       &     &  $[$G85$]$\,5, $[$PGS95$]$\,17, $[$PGS95$]$\,19, FSE\,6 \\

LDN 1105         & 21 37.1 & +57 33.5 & ~99.15 & ~+3.95 &  0.008 & ~$-$8.0~~ & 22 & TDS\,369, $[$B77$]$\,37, DBY\,099.1+04.0, FSE\,5  \\
YMD CO\,19       & 21 37.3 & +67 01.2 & 105.60 & +10.93 &  0.279 & $-$10.2~~ &   1 & TGU\,642, $[$KTK2006$]$ G105.5+10.8, \\
                 &         &          &        &        &        &       &     & $[$KTK2006$]$ G105.7+10.8 \\
LDN 1092         & 21 37.6 & +56 58.5 & ~98.81 & ~+3.48 &  0.014 &       &   1 &  \\
LDN 1112         & 21 38.5 & +58 03.6 & ~99.63 & ~+4.20 &  0.013 & ~$-$0.3~~ &   0 & DBY\,099.7+04.1 \\
LDN 1117         & 21 38.5 & +58 18.6 & ~99.79 & ~+4.39 &  0.004 & ~$-$0.7~~ &   0 & WWC\,114--116 \\
LDN 1088         & 21 38.6 & +56 13.6 & ~98.41 & ~+2.83 &  0.019 & ~~~7.2~~   &   2 & Barnard 160, TGU\,585, \\
	         &         &          &        &        &        &       &     &  DBY\,098.4+02.9 \\
LDN 1135         & 21 39.4 & +60 38.6 & 101.44 & ~+6.06 &  0.054 &       &   0 &  \\
LDN 1110         & 21 39.5 & +57 53.6 & ~99.62 & ~+3.99 &  0.007 & ~$-$0.3~~  &   2 & DBY\,099.7+04.1 \\
LDN 1111         & 21 39.5 & +57 53.6 & ~99.62 & ~+3.99 &  0.002 &  ~$-$0.3~~  &   2 & Barnard 161, DBY\,099.7+04.1, CB\,233 \\
LDN 1123         & 21 39.5 & +58 28.6 & 100.00 & ~+4.43 &  0.003 &       &   0 &  \\
LDN 1126         & 21 39.5 & +58 33.6 & 100.06 & ~+4.49 &  0.006 &       &   0 &  \\
LDN 1101         & 21 39.6 & +56 58.6 & ~99.01 & ~+3.30 &  0.095 &       &   0 &  \\
LDN 1131         & 21 40.0 & +59 33.7 & 100.77 & ~+5.20 &  0.046 & ~$-$0.4~~ &   0 & Barnard 366, TDS\,378, \\
 	         &         &          &        &        &        &       &     & DBY\,100.9+05.3,  TGU\,609 \\
LDN 1140         & 21 40.4 & +60 53.7 & 101.70 & ~+6.16 &  0.240 &       &   0 &  \\
LDN 1121         & 21 40.5 & +58 16.7 & ~99.97 & ~+4.19 &  0.008 &  ~$-$0.2~~  &  $\ga25$~~ & DBY\,100.0+04.2, $[$IS94$]$\,5, $[$G85$]$\,14,  \\
 	         &         &          &        &        &        &       &     & $[$LM99$]$\,379, $[$IS94$]$\,17, IC 1396\,N, \\
 	         &         &          &        &        &        &       &     &  SFO 38, $[$PGS95$]$\,24, FSE\,19, TGU\,599  \\
CB 234           & 21 40.5 & +70 18.6 & 108.09 & +13.15 &  0.044 & ~$-$5.0~~ & 0 & \\
LDN 1127         & 21 40.9 & +58 33.7 & 100.20 & ~+4.37 &  0.004 &       &   0 & TGU\,599 \\
TDS\,395         & 21 40.9 & +66 35.7 & 105.57 & +10.38 &  0.161 & $-$10.8~~ &   1 & $[$KTK2006$]$ G105.5+10.3, YMD\,CO\,18 \\
LDN 1124         & 21 41.5 & +58 13.7 & 100.04 & ~+4.07 &  0.015 & ~$-$0.2~~  &  20 & DBY\,100.0+4.2, IC\,1396\,N \\
LDN 1128         & 21 41.5 & +58 33.7 & 100.26 & ~+4.32 &  0.015 & ~~~1.2~~   &   0 & TGU\,599, WWC\,117 \\
LDN 1134         & 21 41.5 & +60 13.7 & 101.35 & ~+5.58 &  2.630 & ~$-$0.4~~  &   0 & $[$PGH98b$]$\,Cloud 19, DBY 100.9+05.3, \\
 	         &         &          &        &        &        &       &     &  $[$PGH98b$]$\,Cloud 21, \\
 	         &         &          &        &        &        &       &     &  DBY\,101.7+05.0 \\
LDN 1103         & 21 41.7 & +56 43.7 & ~99.07 & ~+2.92 &  0.003 &       &   0 &  \\
LDN 1104         & 21 42.1 & +56 43.7 & ~99.11 & ~+2.89 &  0.006 &  ~~~4.1~~ &   2 & Barnard 163, TDS\,371, WWC\,184, \\
 	         &         &          &        &        &        &       &     & $[$G85$]$\,17, $[$LM99$]$\,381 \\
LDN 1181         & 21 42.1 & +66 08.7 & 105.36 & ~+9.97 &  0.008 & ~$-$9.9~~ &   0 & YMD\,CO\,18, TGU\,645 \\

LDN 1183         & 21 42.1 & +66 13.7 & 105.42 & +10.03 &  0.090 & ~$-$9.9~~ & $>$84 & Ced\,196, GM 1-57, $[$B77$]$\,40, TDS\,395, \\ 
                 &         &          &        &        &        &           &       & $[$FMS2001$]$\,NGC 7129, \\
 	         &         &          &        &        &        &       &     & $[$MPR2003$]$\,HI Ring, $[$MPR2003$]$\,HI Knot, \\
 	         &         &          &        &        &        &       &     & BFS 11, JWT Core 46  \\
LDN 1095         & 21 42.6 & +56 18.8 & ~98.89 & ~+2.52 &  0.012 &       &   0 &  \\
LDN 1106         & 21 42.6 & +56 53.8 & ~99.27 & ~+2.97 &  0.009 & ~$-$0.4~~ &   0 & $[$PGS95$]$\,28 \\
LDN 1113         & 21 44.6 & +57 11.8 &  99.67 & ~+3.02 &  0.002 & ~~~4.5~~   &   1 & Barnard 367, WWC\,186 \\
LDN 1130         & 21 44.6 & +58 18.8 & 100.40 & ~+3.87 &  0.012 & ~$-$3.0~~ &  8 & TGU\,604, $[$PGS95$]$\,32, FSE\,8 \\
LDN 1136         & 21 45.5 & +59 58.9 & 101.57 & ~+5.06 &  0.012 & ~$-$2.9~~ & 0 & DBY\,101.7+05.0 \\
DBY\,100.0+03.0  & 21 46.3 & +57 25.1 & 100.00 & ~+3.03 &  0.015 & ~$-$1.9~~ & 16 & IC\,1396\,E,SFO 39, FSE\,9 \\
LDN 1115         & 21 46.6 & +56 58.9 & ~99.74 & ~+2.68 &  0.002 &       &   0 &  \\
LDN 1118         & 21 46.6 & +57 12.9 & ~99.89 & ~+2.86 &  0.001 & ~$-$2.1~~ &   0 & SFO 42, $[$PGS95$]$\,38 \\
LDN 1129         & 21 46.6 & +57 53.9 & 100.33 & ~+3.38 &  0.041 & ~$-$2.1~~ &   0 & $[$PGS95$]$\,36, DBY\,100.4+03.4, WWC\,40--42 \\
LDN 1132         & 21 46.6 & +58 43.9 & 100.87 & ~+4.02 &  0.003 &       &   0 &  \\
LDN 1120         & 21 47.6 & +57 09.0 & ~99.96 & ~+2.72 &  0.001 &       &   0 &  \\
LDN 1114         & 21 49.7 & +56 24.0 & ~99.69 & ~+1.96 &  0.016 & ~$-$1.8~~ &   0 & TDS\,375, DBY\,099.9+01.8 \\
LDN 1241         & 21 50.0 & +76 44.1 & 113.08 & +17.48 &  1.380 & ~$-$3.7~~ &   0 & YMD\,CO\,72, YMD\,CO\,75, \\
 	         &         &          &        &        &        &       &     & TGU\,728, TGU\,739 \\
LDN 1144         & 21 50.5 & +60 07.1 & 102.14 & ~+4.77 &  0.027 & ~$-$2.2~~ &  1 & Barnard 166, DBY\,102.1+04.8 \\
LDN 1137         & 21 51.6 & +59 04.1 & 101.58 & ~+3.87 &  0.018 & $-$10.5~~ & 0 & $[$G85$]$\,31, DBY\,101.5+03.8\\
LDN 1138         & 21 51.6 & +59 04.1 & 101.58 & ~+3.87 &  0.018 & $-$10.5~~ & 0 & DBY\,101.5+03.8 \\
LDN 1109         & 21 51.7 & +55 49.1 & ~99.55 & ~+1.33 &  0.150 & ~$-$1.5~~ &  0 & DBY\,099.5+01.2 \\
Barnard 167      & 21 52.0 & +60 04.0 & 102.25 & ~+4.61 &  0.005 &       &   1 &  \\
LDN 1139         & 21 55.6 & +58 34.3 & 101.68 & ~+3.15 &  0.015 & ~~~0.1~~  &   15 & Barnard 169, DBY\,101.3+03.00, TGU\,620, \\
 	         &         &          &        &        &        &       &     & $[$LM99$]$\,386, $[$PGH98b$]$\,Cloud 27, FSE 10 \\
LDN 1141         & 21 55.6 & +58 44.3 & 101.78 & ~+3.28 &  0.010 & ~~~0.1~~  &   0 & Barnard 171, DBY\,101.3+03.00, \\
LDN 1142         & 21 56.6 & +59 04.3 & 102.09 & ~+3.47 &  0.002 & ~~~0.1~~  &   0 & CB\,235, DBY\,101.3+03.00 \\
LDN 1143         & 21 57.6 & +58 59.3 & 102.14 & ~+3.32 &  0.020 & ~~~0.1~~  &   1 & Barnard 170, TDS\,380, DBY\,101.3+03.00 \\
LDN 1149         & 21 57.6 & +59 07.3 & 102.22 & ~+3.43 &  0.015 & ~~~0.1~~  &   0 & DBY\,101.3+03.00 \\
LDN 1151         & 21 59.6 & +59 04.4 & 102.40 & ~+3.23 &  0.027 & ~~~0.1~~  &   0 & DBY\,101.3+03.00 \\
LDN 1153         & 22 00.6 & +58 59.5 & 102.45 & ~+3.09 &  0.079 & ~~~0.1~~  &   0 & TDS\,383, DBY\,101.3+03.00 \\
LDN 1133         & 22 02.7 & +56 14.5 & 101.02 & ~+0.72 &  0.248 &       &   0 &  \\
LDN 1160         & 22 04.7 & +58 59.6 & 102.87 & ~+2.78 &  0.019 & ~~~0.1~~  &   0 & DBY\,101.3+03.00 \\

LDN 1166         & 22 05.7 & +59 34.6 & 103.32 & ~+3.18 &  0.001 & ~$-$3.0~~  &   0 & CB 236 \\
LDN 1159         & 22 06.7 & +58 34.7 & 102.84 & ~+2.29 &  2.790 & ~$-$1.0~~ & 17 &  Barnard 174, TDS\,384, TDS\,391, $[$LM99$]$\,389,\\
 	         &         &          &        &        &        &       &     & $[$GA90$]$\,3-36,  DSH J2206.2+5819, \\
        	 &         &          &        &        &        &       &     & DBY\,102.9+02.4, $[$G85$]$\,32, \\
 	         &         &          &        &        &        &       &     &  DBY\,102.8+02.1, DBY\,103.3+02.8, \\ 
 	         &         &          &        &        &        &       &     & DBY\,103.2+01.8, $[$PGH98b$]$\,Cloud 30, \\
 	         &         &          &        &        &        &       &     & DBY\,103.5+02.0, GSH 103+02-66 \\
LDN 1164         & 22 06.7 & +59 09.7 & 103.18 & ~+2.76 &  0.019 & ~$-$2.2~~ & 36 & DBY\,103.3+02.8, FSE\,11 \\
LDN 1165         & 22 07.2 & +59 04.7 & 103.18 & ~+2.66 &  0.019 & ~$-$2.2~~ &  1 & DBY\,103.3+02.8 \\
LDN 1169         & 22 07.2 & +59 44.7 & 103.57 & ~+3.20 &  0.004 & ~$-$3.2~~ &   0 & CB 237 \\
TGU 659          & 22 09.0 & +64 29.7 & 106.53 & ~+6.93 &  1.070 &       &   1 & $[$KTK2006$]$ G106.9+07.1 \\
LDN 1178         & 22 09.1 & +62 19.8 & 105.27 & ~+5.16 &  0.005 &       &   0 &  \\
LDN 1243         & 22 10.6 & +75 20.0 & 113.16 & +15.61 &  0.081 & ~$-$2.9~~~ & 0 & YMD\,C0\,74 \\
LDN 1219         & 22 11.6 & +70 59.9 & 110.58 & +12.06 &  0.003 & ~$-$4.6~~  & 1 &  Barnard\,175, TDS\,414, YMD\,CO\,57 \\
LDN 1182         & 22 13.1 & +61 54.9 & 105.42 & ~+4.55 &  0.004 &       &   0 &  \\
LDN 1217         & 22 13.1 & +70 44.9 & 110.54 & +11.79 &  0.186 & ~$-$4.6~~ &   1 & Ced\,201, YMD\,CO\,57, TDS\,414, TGU\,696 \\
LDN 1186         & 22 13.6 & +62 07.9 & 105.59 & ~+4.70 &  0.005 &       &   0 & TGU\,649 \\
LDN 1175         & 22 13.7 & +60 44.9 & 104.81 & ~+3.55 &  0.015 &       &   0 & TGU\,635 \\
LDN 1191         & 22 14.1 & +62 24.9 & 105.80 & ~+4.90 &  0.004 &       &   0 &  \\
LDN 1193         & 22 14.6 & +62 24.9 & 105.85 & ~+4.87 &  0.002 &       &   0 &  \\
LDN 1235         & 22 14.9 & +73 25.0 & 112.24 & +13.88 &  0.037 & ~$-$4.0~~ &  4 & YMD\,CO\,69, TDS\,426, HCL 1F,  \\
 	         &         &          &        &        &        &       &     & TGU725, $[$BM89$]$\,1-117, $[$LM99$]$\,390 \\
TGU 627          & 22 15.2 & +58 47.6 & 103.87 & ~+1.83 &  0.210 &       &   3 &  \\
LDN 1150         & 22 15.8 & +56 00.0 & 102.37 &  $-$0.53 &  0.007 & ~$-$6.9~~ &   0 & Barnard 369, TDS\,385, TGU\,621 \\
LDN 1188         & 22 16.7 & +61 45.0 & 105.67 & ~+4.18 &  0.398 & $-$10.6~~ & $\ga20$~~ & YMD\,CO\,21, TGU\,652, \\
 	         &         &          &        &        &        &       &     & $[$PGH98b$]$\,Cloud 33, $[$ADM95$]$\,3, \\
 	         &         &          &        &        &        &       &     & $[$ADM95$]$\,4, $[$ADM95$]$\,5, $[$ADM95$]$\,7, \\
 	         &         &          &        &        &        &       &     & $[$ADM95$]$\,13, $[$ADM95$]$\,8, GN 22.15.0 \\
LDN 1154         & 22 16.8 & +56 13.0 & 102.61 &  $-$0.43 &  0.007 &       &   0 &  \\
TGU 636          & 22 16.9 & +60 11.6 & 104.83 & ~+2.87 &  0.240 &       &   2 &  \\
TGU 640          & 22 17.6 & +60 36.5 & 105.13 & ~+3.17 &  0.330 &       &   2 & DG 181, DG 182, GN 22.14.9 \\
LDN 1156         & 22 19.9 & +55 45.1 & 102.71 & $ -$1.05 &  0.043 &       &   0 &  \\

LDN 1161         & 22 19.9 & +56 08.1 & 102.92 &  $-$0.73 &  0.006 &       &   0 & $[$KC97c$]$\,G102.9-00.7 \\
YMD\,CO\,29      & 22 20.2 & +63 53.1 & 107.20 & ~+5.73 & 0.018  & $-$11.0~~ & 13 & Sh2-145, SFO\,44 \\
LDN 1184         & 22 20.7 & +60 45.1 & 105.53 & ~+3.08 &  0.124 &       &   0 & TGU\,643, TGU\,644 \\
LDN 1247         & 22 20.8 & +75 15.2 & 113.66 & +15.17 &  0.167 & ~$-$5.1~~ &   0 & TDS\,430, TGU\,742 \\
LDN 1163         & 22 20.9 & +56 10.1 & 103.05 &  ~$-$0.78 &  0.007 &       &   0 &  \\
LDN 1201         & 22 23.6 & +63 30.2 & 107.31 & ~+5.21 &  0.009 &       &   0 & TGU\,661 \\
LDN 1202         & 22 26.7 & +63 05.3 & 107.38 & ~+4.67 &  0.004 &       &   0 & TGU\,661  \\
LDN 1204         & 22 26.7 & +63 15.3 & 107.47 & ~+4.82 &  2.500 & ~$-$7.6~~ & $\ga100$~~ & YMD\,CO\,27, TDS\,399, TDS\,401, \\
 	         &         &          &        &        &        &       &     & TDS\,403, TDS\,404, TDS\,405, $[$KC97c$]$,  \\ 
 	         &         &          &        &        &        &       &     & $[$PGH98b$]$\,Cloud 32, $[$PGH98b$]$\,Cloud 37, \\   
                 &         &          &        &        &        &       &     & $[$PGH98b$]$\,Cloud 38, $[$G84b$]$\,12, \\ 
 	         &         &          &        &        &        &       &     & DG 185, $[$PGH98b$]$\,Cloud 36, Sh2-140, \\
 	         &         &          &        &        &        &       &     &  GSH 108+05-46, GN 22.21.5, TGU\,661 \\		 
LDN 1180         & 22 26.8 & +59 15.3 & 105.37 & ~+1.41 &  7.000 & ~$-$3.6~~ &  19 & YMD\,CO\,20, YMD\,CO\,24, TDS\,396, TDS\,397, \\
 	         &         &          &        &        &        &       &     & CB\,240, TGU\,631, TGU\,638,  TGU\,655, \\ 
 	         &         &          &        &        &        &       &     & TGU\,658, Min 2-72, $[$PGH98b$]$\,Cloud~34,  \\
 	         &         &          &        &        &        &       &     & $[$KC97c$]$\,G104.6+01.4, KR 47, \\
 	         &         &          &        &        &        &       &     &$[$LM99$]$\,394, $[$KC97c$]$\,G105.6+00.4, \\
 	         &         &          &        &        &        &       &     & Sh\,2-138, $[$GSL2002$]$\,109, $[$GSL2002$]$\,110, \\ 
 	         &         &          &        &        &        &       &     & $[$P85b$]$\,18, $[$LM99$]$\,396, \\ 
 	         &         &          &        &        &        &       &     & DSH J2222.5+5918B, TGU\,657, \\
LDN 1195         & 22 26.8 & +61 15.3 & 106.42 & ~+3.11 &  0.017 &       &   1 & $[$B77$]$\,42, $[$LM99$]$\,391, GN 22.24.9 \\
LDN 1196         & 22 26.8 & +61 15.3 & 106.42 & ~+3.11 &  0.029 &       &   1 &  \\
LDN 1179         & 22 27.3 & +59 02.3 & 105.31 & ~+1.19 &  0.002 &       &   0 &  \\
LDN 1203         & 22 27.7 & +63 00.3 & 107.43 & ~+4.54 &  0.016 &       &   0 &  TGU\,661 \\
TGU 719          & 22 27.8 & +71 21.4 & 111.90 & +11.63 &  0.510 &       &   3 &  \\
LDN 1221         & 22 28.4 & +69 00.4 & 110.67 & ~+9.61 &  0.020 & ~$-$4.9~~  &  3 & TDS 416, $[$KTK2006$]$ G110.6+09.6, \\
 	         &         &          &        &        &        &       &     & $[$LM99$]$\,392, TGU\,702\\
LDN 1206         & 22 28.7 & +64 25.4 & 108.27 & ~+5.70 &  0.083 &       &   0 & TGU\,673, $[$PGH98b$]$\,Cloud 35 \\
LDN 1185         & 22 29.3 & +59 05.4 & 105.56 & ~+1.10 &  0.006 &       &   0 &  \\
LDN 1207         & 22 29.7 & +64 25.4 & 108.36 & ~+5.64 &  0.047 &       &   0 &  \\
LDN 1209         & 22 29.7 & +64 45.4 & 108.53 & ~+5.93 &  0.111 & ~$-$8.8~~ &   3 & Sh2-150, TGU\,680 \\

LDN 1208         & 22 30.7 & +64 25.4 & 108.46 & ~+5.58 &  0.018 &       &   0 &  \\
LDN 1190         & 22 30.9 & +59 08.4 & 105.75 & ~+1.04 &  0.004 &       &   0 &  \\
LDN 1242         & 22 31.2 & +73 15.4 & 113.15 & +13.11 &  0.793 &       &   0 &  \\
LDN 1213         & 22 31.6 & +65 25.5 & 109.06 & ~+6.39 &  0.006 & ~$-$9.2~~ &   0 & YMD\,CO\,48,  TGU\,686, Sh2-150  \\
LDN 1194         & 22 32.9 & +59 05.5 & 105.95 & ~+0.87 &  0.009 &       &   0 &  \\
LDN 1214         & 22 33.6 & +65 45.5 & 109.41 & ~+6.57 &  0.256 & ~$-$8.5~~ &   1 & TDS\,407, YMD\,CO\,48, Sh2-150 \\
LDN 1192         & 22 33.9 & +58 35.5 & 105.81 & ~+0.37 &  0.007 & ~$-$3.6~~ &   0 & CB\,240 \\
TDS\,417         & 22 35.5 & +69 13.1 & 111.33 & ~+9.47 &  0.014 & ~$-$7.2~~ & 1 &  YMD CO\,65, $[$B77$]$\,46, GN 22.33.6 \\
                 &         &          &        &        &        &       &     & $[$KTK2006$]$ G101.9+15.8 \\
TGU 679          & 22 35.9 & +63 35.2 & 108.53 & ~+4.57 &  0.170 &       &   1 &  \\
LDN 1251         & 22 36.1 & +75 15.6 & 114.51 & +14.65 &  0.195 & ~$-$3.8~~ & $\ga20$~~ & YMD\,CO\,79, TGU\,750, $[$SMN94$]$\,B,\\
 	         &         &          &        &        &        &       &     & HCL 1A, $[$SMN94$]$\,C, $[$SMN94$]$\,D, \\
 	         &         &          &        &        &        &       &     &  $[$LM99$]$\,397, $[$NJH2003$]$\,5, $[$TW96$]$\,H2, \\ 
 	         &         &          &        &        &        &       &     &  $[$SMN94$]$\,E, $[$TW96$]$\,H1, \\ 
	         &         &          &        &        &        &       &     &  $[$KTK2006$]$ G114.4+14.6 \\
LDN 1198         & 22 36.9 & +59 25.6 & 106.57 & ~+0.90 &  0.054 & $-$11.91~~ &   2 &  \\
TGU 672          & 22 37.5 & +62 20.6 & 108.07 & ~+3.40 &  0.360 &       &   1 &  \\
LDN 1197         & 22 37.9 & +58 55.6 & 106.43 & ~+0.40 &  0.009 &       &   0 &  \\
LDN 1187         & 22 38.0 & +57 15.6 & 105.62 & ~$-$1.06 &  0.145 &       &   0 &  \\
LDN 1189         & 22 39.0 & +57 15.6 & 105.74 & ~$-$1.12 &  0.473 &       &   0 &  \\
TGU 671          & 22 44.1 & +60 16.1 & 107.77 & ~+1.20 &  0.200 &       &   2 &  \\
LDN 1210         & 22 44.9 & +62 05.8 & 108.70 & ~+2.77 &  0.011 & $-$10.0~~ &   0 & YMD\,CO\,40, TGU\,699  \\
Sh\,2-142        & 22 45.0 & +57 55.8 & 106.76 & ~$-$0.92 &  1.063 & $-$41.0~~ &  14 & Ced\,206, GM 2-42, TGU\,663, SFO 43, \\
 	         &         &          &        &        &        &       &     &  $[$KC97c$]$\,G107.2-01.0, $[$KC97c$]$\,G107.3-00.9 \\ 
LDN 1205         & 22 45.9 & +60 25.8 & 108.04 & ~+1.24 &  0.095 &       &   0 &  \\
LDN 1200         & 22 46.0 & +58 45.8 & 107.27 & ~$-$0.24 &  0.020 & ~$-$3.7~~ &  3 & YMD\,CO\,28, TDS\,402, TGU\,665 \\
TGU 678          & 22 46.5 & +61 12.8 & 108.47 & ~+1.90 &  0.170 &       &   1 &  \\
LDN 1211         & 22 46.9 & +62 10.8 & 108.95 & ~+2.74 &  0.011 & $-$11.1~~ &   4 & YMD\,CO\,40, TGU\,699  \\
LDN 1212         & 22 47.9 & +62 12.9 & 109.07 & ~+2.71 &  0.014 & $-$10.0~~ &   0 & YMD\,CO\,40, TGU\,699 \\
TGU 689          & 22 50.2 & +62 51.5 & 109.60 & ~+3.17 &  0.160 &       &   2 &  \\
TGU 692          & 22 50.7 & +63 15.4 & 109.83 & ~+3.50 &  0.240 &       &   2 &  \\
YMD CO\,38       & 22 51.1 & +60 51.6 & 108.80 & ~+1.33 &  0.028 & $-$8.5 &   1 & TGU\,606, $[$KTK2006$]$\,G100.6+16.2 \\
YMD CO\,49       & 22 51.1 & +62 38.9 & 109.60 & ~+2.93 &  0.014 & $-$9.6 &   1 & TGU\,690 \\

LDN 1215         & 22 51.9 & +62 06.0 & 109.44 & ~+2.40 &  0.021 &       &   0 & TGU\,699 \\
LDN 1216         & 22 51.9 & +62 16.0 & 109.52 & ~+2.55 &  0.142 & $-$9.5~~ &  11 & Cep~F,  TDS\,408,  \\
 	         &         &          &        &        &        &       &     & GN 22.51.3, TGU\,699 \\
LDN 1236         & 22 52.7 & +68 56.0 & 112.56 & ~+8.49 &  0.044 & $-$5.0~~ &   0 & YMD\,CO\,73, TGU\,729 \\
YMD CO\,51       & 22 56.2 & +62 02.7 & 109.87 & ~+2.13 &  0.489 & $-$10.2~~ & $\ga15$~~ & Cep\,A \\
LDN 1223         & 22 56.9 & +64 16.1 & 110.89 & ~+4.11 &  1.010 &       &   1 & TGU\,703, $[$BKP2003$]$\,455 \\
 	         &         &          &        &        &        &       &     & GN 22.58.2, GSH 111+04-105 \\
TGU 715          & 22 57.0 & +65 53.5 & 111.60 & ~+5.57 &  1.280 &       &   0 &  \\
TGU 684          & 22 57.4 & +59 27.5 & 108.90 & ~$-$0.27 &  0.230 & $-$47.0~~ & 6 & $[$KC97c$]$\,G109.1-00.3, $[$G82a$]$\,13 \\
TGU 705          & 22 59.5 & +63 26.0 & 110.80 & ~+3.23 &  0.410 &       &   2 &  \\
LDN 1218         & 23 02.0 & +62 16.2 & 110.58 & ~+2.05 &  1.590 & $-$5.1~~ & $>$400~~ & YMD\,CO\,53, YMD\,CO\,51 \\
 	         &         &          &        &        &        &       &     & YMD\,CO\,56, YMD\,CO\,58, YMD\,CO\,59, YMD\,CO\,62, \\
 	         &         &          &        &        &        &       &     & YMD\,CO\,63, TDS\,418,  Cep A, Cep A west, \\
 	         &         &          &        &        &        &       &     & $[$THR85$]$ Cep A-3, GAL 109.88+02.11, \\
                 &         &          &        &        &        &       &     & $[$HW84$]$ 7d, $[$B77$]$ 44, $[$TOH95$]$ Ridge, \\
                 &         &          &        &        &        &       &     & $[$TOH95$]$ C, $[$TOH95$]$ B, $[$TOH95$]$ A, \\
                 &         &          &        &        &        &       &     & Cep B, GN 22.55.2, $[$KC97c$]$\,G110.2+02.5, Cep E, \\
                 &         &          &        &        &        &       &     & Cep E South, Cep E North, $[$YNF96$]$\,a, \\
                 &         &          &        &        &        &       &     & $[$YNF96$]$ \,b, TGU\,699 \\
YMD CO\,55       & 23 02.1 & +61 29.2 & 110.27 & ~+1.33 &  0.056 & $-$7.3~~ &   2 &  \\
LDN 1224         & 23 04.0 & +63 46.2 & 111.39 & ~+3.33 &  0.016 &       &   0 &  \\
LDN 1220         & 23 04.1 & +61 51.2 & 110.63 & ~+1.57 &  0.006 & $-$10.1~~ &   1 & YMD\,CO\,58, Cep\,E \\
LDN 1222         & 23 05.1 & +61 46.2 & 110.70 & ~+1.45 &  0.002 & $-$10.1~~ &   0 & YMD\,CO\,58, Cep\,E, TGU\,699  \\
TGU 707          & 23 09.0 & +61 04.8 & 110.87 & ~+0.63 &  0.160 &       &   1 &  \\
TGU 700          & 23 09.7 & +60 06.4 & 110.57 & ~$-$0.30 &  0.650 & ~$-$52.8~~ &  10 & BFS 18 \\
LDN 1226         & 23 10.1 & +62 16.3 & 111.44 & ~+1.68 &  0.009 & $-$10.7~~ &   0 & YMD\,CO\,63, TGU\,699 \\
LDN 1227         & 23 10.1 & +62 19.3 & 111.46 & ~+1.73 &  0.008 & $-$10.7~~ &   0 & YMD\,CO\,63, TGU\,699  \\
LDN 1240         & 23 11.0 & +66 26.3 & 113.12 & ~+5.50 &  0.126 & &   0 &  \\
LDN 1225         & 23 12.1 & +61 36.3 & 111.41 & ~+0.97 &  0.036 & $-$10.9~~  &  4 & TDS\,419, CB\,242, TGU\,699 \\
LDN 1239         & 23 12.3 & +66 04.3 & 113.11 & ~+5.11 &  0.005 &  ~$-$7.6~~ &   0 & CB 241 \\
NGC\,7538        & 23 14.1 & +61 29.4 & 111.58 & ~+0.77 &  0.011 & $-$57.0~~ & $\sim$2000~~~ & $[$KSH92$]$\,CS 5, $[$KSH92$]$\,CS\,4, $[$M73$]$\,C, \\
 	         &         &          &        &        &        &       &     & $[$KSH92$]$\,CS 3, Ced\,209, $[$WAM82$]$\,111.543+0.7, \\
 	         &         &          &        &        &        &       &     & $[$M73$]$\,B, $[$M73$]$\,A, $[$KSH92$]$\,CS 2, Sh2-158\\ 
 	         &         &          &        &        &        &       &     & GAL 111.53+00.82, $[$WC89$]$\,111.54+0.78, $[$KSH92$]$\,CS 1  \\ 

LDN 1229         & 23 14.1 & +61 59.4 & 111.78 & ~+1.24 &  0.004 & $-$10.0~~  &   0 & Cep D, TGU\,699  \\
LDN 1230         & 23 14.1 & +62 01.4 & 111.79 & ~+1.27 &  0.002 & $-$10.0~~  &   0 & Cep D, TGU\,699  \\
LDN 1233         & 23 16.5 & +62 20.4 & 112.15 & ~+1.47 &  0.001 & $-$10.0~~  &   0 & Cep D  \\
LDN 1232         & 23 17.2 & +61 46.4 & 112.03 & ~+0.91 &  0.004 & $-$10.6~~ &   0 & YMD\,CO\,68, TGU\,699  \\
LDN 1234         & 23 17.2 & +62 24.4 & 112.25 & ~+1.51 &  0.004 &       &   0 &  \\
TGU 717          & 23 17.3 & +60 48.1 & 111.70 &   ~~0.00 &  0.200 & $-$30.1~~ & $\ga9$~~ &  \\
TGU 757          & 23 17.3 & +69 56.0 & 114.97 & ~+8.53 &  0.270 &       &   1 &  \\
LDN 1231         & 23 18.2 & +61 16.4 & 111.97 & ~+0.40 &  0.013 & $-$11.2~~ &   0 & TDS\,424 \\
LDN 1250         & 23 22.1 & +67 16.5 & 114.45 & ~+5.89 &  2.630 & ~$-$8.6~~ &   1 & YMD\,CO\,78, YMD\,CO\,80, \\ 
 	         &         &          &        &        &        &       &     & TDS\,432, TGU\,747 \\
                 &         &          &        &        &        &       &     & GSH 114+06-47 \\
LDN 1259         & 23 22.9 & +74 16.5 & 116.93 & +12.44 &  0.035 &   ~~~3.9~~ &   0 & YMD\,CO\,101, TGU\,772 \\
LDN 1244         & 23 25.2 & +62 46.5 & 113.25 & ~+1.53 &  0.007 &       &   0 &  \\
LDN 1246         & 23 25.2 & +63 36.5 & 113.52 & ~+2.32 &  0.002 & $-$11.1~~ &   0 & CB\,243, $[$GA90$]$\,3-40, $[$LM99$]$\,400 \\
LDN 1261         & 23 27.0 & +74 16.5 & 117.20 & +12.35 &  0.030 &   ~~~3.9~~ &  2 & YMD\,CO\,101, TDS\,447, TGU\,772, \\ 
 	         &         &          &        &        &        &       &     & TDS\,448, CB\,244, $[$LM99$]$\,401 \\
LDN 1262         & 23 27.0 & +74 16.5 & 117.20 & +12.35 &  0.066 & ~~~3.9~~  &  2 & $[$BM89$]$\,3B-10, $[$BM89$]$\,1-119, TGU\,772 \\
YMD CO\,85       & 23 32.6 & +67 02.0 & 115.33 & ~+5.33 &  0.084 &  ~$-$5.4~~ &   0  \\
YMD CO\,86       & 23 35.8 & +66 26.2 & 115.47 & ~+4.67 &  0.056 & ~$-$7.0~~ &   0  \\
YMD CO\,87       & 23 39.3 & +65 42.4 & 115.60 & ~+3.87 &  0.168 & $-$25.3~~ &   2 \\
TGU 767          & 23 42.9 & +68 52.9 & 116.80 & ~+6.83 &  0.400 &       &   2 &  \\
LDN 1264         & 23 52.5 & +68 16.7 & 117.50 & ~+6.03 &  0.102 &       &   0 &  \\
LDN 1266         & 23 57.5 & +67 16.7 & 117.75 & ~+4.95 &  2.540 & ~$-$6.8~~  &  14 & YMD\,CO\,103, YMD\,CO\,104, YMD\,CO\,105,  \\ 
 	         &         &          &        &        &        &       &     & YMD\,CO\,108, YMD\,CO\,111, TDS\,449, TDS\,451,  \\ 
 	         &         &          &        &        &        &       &     & TGU\,774, $[$LM99$]$\,404, $[$YF92$]$\,C1, \\ 
 	         &         &          &        &        &        &       &     & $[$LM99$]$\,405, $[$YF92$]$\,C2, GN 23.56.1, \\
		 &         &          &        &        &        &       &     & $[$KC97c$]$\,G118.1+05.0, $[$LM99$]$\,406, $[$GA90$]$\,1-1, \\
                 &         &          &        &        &        &       &     & $[$GA90$]$\,1-1a, $[$LM99$]$\,1, $[$KC97c$]$\,G118.4+04.7, \\ 
 	         &         &          &        &        &        &       &     &  DG 1, $[$GA90$]$\,3-41, $[$GA90$]$\,3-42,  \\
 	         &         &          &        &        &        &       &     & SFO 1, SFO 3, DSH J2359.6+6741 \\		 

LDN 1274         & 23 57.5 & +70 56.7 & 118.52 & ~+8.54 &  0.032 & ~$-$2.5~~ &   0 & YMD\,CO\,109 \\
LDN 1268         & 23 59.5 & +67 26.7 & 117.97 & ~+5.07 &  0.158 & ~$-$6.2~~ &  3 & Sh2-171, YMD\,CO\,104, TGU\,774  \\
LDN 1269         & 00 00.6 & +67 09.7 & 118.01 & ~+4.78 &  0.025 &       &   0 & TGU\,774 \\
LDN 1270         & 00 01.6 & +67 09.7 & 118.11 & ~+4.76 &  0.009 &       &   2 &  TGU\,774  \\
LDN 1271         & 00 01.6 & +67 16.7 & 118.13 & ~+4.87 &  0.010 & $-$15.1~~ &  0 & YMD\,CO\,108, TGU\,774 \\
LDN 1272         & 00 02.6 & +67 16.7 & 118.23 & ~+4.85 &  8.690 & ~$-$6.3~~ &  23 & YMD\,CO\,97, YMD\,CO\,98, YMD\,CO\,110,  \\
 	         &         &          &        &        &        &       &     &  YMD\,CO\,112, YMD\,CO\,113, YMD\,CO\,114, \\ 
 	         &         &          &        &        &        &       &     & YMD\,CO\,116, TGU\,779, SFO\,2, $[$GA90$]$\,3-1,\\ 
 	         &         &          &        &        &        &       &     & $[$GA90$]$\,1-1b, GSH 119+05-74 \\
LDN 1273         & 00 02.6 & +68 31.7 & 118.46 & ~+6.08 &  0.199 & ~$-$8.8~~ &   4 & YMD\,CO\,114 \\
LDN 1275         & 00 06.6 & +67 26.7 & 118.64 & ~+4.95 &  0.020 &       &   0 & TGU\,781 \\
\label{Tab_cloud}
\end{longtable} 
\end{center}
\end{landscape}

\normalsize

\section{The Cepheus Flare}

\subsection{Large-scale Studies of the Cepheus Flare}

The term `Cepheus flare' was first used by \citet{Hubble} who recognized that the zone of
avoidance of external galaxies, confined to the Galactic plane, extended to higher
latitudes at certain segments of the plane, suggesting significant obscuration
outside the main Galactic belt. He called these wide segments for Galactic disk {\em flares\/}.
The Cepheus Flare can be found at $100\deg \leq l \leq 120\deg$:
there is a large amount of dense ISM above $b \geq 10\deg$ \citep*{LyndsD,TDS, CB88,DUK}.

\citeauthor{Heiles}' \citeyearpar{Heiles} study of the HI distribution in the region
revealed two kinematically separate sheets of interstellar gas
in the Galactic latitude interval +13\deg $\leq b \leq$ +17\deg,
moving at a velocity of $\sim$ 15~km\,s$^{-1}$ relative to each other.
Heiles speculated that the two sheets probably represent
an expanding or colliding system at a distance interval of 300--500~pc.
\citet{Berkhuijsen} found a giant radio continuum region {\em Loop III}
centered on $l$=124$\pm$2\deg, $b$=+15.5$\pm3$\deg\  and stretching
across 65\deg, and suggested that it was a result of multiple supernova
explosions. The HI shell reported by \citet{Hu} at $l$=105\deg, $b$=+17\deg\  and
$v_{centr}$=+3~km\,s$^{-1}$ also indicates that the interstellar medium in this
region is in a state of energetic motion. The wide range in the velocity thus
may reflect disturbances from various shocks.

\citeauthor{Lebrun}'s \citeyearpar{Lebrun} low-resolution CO survey revealed that,
in this region, the clouds constitute a coherent giant molecular
cloud complex. Based on \citeauthor{Racine}'s \citeyearpar{Racine}
study of reflection nebulae, Lebrun placed the Cepheus Flare molecular clouds
between 300 and 500 pc. \citet{GLADT} extended the CO survey to a region of
490 deg$^2$ in Cepheus--Cassiopeia above $b$=+10\deg. They found
that the clouds could be divided into two kinematically well separated
subsystems around the radial velocities of $v_{LSR} \sim 0$~km\,s$^{-1}$ and
$-$10~km\,s$^{-1}$, respectively. They also detected CO emission around
$v_{LSR} \sim 0$~km\,s$^{-1}$ at higher longitudes ($124\deg  < l < 140\deg$)
in Cassiopeia, and found an area free of CO emission at $118\deg  < l < 124\deg$.
They suggested that the void  between the Cepheus and Cassiopeia clouds is
a supernova bubble. They estimated the age of the bubble as $4 \times 10^4$~years,
and proposed that it may result from a Type~I supernova.

\begin{table}[!p]
\vspace{-5mm}
\caption{Cloud groups in Cepheus, classified by \citet{YDMOF}.}
\label{Tab_groups}
\begin{center}
{\footnotesize
\begin{tabular}{lc@{\hskip1mm}c@{\hskip1mm}c@{\hskip1mm}c
@{\hskip1mm}c@{\hskip1mm}c@{\hskip1mm}r@{\hskip1mm}r@{\hskip1mm}r@{\hskip1mm}r}
\noalign{\smallskip}
\tableline
\noalign{\smallskip}
N & \multicolumn{2}{c}{Longitude} & \multicolumn{2}{c}{Latitude}
& \multicolumn{2}{c}{V$_\mathrm{LSR}$} & D~~ & M~~ &  Associated~~
 & Ref. \\[2pt]
& l$_{l}$ & l$_{u}$ & ~~b$_{l}$ & ~~b$_{u}$ & ~~V$_{l}$ & ~V$_{u}$ & ~~(pc) & ~~(M$_{\sun}$) & objects~~ \\
& (\deg) & (\deg) & (\deg) & (\deg) & \multicolumn{2}{c}{(km\,s$^{-1}$)}   \\
\tableline
\noalign{\smallskip}
\multicolumn{11}{c}{Close Group} \\
\tableline
\noalign{\smallskip}
1 & ~99 & 105 & 13 & 18 & ~~1 & ~~5 & 440 & 3900  & NGC\,7023, L1157 & 1, 2 \\
2 & 102 & 104 & 16 & 18 & $-$4 & $-$1 & 300 & 90 &  & 3 \\
3 & 105 & 110 & $-$1 & 1 & $-$4 & $-$1 & 300 & 110  & L1200 & 3 \\
4 & 106 & 116 & 13 & 19 & $-$8 & $-$1 & 300 & 3600  & Cepheus Flare & 4 \\
5 & 110 & 116 & 4 & 12 & $-$9 & $-$3 & 300 & 730  & L1221, L1250 & 3 \\
6 & 110 & 117 & 19 & 21 & $-$9 & $-$4 & 300 & 680  & L1228 & 3 \\
7 & 115 & 117 & $-$4 & $-$1 & $-$4 & $-$1 & 140 & 20 & L1253, L1257 & 5 \\
8 & 116 & 118 & 12 & 13 & ~~3 & ~~5 & 200 & 26 & L1262 & 6 \\
9 & 117 & 124 & 6 & 10 & $-$3 & $-$1 & 300 & 200 &  L1304 & 3 \\[2pt]
\tableline
\multicolumn{11}{c}{Distant Group} \\
\tableline
\noalign{\smallskip}
10$^*$ & ~95 & ~99 & 3 & 7 & $-$8 & ~~2 & 624 & 4000 & IC\,1396 & 17 \\
13 & 103 & 111 & 9 & 15 & $-$21 & $-$8 & 1000 & 12000  & NGC 7129 & 7, 8 \\
14 & 105 & 110 & 0 & 7 & $-$14 & $-$6 & 910 & 11000  & S\,140 & 9 \\
15 & 108 & 117 & 0 & 4 & $-$17 & $-$4 & 730 & 15000  & Cep OB3 & 10 \\
16 & 115 & 116 & 3 & 5 & $-$26 & $-$23~ & 1000 & 1400 & M115.5+4.0 & 11 \\
17 & 116 & 124 & $-$2 & 7 & $-$22 & $-$2 & 850 & 27000  & Cep OB4 & 12 \\
\noalign{\smallskip}
\tableline
\multicolumn{11}{c}{Clouds in Perseus Arm} \\
\tableline
\noalign{\smallskip}
22 & 111 & 112 & $-$1 & ~1 & $-$31 & $-$30 & 2200 & 2500  & MWC1080 & 13, 14 \\
23 & 112 & 113 & $-$3 & $-$2 & $-$37 & $-$34 & 3000 & 5400  & Cas A & 15 \\
24 & 123 & 124 & $-$7 & $-$6 & $-$32 & $-$30 & 2100 & 2000  & NGC 281 & 16 \\[2pt]
\tableline
\end{tabular}}
\end{center}
{\footnotesize
References. (1) \citet{Viotti}; (2) \citet{SIY}; (3) \citet{GLADT};
(4) \citet{KP93}; (5) \citet{Snell}; (6) \citet{MB83};
(7) \citet{Racine}; (8) \citet{SY89}; (9) \citet{Cramp74};
(10) \citet{Crawford}; (11) \citet{Yang1}; (12) \citet{MacConnell}; (13) \citet{Levreault85};
(14) \citet{Levreault88};  (15) \citet{UT89};
(16) \citet{Hogg}; (17) \citet{deZeeuw}.  \\
$^*$ The $^{13}$CO clouds in the region of IC\,1396 are included in \citet{DBYF},
not in \citet{YDMOF}.}
\end{table}

\begin{figure*}[p!]
\centerline{
\includegraphics[draft=False,width=8cm]{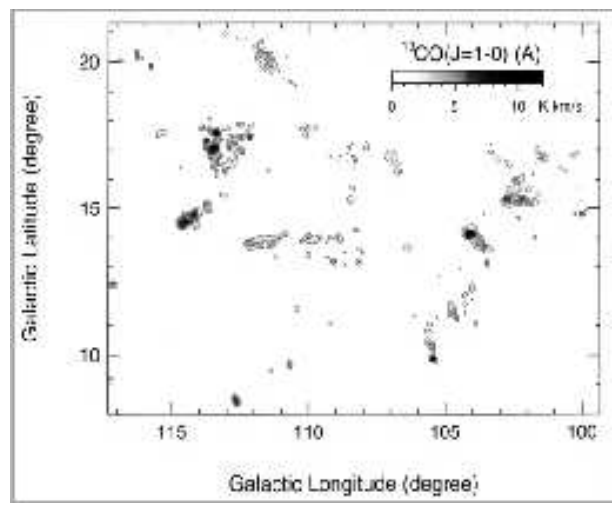}
}
\caption{Distribution of $^{13}$CO clouds in the Cepheus Flare
(adopted from Yonekura et al. 1997)}
\label{Fig_CO}
\end{figure*}

\citet{YDMOF} conducted a large-scale  $^{13}$CO survey of the
Cepheus--Cassiopeia region  at an 8-arcmin grid spacing and with
2\farcm7 beam size. Out of the 188 molecular clouds found in the whole surveyed
region 51  fall in the Cepheus Flare region.  Their surface
distribution, adopted from \citet{YDMOF}, is shown in Fig.~\ref{Fig_CO}.

These clouds, distributed over the velocity interval of ($-$15,+6)~km\,s$^{-1}$,
were classified into three kinematically  different components
by \citeauthor{YDMOF} (see Table~\ref{Tab_groups}).
The high latitude part of the molecular complex is included in \citeauthor{MBM}'s
\citeyearpar{MBM} catalog of high latitude molecular clouds.
Ammonia observations of dense cores have been reported by \citet{BM89} and \citet{TW}.
In addition to the two major catalogs of dark clouds \citep{LyndsD,DUK},
dark cloud cores were catalogued by \citet{LM99}.

\citet{Kun98} determined cloud distances using Wolf diagrams, and
presented a list of candidate YSOs found during an objective  prism H$\alpha$
survey, and selected from the IRAS Catalogs.

\citet{KTK2006} performed a complex study of the visual and infrared properties
of the ISM in the Cepheus Flare region using USNO, 2MASS, DIRBE, IRAS, and ISO
data. Based on the distribution of visual extinction they identified 208 clouds,
and divided them into 8 complexes. They examined the morphology of clouds,
and established several empirical relationships between
various properties of the clouds.

\citet*{Olano} studied the space distribution and kinematics of the interstellar matter
in Cepheus and Cassiopeia, using the Leiden--Dwingeloo HI data and the Columbia Survey CO
data. They found that the broad and often double-peaked spectral line
profiles suggest that the Cepheus Flare forms part of a big expanding shell that
encloses an old supernova remnant. Assuming a distance of 300~pc for the center
of the shell, located at (l,b) $\approx$ (120\deg,+17\deg), they derived a radius of
approximately 50~pc, expansion velocity of 4~km\,s$^{-1}$, and HI mass of
$1.3 \times 10^{4}$~M$_{\sun}$ for the Cepheus Flare Shell. The supernova bubble
proposed by \citet{GLADT}, the radio continuum structure Loop~III
\citep{Berkhuijsen}, and the Cepheus Flare Shell are various observable aspects
of the supernova explosion(s) that shaped the structure of the interstellar medium
and triggered star formation in the Cepheus Flare during the past few million
years.

\begin{table}[!t]
\caption{Distance measured for the clouds within the Cepheus Flare region
 together with their probable lower and upper limits and the method of determination.}
\label{Tab_cepflare_dist}
\begin{center}
{\footnotesize
\begin{tabular}{lc@{\hskip2mm}c@{\hskip2mm}c@{\hskip2mm}l@{\hskip2mm}l@{\hskip2mm}l}
\tableline
\noalign{\smallskip}
Cloud & l & b & d\,($\Delta$d) & Method & Ref. \\
& (\deg) & (\deg) & (pc) &  &  \\
\noalign{\smallskip}
\tableline
\noalign{\smallskip}
L1147/L1158 & 102.0 & 15.0 & 325$\pm$13 & A$_V$ vs. distance & 4 \\
L1167/L1174 & 104.0 & 15.0 & 440$\pm$100 & spectroscopy \& photometry of HD\,200775 & 1 \\
L1167/L1174 & 104.0 & 15.0 & 288$\pm$25 & A$_V$ vs. distance & 4 \\
L1167/L1174 & 104.0 & 15.0 & 430$^{+160}_{-90}$ & {\it Hipparcos\/} parallax of  HD\,200775 & 10 \\
L1177  & 105.14 & 13.12 & 300$\pm$30 & Wolf diagram & 6 \\
L1199  & 106.50 & 12.21 & 500$\pm$100 & Wolf diagram & 6 \\
L1199  & 106.50 & 12.21 & 800 & spectroscopy \& photometry of HD\,206135 & 2 \\
TDS400 & 107.01 & 16.78 & 300$^{+50}_{-10}$  & Wolf diagram  & 6 \\
CB 234 & 108.10 & 13.15 & 300$^{+50}_{-10}$ & Wolf diagram & 6 \\
TDS406 & 108.50 & 18.15 & 300$^{+50}_{-20}$ & Wolf diagram & 6 \\
L1217  & 110.34 & 11.41 & 400$^{+50}_{-20}$ & Wolf diagram  & 6 \\
L1219  & 110.60 & 11.96 & 400$^{+50}_{-20}$ & Wolf diagram  & 6 \\
L1219  & 110.60 & 11.96 & 400  & spectroscopy \& photometry of BD+69\,1231 & 2, 8 \\
TDS420 & 111.57 & 14.25 & 300$^{+50}_{-10}$ &  Wolf diagram  & 6 \\
L1228  & 111.63 & 20.14 & 200$^{+100}_{-20}$ & Wolf diagram  & 6 \\
L1228  & 111.63 & 20.14 & 180$^{+30}_{-10}$ & spectroscopy \& photometry of BD+76\,825 & \\
TDS421 & 111.71 & 13.80 & 250$^{+30}_{-10}$ & Wolf diagram  & 6 \\
L1235  & 112.22 & 13.86 & 200 & A$_V$ vs. distance & 3 \\
L1235  & 112.22 & 13.86 & 300$^{+50}_{-10}$ & Wolf diagram  & 6 \\
L1235  & 112.22 & 13.86 & 400$\pm$80 & spectroscopy \& photometry of HD\,210806 & 2, 8 \\
L1235  & 112.22 & 13.86 & 300$^{+80}_{-40}$ & Hipparcos parallax of HD\,210806 & 7 \\
L1241  & 113.03 & 17.51 & 300$^{+50}_{-10}$ & Wolf diagram  & 6 \\
L1242  & 113.08 & 13.14 & 300$^{+30}_{-10}$ &  Wolf diagram  & 6 \\
L1243  & 113.10 & 15.64 & 300$^{+50}_{-10}$ & Wolf diagram  & 6 \\
L1247  & 113.60 & 15.20 & 300$^{+50}_{-10}$  & Wolf diagram  & 6 \\
L1251  & 114.45 & 14.68 & 300$\pm50$ & A$_V$ vs. distance & 5 \\
L1251  & 114.45 & 14.68 & 300$^{+50}_{-10}$ & Wolf diagram  & 6 \\
L1251  & 114.45 & 14.68 & 337$\pm50$ & star count analysis  & 9\\
MBM163--165 & 116.00 & 20.25 & 200$^{+100}_{-20}$ & Wolf diagram & 6 \\
L1259--1262 & 117.00 & 12.40 & $180^{+40}_{-20}$ & Wolf diagram & 6 \\
\noalign{\smallskip}
\tableline
\end{tabular}}
\end{center}
\smallskip
{\footnotesize
References: (1)  \citet{Viotti}; (2) \citet{Racine}; (3) \citet{Snell}; (4) \citet{SCKM};
(5)  \citet{KP93}; (6) \citet{Kun98}; (7) \citet{Hipparcos}; (8) \citet{KVSz};
(9) \citet{BAKTK}; (10) \citet{vandenA97}}
\end{table}

\subsection{Distance to the Cepheus Flare Clouds}
\label{Sect_dist}

Spectroscopic and photometric studies of stars illuminating reflection nebulae
in the Cepheus Flare
\citep{Racine} indicated, long before the discovery of
the molecular cloud complex, that interstellar dust can be found at several
distances along the line of sight in this region.
The presence of clouds at different velocities also suggests
that there are clouds at various distances \citep{GLADT}.

At the low Galactic latitude boundary of the Cepheus Flare we find the associations
Cep~OB2 and Cep~OB3 at a distance of $\sim$\,800\,pc.
Therefore \citet{GLADT}  propose that the negative velocity component of the
Cepheus flare clouds (v$_\mathrm{LSR} \sim -10$~km\,s$^{-1}$) is an extension of
these local arm features, while the more  positive (v$_\mathrm{LSR} \sim 0$~km\,s$^{-1}$)
velocity component corresponds to a nearby cloud complex at a distance of 300\,pc.

Table~\ref{Tab_groups} suggests a more complicated pattern of cloud distances.
Both the close and the distant components are composed of several complexes,
probably located at different distances. Other distance determinations found in the
literature support this suggestion. Below we list some results and problems
related to the distance of Cepheus Flare clouds.

Distances of individual clouds can reliably be derived by studying the
effects of the clouds on the light of associated stars. \citet{Racine}, in a
spectroscopic and photometric study of stars in reflection nebulae,
obtained a distance of 400$\pm$80\,pc for the Cepheus~R2 association above
the latitude +10\deg. Both velocity components are represented among the clouds
being illuminated by the stars of Cep~R2.

A prominent object of the distant component, in addition to the possible outer
parts of Cep~OB2 and Cep~OB3, is NGC\,7129. Though \citet*{KBT}
and \citet*{ABK} proposed that NGC\,7129 may be associated with
the Cepheus Bubble, and thus with Cep~OB2, other observations suggest that
NGC\,7129 may be farther than Cep~OB2. \citet{Racine} investigated
three member stars, and derived $m-M \approx 12.2$  for BD+65\deg1637, and 10.0 for
both BD+65\deg1638 and LkH$\alpha$\,234, and labeled each value as uncertain.
\citet{SY89}, based on an $A_V$ vs. distance diagram,
derived 1250\,pc for NGC\,7129. \citet{YDMOF} found a group
of clouds at $l \sim 107\deg-111\deg, b \sim +13\deg$ at similar velocity,
and regarded them as an extension of the NGC\,7129 clouds to the northeast (group~13
in Table~\ref{Tab_groups}). This result suggests that a considerable part of the
Cepheus Flare clouds is located at about 1\,kpc.
 The situation is, however, far from clear.
\citet{SVSG} found a large HI cloud coinciding
both in position and velocity  with the molecular clouds of
Yonekura et al.'s group~13. They associated this cloud not
with NGC\,7129, but with the reflection nebulae of Cep~R2 \citep{Racine}
at a distance of some 400\,pc. The Wolf diagrams constructed by \citet{Kun98}
show two layers of extinction towards $b \sim +11\deg - +13\deg$, at 300\,pc
and $\sim$ 450\,pc, respectively, thus it is tempting to identify the two
velocity components with these two layers.

Another direct distance determination within the area of the Cepheus Flare
is that of \citet{Viotti}, who derived $440\pm100$\,pc for the reflection
nebula NGC\,7023, based on high resolution spectroscopy and UBV photometry of
its illuminating star, HD\,200775. In spite of the large uncertainty of
this value, this is the most frequently cited distance of
NGC\,7023. \citet{Alecian08} pointed out that HD\,200775 is a spectroscopic binary with
two nearly identical components, thus its observed luminosity has to be partitioned
on  both components, which suggests a distance of about 350~pc.
The distance from the {\it Hipparcos\/}
parallax of  HD\,200775, 430$^{+160}_{-90}$~pc \citep{vandenA97} has to be treated with
some caution,  since part of the measured displacement of the star resulted
from its orbital motion. The projected separation of the components, estimated by \citet{Alecian08},
$16\pm9$~mas, commensurates with  the measured parallax, and the orbital
period of 1412~days suggests that {\it Hipparcos\/} might have measured the position
of the star at any point of the orbit.

\citet{SCKM} determined the visual extinction $A_V$ vs. distance
using photometric measurements of 79 stars in the Vilnius photometric system
for the L1147/1158 and NGC\,7023 (L\,1167/1174) regions. They obtained a
distance  288$\pm$25\,pc for NGC 7023, and the same method resulted in
325$\pm$13\,pc for the L1147/L1158 group.

\citet{Snell} studied the interstellar reddening towards L\,1235
and pointed out the presence of an absorbing layer at a distance of
200\,pc. This value is frequently assumed to be the  distance of several other
clouds in the region as well \citep[e.g.][]{BM89}.

\begin{figure*}[ht!]
\centerline{
\includegraphics[draft=False,width=12cm]{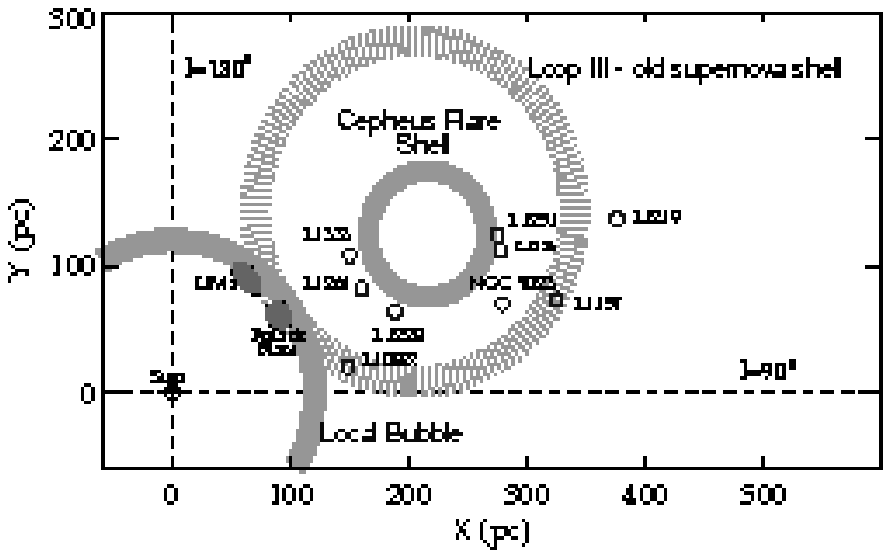}
}
\centerline{
\includegraphics[draft=False,width=12cm]{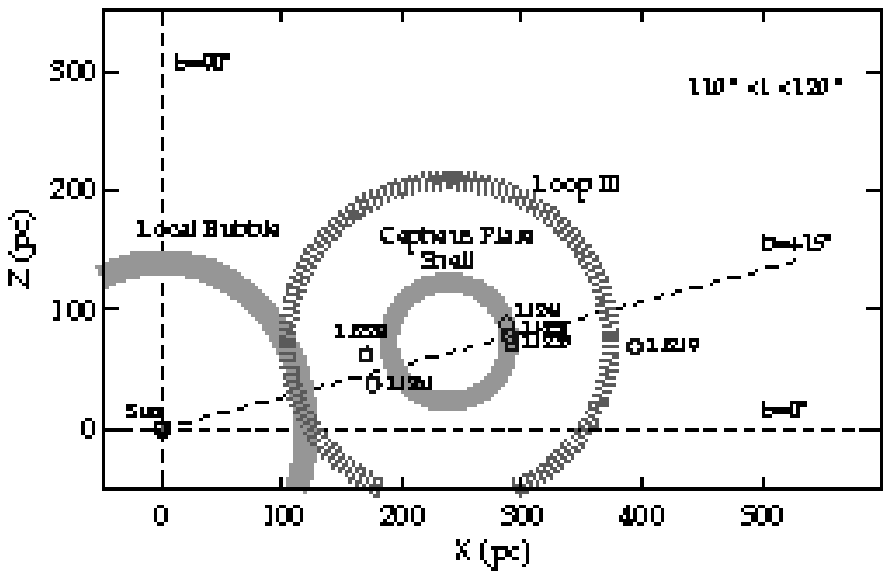}
}
\caption{Upper panel: Distribution of the most prominent molecular clouds of the
Cepheus Flare  and nearby interstellar shells, projected on the Galactic plane, and
viewed from the direction of the North Galactic pole.
Lower panel: the same objects projected onto a plane perpendicular to the Galactic
equator.}
\label{Fig_cf1}
\end{figure*}

\citet{KP93} found a distance of 300$\pm$50~pc for L\,1251 by examining the
interstellar reddening as a function of distance moduli of field stars.
\citet{Kun98} determined distances of dark clouds over the whole area of the
Cepheus Flare using Wolf diagrams. The Wolf diagrams indicated that the
interstellar matter in the Cepheus Flare is concentrated at
three characteristic distances: 200, 300 and 450\,pc. The three components,
though partly overlapping, can be separated along the Galactic latitude.
The three absorbing layers can be identified with  \citeauthor{YDMOF}'s
\citeyearpar{YDMOF} groups 6, 4, and 1, respectively. The overlap of the
layers makes the distance determination of some dark clouds
ambiguous. For instance, both the 300\,pc and 200\,pc layers
can be recognized towards L\,1228. However, L\,1228 differs in radial velocity
from the other clouds of the 300\,pc component. Moreover, the star BD+76\deg825
({\it spectral type: F2\,V, B=11.21, V=10.62\/}) is projected within a compact
group of pre-main sequence stars of L\,1228  and illuminates
a faint reflection nebula \citep{Padgett}. Thus the photometric
distance of this star, 180\,pc, is a good estimate of the distance of L\,1228.
The Cepheus Flare shell, proposed by \citet{Olano}, may explain the existence
of dark clouds at both 200 and 300~pc (see Fig.~\ref{Fig_cf1}).
 Table~\ref{Tab_cepflare_dist}
summarizes the results of distance determinations in the Cepheus Flare region.
The  group of clouds whose distance remains controversial is associated with
the negative velocity component at $l \sim 107\deg-111\deg, b \sim +13\deg$.

\subsection{Star Formation in the Cepheus Flare}

Star formation takes place in dense cores of molecular clouds.  Dense
cores within dark clouds are usually designated with letters appended
to the name of the cloud, e.g. L\,1082\,A, L\,1251\,E.  Several dense
cores and IRAS sources of the Cepheus Flare clouds have been included
in large molecular surveys aimed at studying their various properties
and the associated young stellar objects. Below we list some major
survey papers including key data for Cepheus Flare clouds, cores, and
IRAS sources. \\ \citet{MLB83} -- CO observations (L1152, L1155\,H,
L1155\,D, L1082\,C, L1082\,A, L1082\,B, L1174, L1172\,D, L1172\,B,
L1262\,A); \\ \citet{CB88} -- positions, radial velocities, and IRAS
associations of small, optically selected molecular clouds (CB\,222,
CB\,224, CB\,229 (L1171), CB\,230, CB\,232, CB\,244); \\ \citet{BM89}
-- NH$_3$ observations (L1152, L1155\,B, L1155\,C, L1155\,D, L1155\,G,
L1158, L1082\,C, L1082\,A, L1174, L1174\,B, L1172\,D, L1172\,A,
L1172\,B, L1228\,C, L1228\,B, L1235, L1251\,A, L1262\,A);
\\ \citet{GBFM93} -- velocity gradients (L1152, L1082\,A, L1082\,B,
L1082\,C, L1174, L1172\,A, L1251\,A, L1251\,E, L1262\,A); \\
\citet{BCM98} -- N$_2$H$^+$, C$_3$H$_2$, and CCS observations (L1155,
L1152, L1152\,(IR), L1082\,A, L1082\,C, L1174, L1172\,A, L1228,
L1228\,D, L1221, L1251\,A, L1251\,E, L1262);\\
\citet{Myers88} -- search for outflows (L1152, L1082\,A, L1174,
L1172\,D, L1262); \\ \citet{Fukui} -- search for outflows (PV Cep,
L1228, L1172, NGC\,7129, LkH$\alpha$\,234, L1221, L1251\,A, L1251\,B,
L1262); \\ \citet{Furuya03} -- search for H$_2$O masers (L1082,
L1082\,A, L1082\,B, L1228, L1174, L1172\,D, L1221, L1251\,A, L1251\,B,
L1262);\\ \citet{Mardones} -- search for protostellar infall
(IRAS~20353+6742,\linebreak IRAS~20386+6751, IRAS~21017+6742,
IRAS~22343+7501, IRAS 22376+7455, IRAS 23238+7401).

\begin{table}[!ht]
\caption{Molecular outflows and sources in the Cepheus Flare.}
\label{tab_outflow}
\smallskip
\begin{center}
{\footnotesize
\begin{tabular}{ l c c l l }
\tableline
\noalign{\smallskip}
Cloud & RA(2000) & Dec(2000) & Source & References \\
\noalign{\smallskip}
\tableline
\noalign{\smallskip}
L\,1157  & 20 39 06   & +68 02 13 & IRAS 20386+6751  & 5,17 \\
L\,1157  & 20 45 54   & +67 57 39 & PV Cep & 5,10,12 \\
L\,1082  & 20 47 56.6 & +60 04 14 & IRAS 20468+5953 & 22 \\
L\,1082  & 20 51 30.1 & +60 18 39 & GF9--2 & 22,23 \\
L\,1082  & 20 53 13.6 & +60 14 40 & IRAS 20520+6003 & 22 \\
L\,1228\,A  & 20 57 13 & +77 35 47 & IRAS 20582+7724 & 5,7,9 \\
L\,1228\,B & 20 57 06  & +77 36 56 & HH\,200 IRS & 9 \\
L\,1174\,A & 21 00 22 & +68 12 52 & L\,1174\,A & 21 \\
L\,1172  & 21 02 24 & +67 54 27 & IRAS 21017+6742 & 2, 5,13,\\
L\,1174  & 21 03 02 & +68 06 57 & RNO 131A & 1,3,20 \\
L\,1177  & 21 17 40 & +68 17 32 & IRAS 21169+6804 & 19, 21\\
L\,1183  & 21 42 57 & +66 04 47 & RNO 138  & 1,3,24 \\
L\,1183  & 21 43 01 & +66 03 37 & NGC 7129\,FIRS\,2 & 4,5,8,10 \\
L\,1183  & 21 43 02 & +66 06 29 & LkH$\alpha$ 234 & 4,5,8,10 \\
L\,1183  & 21 43 00 & +66 11 28 & V350 Cep & 24 \\
L\,1219  & 22 14 08 & +70 15 05 & IRAS 22129+7000 & 14,25  \\
L\,1221  & 22 28 03 & +69 01 13 & IRAS 22266+6845 & 6,11,18 \\
TDS\,417  & 22 35 06 & +69 10 53 & IRAS 22336+6855 & 6 \\
L\,1251\,A  & 22 35 24 & +75 17 06 & IRAS 22343+7501 & 5,15 \\
L\,1251\,B & 22 38 47 & +75 11 29 & IRAS 22376+7455 & 5,15,16 \\
L\,1262 & 23 25 46 & +74 17 33 & IRAS 23238+7401 & 5,17,20 \\
\noalign{\smallskip}
\tableline
\end{tabular}
}
\end{center}
\smallskip
{\footnotesize References.  1: \citet{Armstrong}; 2: \citet{BME86};
3: \citet{Cohen}; 4: \citet{ES83};  5: \citet{Fukui}; 6: \citet{Haikala1};
7: \citet{Haikala2}; 8: \citet*{HWJ84}; 9: \citet{Bally95}; 10: \citet{Lada85}; 11: \citet{Lee2002};
12: \citet{Levreault84};  13: \citet{Myers88}; 14: \citet{NK04}; 15: \citet{SF89}; 16: \citet{SFN94};
17: \citet{TVM89}; 18: \citet{UM91}; 19: \citet{Wang95};  20: \citet{WSE92};
21: \citet{YC92}; 22: \citet{Wiesemeyer99}; 23: \citet{Furuya06}; 24: \citet{Liseau83}; 25: \citet{Goicoechea}.}
\end{table}

The Cepheus Flare cloud cores can be found among the targets of infrared, submillimeter
and millimeter continuum surveys of embedded low mass young stellar objects as well:  \\
\citet{Connelley07} -- a K-band atlas of reflection nebulae (IRAS 20353+6742, IRAS 20453+6746,
IRAS~21017+6742, IRAS 22266+6845, IRAS 22376+7455); \\
\citet{Young06} -- submillimeter (450 and 850~\micron) survey of cores included in the Spitzer c2d project (L\,1152, CB\,224,
L\,1157, L\,1082\,C, L\,1082\,A, L\,1228, L\,1177 (CB\,230), L\,1221, L\,1251\,C, L\,1251\,E, L\,1155\,C);\\
\citet{RR98} -- VLA observations of HH exciting sources (L\,1152, L\,1157, L\,1221).

We find these objects in statistical studies of dense cores and young stellar objects: \\
\citet{FM92} -- line width--size relations (L\,1152, L\,1262); \\
\citet{MFGB91} -- shapes of cores (L\,1152, L\,1251, L\,1262); \\
\citet{Wu04} -- properties of outflows (L\,1152, L\,1155, L\,1082, L\,1174, L\,1172,
L\,1228, L\,1177, L\,1221, L\,1251, L\,1262); \\
\citet{Wu07} -- submm (350~\micron) survey of cores included in the Spitzer c2d project (L\,1152,
L\,1157, L\,1148, L\,1177, L\,1228, RNO\,129, L\,1221, L\,1251).

\begin{table}
\caption{Herbig--Haro objects and their sources in the Cepheus Flare.}
\label{tab_cflare_HH}
\smallskip
\begin{center}
{\footnotesize
\begin{tabular}{l@{\hskip3mm}c@{\hskip3mm}c@{\hskip3mm}l@{\hskip3mm}l@{\hskip3mm}c@{\hskip3mm}l}
\noalign{\smallskip}
\tableline
\noalign{\smallskip}
Name & RA(2000) & Dec(2000) & Source & Cloud & D(pc) & Reference \\
\noalign{\smallskip}
\tableline
\noalign{\smallskip}
HH 376B       & 20 35 06.1 & +67 48 47 & IRAS 20359+6745 & L1152    &  440 & 11,13 \\[-1pt]
HH 376A       & 20 36 02.4 & +67 54 28 & IRAS 20359+6745 & L1152    &  440 & 11,13 \\[-1pt]
HH 376        & 20 36 55.3 & +67 59 28 & IRAS 20359+6745 & L1152    &  440 & 15,19 \\[-1pt]
HH 375        & 20 39 06.2 & +68 02 15 & IRAS 20386+6751 & L1157    &  440 & 2,18,19 \\[-1pt]
HH 315C       & 20 45 06.9 & +68 04 50 & PV Cep          & L1158    &  500 & 2,8,13 \\[-1pt]
HH 315        & 20 45 34.0 & +68 03 25 & PV Cep          & L1158    &  500 & 2,8,13 \\[-1pt]
HH 315B       & 20 45 34.0 & +68 03 25 & PV Cep          & L1158    &  500 & 2,8,13 \\[-1pt]
HH 315A       & 20 45 38.4 & +68 00 55 & PV Cep          & L1158    &  500 & 2,8,13 \\[-1pt]
HH 215        & 20 45 53.8 & +67 57 39 & PV Cep          & L1158    &  500 & 10,12,13 \\[-1pt]
HH 415        & 20 46 04.6 & +68 00 28 &                 & L1158    &  500 & 8,11,16 \\[-1pt]
HH 315D       & 20 46 06.4 & +67 54 13 & PV Cep          & L1158    &  500 & 2,8,13 \\[-1pt]
HH 315E       & 20 46 28.1 & +67 52 20 & PV Cep          & L1158    &  500 & 2,8,13 \\[-1pt]
HH 315F       & 20 47 09.9 & +67 50 05 & PV Cep          & L1158    &  500 & 2,8,13 \\[-1pt]
HHL 65        & 20 53 06.0 & +67 10 00 &                 &          &  300 & 7 \\[-1pt]
HH 199R3      & 20 54 49.1 & +77 32 16 & IRAS 20582+7724 & L1228    &  200 & 4 \\[-1pt]
HH 199R2      & 20 54 56.2 & +77 32 21 & IRAS 20582+7724 & L1228    &  200 & 4 \\[-1pt]
HH 200B6      & 20 55 09.4 & +77 31 20 & HH 200 IRS      & L1228    &  200 & 4 \\[-1pt]
HH 199R1      & 20 55 12.2 & +77 33 11 & IRAS 20582+7724 & L1228    &  200 & 4 \\[-1pt]
HH 200B5      & 20 55 22.5 & +77 32 17 & HH 200 IRS      & L1228    &  200 & 4 \\[-1pt]
HH 200B4      & 20 55 33.9 & +77 33 07 & HH 200 IRS      & L1228    &  200 & 4 \\[-1pt]
HH 200B4a     & 20 56 11.0 & +77 34 18 & HH 200 IRS      & L1228    &  200 & 4 \\[-1pt]
HH 200B3      & 20 56 22.2 & +77 35 01 & HH 200 IRS      & L1228    &  200 & 4 \\[-1pt]
HH 200B2      & 20 56 35.9 & +77 35 34 & HH 200 IRS      & L1228    &  200 & 4 \\[-1pt]
HH 200B1      & 20 56 51.2 & +77 36 21 & HH 200 IRS      & L1228    &  200 & 4 \\[-1pt]
HH 199B1      & 20 57 27.2 & +77 35 38 & IRAS 20582+7724 & L1228    &  200 & 4 \\[-1pt]
HH 199B2      & 20 57 31.0 & +77 35 44 & IRAS 20582+7724 & L1228    &  200 & 4 \\[-1pt]
HH 199B3      & 20 57 34.1 & +77 35 53 & IRAS 20582+7724 & L1228    &  200 & 4 \\[-1pt]
HH 199B4      & 20 58 21.7 & +77 37 42 & IRAS 20582+7724 & L1228    &  200 & 4 \\[-1pt]
HH 199B5      & 20 59 08.0 & +77 39 25 & IRAS 20582+7724 & L1228    &  200 & 4 \\[-1pt]
HH 198        & 20 59 09.7 & +78 22 48 & IRAS 21004+7811 & L1228    &  200 & 4,14,16,17 \\[-1pt]
HH 200R1      & 20 59 48.2 & +77 43 50 & HH 200 IRS      & L1228    &  200 & 4 \\[-1pt]
HH 199B6      & 21 00 27.4 & +77 40 54 & IRAS 20582+7724 & L1228    &  200 & 4 \\[-1pt]
HHL 67        & 21 05 00.0 & +66 47 00 &                 &          &  300 & 7 \\[-1pt]
HH 450        & 22 14 24.1 & +70 14 26 & IRAS 22129+7000 & L1219    &  400 & 5 \\[-1pt]
HH 450X       & 22 14 50.1 & +70 13 47 &                 & L1219    &  400 & 5 \\[-1pt]
HH 363        & 22 27 46.7 & +69 00 38 & IRAS 22266+6845 & L1221    &  200 & 1,9 \\[-1pt]
HH 149        & 22 35 24.2 & +75 17 06 & IRAS 22343+7501 & L1251    &  300 & 3,13 \\[-1pt]
HH 373        & 22 37 00.0 & +75 15 16 &                 & L1251    &  300 & 16 \\[-1pt]
HH 374        & 22 37 39.1 & +75 07 31 &                 & L1251    &  300 & 16 \\[-1pt]
HH 374A       & 22 37 39.2 & +75 07 31 &                 & L1251    &  300 & 1 \\[-1pt]
HH 374B       & 22 37 50.0 & +75 08 13 &                 & L1251    &  300 & 1 \\[-1pt]
HH 364        & 22 38 19.2 & +75 13 07 &                 & L1251    &  300 & 16 \\[-1pt]
HH 189C       & 22 38 39.4 & +75 09 49 &                 & L1251    &  300 & 6 \\[-1pt]
HH 189        & 22 38 39.9 & +75 10 41 & IRAS 22376+7455? & L1251    &  300 & 6 \\[-1pt]
HH 189B       & 22 38 40.0 & +75 10 40 & KP 44?          & L1251    &  300 & 6 \\[-1pt]
HH 189E       & 22 38 40.3 & +75 13 52 &                 & L1251    &  300 & 1 \\[-1pt]
HH 189A       & 22 38 40.4 & +75 10 53 &                 & L1251    &  300 & 6 \\[-1pt]
HH 189D       & 22 38 44.2 & +75 13 28 &                 & L1251    &  300 & 1 \\[-1pt]
HH 358        & 23 24 39.0 & +74 12 35 &                 & L1262    &  180 & 1,16 \\[-1pt]
HH 359        & 23 26 29.0 & +74 22 28 &                 & L1262    &  180 & 1,16 \\[-1pt]
\noalign{\smallskip}
\tableline
\end{tabular}
}
\end{center}
\smallskip
{\footnotesize
References. 1: \citet{Alten97}; 2: \citet{AG02}; 3: \citet{BEHKK}; 4: \citet{Bally95};  5: \citet{BR01};
6: \citet{Eiroa94}; 7: \citet{G87}; 8: \citet{Gomez97}; 9: \citet{Lee2002}; 10: \citet*{Moreno95};
11: \citet{MMBA04}; 12: \citet{Neckel87}; 13: \citet{Reipurth97}; 14: \citet{MM2004};
15: \citet{RR98}; 16: \citet{WSE92}; 17: \citet{Brugel90}; 18: \citet*{DRB97}; 19: \citet{DE95}.}
\end{table}

The molecular outflows discovered in the Cepheus Flare and their driving sources
are listed in Table~\ref{tab_outflow}, and the Herbig--Haro objects and driving sources
can be found in Table~\ref{tab_cflare_HH}.

Figure~\ref{fig_yso} shows the surface distribution of pre-main sequence stars and
candidates in the Cepheus Flare. Apparently star formation
occurs in small aggregates, especially along the boundaries of the complex.
The central part of the cloud complex contains clouds of low density and is
avoided by known signposts of star formation.

The samples of T Tauri stars, displayed in Fig.~\ref{fig_yso}, result from several
surveys for H$\alpha$ emission stars conducted in the region
of Cepheus Flare. Only 10 pre-main sequence objects are
catalogued in the Herbig--Bell Catalog \citep[hereinafter HBC]{HBC}.
\citet{OS90} reported on 69 detected
and 49 suspected H$\alpha$ emission stars in a wide environment of L\,1228.
\citet{Kun98} reported 142 H$\alpha$ emission stars, distributed over the
whole area of the Cepheus Flare, identified on objective prism
photographic Schmidt plates and 128 IRAS sources as possible YSOs. Spectroscopic
follow up observations of both samples are underway (Kun et al.,
in preparation). We show in Fig.~\ref{fig_yso} and list in Table~\ref{Tab_pms}
those stars from these two surveys whose pre-main sequence
nature has already been confirmed. The known Herbig~Ae/Be stars are also listed in
Table~\ref{Tab_pms}.

\citeauthor{Tachihara}'s \citeyearpar{Tachihara}  spectroscopic observations
toward the ROSAT X-ray sources resulted in detecting 16 Li-rich stars,
representing weak line T~Tauri stars. The main properties
of these WTTSs are listed in Table~\ref{Tab_wtts}.

The distribution of the WTTSs in the Cepheus Flare differs from that of
other YSOs. In the CO void found by \citet{GLADT}, a group of WTTSs is
separated from the $^{13}$CO cloud by $\ga 10$ pc.
The cloud-to-WTTS separations are significantly larger in Cepheus than in
other nearby SFRs such as Chamaeleon. Because of their grouping, Tachihara et~al.
propose the in-situ formation model for them. From the
total mass of the group of TTSs, a $\sim 800~M_{\sun}$ molecular cloud might have
formed them, while only $\sim 200~M_{\sun}$ molecular gas remained in
their vicinity. An external disturbance might have dissipated the molecular cloud within
several $10^5$ yr. As the distances to the WTTSs are unknown,
\citeauthor{Tachihara} suggest two possible scenarios for the history of the formation
of the WTTS sample isolated from the cloud complex:
(1) The WTTSs in the CO void were formed at 300~pc and affected by the supernova shock
discussed by \citet{GLADT}; or (2) They are at 200 pc and an
unknown supernova explosion has the responsibility for the parent cloud dissipation.
Radial velocity measurements might help to find the relationship between the
stars and the cloud complex. Taking into account the picture of the Cepheus Flare
Shell, the same supernova might have triggered star formation at both 200 and 300~pc.

\subsection{Notes on Individual Objects}

\subsubsection{L\,1147/L\,1158} The cloud group often referred to as the
{\it L\,1147/L\,1158 complex} consists of the clouds Lynds~1147,
1148, 1152, 1155, 1157, and 1158.
L\,1157 harbors a Class~0 object,
L\,1157-mm, with $L_\mathrm{bol} \sim 11\,L_{\sun}$.  It coincides
with IRAS 20386+6751 and drives a spectacular outflow. The L\,1157
outflow has been studied in detail through many molecular lines, such
as CO \citep*{UM92,Gueth96,BPG97,Hirano}, SiO
\citep*{Mikami92,Zhang95,Gueth98,Zhang00, Bachiller01}, H$_2$
\citep{Hodapp,DE95}, NH$_3$ \citep{Bachiller01,TB95, UM99}, and
CH$_3$OH \citep{Bachiller95,Bachiller01,AC96}.  Many other lines have
been detected \citep{BPG97,Bachiller01,Beltran04,Benedettini07,Arce08},
making L\,1157 the prototype of chemically active outflows. Gas phase
shock chemistry models have been used by \citet{Amin} to study the
production of the observed species in the L\,1157 outflow
. \citet{AS06} studied the outflow--envelope interaction on a
$10^4$~AU scale using high angular resolution multiline observations.
\citet{Velusamy} detected spatially resolved methanol emission at 1~mm
from L\,1157. Their results indicate the presence of a warm gas layer
in the infall--disk interface, consistent with an accretion shock.
Regarding the protostar itself, dust continuum observations have been
carried out at 2.7~mm \citep{Gueth96,Gueth97,Beltran04}, 1.3~mm
\citep*{Shirley00,Chini,Gueth03,Beltran04}, 850~\micron \
\citep{Shirley00,Chini,Young06}, 450~\micron \ \citep{Chini}, as well
as 60, 100, 160, and 200~\micron \ \citep{Froebrich03}.  Using the
VLA, \citet{RR98} detected the protostar as a radio continuum source
at 3.6~cm.  \citet{Froebrich03} obtained a far-infrared spectrum of the
L\,1157 protostar using the LWS on board ISO. Deep Spitzer {\it
IRAC\/} images of L\,1157 reveal many details of the outflow and the
circumstellar environment of the protostar. \citet{Looney07} report on
the detection of a flattened structure seen in absorption at
8\,\micron\ against the background emission. The structure is
perpendicular to the outflow and is extended to a diameter of
2~arcmin. This structure is the first clear detection of a flattened
circumstellar envelope or pseudo-disk around a Class 0 protostar.

\begin{table}[!ht]
\caption{Weak-line T Tauri stars in the Cepheus Flare (Tachihara et al. 2005)}
\label{Tab_wtts}
\smallskip
\begin{center}
{\footnotesize
\begin{tabular}{c@{\hskip2mm}c@{\hskip2mm}c@{\hskip2mm}c@{\hskip2mm}c@{\hskip2mm}c@{\hskip2mm}c@{\hskip2mm}c@{\hskip2mm}c}
\tableline
\noalign{\smallskip}
ID & GSC  & RA(2000) & Dec(2000)  & Sp. Type &  $V$  & $V-I_\mathrm{C}$ & M$^{1}$(M$_\odot$) & Age(Myr)$^{1}$ \\
\noalign{\smallskip}
\tableline
\noalign{\smallskip}
4c1 & 0450001478 & 00 38 05.4 & +79 03 21 & K1~ & 10.43 & 0.92 & 1.6 & 2 \\[-1pt]
4c2 &            & 00 38 05.4 & +79 03 21 & K7~ & 13.86 & 1.77 & 0.8 & 15 \\[-1pt]
5c1 & 0450001549 & 00 39 06.1 & +79 19 10 & K6~  & 12.18 & 1.52 & 0.6 & 1 \\[-1pt]
5c2 &            & 00 39 06.1 & +79 19 10 & M2$^{2}$ & 14.12 & 2.10 & 0.4 & 1 \\[-1pt]
19~ & 0458901101 & 20 20 29.3 & +78 07 22 & G8~ &  10.39 & 0.82 & 1.6 & 6 \\[-1pt]
20~~ & 0445900083 & 20 25 15.4 & +73 36 33 & K0~ &  10.62 & 0.84 & 1.6 & 4 \\[-1pt]
28~~ & 0458601090 & 21 11 29.4 & +76 14 29 & G8~ &  11.66 & 0.81 & 1.0 & 25 \\[-1pt]
34~~ & 0460801986 & 22 11 11.0 & +79 18 00 & K7~ &  13.09 & 1.55 & 0.7 & 3 \\[-1pt]
36c1 & 0427200174 & 22 27 05.3 & +65 21 31 & K4~  & 12.92 & 1.13 & 0.9 & 20 \\[-1pt]
36c2 &            & 22 27 05.3 & +65 21 31 & M4$^{2}$ & 15.55 & 2.50 & 0.2 & 0.2 \\[-1pt]
37c1 & 0448000917 &  22 33 44.9 & +70 33 18 & K3~ &  11.63 & 1.65 & 1.0 & 0.4 \\[-1pt]
38~~ & 0460400743 &  22 39 58.1 & +77 49 40 & K2~ & 11.88 & 1.03 & 1.2 & 8 \\[-1pt]
40~~ & 0460502733 &  23 00 44.4 & +77 28 38 & K0~ & 10.98 & 0.88 & 1.4 & 6 \\[-1pt]
41~~ & 0460500268 & 23 05 36.1 & +78 22 39 & K6~ & 13.19 & 1.59 & 0.8 & 7 \\[-1pt]
43c1 & 0448900036 &  23 09 43.4 & +73 57 15 & K7~ & 13.21 & 1.86 & 0.3 & 4 \\[-1pt]
43c2 &           &  23 09 43.4 & +73 57 15 & M3$^{2}$ & 15.55 & 2.50 & 0.7 & 3 \\[-1pt]
44~~ & 0460500170 & 23 16 18.1 & +78 41 56 &  K4~ & 11.77 & 1.24 & 1.0 & 2 \\[-1pt]
45~~ & 0447900348 & 23 43 41.9 & +68 46 27 & K2~ & 12.64 & 1.00 & 0.9 & 25 \\[-1pt]
46~~ & 0461001318 & 23 51 10.0 & +78 58 05 & K1~  & 11.34 & 0.93 & 1.3 & 6 \\[-1pt]
\noalign{\smallskip}
\tableline
\noalign{\smallskip}
\multicolumn{9}{l}{\parbox{0.9\textwidth}{\footnotesize
$^{1}$Derived from the evolutionary tracks by D'Antona \& Mazzitelli (1994).\\
$^{2}$Derived only from the $V-I_\mathrm{C}$ color.}}
\end{tabular}}
\end{center}
\vspace{-8mm}
\end{table}

The K$^\prime$ image of the outflow source, presented by \citet{Hodapp},
is dominated by  nebulosity of bipolar morphology, indicative of
an in-plane bipolar outflow. Knots of nebulosity extend to the north and south of the
outflow position. Both the northern and the southern lobes contain bow shock fronts.

X-ray observations of L\,1157, performed by the ASCA satellite, have been published
by \citet{Furusho}.

The molecular cloud L\,1155 was mapped by \citet{Harjun91} in the lines
of C$^{18}$O, HCO$^{+}$, and NH$_3$. The observations revealed that L\,1155 consists of
two separate clumps, L1155\,C1 and L1155\,C2. The optically visible pre-main sequence
star associated with L\,1152 is HBC\,695 (RNO\,124), studied in detail by \citet{MMBA04}.

Recently \citet{Kauffmann05}, using the data base of the Spitzer Space Telescope
Legacy Program {\it From Molecular Cores to Planet Forming Disks\/} \citep[c2d,][]{Evans} found a candidate
sub-stellar ($M_{\*} \ll 0.1 M_{\sun}$) mass protostellar object in L\,1148.
The object L\,1148--IRS coincides with IRAS~F20404+6712.

\subsubsection{PV Cep}

The highly variable pre-main sequence star PV~Cep lies near the northeastern edge of the
dark cloud complex L\,1147/L\,1158. It is a bright IRAS source, and
has been detected in radio continuum \citep{Anglada92}.
It illuminates a reflection nebula, known as GM--29 \citep{GM77}
and RNO~125 \citep{Cohen}.  A dramatic brightening of the star
(an EXor-like outburst) was observed in the period 1976--1978, and at the same
time the shape of the associated nebula changed drastically \citep*{Cohen77}.

The stellar parameters of PV~Cep are somewhat uncertain. Most of its
optical spectrograms available show no photospheric absorption features.
\citet{Cohen77}, based on measurements of narrow-band continuum indices, estimated
a spectral type about A5. \citet{Cohen81} found the same spectral
type based on the strength of the H$\delta$ absorption line, apparent in two
blue spectra. They note, however, that the hydrogen features probably represent
merely a shell spectrum. \citet{Magakian01} estimated a spectral type of
G8--K0, based on a spectrum taken in July 1978, when the star was some
2~magnitudes fainter than during the outburst. No other estimate of spectral type
can be found in the literature \citep[see][for a review]{Hernan04}.
The spectral type of F, quoted by \citet{Staude86} and \citet{Neckel87} is also
based on the spectral information presented by \citet{Cohen81}.

\citet{Cohen81} found a distance of about 500~pc for PV~Cep. Their estimate was
based on three independent arguments. (1) PV~Cep is probably related to NGC\,7023.
The spectroscopic and photometric data of its illuminating star, HD\,200775, suggest a
distance of 520~pc. (2)  The spectroscopic and photometric data of the nebulous star RNO~124,
located in the same cloud complex as PV~Cep, suggest the same distance.
(3) A similar distance can be obtained from the light travel time from the star to a
nebular spike which brightened about a year after the outburst of the star.
On the contrary, \citet{SCKM} obtained a distance of 325~pc for the  L\,1147/L\,1158
dark cloud complex (see Table~\ref{Tab_cepflare_dist}).

The environment of the star shows a bipolar and rapidly changing
optical morphology
\citep{Cohen81,Neckel84,Staude86,Neckel87,Levreault87,Scarrott-a,Scarrott-b},
as well as a bipolar CO outflow parallel to the symmetry axis of the
reflection nebula \citep{Levreault84}.  \citet{Neckel87} detected
several HH-knots, known as HH~215, emanating from PV~Cep.
\citet*{Reipurth97} and \citet*{Gomez97} discovered a giant ($\sim
2.3$~pc long) Herbig-Haro flow, HH~315, consisting of 23 knots.
HH~215 is also part of this giant flow. \citet{Reipurth97} detected a
further small knot, HH~415, located north-east of PV~Cep, but this may
be a dwarf galaxy with H$\alpha$ redshifted into the [SII] passband
(Bally, priv. comm.).  Near-infrared spectroscopy of PV~Cep is
presented by \citet{HP94} and \citet{GL96}, and optical spectroscopy
by \citet{Corcoran97}.

Far-infrared data obtained by ISOPHOT \citep{Abraham00} indicate the presence of an extended dust
component on arcminute scale around PV~Cep. The existence of a dust core close to PV~Cep
is also supported by the $^{13}$CO (J=1--0) mapping of the region by \citet{Fuente98b},
who found a molecular core with a size of  some 60\arcsec.

\begin{figure*}[!ht]
\centering
\includegraphics[draft=False,width=5.2in,angle=180]{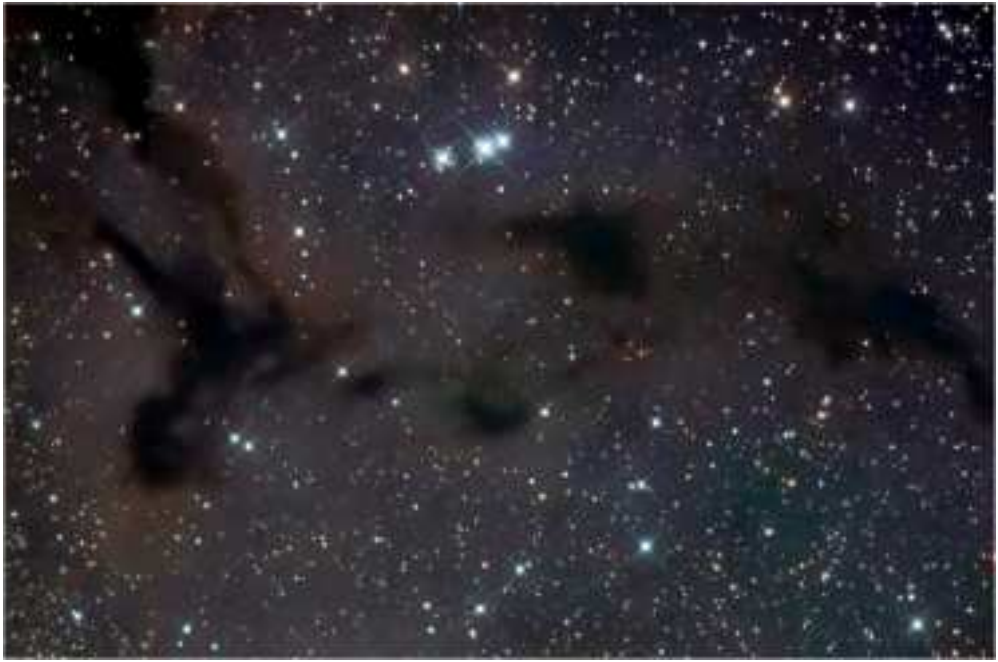}
\caption{Optical image of a field of $36\arcmin \times 24\arcmin$ of L\,1082, obtained
by Giovanni Benintende (http://www.astrogb.com/).}
\label{Fig_B150}
\end{figure*}

\subsubsection{L\,1082} 
\begin{figure*}[!ht]
\centering
\includegraphics[draft=False,width=\textwidth]{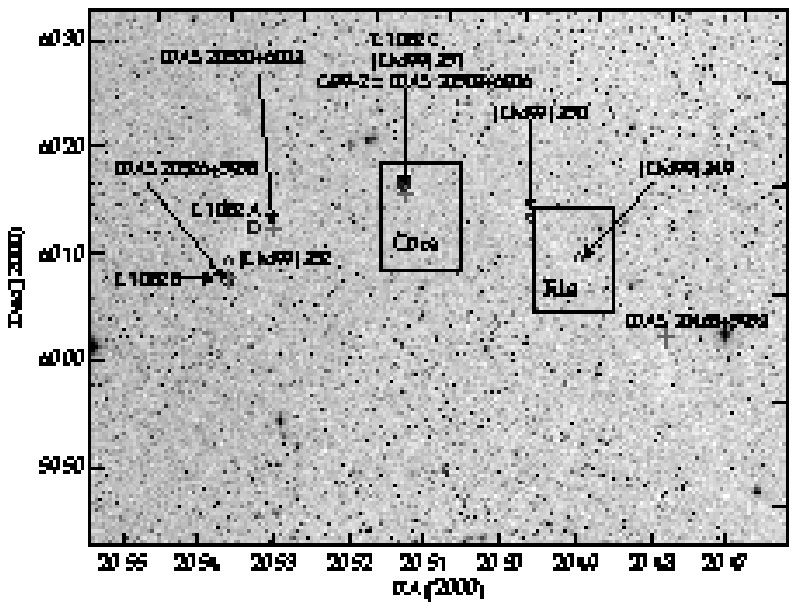}
\caption{Finding chart for the structure of L\,1082, based on \citeauthor{Poidevin06}'s
\citeyearpar{Poidevin06} Fig.~1 and \citeauthor{Wiesemeyer98}'s \citeyearpar{Wiesemeyer98} Fig.~1.
Crosses indicate the positions of the IRAS and ISOCAM point sources,
open cirles mark the dense cores L\,1082\,A, B, and C \citep{BM89},
diamonds show the dense cores catalogued by \citet{LM99}. Two rectangles show the
regions GF\,9~Core and GF\,9 Fila, studied in detail by \citeauthor{Ciardi98}
\citeyearpar{Ciardi98,Ciardi00}. The size of the underlying DSS red image is
$70\arcmin \times 50\arcmin$.}
\label{Fig_L1082}
\end{figure*}

is a remarkable filamentary cloud (see Fig.~\ref{Fig_B150}), first
catalogued by E. E. Barnard \citeyearpar{Barnard} as Barnard~150. It appears as GF~9 in the catalog of globular
filaments by \citet{GF}. Several dense cores, namely L\,1082\,A,B,C \citep{MLB83},
and LM99~349, 350, 351 ($\equiv$ L\,1082\,C), 352 \citep{LM99}, as well as four IRAS
sources, IRAS~20468+5953, 20503+6006, 20520+6003, and 20526+5958 are found along the filament.
A finding chart for the objects in L\,1082 is given in Figure~\ref{Fig_L1082}.

No distance determination is available in the literature for L\,1082. Several authors
assume that L\,1082 is close to NGC\,7023 not only on the sky, but also in space,
and thus accept 440\,pc as its distance \citep[e.g.][]{Ciardi98}. We note that the
angular separation of 10\deg \ between NGC\,7023 and L\,1082 corresponds to
70~pc at the distance of NGC\,7023. If both objects belong to the same complex,
a similar difference can be expected between their distances.
\citet{Wiesemeyer97}, based on statistical arguments, assume a distance of 100$\pm$50~pc.
\citet{Furuya03} refer to 150~pc, and \citet{Furuya06} derive the physical parameters
of GF\,9--2 using 200~pc. \citet{Kun07} speculates that L\,1082 may lie at the interaction
region of the Local Bubble and Loop~III (Cepheus Flare Shell). In this case its likely
distance is about 150~pc.

\citet{Ciardi98} performed near-infrared observations of a core and a filament region within GF~9
(GF\,9-Core and GF\,9--Fila, see Fig.~\ref{Fig_L1082}). They found that neither
the core nor the filament contains a Class~I or Class~II YSO. The extinction
maps of the two $7\arcmin \times 7\arcmin$ fields observed reveal masses 26 and
22~M$_{\sun}$ in the core and the filament, respectively (at a distance of 440\,pc).
The core contains a centrally condensed extinction maximum that appears
to be associated with IRAS 20503+6006, whereas GF\,9--Fila does not show centrally peaked dust
distribution.

\citet{Wiesemeyer98} presented  mid-infrared ISOCAM observations of L\,1082. They identified
9 sources along the filament, and designated them as GF\,9--1, 2, 3a, 3b, 4, 5, 6, I6, and  7.
The designations probably follow \citet{Mezger94} who labeled dense cores along
the filament by the same numbers.
GF\,9--2 coincides in position with IRAS~20503+6006, GF\,9--3b with IRAS 20520+6003,
GF\,9--I6 with IRAS 20468+5953, and GF\,9--4 lies at 12\arcsec \ west of IRAS 20526+5958.
They find that these latter two sources are Class~0
protostars, both associated with CO outflows, whereas GF\,9--6 and  GF\,9--7 are most
likely reddened background stars. They speculate that the other ISOCAM sources, not
associated with outflows, may be transitional objects between prestellar (Class~$-1$) and
Class~0 evolutionary stages. \citet{Wiesemeyer99} presented far-infrared
ISOPHOT and millimeter continuum measurements for GF\,9--2, GF\,9--3a/3b, and GF\,9--4.

\citet{Ciardi00} performed CO, $^{13}$CO, and CS observations of the GF\,9--Core and
GF\,9--Fila regions, determined excitation temperatures, densities and masses. The CS
observations reveal that both regions contain centrally condensed,
high-density gas cores. The temperatures and masses of the two regions and of the
cores contained within the regions are similar, but the densities in GF\,9--Core are
twice those of GF\,9--Fila.

\Citet{Gregorio} detected CCS emission from GF\,9--2.
The structure of the magnetic field associated with GF\,9 was studied by
\citet{Jones03} and \citet{Poidevin06}.

\citet{Furuya03} detected H$_2$O maser emission from GF\,9--2.
\citet{Furuya06} studied in detail the spatial and velocity structure of GF\,9--2,
using several molecular transitions obtained by single-dish and interferometric
radio observations and 350~\micron \ continuum data.
The observations revealed a dense core with a diameter of $\sim 0.08$~pc and mass
of $\sim 3$~M$_{\sun}$. Within the core a protostellar envelope with a size of
$\sim 4500$~AU and mass of $\sim 0.6$~M$_{\sun}$ could be identified.
The radial column density profile of the core can be well fitted by a
power-law form of $\rho(r) \propto r^{-2}$ for the $0.003 < r/pc < 0.08$ region.
The power-law index of $-2$ agrees with the expectation for an outer part
of the gravitationally collapsing core. They found no jet-like outflow, but a compact,
low-velocity outflow may have formed at the center.
They discovered a potential protobinary system with a projected separation of
$\sim 1200$~AU, embedded in a circumbinary disk-like structure with $\sim 2000$~AU
radius at the core center. The binary consists of a very young protostar and
a pre-protostellar condensation. The studies led to the conclusion that
GF\,9--2 is very likely at an extremely early stage of low-mass
star formation before arriving at its most active outflow phase.

\citet*{Stecklum} discovered 14 Herbig--Haro objects in the GF\,9 region which
apparently belong to at least three large HH-flows. Five HH-objects and GF\,9--2
are linearly aligned, suggesting that they constitute an HH-flow driven
by IRAS~20503+6006. Its overall length amounts to 43.5~arcmin, which corresponds
to 2.3\,pc for an assumed distance of 200\,pc. The presence of a well-developed,
parsec-scale outflow from GF\,9--2 indicates a more advanced evolutionary stage of this
source than previously believed.

\citet{Furuya08} mapped GF\,9 in the NH$_3$~(1,1) and (2,2) inversion lines,  using
the Nobeyama 45-m telescope, with an angular resolution of  73\arcsec. The large-scale
map reveal that the filament contains at least 7 dense cores, as well as 3
candidates, located at regular intervals of $\sim0.9~$pc (at an assumed distance of 200~pc).
The cores have kinetic temperatures
of  $\la$~10~K and LTE-masses of 1.8 -- 8.2 M$_{\sun}$, making them typical
sites of low-mass star formation, probably  formed via the gravitational
fragmentation of the natal filamentary cloud.

\subsubsection{NGC\,7023} 
\begin{figure*}[htb]
\centering
\includegraphics[draft=False,width=5.2in]{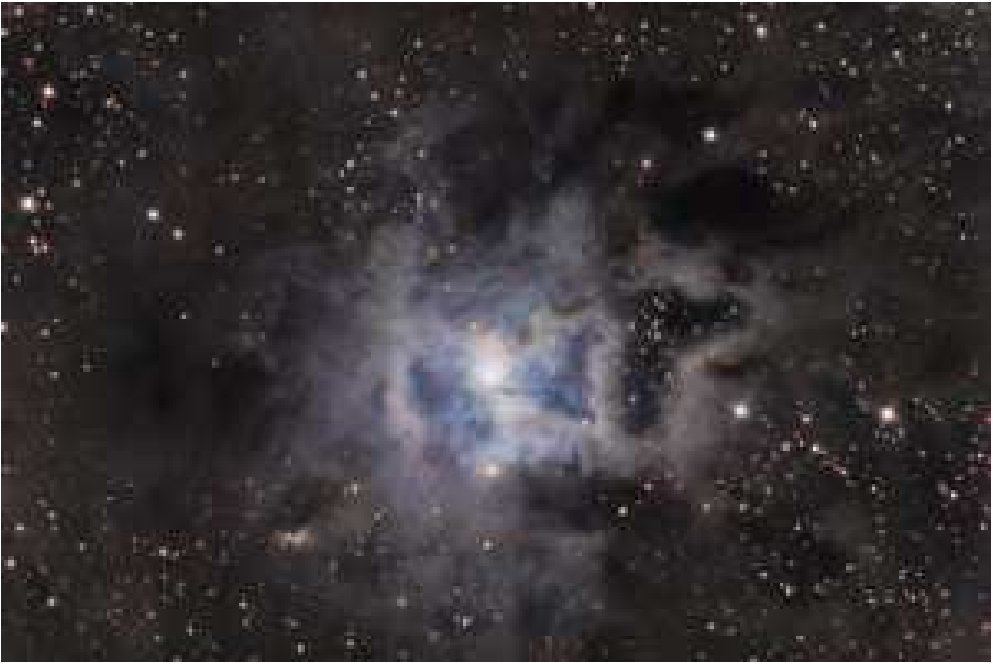}
\caption{Optical image of NGC\,7023, illuminated by the Herbig~Be star HD\,200775.
The size of the field is about 30$\arcmin \times 20 \arcmin$.
Courtesy of Richard Gilbert of the Star Shadows Remote Observatory.}
\label{Fig_n7023}
\end{figure*}

is a reflection
nebula, illuminated by the young massive star \linebreak
HD\,200775 and a group of fainter stars.  It was discovered by William Herschel
in  1794. HD\,200775 (also known as V380~Cep and HBC~726) is a Herbig Be star
that has been extensively studied  \citep[e.g.][]{Herbig60,Altamore80,Pogodin04,Alecian08}.
The surrounding reflection nebula has been observed in detail
\citep[e.g.][]{Slipher1918,Witt80,Witt82,Rogers95,Laureijs96,Fuente00,Werner04,Berne08}.
Figure~\ref{Fig_n7023} shows an optical image of the reflection nebula.

\citet{Weston} found that, centered on the reflection
nebula, there is a small cluster consisting of stars which are variable and show
H$\alpha$ line in emission. Some two dozens of variable stars of the region
were studied by \citet{Rosino}.  Nevertheless, the HBC lists only
four of these stars (HBC~726, 304, 306, 307, see Table~\ref{Tab_pms})
as confirmed T~Tauri type stars. Recently \citet{Goodman} speculated that
the young Herbig~Ae star PV~Cep, located more than 10\,pc to the west of the cluster,
might have been ejected from NGC\,7023 at least 100,000 years ago.
HD\,200775 is located at the northern edge of an elongated molecular cloud,
corresponding to the  dark clouds L\,1167,
1168, 1170, 1171,
1172, 1173 and 1174,
referred to as the {\it L\,1167/L\,1174 complex\/}. The cloud complex has
been mapped in CO by \citet{Elmegreen}. They found that the
size of the cloud is 0.5\deg$\times$1.0\deg, or 3.9\,pc$\times$7.7\,pc, and
the mass of the molecular hydrogen is some 600\,M$_{\sun}$.



\begin{landscape}
\begin{center}
{\footnotesize
\begin{longtable}{@{\extracolsep\fill}lr@{\hskip1mm}c@{\hskip1mm}c@{\hskip1mm}r@{\hskip3mm}
r@{\hskip1mm}r@{\hskip1mm}r@{\hskip1mm}r@{\hskip1mm}r@{\hskip1mm}r@{\hskip1mm}r}

\caption{Pre-main sequence stars in the Cepheus Flare -- (A)  Classical T Tauri stars}\\
\noalign{\smallskip}
\tableline
\noalign{\smallskip}
Names  & IRAS~~~ & 2MASS J/ & \multicolumn{3}{c}{2MASS magnitudes} & \multicolumn{4}{c}{IRAS fluxes} \\[3pt]
 & & RA,Dec(J2000) &  J~  &  H~~ &  K~~  &  F(12)  &   F(25)  &  F(60)  &  F(100) \\
\noalign{\smallskip}
\tableline
\noalign{\smallskip}
\endfirsthead

\caption{Pre-main sequence stars in the Cepheus Flare -- (A)  Classical T Tauri stars  (cont.)}\\
\noalign{\smallskip}
\tableline
\noalign{\smallskip}
Names  & IRAS~~~ & 2MASS J/  & \multicolumn{3}{c}{2MASS magnitudes} & \multicolumn{4}{c}{IRAS fluxes} \\[3pt]
 & & RA/Dec(J2000) & J~~  &  H~~ &  K~~  &  F(12)  &   F(25)  &  F(60)  &  F(100) \\
\noalign{\smallskip}
\tableline
\noalign{\smallskip}
\endhead

\noalign{\smallskip}
\noalign{\smallskip}
\multicolumn{10}{l}{\parbox{20cm}{\footnotesize \hskip 7mm $^*${\it No 2MASS counterpart.}}}
\endfoot

\noalign{\smallskip}
\tableline
\noalign{\smallskip}
\multicolumn{10}{l}{\parbox{20cm}{\footnotesize \hskip 7mm $^*${\it No 2MASS counterpart.}}}
\endlastfoot

HBC\,695n, RNO\,124, K98\,6 &  20359+6745 & 20361986+6756316 & 11.364 & 9.739 &  8.781 & 0.44 & 1.05 & 1.81 \\  
GSC 04472-00143  & 20535+7439 & 20530638+7450348    & 10.149 & 9.405 & 8.861  & 0.42 & 0.63  & 0.68  \\   
HH\,200 IRS, L\,1228 VLA~4 &  & 20570670+7736561$*$ & \\
FT Cep, K98\,26 & 20587+6802 & 20592284+6814437   & 10.588 &  9.342 & 8.532 & 0.55 & 0.92  & 0.89 \\
K98\,30, OSHA\,42 & F20598+7728 & 20584668+7740256    & 11.513 & 10.365 & 9.699 & 0.09 & 0.16   \\
RNO\,129\,S1, OSHA\,44, K98\,32  & 21004+7811 & 20591409+7823040   &  9.437 & 7.530  & 6.319 & 6.23 & 11.12 & 36.30 & 76.60: \\
RNO\,129\,S2, OSHA\,44, K98\,32  & 21004+7811 &  20591256+7823078 & 10.993:& 12.060 &  9.174 \\
RNO\,129\,A  &  &   20590373+7823088 &  12.565 & 11.280 & 10.726 \\	
HBC\,304, FU Cep, LkH$\alpha$ 427 & 21009+6758 & 21014672+6808454   & 11.792 & 10.798 & 10.159 \\
K98\,35, PRN S5   & F21016+7651 &  21005285+7703149   & 11.290 & 10.288 & 9.773 & 0.12 & 0.15 & 0.34  \\
 & F21022+7729 &  21011339+7741091  & 12.676 & 11.202 & 10.327 & 0.08 & 0.13 & 0.17 \\
NGC\,7023 RS\,2 & & 21012706+6810381 & 12.323 & 11.150 & 10.417 \\
NGC\,7023 RS\,2 & & 21012637+6810385 & 11.107 & 10.084 &  9.571 \\
PW Cep, LkH$\alpha$ 425, NGC\,7023 RS\,3 & & 21013590+6808219 &  12.336 & 11.564 & 11.052  \\
NGC\,7023 RS\,5 & & 21014250+6812572   & 11.911 & 10.892 & 10.421 \\
LkH$\alpha$ 428, NGC\,7023 RS\,8 & & 21022829+6803285 &  11.141  & 10.457 & 9.723 \\
HZ Cep, NGC\,7023\,RS\,S3 & &  21014358+6809361 &  11.218 & 10.415 & 10.156 \\  
HBC\,306, FV Cep, LkH$\alpha$ 275, K98\,38 & F21017+6813  & 21022039+6825240  & 11.513 & 10.529 & 9.880 & 0.12 & 0.14 \\
PRN S1(b) &  & 21012508+7706540  & 17.212: & 14.640 & 13.156 & \\
PRN\,S1(a) & & 21012638+7707029   & 16.460: & 14.622 & 13.244 & \\
OSHA\,48, PRN S6 &  & 21012919+7702373   &  9.919 & 9.093  & 8.563   \\
OSHA\,49, K98\,40, PRN S7 & F21023+7650  &  21013097+7701536    & 11.669 & 10.910 & 10.614 & 0.17 & 0.23 & 0.72 \\
OSHA\,50, K98\,41, PRN S8  & &  21013267+7701176 & 11.993 & 11.060 & 10.419  & 0.08 & 0.08 & 0.09 \\
PRN S4 & & 21013505+7703567 & 13.217 & 12.018 & 11.084 & \\
PRN S2 & & 21013945+7706166  & 15.399 & 14.007 & 12.973  \\
PRN S3 & & 21014960+7705479 & 12.449 & 11.272 & 10.802 & \\
OSHA\,53, K98\,43, PRN S9 & F21028+7645 &  21020488+7657184  & 11.108 & 10.352 & 10.027 & 0.09 & 0.20 & 0.24 \\
FW Cep, NGC\,7023 RS\,9 &   & 21023299+6807290   & 11.559 & 10.713 & 10.411  \\ 
NGC\,7023 RS\,10 & 21023+6754 &  21025943+6806322    & 13.871 & 13.083 & 12.357 & 0.27 & 0.39    \\   
K98\,46 & F21037+7614 &  21030242+7626538  & 11.585 & 10.843 & 10.471 & 0.13 & 0.18 \\
HBC\,307, EH Cep, LkH$\alpha$ 276, K98\,42 & 21027+6747 &  21032435+6759066  &  9.538 & 8.767  & 8.196 & 0.59 & 0.71   \\    

OSHA\,59, K98\,49 & F21066+7710 & 21055189+7722189   & 10.689 & 9.755 & 9.086 & 0.17 & 0.24 & 0.18 \\
K98\,53  &  & 21153595+6940477   & 11.585 & 10.843 & 10.471 & 0.13 & 0.18 \\
K98\,58 & F21202+6835   &  21205785+6848183   & 14.318 & 13.035 & 11.866 &  & 0.20 &  &  5.49: \\
RNO\,135, K98\,61 & 21326+7608 &  21323108+7621567 & 11.101 & 10.072 &  9.688   & &  & 1.46 &  5.01  \\
K98\,66 &  & 21355434+7201330  & 11.323 & 10.724 & 10.463 \\ 
K98\,71 & F21394+6621 &  21402754+6635214  & 11.344 & 10.540 & 10.046 & 0.34 & 0.64 & 2.56 \\
HBC\,731, SVS 6 & & 21425961+660433.8  & 12.886 & 11.624 & 10.948 & \\	
HBC\,732, V350\,Cep, MMN\,13 & & 21430000+6611279 & 12.714 & 11.691 & 11.008 & \\
NGC\,7129 S V1 & & 21401174+6630198   & 13.161 & 12.173 & 11.591 \\
NGC\,7129 S V2 &  & 21402277+6636312   &  13.882  & 12.671  & 11.890 \\
NGC\,7129 S V3 &  & 21403852+6635017   &  13.108 & 11.888 & 11.267  \\
V391 Cep, K98\,72  & F21404+6608 &  21413315+6622204  & 11.680 & 10.555 & 9.750  & 0.36 & 0.40  \\      
NGC\,7129 MMN\,1 & & 21422308+6606044   & 15.059 & 14.050 & 13.394 \\
NGC\,7129 HL85 14, MMN2\,2  &  &  21423880+6606358  &  14.796  &  13.479  &  12.514 \\
NGC\,7129 MMN\,3 & & 21424194+6609244   & 15.179 & 14.340 & 13.987  \\
NGC\,7129 MMN\,5 & & 21425142+6605562   & 15.199 & 14.128 & 13.573 \\
NGC\,7129 MEG 1 &  & 21425177+6607000   & 16.679 & 15.546 & 14.675 \\ 
NGC\,7129 MMN\,6 & & 21425262+6606573   & 13.821 & 12.650 & 11.803 \\
NGC\,7129 MMN\,7 & & 21425314+6607148   & 14.338 & 13.193 & 12.679 \\
NGC\,7129 MMN\,8 & & 21425346+6609197   & 16.990 & 15.497 & 14.578 \\
NGC\,7129 MMN\,9 & & 21425350+6608054   & 13.214 & 12.365 & 12.124 \\
NGC\,7129 MMN\,10 & & 21425481+6606128  & 14.192 & 13.254 & 12.907 \\
NGC\,7129 MEG 2 & & 21425476+6606354    & 14.183 & 13.179  & 12.564 \\ 
NGC\,7129 MMN\,11 & & 21425626+6606022  & 12.423 & 11.657 & 11.406  \\
RNO 138, V392 Cep & & 21425771+6604235  & 14.567: & 14.478 & 13.658 \\
NGC\,7129 MMN\,12 & & 21425810+6607394   & 14.299 & 13.532 & 13.030 \\
NGC\,7129 MEG 3 & & 21425878+6606369  & 13.487 & 12.561  &  12.337 \\
NGC\,7129 MMN\,14 & & 21430024+6606475  & 14.161 & 12.778 & 12.085 \\
NGC\,7129 MMN\,15 & & 21430246+6607040  & 14.004 & 13.109 & 12.256 \\
GGD 33A  &  &  21430320+6611150 & 16.619 & 15.461 & 14.337 \\

NGC\,7129 MMN\,17 & & 21431162+6609115   & 12.605 & 11.800 & 11.487  \\	
NGC\,7129 MMN\,18 & & 2143124+661238$^*$   & \\ 
NGC\,7129 MMN\,19 & & 21431683+6605487   & 14.184 & 13.300 & 13.123  \\
NGC\,7129 MMN\,20 & & 21433183+6608507   & 13.935 & 12.894 & 12.295 \\
NGC\,7129 MMN\,21 & & 21433271+6610113   & 14.986 & 13.792 & 13.547 \\
NGC\,7129 MMN\,22 & & 21434345+6607308   & 13.655 & 12.609 & 12.146 \\ 	
K98\,73  &   &  21443229+7008130  & 11.777 & 10.939 & 10.535 &  & 0.14 & 0.25 \\
K98\,95   &  &  22131219+7332585   & 10.987 & 10.048 & 9.498 & 0.3  & 0.26 & 0.32 & 0.28 \\
GSC 04467-00835 & 22129+6949 & 221406ö8+7005043   &  9.620  & 9.072  & 8.753 & 0.254 & 0.567 & 0.484 &  0.50: \\
K98\,108  &  &  22190203+7319252   & 10.666 & 9.947  & 9.561 & 0.15 & 0.26 & 0.3 \\
K98\,109  &  &  22190169+7346072   & 11.677 & 10.738 & 10.234 & 0.06 & 0.12 & 0.16 & 0.4 \\
K98\,110  &  &  22190343+7349596   & 12.848 & 12.264 & 12.162 & 0.15 & 0.19 & 0.12 & 0.16 \\
K98\,119 & 22256+7102 &  22265660+7118011   & 10.967 & 9.646  & 8.525  & 1.74 & 2.28 & 2.48 & 4.17 \\ 
KP 1, XMMU J223412.2+751809 & 22331+7502 & 22341189+7518101   &  10.095  & 8.808 & 7.827 & \\
KP 39, XMMU J223516.6+751848 & & 22351668+7518471   &  11.808 & 10.589 & 9.858 \\
KP 2, XMMU J223605.8+751831 & 22350+7502 & 2236059+7518325  &  11.957 & 10.894 & 10.253 & \\
KP \#10  & 22355+7505 &  2236345+7521352  & 11.984 & 10.628 & 10.026 & 0.11 & 0.24 & 0.23 &  \\
KP 43, XMMU J223727.7+751525 & & 22372780+7515256  & 11.289  &  10.202 & 9.854 \\
KP 3, XMMU J223750.1+750408 & & 22374953+7504065    & 11.785 & 10.974 & 10.680 \\
ETM Star 3, XMMU J223818.8+751154 & & 22381872+7511538  & 11.248 & 9.815 & 8.912 \\
KP 44, ETM Star 1, XMMU J223842.5+751146 & & 22384249+7511455   & 12.166 & 11.313 & 11.028 \\
KP 45, XMMU J22397.3+751029 & & 22392717+7510284  & 11.748 & 10.851 & 10.377 \\
KP 46, XMMU J223942.9+750644 & 22385+7457 & 22394030+7513216  & 10.920 & 9.481 & 8.698 \\
GSC 04601-03483 & F22424+7450 & 22433926+7506302  & 10.958 & 10.379 & 10.216 & 0.080 & 0.240 & 0.522 \\
K98\,128 & &  22490470+7513145  & 12.011 & 11.208 & 10.814  \\ 
TYC 4601-1543-1 & 22480+7533  &  22491626+7549438  &  9.938 & 9.256  & 8.727 & 0.61 & 0.90 & 0.81 \\ 
HBC\,741, AS\,507, K98\,140 & 23189+7357 & 23205208+7414071  & 8.308 & 7.754 & 7.480 \\ 
\label{Tab_pms}
\end{longtable}
}
\end{center}

\addtocounter{table}{-1}

\begin{table}[!ht]
\caption{Pre-main sequence stars the Cepheus Flare -- (B) Herbig Ae/Be stars }
\begin{center}
{\footnotesize
\begin{tabular*}{7.6in}{@{\extracolsep\fill}lr@{\hskip1mm}c@{\hskip1mm}r@{\hskip3mm}
r@{\hskip1mm}r@{\hskip1mm}r@{\hskip1mm}r@{\hskip1mm}r@{\hskip1mm}r@{\hskip1mm}r}
\noalign{\smallskip}
\tableline
\noalign{\smallskip}
Names  & IRAS~~~ & 2MASS J  & \multicolumn{3}{c}{2MASS magnitudes} & \multicolumn{4}{c}{IRAS fluxes} \\[3pt]
 & & RA/Dec(J2000) &  J~~  &  H~~ &  K~~  &  F(12)  &   F(25)  &  F(60)  &  F(100) \\
\noalign{\smallskip} 
\tableline
\noalign{\smallskip}
HBC\,696n, PV Cep, K98\,9 & 20453+6746 & 20455394+6757386  & 12.453 &  9.497 & 7.291 & 12.82 & 32.93 & 48.85 & 57.92 \\
HBC\,726, HD\,200775, MWC\,361 & 21009+6758 & 21013691+6809477  & 6.111 & 5.465 & 4.651 & 26.700 & 76.800 & 638.000 & 1100.000 \\
HD\,203024 & 21153+6842 & 21160299+6854521  & 8.377 & 8.209 & 8.120 & 3.680 & 10.800 & 4.260 & \\
GSC\,04461-01336, BD+68\deg1118 & 21169+6842 & 21173917+6855098  & 9.269 & 8.741 & 8.105 & 1.570 & 3.480 & 4.130 & 2.59 \\
HBC\,730, V361 Cep, AS\,475, BD +65\deg1637 & & 21425018+6606352  & 8.973 & 8.729 & 8.474 	\\
HBC\,309, LkH$\alpha$ 234, V373 Cep & 21418+6552 & 21430682+6606542 & 9.528 & 8.201 & 7.081 & 14.78 & 78.96 & 687.70 & 1215.00 \\
HBC\,734, BH Cep, K98\,83 & 22006+6930 & 22014287+6944364 & 9.686 & 8.993 & 8.310 & 0.524 & 1.200 & 1.390 \\ 
HBC\,735, BO Cep, K98\,100 & F22156+6948 & 22165406+7003450 &  10.319 & 9.849 & 9.581 & 0.285 & 1.428  \\ 
HBC\,736, SV Cep, K98\,113 & 22205+7325 & 22213319+7340270 &  9.350  & 8.560 & 7.744 & 4.22 & 5.22 & 2.66 & 1.76 \\
GSC\,04608-02063 & 22219+7908  &  22220233+7923279 &  11.509 & 10.920 & 10.266  \\
\noalign{\smallskip}
\tableline
\noalign{\smallskip}
\end{tabular*}}
\end{center}
\smallskip
{\footnotesize 
{\it References to star names:\/} NGC\,7023\,RS -- variable stars from \citet{Rosino}; OSHA -- H$\alpha$ emission stars from 
\citet{OS90}; KP -- H$\alpha$ emission star from \citet{KP93}, Table~2; KP\# -- IRAS point source 
from \citet{KP93}, Table~3; ETM -- \citet{Eiroa94}; K98 -- H$\alpha$ emission stars from \citet{Kun98}; 
NGC\,7129\,MEG -- \citet{MEG}; NGC\,7129\,MMN -- \citet{MMN04}; PRN -- \citet{Padgett}; NGC\,7129\,S -- \citet{Semkov};
XMMU -- \citet{Simon}}
\end{table}
\end{landscape}

\begin{figure*}[!ht]
\centerline{
\includegraphics[draft=False,width=10cm]{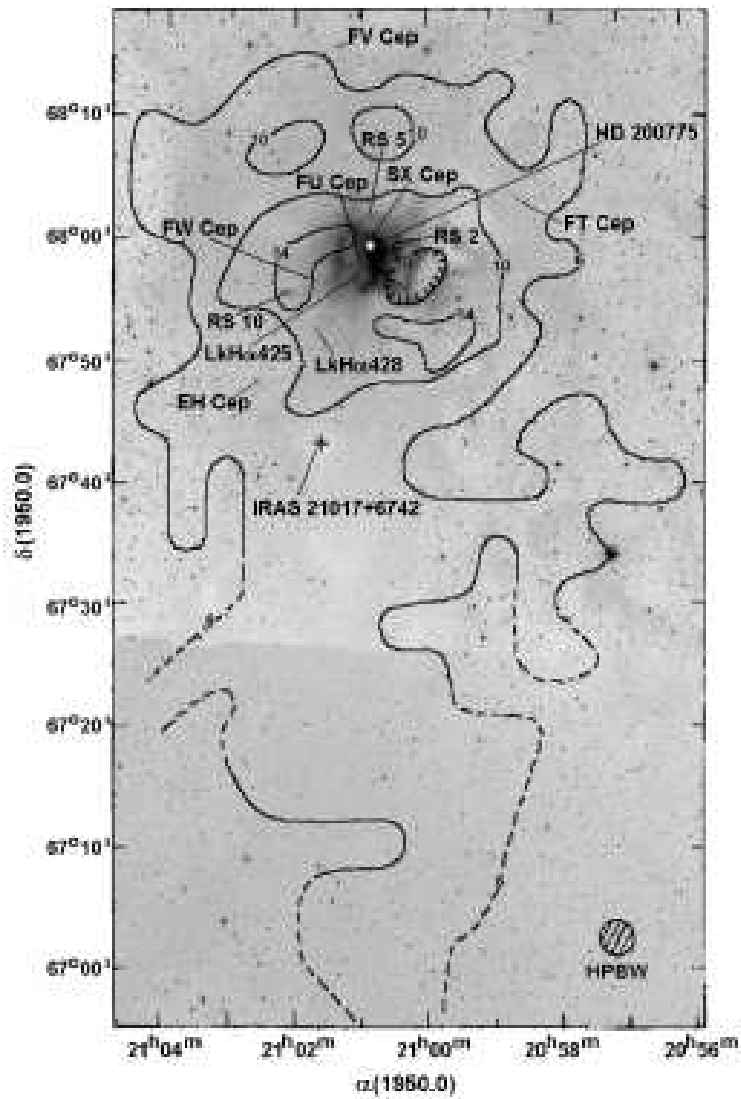}
}
\caption{$^{13}$CO contours of the molecular cloud associated with NGC\,7023,
overplotted on the DSS red image \citep{Elmegreen}. Positions of HD\,200775
and lower mass pre-main sequence stars, as well as
the protostar IRAS 21017+6742 are indicated.}
\label{Fig_l1172}
\end{figure*}

\citet{Watt} found a bipolar outflow associated with HD\,200775.
The region of the outflow has been mapped in $^{13}$CO(1-0)
by \citet{Fuente98}. These observations show that the star
is located within a biconical cavity, which has probably been excavated by
a bipolar outflow. However, Fuente et al. found no evidence for
current high-velocity gas within the lobes of the cavity.

\citet{Myers88} detected another molecular outflow centered on the
IRAS source IRAS 21017+6742.  \citeauthor{Hodapp}'s \citeyearpar{Hodapp}
K$^{\prime}$ image of the L\,1172 outflow
shows four stars associated with localized nebulosity. None of them are close
to the nominal outflow position. \citet*{Visser} detected three submillimeter
sources at the position of IRAS~21017+6742: L\,1172\,SMM\,1--SMM\,3.
They found that  L\,1172\,SMM\,1, located at
RA(2000)=21$^\mathrm{h}02^\mathrm{m}21.5^\mathrm{s}$, Dec(2000)=$+67\deg54\arcmin14\arcsec$
is a protostar and the driving source of the outflow, whereas
SMM\,2 and SMM\,3 are starless dust clumps. Figure~\ref{Fig_l1172} shows the
$^{13}$CO contour map of the L\,1167/L\,1174 complex, adopted from \citet{Elmegreen}.
Known pre-main sequence stars of the region are also indicated.

\subsubsection{L\,1177} (CB~230) contains a molecular outflow driven by the IRAS source 21169+6804
\citep{YC94}. Near infrared observations by \citet{Yun96} revealed this source to be a binary
protostar with a projected separation of 12\arcsec, and embedded in a common infrared
nebula. The stars can be found near the center of a dense core whose size is about 360\arcsec \
(0.5\,pc at 300\,pc). Further CO and infrared studies can be found in \citet*{CYH91}.
Submillimeter polarization measurements by  \citet*{WLH03}
reveal a magnetic field strength of 218\,$\mu$G for the envelope of CB\,230.
\citeauthor{WLH03} found that the outflow is oriented almost perpendicular to the symmetry
axis of the globule core, whereas the magnetic field is parallel to the same axis.
They discuss the possibility that the orientation of the
magnetic field relative to the outflow directions reflects the evolutionary stage of the globule.
Two A-type emission line stars, HD\,203024 and BD\,+68\deg 1118
can be found to the north of the globule, at the edge of the diffuse outer part
of the cloud. Their formation history may be connected to each other \citep{Kun98}.
\citet{MBK97} and \citet*{KVSz} classify these objects as candidate Herbig Ae/Be stars,
whereas \citet{Mora} state that they are main sequence stars.

\subsubsection{L\,1228} is a small cloud stretching some 3\deg \ along a north-south
direction. Its most probable distance is 180~pc (see Sect.~\ref{Sect_dist}). L\,1228 differs kinematically
from the rest of the Cepheus Flare molecular clouds, suggesting that the cloud is located
on the near side of the Cepheus Flare shell. Numerous H$\alpha$ emission
stars have been found around this cloud  \citep{OS90,Kun98}, as well as several
molecular outflows \citep{Haikala2} and Herbig--Haro
objects \citep{Bally95}.

The elongated cloud consists of three centers of star formation.

(1) The northernmost part is a small, nebulous group of stars, {\it RNO\,129\/},
associated with IRAS~21004+7811. \citet{Bally95} found Herbig--Haro emission from RNO\,129.
A detailed study of RNO\,129 can be found in \citet{MM2004}. \citet{AS06}
included RNO\,129 in their study of the  evolution of outflow-envelope interactions
in low-mass protostars.

\begin{figure*}[!t]
\centerline{
\includegraphics[draft=False,width=5.2in]{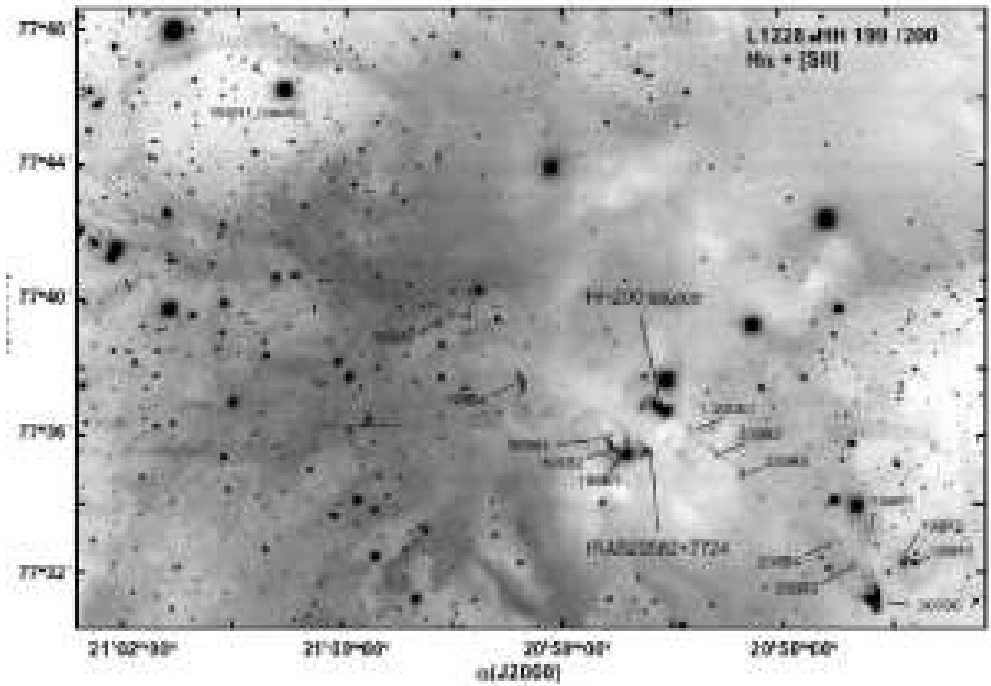}
}
\caption{H$\alpha$ + [SII] image of  L\,1228~A, based on KPNO 4\,m images obtained
with the Mosaic~1 prime focus CCD camera through narrow-band H$\alpha$
and [SII] filters (80A passband). HH objects discussed in \citet{Bally95} are marked.
(Courtesy of John Bally).}
\label{fig_L1228}

\end{figure*}

(2) The  {\it L\,1228 core\/} or {\it L\,1228\,A\/} contains the Class~I source IRAS 20582+7724.
A $^{13}$CO map of the cloud is presented in \citet{MB94}. A dense core, mapped in ammonia
by \citet*{ASG97}, contains at least two sources driving molecular outflows as well as
two Herbig-Haro flows, HH~199 and HH~200, revealed by the H$\alpha$ image of L\,1228\,A,
obtained by \citet{Bally95} and shown in  Fig.~\ref{fig_L1228}. HH~199 emerges from
IRAS~20582+7724, associated with an east--west oriented infrared
reflection nebula \citep{Hodapp,Reipurth2000}. Whereas the molecular outflow
and the HH~199 flow have a position angle of about 60\deg, \citet{Hodapp} and
\citet{Bally95} found a well-collimated H$_2$ emission flow at a position angle of
about 100\deg. Either IRAS~20582+7724 is a possibly wide binary where each component is
launching a separate flow, or one of the components of a very close binary is precessing
rapidly, giving rise to the two very different flow angles. HH~200 is driven by an embedded
T Tauri star about 1.5~arcmin further to the northwest \citep{Bally95}.
A low-resolution 3.6~cm survey of the L\,1228 cloud by \citet{RR96}
revealed two sources. L\,1228~VLA~1 is associated with the IRAS source, and the other, VLA~2,
has no known counterpart but is located in the direction of the high extinction part of the
L\,1228 core. \citet{Reipurth2004} detected two further 3.6~cm sources, VLA~3 and VLA~4.
VLA~4 is supposed to be the driving source of the HH~200 flow.
The environment of  IRAS~20582+7724 was studied in detail by
\citet{TM97}, \citet{AS04}, and \citet{AS06}.

The K$^{\prime}$ image of L\,1228\,A, presented by \citet{Hodapp}, shows a star associated with a
parabola-shaped nebula, located near the molecular outflow position.
Two other stars further north at offsets  ($-$21\arcsec,72\arcsec) and
(12\arcsec,75\arcsec) are also associated with some
less extended nebulae. The relatively bright stars in this region clearly
stand out against the faint background stars, so that Hodapp classified this region
as a cluster.

(3) {\it L\,1228 South\/} contains a small aggregate of low-mass pre-main sequence stars.
\citet{Padgett} identified 9 infrared sources in the images taken with IRAC on board the
{\it Spitzer Space Telescope\/} (see Table~\ref{Tab_pms}).

\subsubsection{L\,1219 (B\,175)} 
\begin{figure*}[!ht]
\centering
\includegraphics[draft=False,width=7.6cm]{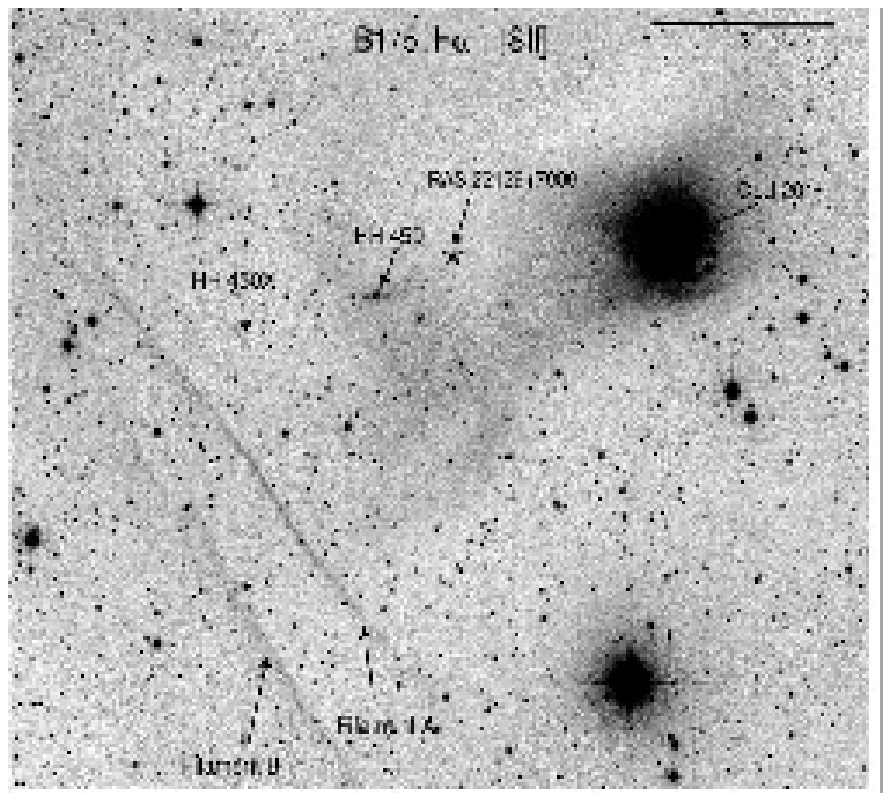}
\includegraphics[draft=False,width=5.6cm]{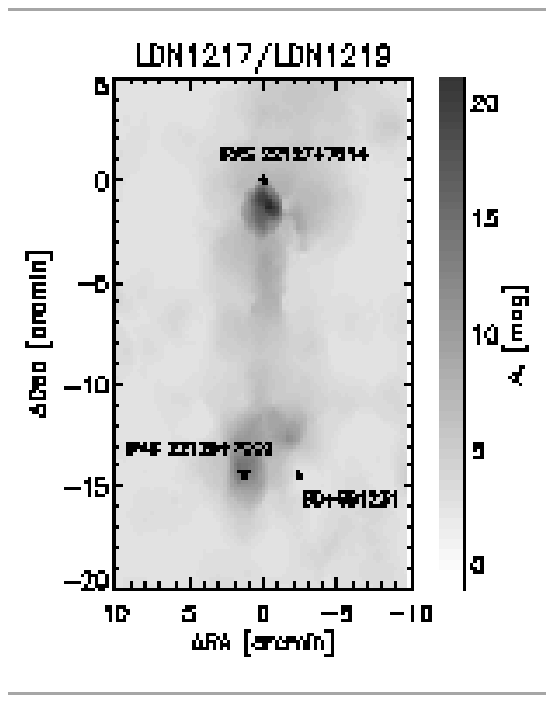}
\caption{Left: An image of B\,175 (L\,1219) from \citet{BR01}. Right: the map of
visual extinction, obtained from the 2MASS data using the NICER algorithm of
\citet{Lombardi}, shows the northern core of the cloud centered on IRAS 22127+7014.}
\label{fig_l1219}
\end{figure*}

is a small cometary shaped cloud at the southernmost edge of the
Cepheus Flare cloud complex. The cloud is illuminated by the B9.5V
type star BD~+69$^\circ$ 1231,
associated with the reflection nebula Ced~201
\citep[see][and references therein]{Cesarsky00}.  Two cold IRAS
sources, 22129+7000 and 22127+7014, are projected within the dark
cloud.  By an imaging and spectroscopic study \citet{BR01} discovered
a Herbig-Haro object, HH~450, emerging from IRAS
22129+7000. Furthermore, they found several parsec-scale filaments of
emission that trace the rim of a new supernova remnant, G\,110.3+11.3,
which appears to be approaching the globule (see
Figure~\ref{fig_l1219}).  At 400~pc, G\,110.3+11.3 is one of the
closest known supernova remnants.  The supernova remnant and the HH
flow appear to be heading toward a frontal collision in about
1000~yr. \citet{NK04} discovered a CO outflow from IRAS 22129+7000.
\citet{Goicoechea} present Spitzer IRAC and MIPS data, 1.2-mm dust
continuum map, as well as observations of several molecular lines for
IRAS 22129+7000.  They detected a collimated molecular outflow in the
CO $J=3-2$ line, whereas the profile of the HCO$^{+}$ $J=1-0$ line
suggested inward motion. Based on the SED they classified the object
as either a transition Class~0/I source or a multiple protostellar
system.  They discuss the role of the photodissociation region
associated with Ced~201 in triggering the star formation in L\,1219.

Three known pre-main sequence stars can be found to the south of L\,1219:
the Herbig~Ae stars BH~Cep and BO~Cep (HBC 734 and 735, respectively)
and a T~Tauri star, associated with IRAS~22129+6949. The head of the cometary globule
with IRAS 22129+7000 is pointing toward south. The embedded source IRAS 21127+7014,
located to the north of IRAS 22129+7000, may be the youngest object associated with this cloud.
The left panel of Fig.~\ref{fig_l1219} shows the head of the globule B\,175 and the supernova
filaments, adopted from \citet{BR01}, and the right panel shows the extinction map of the
whole cloud, revealing another core to the north.

\subsubsection{L\,1221} is a small, isolated cometary dark cloud to the south of the main
body of the Cepheus Flare cloud complex. No distance determination
has been published for this cloud.
A frequently assumed distance is 200~pc \citep[e.g.][]{Fukui,UM91}.
Inside the cloud, \citet{UM91} found an unusual U-shaped CO outflow
associated with a low-luminosity ($2.7 L_{\sun}$) Class~I source, IRAS 22266+6845.
More recent CO observations at high resolution showed that the U-shaped
outflow may actually consist of two bipolar outflows, an east-west outflow associated
with the IRAS source and a north-south outflow about 25\arcsec \ to the east of the
IRAS source, interacting with each other \citep{Lee2002}.
To the south of the IRAS source, a fairly bright
compact object, HH~363, is detected in H$\alpha$ and [S\,II] \citep{Alten97}.
There are three infrared sources within the error ellipse of the IRAS~source:
a close binary consisting of an east source and a west source around the IRAS
source position and another source 45\arcsec \ to the southeast.
 The east source is identified as the IRAS source.
\citet{Furuya03} detected H$_2$O maser emission associated with IRAS~22266+6845.
\citet{Lee05} mapped IRAS~22266+6845 in 3.3~mm continuum, CO, HCO$^+$, and N$_2$H$^+$.
Continuum emission is seen around the east source and the southeast source at 3.3~mm,
probably tracing the dust around them. Assuming a temperature of 40~K, the masses of
the dust plus gas are estimated to be 0.02 and 0.01~M$_{\sun}$ around the east source
and southeast source, respectively. No continuum emission is seen toward the west
source. The east--west outflow is likely powered by the east source, which shows a southeast
extension along the outflow axis in the K$^{\prime}$ \ image \citep{Connelley07}.
\citet{Wu07} detected two submillimeter sources in the cloud, L\,1221~SMM~1 and
L\,1221~SMM~2, apparently coinciding with the binary and the southeast source,
respectively.

\subsubsection{L\,1251} is a cloud elongated east--west at the eastern boundary of the
Cepheus Flare molecular complex. Its cometary shape suggests interaction with the supernova bubble
described by \citet{GLADT}. Recent star formation is indicated by two molecular outflows,
driven by IRAS 22343+7501 and  IRAS 22376+7455, respectively \citep{SF89}.

The distance of L\,1251 was determined by three different methods
(see Table~\ref{Tab_cepflare_dist}). The cloud has been
mapped in several molecular lines, such as $^{13}$CO, C$^{18}$O, H$^{13}$CO$^{+}$,
SiO \citep{SFN94}, NH$_3$ \citep{BM89,TW}, HNC, HCN, HCO$^{+}$, CS \citep*{NJH03}.
\citet{KP93} studied the YSO population and reported on 12 H$\alpha$ emission
stars and IRAS point sources as YSO candidates.
\citet{BEHKK} discovered an optical jet, HH\,149, originating from IRAS~22343+7501.
\citet{RD95} found that this IRAS source  is associated with a cluster of five near-infrared
sources spread over a $10\arcmin\times 10\arcmin$ area (sources A--E). \citet{Meehan} found two
thermal radio continuum sources, VLA~A and VLA~B, coinciding with the near infrared
sources D and A, respectively.
\citet{Beltran01} found 9 radio continuum sources around IRAS~22343+7501,
two of them, VLA~6 and VLA~7 separated by 7\arcsec, are located within the error
ellipse of the IRAS source and identical with \citeauthor{Meehan}'s VLA~A and VLA~B,
respectively. \citeauthor{Beltran01} found a third source, VLA~5 to be
a probable YSO, based on the positive spectral index. \citet{NJH03} concluded that
both VLA~6 and VLA~7 are protostars driving their own outflow.

The high resolution VLA observations by \citet{Reipurth2004} revealed
four radio continuum sources in the region around IRAS~22343+7501,
three of which were known from previous studies. The high resolution VLA~A
map has revealed a new source, VLA~10, close to VLA~6, with which it was blended in the
earlier low-resolution data of \citet{Meehan}. The designations VLA~10
and 11 is a continuation of the numbering scheme of \citet{Beltran01}.
\citet{Meehan} suggest that VLA~6 corresponds to the very red and embedded source
IRS~D, while VLA~7 is the brighter source IRS~A.

\citet{Eiroa94} discovered a chain of Herbig--Haro objects, HH\,189A,B,C near
IRAS~22376+7455 (L\,1251\,B).  The Spitzer Space Telescope observed
L\,1251\,B as part of the Legacy Program {\em From Molecular Cores to Planet Forming
Disks\/} \citep{Evans} at wavelengths from 3.6 to 70~\micron. The observations revealed a small cluster
of protostars, consisting of 5 Class~0/I and 14 Class~II objects \citep{Lee06}. Three
Class~0/I objects are projected on IRAS~22376+7455, the most luminous is located 5\arcsec \ north
of the IRAS position. Thus the molecular outflow observed from IRAS~22376+7455 \citep{SF89}
is probably a combined effect of more outflows. \citet{Lee07} studied the complex
motions in the region, based on both single-dish and interferometric molecular line
observations. The data have shown very complex kinematics including infall, rotation, and
outflow motions. The well-known outflow, associated with L\,1251\,B, was resolved into
a few narrow and compact components.
They detected infall signatures in the shape of HCO$^{+}$, CS, and HCN
lines to the east of L\,1251\,B, where no infrared object has been detected, and an
extended emission has been found at 850~\micron. This result shows that, in addition
to the Class~0--Class~II objects, the young cluster contains a pre-protostellar
core as well.
Results of spectroscopic follow-up observations of the optically visible candidate YSOs
reported by \citet{KP93} are given in \citet{EK03}.
\citet{Simon} detected 41 X-ray sources in the image obtained with the XMM-Newton
telescope. The list of X-ray sources contains both outflow sources and 8 optically
visible T~Tauri stars.

The structure of the cloud and the properties of its dust grains were studied,
based on optical extinction maps, by \citet{Kandori} and \citet{BAKTK}.

\citet{Young06} and \citet{Wu07} observed submm sources associated
with the dense cores L\,1251\,A, B, and C.

\subsubsection{L\,1261/L\,1262 (CB\,244)} are small clouds to the east of the main body of
the Cepheus Flare cloud complex, at a probable distance of 180\,pc from the Sun \citep{Kun98}.
Two young, low luminosity objects, the G2 type classical T~Tauri star HBC\,741
(AS\,507) and the cold IRAS source IRAS~23238+7401 are projected on the cloud.
A CO outflow centered on IRAS~23238+7401 was found by \citet{Parker88}.
\citeauthor{WLH03}'s \citeyearpar{WLH03} submillimeter polarization measurements
resulted in a magnetic field strength of  257\,$\mu$G for the envelope of CB\,244.

\subsection{NGC 7129}

\begin{figure*}[!ht]
\centerline{
\includegraphics[draft=False,width=5.25in]{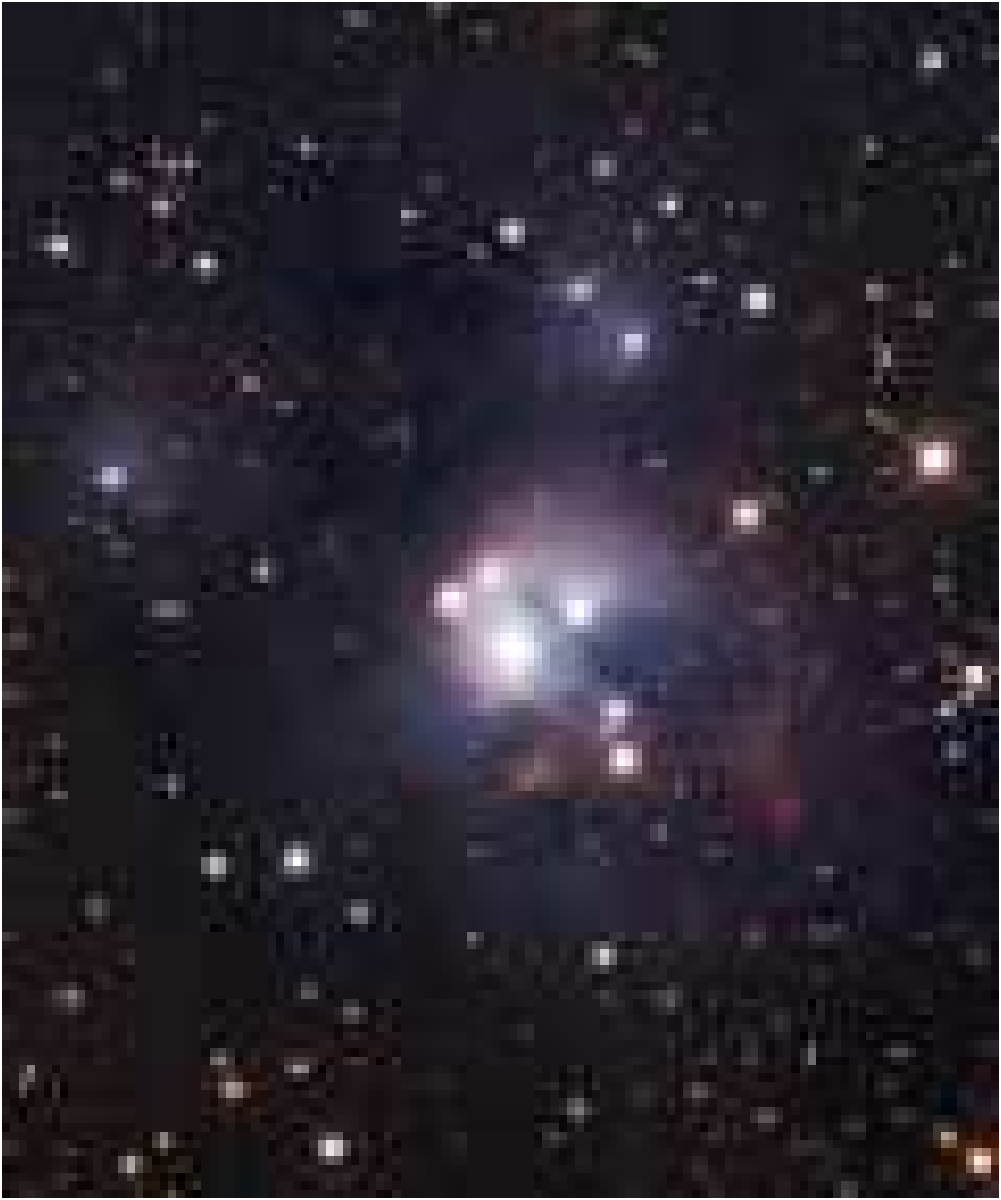}
}
\caption{An optical image of NGC~7129, displaying the young cluster, embedded in
a reflection nebula, as well as several HH objects. The size of the area is
about $15\arcmin \times 18\arcmin$. Photograph by Robert Gendler.}
\label{n7129_image}
\end{figure*}

\begin{figure*}[!ht]
\centerline{
\includegraphics[draft=False,width=5.25in]{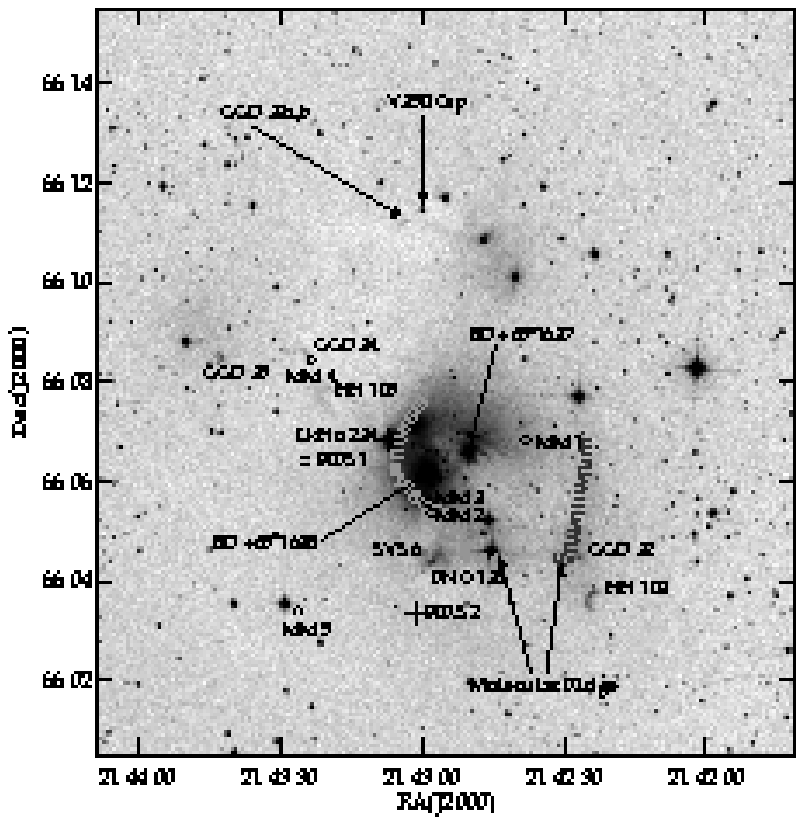}
}
\caption{A DSS\,2 red image of NGC\,7129 in which the crude position of the molecular ridge,
bordering the cluster, is overplotted by white cross-dashed lines. The B-type cluster members,
HH objects, and some interesting low-mass young stars are labeled.
Crosses indicate the far-infrared sources FIRS~1 and FIRS~2, and diamonds stand for the compact
millimeter sources detected by \citet{Fuente01} outside the FIRS~1 and FIRS~2 regions.}
\label{map_n7129}
\end{figure*}

Though NGC\,7129 lies in the Cepheus Flare, it is more distant than the clouds
discussed above. NGC\,7129 (Ced~196) is a reflection nebula in the region of a young cluster,
containing three B-type stars, namely BD\,+65\deg1637, BD\,+65\deg1638, and LkH$\alpha$\,234,
as well as several low-mass pre-main sequence stars
\citep*[e.g.][see Table~\ref{Tab_pms}]{Herbig60,SVS76,Cohen83,MMN04}. Whereas BD\,+65\deg1638
is regarded a young main sequence star (but see \citeauthor{MPR} \citeyear{MPR}),  BD\,+65\deg1637 and
LkH$\alpha$\,234 are  pre-main sequence stars \citep{Herbig60,Hernan04}.
An optical image of NGC\,7129, displaying several spectacular signposts of
the interactions between the young stars and their environments, is shown in
Fig.~\ref{n7129_image}. A finding chart for the most prominent objects,
related to star formation, is displayed in Fig.~\ref{map_n7129}.

\subsubsection{LkH$\alpha$~234 and its environments}
 It has been suggested that
LkH$\alpha$~234 is the youngest among the three B-type stars \citep{Hillen92}.
This star and its environment have been studied
extensively at optical, infrared, radio, centimeter to submillimeter wavelengths.
Photometric and spectroscopic variability of LkH$\alpha$\,234
have been studied by \citet{Shev91} and \citet{Chakra04}, respectively.

\citet{Wilking86} presented high-resolution continuum and molecular-line observations of
the circumstellar environment of LkH$\alpha$\,234.
\citet{Sandell81} and \citet{Tofani95} detected
three H$_2$O maser sources from the environment of LkH$\alpha$\,234.
\citet{Ray90} identified an optical jet originating from this region. \citet{Mitchell94} detected a
molecular jet associated with LkH$\alpha$\,234, and  \citet{Schultz95} observed
shocked molecular hydrogen in the LkH$\alpha$\,234 region.
VLA observations by \citet{Trinidad04} of water masers and radio continuum emission at 1.3 and 3.6~cm
show that the LkH$\alpha$\,234 region contains a cluster of YSOs. In a field of $\sim 5\arcsec$
they detected five radio continuum sources (VLA~1, VLA~2, VLA~3A, VLA~3B, and LkH$\alpha$\,234) and
21 water maser spots. These water masers are mainly distributed in three clusters associated
with VLA~1, VLA~2, and VLA~3B. The VLA observations suggest that there are at least four
independent, nearly parallel  outflows in the LkH$\alpha$\,234 region.
Probably all sources observed in this region ($\sim 5\arcsec$ in diameter) form a cluster of YSOs,
which were born inside the same core in the NGC~7129 molecular cloud.
This fact could explain that the major axes of the outflows have nearly the same orientation.
\citet{Marvel05} performed VLBI observations of maser sources around LkH$\alpha$~234, and
detected maser emission associated with LkH$\alpha$~234--VLA~2 and LkH$\alpha$~234--VLA~3b.
No maser source associated with LkH$\alpha$~234 itself has been detected.

\citet{Tommasi} obtained far-infrared spectra of the LkH$\alpha$\,234 region, using the
Long Wavelength Spectrograph of ISO. The observed spectra are consistent
with a photodissociation region, associated with not LkH$\alpha$\,234, but with
BD\,+65\deg1637.
\citet{Morris04}  have obtained mid-IR spectroscopy of regions around
LkH$\alpha$~234, with the Spitzer Space Telescope Infrared Spectrograph (IRS). They detected
warm material at 16~\micron \ around BD\,+65\deg1638 which clearly shows that this region
is not free of gas and dust. \citet{Wang07} identified a group of low-mass young stars
around LkH$\alpha$~234 using the {\it Spitzer\/} data base.

\subsubsection{Interstellar matter associated with NGC 7129}
Molecular line observations of the region  \citep*{Bechis,FMS01,Miskolczi,Ridg03}
revealed a kidney-shaped molecular cloud of about 11~pc in extent to the east and south of
the cluster.
BD\,+65\deg1637 and most of
the fainter cluster members are found in a cavity of the cloud, bordered by a
prominent molecular ridge, while LkH$\alpha$\,234, located  to the east of
the main cluster, is associated with a peak of $^{13}$CO emission.
The optical jet detected by \citet{Ray90} is pointing southwest into the cavity.
\citet{Torrelles83} and \citet{Gusten86} presented ammonia observations of NGC\,7129.

\citet{MPR} observed the region in the 21~cm line of H\,I
with an angular resolution of 1\arcmin.  The observations revealed a ring of H\,I emission
about 30\arcmin \ in extent.  The H\,I ring appears to be part of the surface of a molecular cloud
and is centered on a relatively dense concentration of H\,I with
unusually wide line profiles and positionally coincident with BD\,+65\deg1638.
An infrared point source, IRAS 21418+6552, coincides within the
positional errors with the H\,I knot.

A continuum source coincident with BD\,+65\deg1638 has also been detected at 1420 MHz,
which shows a significant extension to the northeast overlapping the position of LkH$\alpha$~234.
Comparing the radio continuum data with other radio observations of BD\,+65\deg1638,
\citet{MPR} found that BD\,+65\deg1638 has a flat centimeter-wave spectrum, consistent
with an optically thin H\,II region around the star.
The authors conclude that the physical association of the star with
the H\,I knot indicates that BD\,+65\deg1638 belongs to
a rare class of ``dissociating stars'', having an extremely young age of not more than
a few thousand years. BD\,+65\deg1638 itself is found to be a 6 M$_{\sun}$ star
that has just emerged from  its cocoon and lies on the birthline.

\subsubsection{Embedded young stellar objects in NGC\,7129\/}
\citet{Bechis} identified two far-infrared sources in NGC\,7129: FIRS\,1 coincides with
LkH$\alpha$~234, while FIRS\,2, a  deeply embedded protostellar object is located
three arcminutes to the south of the cluster, at the primary peak of
the $^{13}$CO emission. It was found to coincide with an H$_2$O maser
\citep{Cesarsky78,Rodriguez80b,Sandell81}.
NGC\,7129 FIRS~2 has not been detected in the optical and near infrared. From the
far-infrared and submillimeter observations, \citet*{Eiroa98} derived a luminosity of
430~$L_{\sun}$, a dust temperature of 35~K, and a mass of 6~$M_{\sun}$. The low dust
temperature and the low $L_{\rm bol}/L_{\rm 1.3~mm}$ ratio of this source suggest
that it is an intermediate mass counterpart of Class~0 sources.
\citet{Fuente05a} detected a hot molecular core associated with FIRS\,2, and
\citet{Fuente05b} carried out a molecular survey of FIRS\,2 and LkH$\alpha$\,234 with
the aim of studying the chemical evolution of the envelopes
of intermediate-mass young stellar objects.
\citet{Fuente08}  present high angular resolution imaging of the hot core of NGC\,7129 FIRS~2,
using the Plateau de Bure Interferometer. This is the first chemical study of an
intermediate-mass hot core and provides important hints to understand the
dependence of the hot core chemistry on the stellar luminosity.

Two molecular outflows were found in NGC\,7129 by
\citet{ES83}. They seem to be associated with LkH$\alpha$\,234 and FIRS\,2.
\citet{Wein94}, via near infrared polarimetry, identified a deeply embedded source,
NGC\,7129~PS\,1, located about  3\arcsec \ northwest of LkH$\alpha$\,234.
This source was not identified in the direct near-infrared images of LkH$\alpha$\,234,
which revealed 5 sources (IRS\,1--IRS\,5, IRS\,1 $\equiv$ LkH$\alpha$\,234) in a
$20\arcsec \times 20\arcsec$ field centered on  LkH$\alpha$\,234.
\citet{Wein96} detected  NGC\,7129~PS\,1 at 3.8~\micron, and proposed that it was the
actual outflow source instead of LkH$\alpha$\,234.

\citet{Cabrit97}  present high-resolution imaging of the region around LkH$\alpha$\,234 in the
10~\micron \ and 17~\micron \  atmospheric windows and in the H$_2$  v=1-0 S(1) line
and adjacent continuum. The cold mid-infrared companion, detected at 2.7\arcsec\ to the
north-west of the optical star, corresponds to  NGC\,7129~PS\,1. The companion illuminates
an arc-shaped reflection nebula with very red colors, and is associated with a
radio continuum source, H$_2$O masers, and a bright extended H$_2$ emission knot,
indicating that it is deeply embedded and has strong outflow activity. \citet{Cabrit97}
refer to this star as IRS\,6,  extending \citeauthor{Wein94}'s notation.

\citet{Fuente01} obtained single-dish and
interferometric continuum images at 2.6~mm and 1.3~mm of both FIRS~2 and LkH$\alpha$\,234.
They identified two millimeter sources associated with FIRS~2: FIRS~2--MM1, apparently associated
with the CO outflow, and a weaker source, FIRS~2--MM2, which does not present any sign of stellar activity.
The interferometric 1.3~mm continuum image of FIRS~1 reveals that LkH$\alpha$234 is a
member of a cluster of embedded objects. Two millimeter clumps are associated with this far-infrared
source. The stronger is spatially coincident with IRS\,6. A new millimeter clump,
FIRS~1--MM1, is detected at an offset ($-$3.23\arcsec, 3.0\arcsec)
from LkH$\alpha$\,234. The extremely young object FIRS~1--MM1
(it has not been detected in the near- and mid-infrared) is the likely driving source
of the H$_2$ jet. There is no evidence for the existence of a bipolar outflow associated
with LkH$\alpha$\,234. In addition to FIRS~1 and FIRS~2,
six other compact millimeter clumps are detected in the region, NGC\,7129~MM1 to MM5
(see Fig.~\ref{map_n7129}),  and the sixth coincides with the bipolar nebula RNO~138.

\begin{figure*}[!ht]
\centerline{
\includegraphics[draft=False,width=8cm]{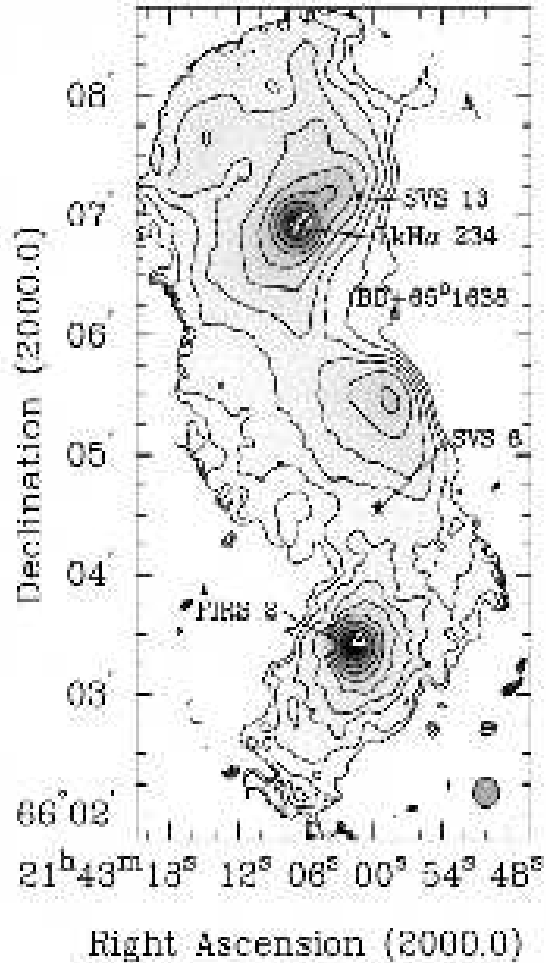}
}
\caption{Dust continuum emission of NGC\,7129, observed at 850\,\micron \ by \citet{FMS01}.
Near-infrared sources are marked by star symbols, and H$_2$O maser sources by triangles.}
\label{fig_ngc7129_SST}
\end{figure*}

Submillimeter continuum observations by \citet{FMS01} revealed three
compact sources: LkH$\alpha$\,234~SMM\,1, LkH$\alpha$\,234~SMM\,2 and
FIRS\,2.  SMM\,1 coincides with IRS\,6, which, according to the
submillimeter observations, may be a deeply embedded Herbig~Be star,
whereas SMM\,2 is a newly discovered source (see Figure~\ref{fig_ngc7129_SST}).
Table~\ref{protostar_NGC7129} shows the coordinates, wavelengths of
detection, measured fluxes and sizes of the deeply embedded young
stellar objects in NGC\,7129, observed in submillimeter, millimeter,
and centimeter continuum.

\begin{table}[!t]
\caption{Embedded YSOs in NGC\,7129.}
\label{protostar_NGC7129}
\smallskip
\begin{center}
{\footnotesize
\begin{tabular}{ l c c c r r c }
\tableline
\noalign{\smallskip}
Name          &  RA(2000) & Dec(2000)   & $\lambda$(mm) & Flux(mJy) & Ref. \\
\noalign{\smallskip}
\tableline
\noalign{\smallskip}
NGC 7129 FIRS1 IRS6   &	21 43 06.4 &	66 06 55.6 & 2.6 &     91  &   1    \\
NGC 7129 FIRS1 IRS6   &	21 43 06.5 &	66 06 55.2 & 1.3 &    313  &   1  \\
NGC 7129 FIRS1 MM\,1  &	21 43 06.3 &	66 06 57.4 & 1.3 &    180  &  1     \\
LkH$\alpha$ 234	&	21 43 06.8 &	66 06 54.4 & 1.3 &  $<$ 20 &  1 \\
NGC 7129 FIRS2 MM\,1 &	21 43 01.7 &	66 03 23.6 & 2.6 &     72  &   1    \\
  	 	&	21 43 01.7 &	66 03 23.6 & 1.3 &     381 &  1   \\
NGC 7129 FIRS2 MM\,2 &	21 43 01.6 &	66 03 26.1 & 2.6 &     22  & 1      \\
  	 	&	21 43 01.7 &	66 03 24.7 & 1.3 &     137 & 1       \\
NGC 7129 FIRS2 IR &	21 43 01.8 &	66 03 27.4 & 2.6 &     $<$15 &   1 \\
  	 	&	21 43 01.8 &	66 03 27.4 & 1.3 &     $<$19 &   1 \\
NGC 7129 FIRS 1	&	21 43 06.5 &	66 06 52.3 & 1.3 &	 690 &	1 \\
  	       	&	21 43 06.5 &	66 06 52.3 & 1.3 &	4283 & 1	 \\
NGC 7129 FIRS 2 &	21 43 01.4 &	66 03 22.3 & 1.3 &	597  & 1   \\
  	        &	21 43 01.4 &	66 03 22.3 & 1.3 &	2613 & 1	  \\
NGC 7129 MM\,1 	&	21 42 57.9 &	66 05 21.4 & 1.3 & 	96 &	1 \\
NGC 7129 MM\,2 	&	21 42 58.7 &	66 05 34.5 & 1.3 & 	89 &	1 \\
NGC 7129 MM\,3 	&	21 42 38.1 &	66 06 50.2 & 1.3 & 	62  &	1\\
NGC 7129 MM\,4 	&	21 43 23.7 &	66 08 29.0 & 1.3 & 	60 &	1\\
NGC 7129 MM\,5 	&	21 43 26.3 &	66 03 24.7 & 1.3 & 	29 &	1\\
RNO 138 	&	21 42 57.6  &   66 04 26   & 1.3 & 	22  &	1\\
LkH$\alpha$ 234 SMM 1 &	21 43 06.76 &	66 06 56.0 & 0.85 & 3120  & 2 \\
LkH$\alpha$ 234 SMM 1 &	21 43 06.76 &	66 06 56.0 & 0.45 & 20700 & 2 \\
LkH$\alpha$ 234 SMM 2 &	21 43 03.20 &	66 07 13.1 & 0.85 & 730  & 2 \\
LkH$\alpha$ 234 SMM 2 &	21 43 03.20 &	66 07 13.1 & 0.45 & 6200   & 2 \\
NGC 7129 FIRS 2 & 21 43 01.51 &	66 03 24.2 & 0.85 & 3350  & 2  \\
NGC 7129 FIRS 2 & 21 43 01.51 &	66 03 24.2 & 0.45 & 18100 &  2 \\
NGC 7129 VLA 1          & 21 43 06.093  & 66 06 58.13 & 36.0 & & 3 \\
NGC 7129 VLA 1          &               &             & 13.0 & & 3 \\
NGC 7129 VLA 2          & 21 43 06.321  & 66 06 55.95 & 36.0 & 0.1  & 3 \\
NGC 7129 VLA 2          &               &             & 13.0 & $<$0.33 &  3 \\
NGC 7129 VLA 3A         & 21 43 06.479  & 66 06 55.02 & 36.0 & 0.67 & 3  \\
NGC 7129 VLA 3A         &               &             & 13.0 & 1.96 & 3 \\
NGC 7129 VLA 3B         & 21 43 06.462  & 66 06 55.22 &  36.0 & 0.61  & 3 \\
NGC 7129 VLA 3B         & & & 13.0 & 1.19 &  3 \\
\noalign{\smallskip}
\tableline
\end{tabular}}
\end{center}
\smallskip
{\footnotesize References: 1: \citet{Fuente01}; 2: \citet{FMS01}; 3: \citet{Trinidad04}}
\end{table}

\subsubsection{Low-mass pre-main sequence members of NGC~7129\/} were identified as H$\alpha$
emission objects by \citet{HL85}, \citet{MEG}, and \citet{MMN04},
as variable stars \citep{Semkov} and near-infrared sources \citep{SVS76,Cohen83}.
\citet{Muzerolle} presented observations of NGC\,7129 taken with the
{\it Multiband Imaging Photometer for Spitzer\/} (MIPS). A significant population
of sources, likely pre-main sequence members of the young stellar cluster,
have been revealed outside the central photoionization region. Combining
{\it Infrared Array Camera\/} (IRAC) and ground-based near-infrared images,
\citet{Gutermuth} obtained colors and spectral energy distributions for some 60 objects.
Most of the pre-main sequence candidates are associated with the densest part of the
molecular cloud, indicating active star formation over a broad (some 3~pc) area
outside the central cluster.

\begin{figure*}[!ht]
\centerline{
\includegraphics[draft=False,width=12cm]{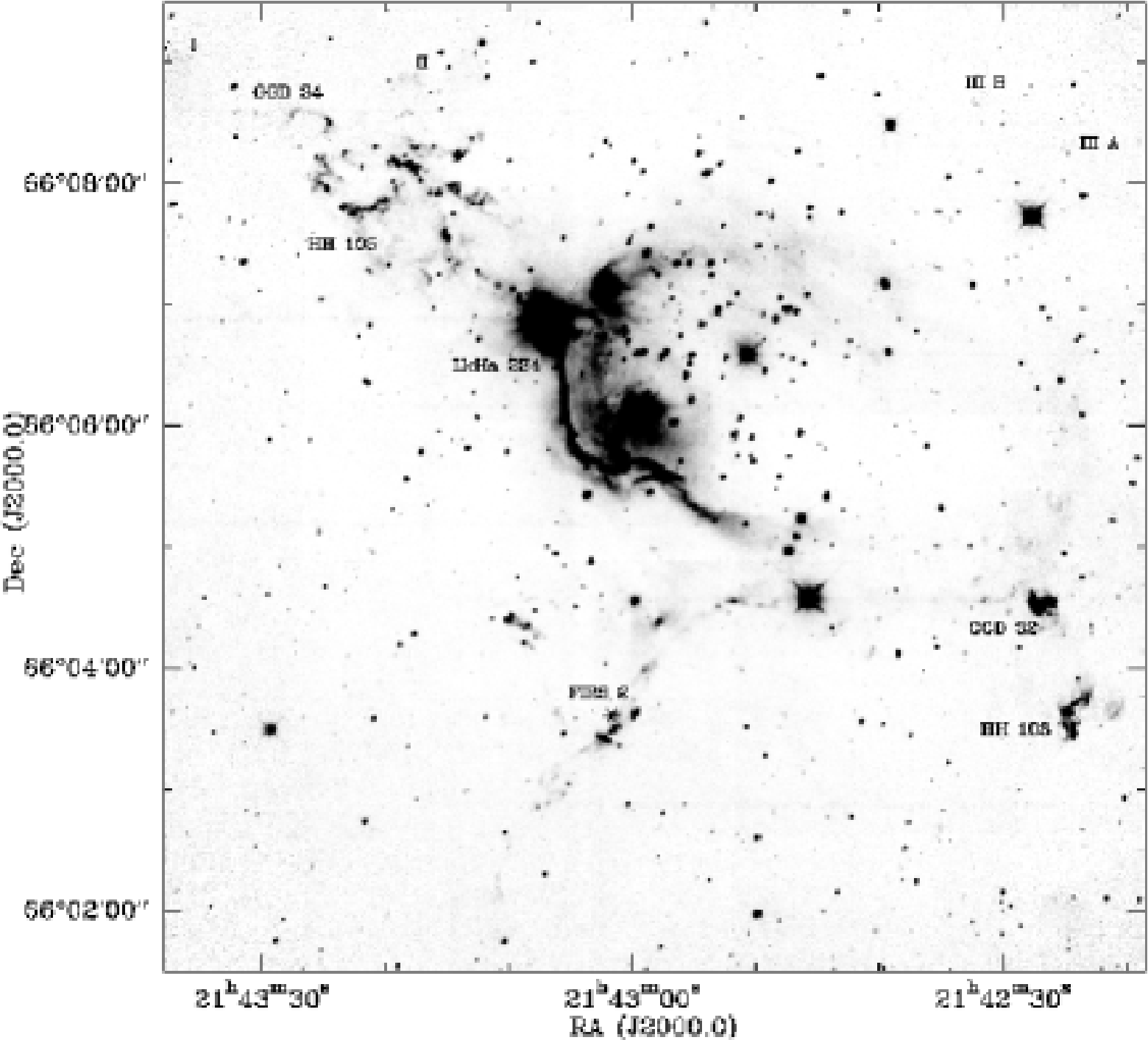}
}
\caption{Mosaic of the NGC\,7129 region in the 1-0 S(1) line of H$_2$ + cont. at 2.12~\micron,
adopted from Eisl\"offel (2000). Emission from several molecular outflows,
as well as from a probable photodissociation region to the east and south of
the stellar cluster can be seen. Some H$_2$ outflows and H$_2$ counterparts of
optical Herbig--Haro objects are labeled.}
\label{fig_ngc7129_HH}
\end{figure*}

A remarkable object is the small bipolar nebula RNO~138, located at the southern edge of
NGC\,7129. Several authors \citep{Cohen83,Draper84} suggest that the T~Tauri star SVS~6,
located 15\arcsec \ NE of RNO~138, is the illuminating source.  However, \citet{Miranda}
report that a star within the nebulosity, RNO~138\,S, appeared as an optically visible
object in 1993. It brightened by $\sim$~1.2~mag between 1988 and 1993. The Li\,I absorption
characteristic of young low-mass stars and P~Cygni profile of the H$\alpha$ line
were recognized in the optical spectrum of RNO~138\,S. \citet{Miranda} suggest that
RNO~138\,S, probably a FUor, is the illuminating source of RNO~138.
Another interesting star is V350~Cep  \citep[denoted as IRS\,1 by][]{Cohen83},
which brightened about 4 mag in the 1970's
\citep{Semkov04}, and has been staying at the high level since then. \citet{Liseau83}
detected CO outflows associated with both RNO~138 and V350~Cep. \citet{Herbig08} has shown
that V350~Cep does not belong to the class of the EXor-type eruptive young stars.
The list of low-mass pre-main sequence stars in
NGC\,7129, based on a literature search, is given in Table~\ref{Tab_pms}.

\subsubsection{Herbig--Haro objects} In the optical, a large number of HH-objects in
and near NGC 7129 have been reported
\citep*[][see Table~\ref{HH_NGC7129}]{GGD,HL85,EGM92,GME93,Miranda,GR99}.
Molecular hydrogen emission in the near-infrared has been detected
from several of the optical Herbig--Haro objects \citep{Wilking90}.
An infrared search for the exciting sources of the optical HH~objects was presented
by \citet{Cohen83}. Spectroscopic observations of HH objects are presented by
\citet{CF85}. \citet{Fuente01} suggest that  NGC\,7129~MM4  is the illuminating
star of the nebular object GGD~34.
\citet{Eisloffel2000} used deep imaging in the near-infrared 1-0 S(1) line of H$_2$
at 2.12~\micron \  to search for parsec-scale outflows in NGC\,7129 (Fig.~\ref{fig_ngc7129_HH}).
They identified numerous outflows (see Table~\ref{HH_NGC7129}),
but likely driving sources could be identified for only three of them. For most of
the other emission-line knots and molecular flows no evident sources could be identified.
The Spitzer observations revealed several distinct outflow arcs, traced by
4.5~\micron \  bright knots, associated with FIRS\,2 \citep{Muzerolle}.
The multipolar nature of this outflow system, in general agreement with the outflow
analysis of \citet{Fuente01}, supports the claim by
\citet{Miskolczi} that FIRS\,2 is a multiple protostellar system.

\begin{table}[!ht]
\caption{Herbig--Haro objects in NGC\,7129.}
\label{HH_NGC7129}
\smallskip
\begin{center}
{\footnotesize
\begin{tabular}{ l c c l l }
\tableline
\noalign{\smallskip}
name          &  RA(2000) & Dec(2000)   & Source         & Reference \\
\noalign{\smallskip}
\tableline
\noalign{\smallskip}
HH 822        & 21 41 42.1 & +66 01 45 & LkH$\alpha$\,234        & 5 \\
HH 103A       & 21 42 23.8 & +66 03 47 & LkH$\alpha$\,234        & 5,13 \\
HH 103B       & 21 42 24.7 & +66 03 40 & LkH$\alpha$\,234  & 1,13 \\
HH 232, GGD 32 & 21 42 26.9 & +66 04 27 &                 & 6,8 \\
HH 242        & 21 42 38.7 & +66 06 36 &                   & 6,10,12 \\
HH 825        & 21 42 39.2 & +66 10 56 & IRAS 21416+6556 & 5 \\
HH 238        & 21 42 40.3 & +66 05 41 & LkH$\alpha$\,234 & 6,10 \\
HH 237        & 21 42 42.3 & +66 05 23 & LkH$\alpha$\,234  & 6,10 \\
HH 239        & 21 42 44.1 & +66 06 37 & LkH$\alpha$\,234  & 6,10 \\
HH 824        & 21 42 56.9 & +66 09 10 & IRAS 21416+6556 & 4,10 \\
HH 236        & 21 42 59.2 & +66 07 39 & LkH$\alpha$\,234  & 6,10 \\
HH 233, GGD 33 & 21 43 00.0 & +66 12 00 & GGD\,33a        & 1,3,13 \\
HH 167        & 21 43 06.7 & +66 06 54 & LkH$\alpha$\,234   & 4 \\
HH 105B       & 21 43 19.8 & +66 07 53 & LkH$\alpha$\,234  & 1,6,9 \\
HH 105A       & 21 43 22.1 & +66 07 47 & LkH$\alpha$\,234  & 1,6,9 \\
HH 823        & 21 43 27.9 & +66 11 46 &                 & 4 \\
HH 234        & 21 43 29.7 & +66 08 38 &                 & 6 \\
GGD 34        & 21 43 30.4 & +66 03 43 & NGC\,7129~MM\,4 & 1,2,7,11 \\
HH 235, GGD 35 & 21 43 42.0 & +66 09 00 &                 & 6,8,13 \\
HH 821        & 21 43 43.4 & +66 08 47 & LkH$\alpha$\,234   & 5 \\
HH 820        & 21 43 47.9 & +66 09 50 & LkH$\alpha$\,234   & 5 \\
HH 818        & 21 43 57.7 & +66 10 26 & LkH$\alpha$\,234   & 5 \\
HH 819        & 21 44 01.0 & +66 09 52 & LkH$\alpha$\,234   & 5 \\
HH 817        & 21 44 13.3 & +66 10 55 & LkH$\alpha$\,234   & 5 \\
HH 816        & 21 44 26.4 & +66 10 58 & LkH$\alpha$\,234    & 5 \\
HH 815        & 21 44 29.9 & +66 13 42 & LkH$\alpha$\,234  & 5 \\
\noalign{\smallskip}
\tableline
\end{tabular}}
\end{center}
\smallskip
{\footnotesize References: 1 -- \citet*{EGM92};  2 -- \citet{GR99};
 3 -- \citet*{Miranda};  4 -- \citet{Moreno95};  5 -- \citet{MRB04};
 6 -- \citet{Wu2002}; 7 -- \citet{GME93}; 8 -- \citet{Cohen83}; 9 -- \citet{HL85}; 10 -- \citet{MEG};
 11 -- \citet{Fuente01}; 12 -- \citet{Avila01}; 13 -- \citet{CF85}.}

\end{table}
\section{Star Formation in the Association Cep OB2}

\begin{figure*}[!ht]
\centerline{
\includegraphics[draft=False,width=5.25in]{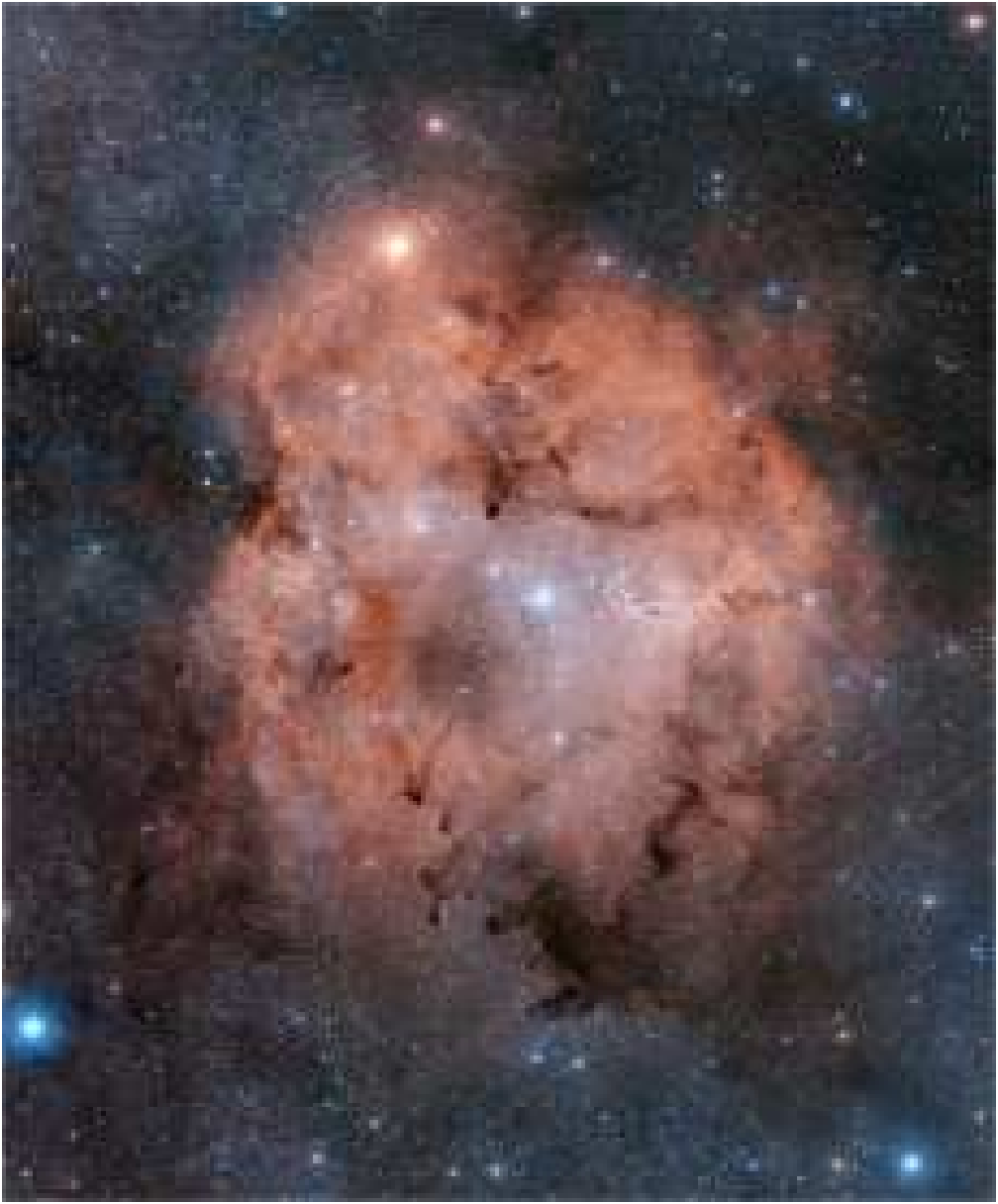}
}
\caption{Optical image of the HII region IC\,1396. Color composite
by Davide De Martin.}
\label{Fig_IC1396}
\end{figure*}

The association Cep~OB2 was discovered by \citet{Ambartsumian}. \citet{Simonson}
identified 75 bright members, including the runaway O6\,Iab star
$\lambda$\,Cep (HIP~109556). The clusters NGC\,7160 and Trumpler~37
(Tr~37), with its associated H\,II region IC\,1396 (Fig.~\ref{Fig_IC1396}),
have similar distances as Cep\,OB2, about 800~pc. \citet{SVSG}  suggested the division
of Cep\,OB2 into two subgroups. The younger subgroup, Cep~OB2b is Tr\,37, one of the youngest known
open clusters, with an age of 3.7~Myr \citep*[e.g.,][]{MKC}.
\citet{GK} suggested that the bright star $\mu$~Cep
(HIP~107259, M2\,Ia) is a member of Tr\,37. The main source of excitation of
IC\,1396 is the O6 star HD\,206267 (HIP~106886), a Trapezium-like system \citep{Harvin04}.
IC\,1396 and neighboring areas contain a large number of H$\alpha$ emission objects
 \citep[e.g.][]{Kun86,KP90,Balazs96}, classical T~Tauri stars \citep{SA1,SA2} and several
globules containing embedded young stellar objects  \citep*[e.g.,][]{Duvert,SWG}.
The other subgroup, Cep~OB2a, contains a large number of evolved massive stars that
are spread over a large area, between  $100\deg \leq l \leq 106\deg$ and
$+2\deg \leq b \leq +8\deg$. The age of this subgroup is about 7-8~Myr,
and contains NGC\,7160. This subgroup is surrounded
by a 9\deg \ diameter infrared emission ring, the Cepheus Bubble (see Fig.~\ref{Fig_DUK}),
which possibly resulted from a supernova explosion \citep{KBT}. This supernova might
have triggered star formation in the ring, as suggested by the presence
of several H\,II regions, and a number of infrared sources that have the
characteristics of embedded young stellar objects \citep{BK89}.

\Citet{deZeeuw} determined the distance of Cep\,OB2, based on Hipparcos results
on 76 members (one O, 56 B, 10 A, five F, one G, two K, and
one M type stars). They obtained a mean distance of 615$\pm$35 pc.

\citet{Daflon99} studied the chemical abundances of OB stars in Cep~OB2. Their
results indicate that this association is slightly metal-poor.
The highest mass association members have been involved in several studies of the
interstellar matter in the region of the association. \citet{Clayton87} detected
anomalous dust in the region of Tr~37, by studying the far-ultraviolet extinction
curves of 17 early-type stars. In order to measure the value of  $R_V = A_V / E_{B-V}$,
\citet{Roth88} combined near-infrared photometry of OB stars with existing optical
and ultraviolet data of the same sample. The results suggest
anomalous grain size distribution with respect to an average Galactic extinction curve.
\citet{Morbidelli97} performed {\it VRIJHK\/} photometry of 14 bright cluster members.
They obtained the normal Galactic value of  $R_V =3.1$.

In order to establish the velocity structure and column density of the interstellar matter
in the region of Cep~OB2, \citeauthor{Pan04} \citeyearpar{Pan04,Pan05} studied several
interstellar absorption lines in high resolution spectra of early-type association members.
Their results are consistent with the large-scale structure
suggested by radio molecular maps, and suggest significant variations in the
column density on small ($\sim 10000$~AU) scales.

Cep~OB2 was mapped in the 21~cm line of  HI by \citet{SVSG}. They detected an HI
concentration of $2\times10^{4}$~M$_{\sun}$ surrounding the HII region and bright-rimmed
dark clouds associated with Cep~OB2b, and found no significant amount of neutral hydrogen
associated with Cep~OB2a.
\citet{Wendker80} constructed  a three-dimensional model of the ionized region based on
a 2695~MHz radio continuum map.
The first molecular study of IC\,1396 was performed by \citet*{Loren75}.
\citet{Heske85} surveyed the dark clouds of IC\,1396 in the
6~cm absorption line of the H$_2$CO. Large-scale $^{12}$CO and $^{13}$CO mapping
of the dark clouds in IC\,1396 are presented by \citet{Patel95} and \citet{Weikard}.
Both IC\,1396 and NGC\,7160 have been included in the CO survey of regions around 34
open clusters performed by \citet*{Leisawitz}.

The low and intermediate mass members of Cep\,OB2 belong to different populations,
born in various subgroups during the lifetime of the association. Significant
low mass populations, born together with and thus nearly coeval with the high luminosity
members, are expected in both Cep~OB2b and Cep~OB2a. Younger subgroups
are being born in the clouds bordering the HII region IC\,1396, due to  triggering
effects by the luminous stars of Cep~OB2b.

\subsection{Pre-main Sequence Stars in the Open Cluster Tr\,37}

\citet{MVA} derived kinematic membership probabilities for stars of Tr\,37, and
identified 427 probable members. \citet{MKC} performed $UBV(RI)_C$ photometry of 120
members, most of them brighter than $V \approx 13.5$. Their HR diagram indicated a
number of probable pre-main sequence members.

During a search for intermediate mass pre-main sequence members of Tr\,37,
\citet{Contreras} found three emission-line stars (MVA~426, MVA~437, and
Kun~314S). They suggest that the low frequency of emission-line activity
in the sample of B--A type stars indicates that inner disks around intermediate-mass
stars evolve faster than those of low-mass stars.

\citet{SA1}, based on spectroscopic and photometric observations
of candidate objects, presented the first identifications of low-mass (spectral types K--M)
pre-main sequence members of Tr\,37 and NGC\,7160. They expanded the studies
in a second paper \citep{SA2}. In all, they identified and studied 130 members
of Tr\,37, and $\sim$30 for NGC\,7160.
They confirmed previous age estimates of 4~Myr for Tr\,37 and 10~Myr for NGC\,7160, and
found active accretion in $\sim$40\% of the stars in Tr\,37, with average accretion
rates of $10^{-8}\,M_{\sun}$yr$^{-1}$, derived from their U-band excesses. These results
expand the existing samples of accreting stars. Only 1 accreting star was detected
in the older cluster NGC\,7160, suggesting that disk accretion ends before the age of
10~Myr.

In order to follow the evolution of protoplanetary accretion disks through
the ages $\sim$3--10~Myr, \citet{SA3} utilized the wavelength
range 3.6--24~\micron, offered by the {\it IRAC\/} and {\it MIPS\/} instruments of the
{\it Spitzer Space Telescope\/}. They found detectable disk emission in the
{\it IRAC\/} bands from 48\% of the low mass stars of Tr\,37. Some 10\% of these disks
have been detected only at wavelengths $> 4.5~\mu$m, indicating optically thin
inner disks. Comparison of the SEDs of Tr\,37 members with those of the younger
Taurus region indicates that the decrease of infrared excess is larger at
6--8~\micron \  than at 24~\micron, suggesting that disk evolution is faster at
smaller radii.

\citet{SA1,SA2} also investigated the spatial asymmetries in Tr\,37 and the possible
presence of younger populations triggered by Tr\,37 itself.
They found a  spatial east-west asymmetry in the cluster that cannot
yet be fully explained. The low-mass Tr\,37 members
are concentrated on the western side of the O6 star
HD\,206267. In contrast, the B-A stars \citep{Contreras} are more uniformly
distributed, with a small excess to the east. Both the high and
intermediate-mass stars and the
low-mass stars show a clear ``edge'' to the east of HD\,206267.
The youngest stars are found preferentially on the western side of HD\,206267.
The presence of dense globules in this region suggests that the expansion of
the HII region into an inhomogeneous environment triggered this later epoch
of star formation. Larger amounts of interstellar material on the western side may also
have helped to shield disk systems from the photoevaporating effects of the central O6 star.

\citet{SA4} obtained high-resolution spectra of a large number of stars in the
region of Tr~37. They derived accretion rates from the H$\alpha$ emission line, and found
lower average accretion rate in Tr~37 than in the younger Taurus.
They used radial velocities as membership criterion, and thus confirmed the membership
of 144 stars and found 26 new members. They also calculated
rotational velocities, and found no significant difference between the rotation of accreting
and non-accreting stars.
In order to study the dust evolution in protoplanetary disks, \citet{SA5} studied the
10~\micron \ silicate feature in the Spitzer IRS spectra of several members of Tr~37.
GM~Cep, a solar-type member of Tr~37 exhibiting EXor-like
outbursts, was studied in detail by \citet{SA6}.

Tr\,37 was searched for X-ray sources as possible WTTSs by \citet*{SBZ97}.
Soft X-ray observations with the ROSAT  PSPC revealed X-ray emission from an area
of 30\arcmin \ radius around the center of globule IC\,1396A, which was resolved
into 85 discrete sources of which 13 sources were identified as foreground objects.
Most of the detected X-ray sources, except HD\,206267,
are very weak, which causes the measured luminosity function to be cut off at
$log L_x < 30.3\,$\,erg\,s$^{-1}$. X-ray sources are located not only in Tr\,37 but are also
scattered around the molecular globules IC\,1396A and B (see Sect.~\ref{Sect_glob}).
Their X-ray spectra appear hard with luminosities between log\,L$\sim$\,30 and 31.
LkH$\alpha$\,349, a $10^5$ yr old pre-main sequence star at the very center of
globule~A, appears very luminous with $ L_x = 5.1\times10^{30}$erg\,s$^{-1}$.
The source density within 5\arcmin \ of the center of emission is 270 sources per square degree.

\citet{Getman07} detected 117 X-ray sources in a field centered on the globule IC\,1396N
(see Sect.~\ref{Sect_glob}), of which ~50-60 are likely members of Trumpler~37.

\begin{table*}[!ht]
\caption{A. Bright rimmed globules and dark clouds associated with IC\,1396.}
\label{Tab_globule}
\begin{center}
{\footnotesize
\begin{tabular}{l@{\hskip2mm}l@{\hskip2mm}l@{\hskip2mm}l@{\hskip2mm}l}
\noalign{\smallskip}
\tableline
\noalign{\smallskip}
Names &  RA(J2000) &  D(J2000) & IRAS Source & Ref.\\[-1pt]
  & [h\,m] & [\deg\,\arcmin\,] \\[-1pt]
\noalign{\smallskip}
\tableline
\noalign{\smallskip}
FSE\,12 & 21 25 & 57 53 &   \\[-1pt]
FSE\,13 & 21 25 & 58 37 &   \\[-1pt]
FSE\,1, IC\,1396\,W &  21 26 & 57 58 & 21246+5743 & 4,14,16,17 \\[-1pt]
FSE\,14, LDN\,1086  & 21 28 & 57 31 &   \\[-1pt]
BRC\,32 & 21 32 24 & 57 24 08 & 21308+5710 \\[-1pt]
BRC\,33, Pottasch IC\,1396C & 21 33 12 & 57 29 33 & 21316+5716 \\[-1pt]
BRC\,34, GRS 3, Pottasch IC\,1396D & 21 33 32 & 58 03 29 &   21320+5750 \\[-1pt]
GRS 1  &  21 32 25 & 57 48 44 &\\[-1pt]
FSE\,2, GRS 2 & 21 33 54 & 57 49 44 & 21312+5736  \\[-1pt]
FSE\,15  & 21 33 & 59 30 & \\[-1pt]
FSE\,16, LDN\,1102   & 21 33 & 58 09 &   \\[-1pt]
FSE\,3, LDN\,1093, LDN\,1098,   & 21 34 11 & 57 31 06 &  21324+5716 & 29 \\[-1pt]
~~~GRS 4, Pottasch IC\,1396B \\[-1pt]
Weikard Rim I & 21 34 35 & 58 19.5 \\[-1pt]
FSE\,4, Pottasch IC\,1396A, &  21 36 12 & 57 27 34 &  21346+5714 & 5,6,7,14,19,\\[-1pt]
~~~GRS 6, BRC\,36  & & & & 29,32,33  \\[-1pt]
GRS 5, BRC\,35 &  21 36 05 & 58 32 17 & 21345+5818 \\[-1pt]
FSE\,5, LDN\,1099, LDN\,1105, GRS 6 &  21 36 54 & 57 30 &  21352+5715 &  \\[-1pt]
FSE\,6, LDN\,1116 &  21 37 & 58 37 &  21354+5823 & \\[-1pt]
FSE\,17, LDN\,1088, GRS\,9 & 21 38 56 & 56 07 36   \\[-1pt]
GRS 7 & 21 37 56 & 57 47 57 \\[-1pt]
Weikard Rim J & 21 38 56 & 56 21.3 \\[-1pt]
FSE\,7, GRS 12, BRC\,37, & 21 40 25 & 56 35 52 &  21388+5622 & 8,9,10,11,14,20, \\[-1pt]
~~~Pottasch IC\,1396H  & & & & 25,28,31,33,34 \\[-1pt]
WB89 108 & 21 40 38 & 56 48 & 21390+5634 \\[-1pt]
GRS 13 & 21 40 30  &  57 46 28 & \\[-1pt]
GRS 14 & 21 40 41 &  58 15 52 &  \\[-1pt]
FSE\,19, IC\,1396\,N(orth), LDN 1121,  &  21 40 43 & 58 20 09 & 21391+5802 & 1,2,3,13,14,15,\\[-1pt]
~~~GRS\,14, WB89 110, & & & & 18,25,30,32 \\[-1pt]
~~~BRC\,38, Pottasch IC\,1396E\\[-1pt]
FSE\,8, GRS 20, LDN\,1130, & 21 44 00 & 58 17 00 &  21428+5802 & 29 \\[-1pt]
~~~Pottasch IC\,1396F \\[-1pt]
GRS 26 & 21 44 51 &  57 08 00  &  \\
GRS 23 &  21 45 05 & 56 59 22 & 21443+5646 \\[-1pt]
GRS 24 & 21 45 09 &  56 47 52 &  \\[-1pt]
LDN\,1132 & 21 45 45 & 58 29.3 & & \\[-1pt]
GRS 25, WB89 122 &  21 45 58  &  57 13 54 & 21436+5657 \\[-1pt]
GRS 27 &  21 46 03 &  57 08 41 &  \\[-1pt]
GRS 28 & 21 46 27 & 57 18 07 &  \\[-1pt]
FSE\,9, IC\,1396\,E(ast), & 21 46 38 & 57 25 55 & 21445+5712 & 14,21,25,33 \\[-1pt]
~~~WB89 123, LDN\,1118, GRS 29, \\[-1pt]
~~~BRC\,39, Pottasch IC\,1396\,G \\[-1pt]
BRC\,40  & 21 46 14 & 57 08 59 & 21446+5655 \\[-1pt]
BRC\,41 & 21 46 29 & 57 18 41 & 21448+5704 \\[-1pt]
BRC\,42 & 21 46 37 & 57 12 25 & 21450+5658 \\[-1pt]
FSE\,21, LDN\,1129  & 21 46 27 & 57 46 37 &  \\[-1pt]
FSE\,22 &  21 49 & 56 43 &    \\[-1pt]
\noalign{\smallskip}
\tableline
\end{tabular}}
\end{center}
\smallskip
{\footnotesize {\it References to globule names and column~5 can be found under
Table~\ref{Tab_globule}~B.}}
\end{table*}

\addtocounter{table}{-1}
\begin{table*}[!ht]
\caption{B. Other clouds in the IC\,1396 region whose relation to Cep~OB2 is uncertain.}
\begin{center}
{\footnotesize
\begin{tabular}{l@{\hskip2mm}l@{\hskip2mm}l@{\hskip2mm}l@{\hskip2mm}l}
\tableline
\noalign{\smallskip}
Names & RA(J2000) & D(J2000) & IRAS source & Ref.\\[-1pt]
  & [h\,m] & [\deg\,\arcmin\,] \\[-1pt]
\noalign{\smallskip}
\tableline
\noalign{\smallskip}
FSE\,18, LDN\,1131  & 21 40 & 59 34 &  \\[-1pt]
FSE\,20, LDN\,1131 & 21 41 & 59 36 &   \\
FSE\,10,  LDN\,1139 & 21 55 36 & 58 35 & 21539+5821 & 14 \\[-1pt]
FSE\,23, LDN\,1153  & 22 01 & 58 54 &  \\[-1pt]
FSE\,11, LDN\,1165, LDN\,1164  &  22 07 00 & 59 02 &  22051+5848 & \\[-1pt]
GRS 32, LDN\,1165 &  22 07 00 & 59 00 & 22051+5848 & 12,14,15,22,23,\\[-1pt]
&&&& 24,25,26,27,28 \\[-1pt]
FSE\,24, LBN 102.84+02.07 & 22 08 & 58 23 &   \\[-1pt]
FSE\,25, LBN 102.84+02.07 & 22 08 & 58 31 &  \\[-1pt]
\noalign{\smallskip}
\tableline
\end{tabular}}
\end{center}
\smallskip
{\footnotesize {\it References to globule names:\/} Pottasch--\citet{Pottasch}; GRS--\citet{GRS};
 WB89--\citet{WB89}; BRC--\citet{SFO}; Weikard--\citet{Weikard}; FSE--\citet{FSEM}. \\
{\it References to column 5.\/}  1. \citet{Beltran02}; 2. \citet{Nisini}; 3. \citet{Codella};
4. \citet{FS2003};  5. \citet{Nakano}; 6. \citet{Hessman}; 7. \citet{Reach};
8. \citet{Duvert};  9. \citet{SFO}; 10. \citet{deVries02};
11. \citet{OSP}; 12. \citet{RB2001}; 13. \citet{RARB}; 14: \citet{SWG}; 15: \citet{Reipurth97};
16: \citet{Froebrich03}; 17: \citet{Zhou06}; 18: \citet{Getman07}; 19: \citet{SA3};
20: \citet{Sugitani97}; 21: \citet{Serabyn93}; 22: \citet{RA97}; 23: \citet{Tapia97}; 24: \citet{Parker91};
25: \citet{Connelley07}, 26: \citet{Visser}; 27: \citet{Slysh97}; 28: \citet{Bronfman96};
29: \citet{Moriarty96}; 30: \citet{Neri07}; 31: \citet{Ogura07}; 32: \citet{Valde05};
33: \citet{Valde08}; 34: \citet{Ikeda08}.}
\end{table*}
\nopagebreak

\begin{figure*}[!hb]
\centerline{
\includegraphics[draft=False,width=\textwidth]{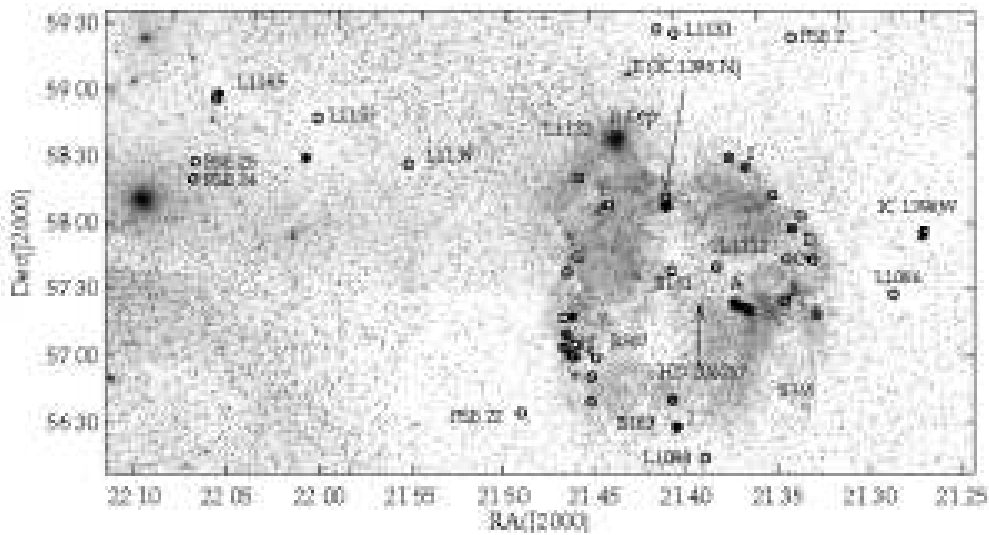}
}
\caption{Distribution of the globules listed in Table~\ref{Tab_globule} and
other dark clouds projected near IC~1396, overplotted on a DSS red image of the region
(open circles).  Star symbols show the embedded protostars, and plusses mark the
IRAS sources of uncertain nature \citep{SWG}.}
\label{fig_glob_yso}
\end{figure*}

\subsection{Star Formation in Globules of IC\,1396}
\label{Sect_glob}

The HII region IC\,1396 is powered by the O6.5\,V star HD\,206267.
It appears that the expansion of this HII region has resulted in sweeping
up a molecular ring of radius 12\,pc \citep{Patel98}. \citeauthor{Patel98} derive
an expansion age of the molecular ring of about 3~Myr.

The ring-like HII region, shown in Fig.~\ref{Fig_IC1396}, is some 3\deg \
in diameter and is surrounded by a number of bright-rimmed globules
which are probable sites of triggered star formation due to
compression by ionization/shock fronts and radiation pressure. Many bright
rimmed clouds harbor IRAS point sources of low dust temperature.
They also frequently contain small clusters of near-IR stars.
The most prominent globules are located in the western and northern
portions of the H II region.

The globules of IC\,1396 received different designations during various
studies. \citet{Pottasch} labeled the most prominent bright rimmed
globules with letters from A to H, in the order of their
increasing distance from the exciting star. \citet{Weikard} supplemented this
list by rims I and J. IC\,1396A corresponds to the famous Elephant Trunk Nebula.
\citet{GRS}  identified 32 globules in the region of Cep~OB2, and designated
them as {\it GRS\/} (globules of radial systems) 1--32.
Four radial systems of globules have been identified near IC\,1396.
One system, consisting of 16 globules, is centered on IC\,1396.
Another system of 12 globules, slightly south of IC\,1396, appears to be
associated with BD\,+54\deg2612, whereas two further radial systems have been
identified to the east and south-east of the main system of globules
surrounding HD\,206267.
The system associated with HD\,206267 is dominated by bright-rimmed globules with
diffuse tails generated by the radiation field of HD\,206267. The other systems
appear as opaque globules without rims. The systems partially overlap
spatially. \citet*{GRC} have surveyed the GRS globules for CO emission.
They found that the radial systems separate in radial velocity.
The mean LSR velocity of the HD\,206267 system is $-2.8\pm2.4$~km\,s$^{-1}$,
whereas the same for the BD+54\deg2612 system is $+6.5\pm1.0$~km\,s$^{-1}$.
Two of the globules, GRS\,12 and GRS\,14, are associated with H$_2$O masers
\citep*{GRC2}.

\citet{SWG} used the IRAS data base
to locate young stellar object candidates associated with the globules
of IC\,1396. They found that only six globule-related sources have point-like
structure and luminosities considerably in excess of that which can be
caused by external heating. Most of the
IRAS point sources associated with the globules are probably externally heated
small-scale dust structures not related to star formation.
Eleven globules of IC\,1396 can be found in the catalog of bright rimmed clouds by
\citet*{SFO} (BRC\,32--42).
\citet{FSEM} presented a large-scale study of the IC\,1396
region using new deep NIR and optical images, complemented by 2MASS data.
They  identified  25 globules (FSE~1--25) using extinction
maps and the list of \citet{SWG}. Four of them were previously
uncatalogued in the SIMBAD database. In all but four cases the masses (or
at least lower limits) of the globules could be determined, and the size
could be measured properly for all but seven objects.
For ten globules in IC\,1396 they determined (J$-$H, H$-$K) color-color diagrams
and identified the young stellar population. Five globules contain
a rich population of reddened objects, most of them probably young stellar objects.
The five globules with many red objects include the targets with the
highest extinction values, suggesting a correlation of the strength of
the star formation activity with the mass of the globule.

\citet{Moriarty96} have made the first arcminute resolution images of atomic
hydrogen toward IC\,1396, and have found remarkable ``tail''-like
structures associated with the globules IC~1396A, B and F, extending up to 6.5~pc
radially away from the central ionizing star. These H\,I ``tails'' may be material
which has been ablated from the globule through ionization and/or photodissociation
and then accelerated away from the globule by the stellar wind, but which has
since drifted into the ``shadow'' of the globules.

Star formation in small globules is often thought to be strongly
influenced by the radiation pressure of a nearby bright star. Froebrich et al.
therefore investigated how the globule properties in IC\,1396 depend on
the distance from the O star HD\,206267. The masses of the globules have
clearly shown positive correlation with the distance from this star, suggesting
that evaporation due to photo-ionization affects the mass distribution
of the globules around HD\,206267. Their data are consistent with a scenario
in which the radiation pressure from the O-type star regulates the star forming
activity in the globules, in the sense that the radiation pressure compresses the
gas and thus leads to enhanced star formation.

The names and coordinates of the known globules of IC\,1396, as well as their
associated IRAS sources are listed in Table~\ref{Tab_globule}, and the distribution
of the same objects with respect to the HII zone is shown in Fig.~\ref{fig_glob_yso}.
Some of the globules in IC\,1396 were already investigated in detail
and/or are known to harbor outflow sources. We give references for the
works on individual globules in the last column of the table.

\begin{figure*}[p]
\centering
\includegraphics[draft=False,width=\textwidth]{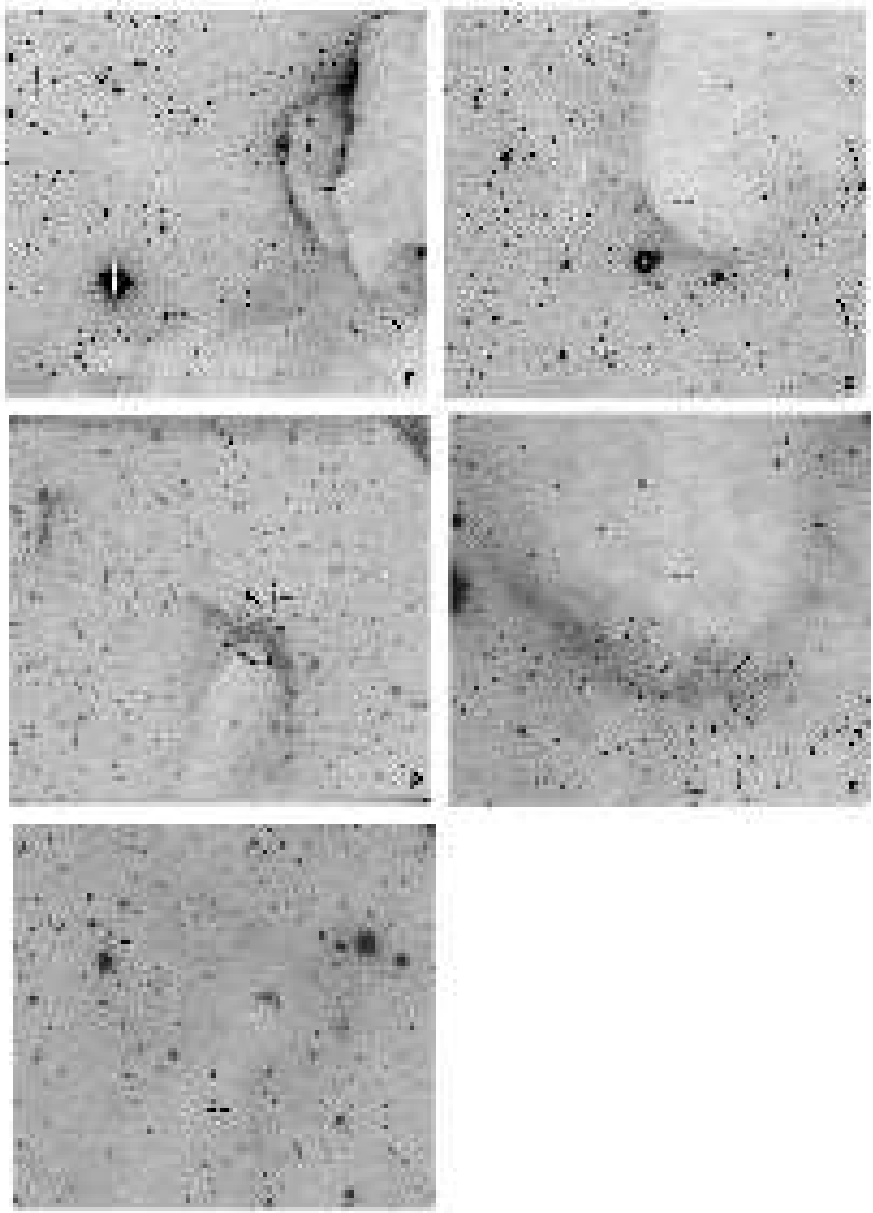}
\caption{H$\alpha$ emission stars in bright rimmed globules of IC\,1396, found by \citet{OSP}.
Top: BRC\,33, BRC\,34; middle: BRC\,37, BRC\,38($\equiv$ IC\,1396\,N); bottom: BRC\,39.
The position of the IRAS source associated with the globule is drawn by a pair of thick
tick marks.}
\label{fig_brc1396}
\end{figure*}

\citet{OSP} performed H$\alpha$ grism spectroscopy
and narrow band imaging observations of the BRCs listed by
\citet{SFO} in order to search for candidate pre-main sequence stars
and Herbig-Haro objects. They have detected a large number of H$\alpha$
emission stars down to a limiting magnitude of about $R = 20$.
Their results for IC\,1396 are reproduced in Table~\ref{tab_brc_ic1396},
and the finding charts for the H$\alpha$ emission stars are shown in
Fig.~\ref{fig_brc1396}.
Submillimeter observations of bright rimmed globules are presented by
\citet{Morgan08}.

\begin{table}[!ht]
\caption{H$\alpha$ emission stars  associated with bright
rimmed clouds in IC\,1396, identified by \citet{OSP}, and revised by \citet{Ikeda08}}
\label{tab_brc_ic1396}
\begin{center}
{\footnotesize
\begin{tabular}{rc@{\hskip1mm}c@{\hskip2mm}r@{\hskip8mm}rc@{\hskip1mm}c@{\hskip2mm}r}
\noalign{\smallskip}
\tableline
\noalign{\smallskip}
N & RA(J2000.0) & Dec(J2000.0) & EW  & N & RA(J2000.0) & Dec(J2000.0) & EW \\
\noalign{\smallskip}
\tableline
\noalign{\smallskip}
\multicolumn{4}{c}{BRC 33} & \multicolumn{4}{c}{BRC 38}  \\[3pt]
1 & 21 34 19.8 & 57 30 01 & 53.4 & 4 & 21 40 31.7 & 58 17 55 & $\cdots$  \\
2 & 21 34 20.8 & 57 30 47 & 3.3 &  5 & 21 40 36.7 & 58 13 46 & 4.0 \\
3 & 21 34 49.2: & 57 31 25: & 66.2 & 6 & 21 40 37.0 & 58 14 38 & 63.3  \\
\multicolumn{4}{c}{BRC 34} &  7 & 21 40 37.2 & 58 15 03 & 29.8 \\
1 & 21 33 29.4 & 58 02 50 & 43.0 &  8 & 21 40 40.5 & 58 13 43 & $\cdots$  \\
2 & 21 33 55.8 & 58 01 18 & $\cdots$ & 9 & 21 40 41.3 & 58 15 11 & 26.1  \\
\multicolumn{4}{c}{BRC 37}&  10 & 21 40 41.7 & 58 14 25 & 14.8 \\
1 & 21 40 25.3 & 56 36 43 & $\cdots$  &  11 & 21 40 45.0 & 58 15 03 & 75.7 \\
2 & 21 40 26.1 & 56 36 31 & 18.4 &  12 & 21 40 48.1 & 58 15 38 & 19.0  \\
3 & 21 40 26.8 & 56 36 23 & 40.9 & 13 & 21 40 48.9 & 58 15 00 &  $\cdots$ \\
4 & 21 40 27.2 & 56 36 30 & $\cdots$ & 14 & 21 40 49.0 & 58 15 12 &  $\cdots$ \\
5 & 21 40 27.4 & 56 36 21 & $\cdots$  &  15 & 21 40 49.2 & 58 17 09 & 22.2 \\
6 & 21 40 28.2 & 56 36 05 & $\cdots$  &  16 & 21 41 02.0 & 58 15 25 &  $\cdots$  \\
7 & 21 40 28.8 & 56 36 09 & 78.8 & \multicolumn{4}{c}{BRC 39}  \\
8 & 21 40 32.4 & 56 38 39 & 14.4  & 1 & 21 45 50.3 & 57 26 49 & $\cdots$   \\
\multicolumn{4}{c}{BRC 38} & 2 & 21 46 01.6 & 57 29 38 &  3.1\\
1 & 21 40 26.2 & 58 14 24 & 22.2 & 3 & 21 46 07.1 & 57 26 31 & 13.0 \\
2 & 21 40 27.4 & 58 14 21 & 59.3 & 4 & 21 46 26.0 & 57 28 28 & $\cdots$   \\
3 & 21 40 28.1 & 58 15 14 & 20.7 & 5N & 21:45:54.08 & 57:28:18.5 &   9.3 \\
\noalign{\smallskip}
\tableline
\end{tabular}}
\end{center}
\end{table}

The primary indicators of star formation in the globules are the embedded IRAS
point sources, molecular outflows and Herbig--Haro objects. Table~\ref{Tab_HH_ic1396}
lists the Herbig--Haro objects found for the IC\,1396 region.

\subsection{Notes on Individual Globules}

\subsubsection{IC 1396\,W}  lies about 1\fdg75 W--NW of  HD\,206267. In the
center of the small (about 6\arcmin) dark cloud, the IRAS source 21246+5743
can be found. This source is not detected at 12~\micron. The very
red IRAS colors and the extended appearance in the 100~\micron \  IRAS image
suggest a young, deeply embedded source. Observations of
this object with the photometer ISOPHOT confirmed that IRAS 21246+5743 is a
Class~0 source that will reach about one solar mass on the main
sequence \citep{FS2003}. The ISOPHOT maps at 160 and 200~\micron \
show two further cold objects (2.5~arcmin SW and NE,
respectively) in the vicinity of the central source. This might be an indication
of other newly forming stars or cold dust in the IC\,1396\,W globule.

\citet{FS2003} have observed the IC\,1396\,W globule in J, H, K$^{\prime}$, and a
narrow band filter centered on the 2.122 $\mu$m 1-0 S(1) line of molecular hydrogen.
They detected three molecular outflows in the field. The flow axes are parallel within
3\deg \ in projection. Magnetic fields cannot consistently explain
this phenomenon. A parallel initial angular momentum of these objects,
caused by the fragmentation of small clouds/globules, might be the reason
for the alignment. NIR photometry, IRAS and ISOPHOT observations
\citep{Froebrich03} led to the discovery of
the driving sources of the outflows.  The brightest outflow is driven by the
Class~0 source IRAS 21246+5743. Two flows are driven by
more evolved Class~I/II objects.
The JHK photometry of the globule also revealed a population of young stars,
situated mainly in a dense embedded subcluster, about 2.5~arcmin
south-west of IRAS 21246+5743. This cluster coincides with a clump of
dense gas. The other young stars are almost uniformly distributed in
the observed field.

\citet{Zhou06} mapped IC~1396\,W in the CO(1-0) line, and found that its CO
molecular cloud may consist of three physically distinct components with
different velocities. They detected neither molecular outflows nor the dense
cores associated with candidate driving sources. One possible reason is that
CO(1-0) and its isotopes cannot trace high density gas, and another is that the
beam of the observation was too large to observe them. The CO cloud may be
part of the natal molecular cloud of IC~1396\,W, in the process of disrupting
and blowing away. The CO cloud seems to be in the foreground of the H$_2$ outflows.

\subsubsection{IC\,1396A, Elephant Trunk nebula} contains an intermediate mass
\citep[$M \sim 3\,M_{\sun}$, Sp. type: F9 -- ][]{Hernan04} pre-main sequence star, LkH$\alpha$\,349
\citep{HRC,Hessman} and a K7 type T~Tauri star LkH$\alpha$\,349c
\citep{CK79,HBC}. No other YSOs were known before the {\it Spitzer Space Telescope\/}.
Radio continuum maps of IC~1396A at 6~cm and 11~cm were obtained by \citet{Baars76}.
The maps suggested the presence of an HII region within the globule.

\begin{table}[!ht]
\caption{Herbig--Haro objects in IC 1396}
\label{Tab_HH_ic1396}
\begin{center}
{\footnotesize
\begin{tabular}{ l c c l l }
\noalign{\smallskip}
\tableline
\noalign{\smallskip}
name          &     RA(J2000)  & Dec(J2000)  & source     & Ref. \\
\noalign{\smallskip}
\tableline
\noalign{\smallskip}
HH 864 A      & 21 26 01.4  & 57 56 09 & IRAS 21246+5743 & 1 \\[-1pt]
              & 21 26 02.0  & 57 56 09 & IRAS 21246+5743 & 1 \\[-1pt]
HH 864 B      & 21 26 07.9  & 57 56 03 & IRAS 21246+5743 & 1 \\[-1pt]
HH 864 C      & 21 26 21.3  & 57 57 40 & IRAS 21246+5743 & 1 \\[-1pt]
              & 21 26 18.6  & 57 57 12 & IRAS 21246+5743 & 1 \\[-1pt]
HH 588SW2D    & 21 40 10.5  & 56 33 46 & IRAS 21388+5622 & 2 \\[-1pt]
HH 588SW2C    & 21 40 12.2  & 56 34 08 & IRAS 21388+5622 & 2 \\[-1pt]
HH 588SW2A    & 21 40 16.7  & 56 33 55 & IRAS 21388+5622 & 2 \\[-1pt]
HH 588SW2B    & 21 40 18.4  & 56 34 16 & IRAS 21388+5622 & 2 \\[-1pt]
HH 777        & 21 40 21.6  & 58 15 49 & IRAS 21391+5802 & 3 \\[-1pt]
HH 778        & 21 40 22.8  & 58 19 19 &                 & 3 \\[-1pt]
HH 588SW1A    & 21 40 24.6  & 56 35 07 & IRAS 21388+5622 & 2 \\[-1pt]
HH 588SW1B    & 21 40 26.6  & 56 34 40 & IRAS 21388+5622 & 2 \\[-1pt]
HH 588SW1C    & 21 40 27.5  & 56 34 55 & IRAS 21388+5622 & 2 \\[-1pt]
HH 588        & 21 40 29.1  & 56 35 55 & IRAS 21388+5622 & 2 \\[-1pt]
HH 588NE1B    & 21 40 32.1  & 56 36 25 & IRAS 21388+5622 & 2 \\[-1pt]
HH 588NE1C    & 21 40 32.1  & 56 36 30 & IRAS 21388+5622 & 2 \\[-1pt]
HH 588NE1A    & 21 40 33.5  & 56 36 16 & IRAS 21388+5622 & 2 \\[-1pt]
HH 588NE1D    & 21 40 33.5  & 56 36 32 & IRAS 21388+5622 & 2 \\[-1pt]
HH 588NE1E    & 21 40 34.6  & 56 36 33 & IRAS 21388+5622 & 2 \\[-1pt]
HH 589C       & 21 40 35.0  & 58 14 37 & IRAS 21391+5802 & 2 \\[-1pt]
HH 590        & 21 40 35.1  & 58 17 52 &                 & 2 \\[-1pt]
HH 591        & 21 40 35.8  & 58 18 21 &                 & 2 \\[-1pt]
HH 592        & 21 40 36.8  & 58 17 02 &                 & 2 \\[-1pt]
HH 589A       & 21 40 37.5  & 58 14 45 & IRAS 21391+5802 & 2 \\[-1pt]
HH 589B       & 21 40 37.7  & 58 14 25 & IRAS 21391+5802 & 2 \\[-1pt]
HH 593        & 21 40 45.2  & 58 16 09 &                 & 2 \\[-1pt]
HH 588NE2E    & 21 40 45.6  & 56 37 15 & IRAS 21388+5622 & 2 \\[-1pt]
HH 588NE2D    & 21 40 47.8  & 56 37 02 & IRAS 21388+5622 & 2 \\[-1pt]
HH 779        & 21 40 47.9  & 58 13 35 &                 & 3 \\[-1pt]
HH 588NE2C    & 21 40 49.0  & 56 37 07 & IRAS 21388+5622 & 2 \\[-1pt]
HH 588NE2B    & 21 40 49.3  & 56 37 09 & IRAS 21388+5622 & 2 \\[-1pt]
HH 588NE2A    & 21 40 49.7  & 56 37 27 & IRAS 21388+5622 & 2 \\[-1pt]
HH 780        & 21 40 53.1  & 58 14 16 &                 & 3 \\[-1pt]
HH 594        & 21 40 53.8  & 58 17 02 &                 & 2 \\[-1pt]
HH 595        & 21 41 00.2  & 58 16 52 &                 & 2 \\[-1pt]
HH 588NE3     & 21 41 00.0  & 56 37 19 & IRAS 21388+5622 & 1 \\[-1pt]
              & 21 41 01.0  & 56 37 25 & IRAS 21388+5622 & 1 \\[-1pt]
HH 865A       & 21 44 28.5  & 57 32 01 & IRAS 21445+5712 & 1 \\[-1pt]
              & 21 44 29.3  & 57 32 24 & IRAS 21445+5712 & 1 \\[-1pt]
HH 865B       & 21 45 10.5  & 57 29 51 & IRAS 21445+5712 & 1 \\[-1pt]
GGD 36        & 21 58 30.0  & 58 56 00 &                 & 4 \\[-1pt]
HH 354        & 22 07 42.5  & 59 11 53 & IRAS 22051+5848 & 1,5 \\[-1pt]
\noalign{\smallskip}
\tableline
\end{tabular}
}
\end{center}
\smallskip
{\footnotesize
References. 1 -- \citet{FSEM}; 2 -- \citet{OSP}; 3 -- \citet{RARB};
4  -- \citet{G87}; 5 -- \citet{Reipurth97}.}
\end{table}

{\it Spitzer Space Telescope\/} images  at 3.6, 4.5, 5.8, 8, and 24~\micron \  \citep{Reach}
revealed this optically dark globule to be infrared-bright and to contain
a set of previously unknown protostars. The mid-infrared colors of the
sources detected at 24~\micron \ indicate several very young (Class~I or 0)
protostars and a dozen Class~II stars. Three of the new sources
(IC 1396A\,$\gamma$, 1396A\,$\delta$, and 1396A\,$\epsilon$) emit over 90\%
of their bolometric luminosities at wavelengths longer than 3~\micron, and
they are located within 0.02~pc of the ionization front at the edge of
the globule. Many of the sources have spectra that are still rising at 24~\micron.
The two previously known young stars LkH$\alpha$\,349 and 349$c$ are both detected,
with component $c$ harboring a massive disk and LkH$\alpha$\,349 itself being bare.
About 5\% of the mass of the globule is presently
in the form of protostars in the 10$^{5}$--10$^{6}$~yr age range.

The globule mass was estimated to be 220\,M$_{\sun}$ from a
high-resolution CO map \citep{Patel95},  much smaller than the virial mass,
estimated as 300--800 M$_{\sun}$  \citep{Patel95,Weikard}.

\begin{figure*}[!ht]
\centerline{
\includegraphics[draft=False,width=8cm]{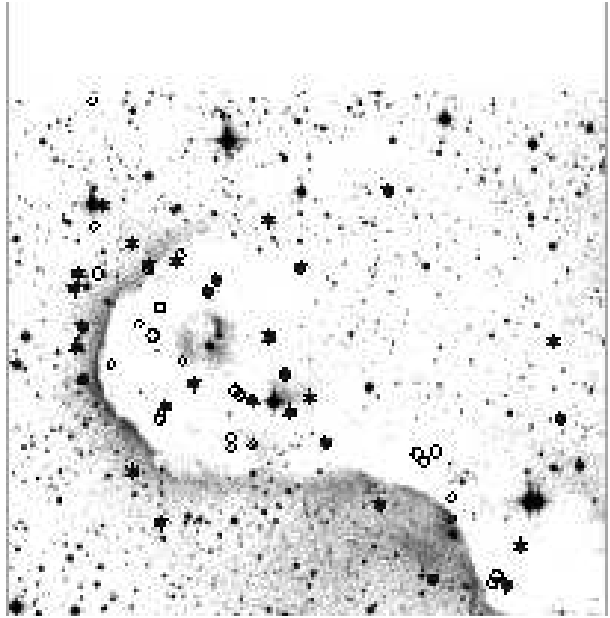}
}
\caption{YSOs in IC\,1396A, discovered by Spitzer Space Telescope \citep{SA3},
overplotted on the DSS red image of the globule. Circles mark Class~I, and star symbols
Class~II sources. Diamonds mark uncertain members.}
\label{fig_IC1396A}
\end{figure*}

\citet{SA3}, based on {\it IRAC\/} and {\it MIPS\/} photometry, identified 57 YSOs
born in the Elephant Trunk. Most of them have no optical counterparts. Based on the color
indices and the shape of the SEDs, \citeauthor{SA3} identified 11 Class~I and 32 Class~II
objects. Their average age is about 1~Myr. The surface distribution of these objects
is displayed in Fig.~\ref{fig_IC1396A}, adopted from \citet{SA3}.

\citet{Valde05,Valde08} detected H$_2$O maser emission from the direction of IRAS~21345+5714,
associated with IC\,1396A. Probably each protostar observed by {\it Spitzer\/}
contribute to the fluxes of the IRAS source.

\subsubsection{BRC 37, IC\,1396H}

High resolution $^{12}$CO, $^{13}$CO and CS observations of this globule have been performed
by \citet{Duvert}. They detected a bipolar outflow and identified the possible optical
counterpart of the driving source IRAS 21388+5622. \citet{Sugitani97} reported on
interferometric $^{13}$CO observations of  BRC\,37. They found evidence of
interaction with the UV radiation from the exciting star of IC\,1396.
\citet{Bronfman96} included IRAS~21388+5622 in their CS(2--1) survey
of IRAS point sources with color characteristic of ultracompact H\,II regions.
In order to study the age distribution of stars \citet{Ogura07} undertook
$BVI_\mathrm{c}JHK_\mathrm{s}$ photometry of stars in and around some bright rimmed
globules including BRC~37. Their results indicate that star formation proceeds from
the exciting star outward of the HII region. \citet{Ikeda08} carried out
near-IR/optical observations of BRC\,37 in order to study the sequential star formation
in the globule. Several results published by \citet{OSP} are revised in the paper.
\citet{Valde08} detected H$_2$O maser emission from IRAS~21388+5622.

\subsubsection{IC\,1396N}

\citet{Serabyn93} estimated the density and temperature structure of
this globule (they use the designation IC~1396E), and found evidence
of the possibility that recent internal star formation was triggered
by the ionization front in its southern surface.  On the basis of
NH$_3$ data, gas temperatures in the globule are found to increase
outward from the center, from a minimum of 17~K in its tail to a
maximum of 26~K on the surface most directly facing the stars ionizing
IC~1396.  \citet{Sugitani00} performed 2~mm continuum observations of
IC\,1396N (BRC\,38).  \citet{Codella} reported mm-wave multiline and
continuum observations of IC 1396N. Single-dish high resolution
observations in CO and CS lines reveal the cometary structure of the
globule with unprecedented detail.  The globule head contains a dense
core of 0.2 pc, whereas the tail, pointing away from the exciting
star, has a total length of 0.8~pc. Two high velocity bipolar outflows
have been identified in the CO maps: the first one is located around
the position of the strong infrared source IRAS~21391+5802 in the head
of the globule, and the second one is located in the northern
region. The outflows emerge from high density clumps which exhibit
strong line emission of CS, HCO$^{+}$, and DCO$^{+}$. The sources
driving the outflows have been identified by mm-wave continuum
observations (e.g. Beltr\'an et al. 2002). The globule head harbors
two YSOs separated by about 10$^{4}$~AU. SiO line observations of the
central outflow unveil a highly collimated structure with four clumps
of sizes pc, which are located along the outflow axis and suggest
episodic events in the mass loss process from the central star.

\begin{figure*}[!ht]
\centerline{
\includegraphics[draft=False,width=8cm]{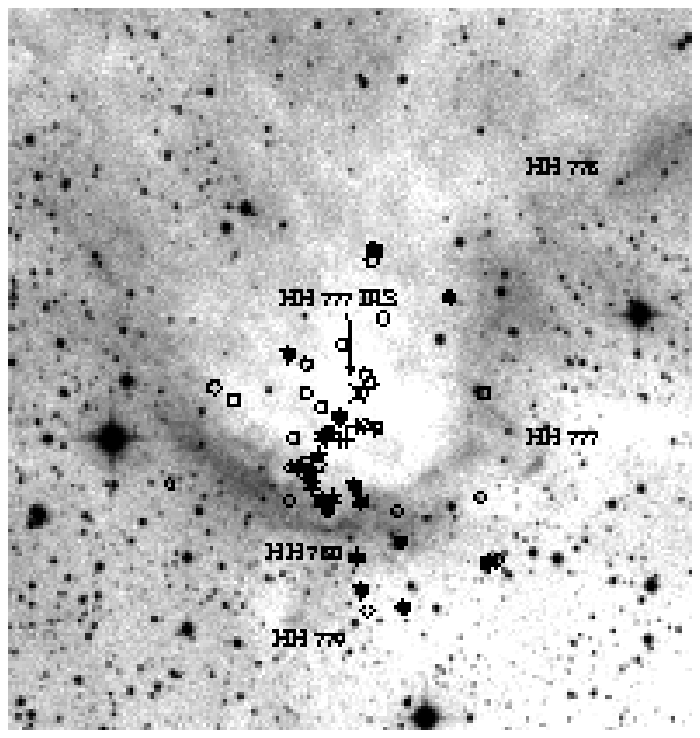}
}
\caption{Members of the young cluster born in IC~1396N, overplotted on the DSS red image of the
globule. Star symbols show the X-ray sources associated with Class~II objects,
plusses mark those corresponding to Class~I objects \citep{Getman07}, diamonds show the H$\alpha$
emission stars detected by \citet{OSP}, and open circles indicate the near-infrared sources found by \citet{Nisini}.
Positions of the HH objects, discovered by \citet{RARB}, are indicated and a large thick cross
shows the position of HH~777~IRS.}
\label{fig_ic1396n}
\end{figure*}

\citet{Nisini} presented near infrared images of
IC\,1396N in the H$_2$ 2.12~\micron \  narrow band filter as well as in broad band
J, H, and K filters. They detected several chains of collimated H$_2$ knots inside
the globule, having different luminosities but similar orientations in the sky.
Most of the knots are associated with peaks of high velocity CO emission, indicating
that they trace shocked regions along collimated stellar jets. From the morphology
and orientation of the H$_2$ knots, they identify at least three different jets:
one of them is driven by the young protostar associated with IRAS~21391+5802,
while only one of the two other driving sources could be identified by means of
near infrared photometry. The NIR photometry revealed the existence of a cluster of
young embedded sources located in a south-north line which follows the
distribution of the high density gas and testifies to a
highly efficient star formation activity through all the globule.
\citet{Valde05} and \citet{Furuya03} detected H$_2$O maser emission from IC~1396N.

\begin{table}[!ht]
\caption{Young stellar objects in IC~1396N detected by Chandra, and their infrared counterparts \citep{Getman07}}
\label{tab_xray_IC1396N}
\begin{center}
{\footnotesize
\begin{tabular}{c@{\hskip2mm}c@{\hskip1mm}r@{\hskip1mm}r@{\hskip1mm}r@{\hskip1mm}c@{\hskip1mm}r@{\hskip1mm}r@{\hskip1mm}c@{\hskip2mm}l}
\noalign{\smallskip}
\tableline
\noalign{\smallskip}
\multicolumn{2}{c}{Source} & \multicolumn{3}{c}{NIR(2MASS)} &	\multicolumn{4}{c}{MIR}  \\
\noalign{\smallskip}
\tableline
\noalign{\smallskip}
No.& CXOU J & J~~~ & H~~~ & K$_s$~~ & [3.6]~ & [4.5]~ & [5.8]~ & Class & Other Id.$^*$  \\
 & &  (mag) & (mag) & (mag) &  (mag) & (mag) & (mag) & \\
\noalign{\smallskip}
\tableline
\noalign{\smallskip}
41 & 	214027.31+581421.1 &   14.30 & 13.30 & 12.88 & 11.62  & 11.09 &     10.74 &  II      &  OSP\,2   \\
49 & 	214031.58+581755.2 &   14.03 & 12.89 & 12.39 & 11.51  & 11.16 &     10.23 & II       &  OSP\,4   \\
53 & 	214036.57+581345.8 &   13.51 & 12.58 & 12.24 & 12.13  & 11.94 &     12.01 & III      &  OSP\,5   \\
55 & 	214036.90+581437.9 &   11.90 & 10.89 & 10.23 & 9.38  &   9.10  &     8.76 & II       &  OSP\,6   \\
60 & 	214039.62+581609.3 &	     &       &       &    11.29  &  9.42  &  8.31 & I$^a$    &  NMV\,2	 \\
61 & 	214039.87+581834.8 & $>$18.29 & 15.30 & 13.36&  11.63  & 10.95  &   10.21 &  II      &  NMV\,3	 \\
62 & 	214041.12+581359.0 &   12.96 & 12.08 & 11.77 &     11.61  & 11.72 & 11.49 &  III     &  	 \\
63 & 	214041.16+581511.2 &   12.97 & 11.61 & 10.68 &  9.15  &  8.61  &     8.05 & II       &  OSP\,9   \\
65 & 	214041.56+581425.5 &   13.65 & 12.62 & 12.17 & 11.40  & 11.20  &    10.68 & II       &  OSP\,10  \\
66 & 	214041.81+581612.3 &	     &       &       &     11.30  & 8.90  &  7.50 & 0/I$^b$  &  	 \\
67 & 	214041.91+581523.1 &   15.68 & 14.30 & 13.65 & 12.69  & 12.49  &  $>$9.82  & II      &  	  \\
68 & 	214042.89+581601.0 &	     &       &       &    10.60  & 8.78  &   7.60 & I$^c$    &  	 \\
70 & 	214043.47+581559.7 &	     &       &       & 12.89  & 11.95  &   $>$9.85 & I/II    &  	  \\
71 & 	214043.64+581618.9 & $>$17.89 & $>$16.09 & 13.51& 9.97  &  8.72  & 8.00   & I        & NMV\,10	  \\
72 & 	214044.34+581513.3 &  16.05 & 14.59 & 13.60  &   12.42  & 11.64  &  10.60 & II       &  	 \\
73 & 	214044.84+581605.1 & $>$15.85 & 14.28 & 12.89& 11.73  & 11.15	&   10.68  & II      & NMV\,11	  \\
74 & 	214044.85+581503.4 &   14.62 & 13.35 & 12.66 & 12.29  & 11.40  &    10.96 & II       & OSP\,11   \\
76 & 	214045.18+581559.8 &	     &       &       & 11.95  & 10.82  &     10.03 &  I      &  	  \\
77 & 	214045.51+581511.4 &   14.65 & 13.71 & 13.11 & 12.54  & 12.21  &     11.82 & II      &  	  \\
78 & 	214045.53+581602.9 &   15.68 & 13.73 & 12.85 & 12.23  & 11.84  &  $>$10.37 & II      &  	  \\
80 & 	214045.79+581549.0 &	     &       &       & 12.59  & 11.19  &     10.04 & I       &  	  \\
81 & 	214046.49+581523.2 &   12.81 & 11.95 & 11.65 & 11.37  & 11.33  &    11.39 & III      &  	 \\
82 & 	214046.89+581533.3 &   15.30 & 13.55 & 12.63 & 11.97  & 11.80  &    11.15 & II       &  	 \\
85 & 	214048.03+581537.9 &   13.89 & 12.95 & 12.67 & 12.08  & 11.91  &    10.77 & II       &  OSP\,12  \\
87 & 	214049.09+581709.3 &   14.14 & 12.86 & 12.13 & 10.77  & 10.19  &     9.69 & II       &  OSP\,15  \\
\noalign{\smallskip}
\tableline
\end{tabular}}
\end{center}
\smallskip
{\footnotesize
$^*$OSP: \citet{OSP}; NMV: \citet{Nisini} \\
 $^a$ X-ray source 60 is offset by 3.2\arcsec \ to the north of the radio continuum source VLA 1 and by 5.0\arcsec \ to the
 northwest of the millimeter source BIMA~1 of \citet{Beltran02}. \\
$^b$ X-ray source 66 is within 0.5\arcsec \ of the radio continuum source VLA~2 and millimeter source BIMA~2 of \citet{Beltran02}.
This is also millimeter source A of \citet{Codella}. \\
$^c$ X-ray source 68 is within 1.0\arcsec \ of the radio continuum source VLA~3 and millimeter source BIMA~3 of  \citet{Beltran02}.}
\end{table}

\citet{Saraceno96} present a far-infrared spectrum of IRAS~21391+5802, together with
submillimeter and millimeter photometry. A rich spectrum of CO, OH, and H$_2$O lines are
detected in the ISO-LWS spectrum, indicative of a warm, dense region around the source.
They also obtained an accurate measure of the bolometric luminosity and an estimate
of the total envelope mass.

\citet{RARB} identified a major Herbig-Haro flow, HH~777, that is
bursting out of the IC~1396N cometary cloud core. Near- and
mid-infrared images reveal a very red object embedded in the center of
the core, located on the symmetry axis of the large HH~777 flow,
suggesting that this is likely the driving source. The projected
separation of the working surface from the source is
0.6~pc. Additionally, 0.4~pc to the east of the source and on the flow
axis, there is a faint, previously known HH object (HH~594) that may
be part of the counterflow (Ogura et al. 2002). It thus appears that
we are seeing a blowout of a parsec-scale flow into the surrounding
H\,II region.

IC~1396N has been observed with the ACIS detector on board the Chandra
X-Ray Observatory \citep{Getman07}. 25 of the 117 detected X-ray
sources are associated with young stars formed within the
globule. Infrared photometry (2MASS and Spitzer) shows that the X-ray
population is very young: 3 older Class~III stars, 16 classical
T~Tauri stars, and 6 protostars including a Class 0/I system. The
total T~Tauri population in the globule, including the undetected
population, amount to $\sim$30 stars, which implies a star formation
efficiency of 1\%-4\%.  Four of the X-ray-selected members coincide
with near-infrared sources reported by \citet{Nisini}, and 9 of them
correspond to H$\alpha$ emission stars detected by \citet{OSP}.  An
elongated spatial distribution of sources with an age gradient
oriented toward the exciting star is discovered in the X-ray
population. The geometric and age distribution is consistent with the
radiation-driven implosion model for triggered star formation in
cometary globules by H\,II region shocks.  The large number of
X-ray-luminous protostars in the globule suggests either an unusually
high ratio of Class I/0 to Class II/III stars or a nonstandard initial
mass function favoring higher mass stars by the triggering
process. The Chandra source associated with the luminous Class 0/I
protostar IRAS 21391+5802 is one of the youngest stars ever detected
in the X-ray band. Table~\ref{tab_xray_IC1396N} shows the list of
young stars identified by their X-ray emission, together with their
NIR (2MASS) and MIR (Spitzer IRAC) magnitudes. Positions of the young
stars born in this globule and some of the associated HH objects are displayed
in Fig.~\ref{fig_ic1396n}.

\citet{Neri07} investigated the mm-morphology of IC~1396N at a scale of $\sim$250~AU.
They have mapped the thermal dust emission at 3.3 and 1.3~mm,
and the emission from the J=13$_k \rightarrow 12_k$ hyperfine transitions of methyl cyanide (CH$_3$CN)
in the most extended configurations of the IRAM Plateau de Bure interferometer.
The observation revealed the existence of a sub-cluster of hot cores in IC~1396\,N, consisting of
at least three cores, and distributed in a direction perpendicular to the emanating outflow.
The cores are embedded in a common envelope of extended and diffuse dust emission.
The CH$_3$CN emission peaks towards the most massive
hot core and is marginally extended in the outflow direction. The protocluster IC\,1396N
has been included in a high angular resolution imaging survey of the circumstellar material
around intermediate mass stars conducted by \citet{Fuente08}, as well as in a study
of clustering properties of Class~0 protostars by \citet{Fuente07}.

\subsubsection{IC\,1396 East (IC\,1396G)}
IRAS~21445+5712, associated with this globule, coincides with a faint, red,
nebulous star \citep{SWG}. \citet{OSP} detected H$\alpha$ emission in the spectrum
of this star (BRC~39 No.~3). \citet{Fukui} detected a molecular outflow associated with IRAS~21445+5712.
\citet{Connelley07} found a small, elongated near-infrared nebula around the star.
IRAS~21445+5712 has been included in several surveys for H$_2$O maser sources
\citep{Felli92,WBF93,Furuya03,Valde08}. High-resolution VLA observations by \citet{Valde08}
resulted in the first detection of water maser emission associated with IRAS~21445+5712.

\subsubsection{L\,1165}
This cloud, harboring IRAS~22051+5848, is included in several studies
of the globules associated with IC~1396 \citep[e.g.][]{SWG,GRS,FSEM},
though it lies at 2.6\deg \ east of the H\,II zone (corresponding to
some 30~pc at the distance of IC\,1396).  IRAS~22051+5848 is
associated with a small reflection nebula, catalogued as Gy~2--21 by
\citet{Gy2}. \citet{SWG} note that, according to its CO radial
velocity \citep{GRC}, this globule may be foreground to
IC~1396. \citet*{Parker91} observed a bipolar CO outflow originating
from IRAS~22051+5848. \citet{Tapia97} presented near-infrared, {\it
IJHK\/}, images of the globule.  They identified the NIR counterpart
of the IRAS source and an extended infrared nebula around
it. \citet{Reipurth97} detected a giant Herbig--Haro flow, HH~354,
associated with IRAS~22051+5848. \citet{RA97} obtained a near-infrared
spectrum of the source, also known as HH~354~IRS. They concluded that
the detected CO absorption and the high luminosity of the star suggest
that HH~354~IRS is probably a FUor.  \citet{Visser} detected a
submillimeter source, L\,1165~SMM\,1, associated with the
globule. \citet{Slysh97} observed an OH maser emission at the position
of the IRAS source. L\,1165 is included in the CS(2--1) survey of IRAS
point sources with colors characteristic of ultracompact H\,II regions
published by \citet{Bronfman96}.  The distance of L\,1165 is
uncertain. Several authors \citep[e.g.][]{Reipurth97,FSEM} associate
this cloud with IC~1396, whereas others, e.~g. \citet{Tapia97} assume
a kinematic distance of 200~pc, and \citet{Visser} use 300~pc, with a
reference to \citet{DBYF}. This value is based on the assumption that
the cloud is part of the Lindblad ring. \citet{GRC} associate L\,1165
with a radial system of globules centered on the A0 type giant
HD~209811. The Hipparcos parallax of this star suggests a distance of
about 400~pc.

\section{Star Formation along the Cepheus Bubble}

\begin{figure*}[!tb]
\centerline{
\includegraphics[draft=False,width=5.25in]{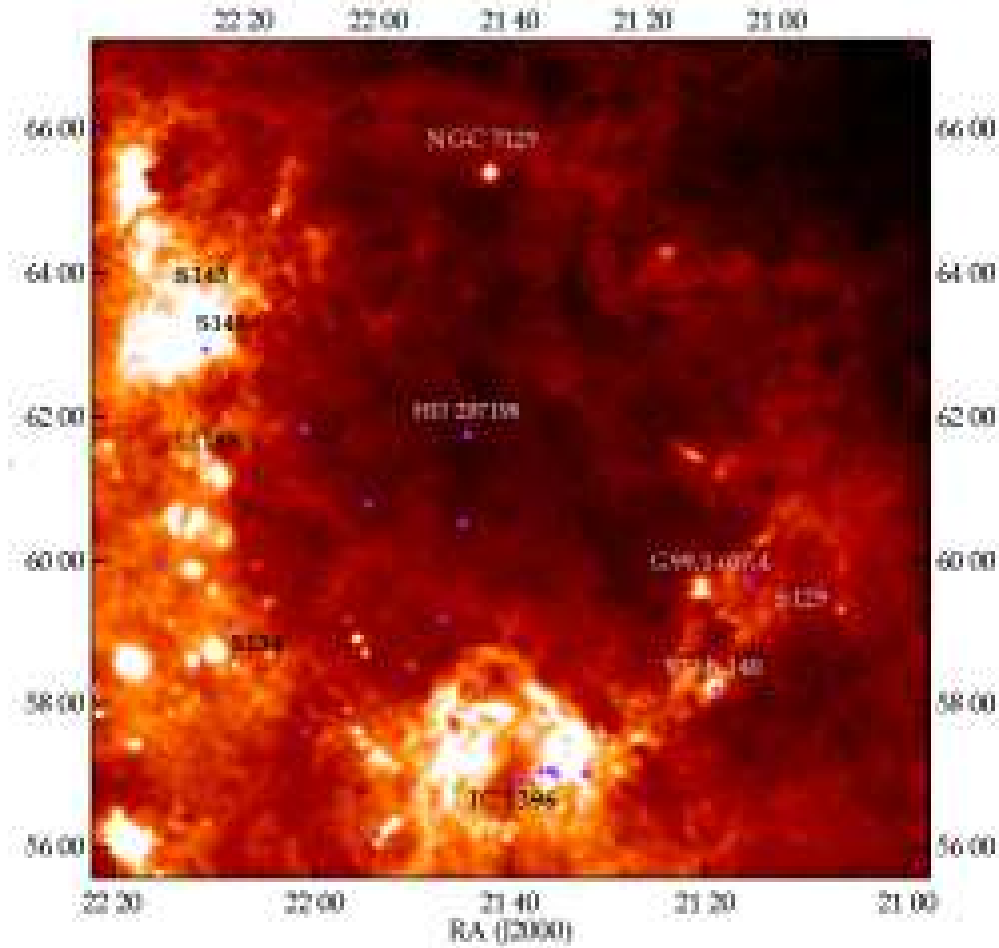}
}
\caption{IRAS 100~\micron \  (IRIS) image of the Cepheus Bubble. The HII regions and reflection
nebulae, probably associated with the Bubble \citep{KBT}, are indicated. Star symbols show the
O-type and supergiant members of Cep~OB2 \citep{Humphreys}.}
\label{fig_bubble}
\end{figure*}

The Cepheus Bubble is a giant far infrared ring-like structure around Cep~OB2a,
described by \citet{KBT}, and identified as an HI shell by  \citet{Patel98}
and \citet{ABK}. Several HII regions, such as IC\,1396, S\,140, S\,134,
S\,129, and G\,99.1+07.4 \citep{KC97} are located on the periphery of the ring.
The similarity of the distances of these
HII regions (700--900~pc) and the presence of the infrared ring-like structure apparently
connecting them suggest that the infrared ring is a real feature and is physically
connected with the HII regions. It was probably created by the stellar
winds and supernova explosions of the evolved high-mass members of Cep~OB2a.
In particular, the O9\,IIe type star HD\,207198, located near the center of
the Bubble, may be a major source of powerful stellar wind \citep{ABK}.
Figure~\ref{fig_bubble} shows the distribution of the  IRAS 100~\micron \  emission
over the area of the Bubble. The known HII regions and reflection nebulae, as
well as the O-type and supergiant members of Cep~OB2  are indicated
in the figure. CO observations  performed by \citet{Patel98} over the $10\deg\times10\deg$
area of the Cepheus Bubble revealed the molecular clouds associated  with it.
They found a total molecular mass of $1 \times 10^{5} M_{\sun}$. Most of the molecular
mass is associated with L\,1204/S\,140 and IC\,1396, but there are further
molecular clouds whose star forming activity has not yet been studied.
\citeauthor{Patel95} have shown that the shapes and kinematic properties of the IC~1396
globules indicate their interaction with the Bubble.
The most comprehensive list of clouds and star forming regions associated with the
Cepheus Bubble can be found in \citeauthor*{KMT04}'s \citeyearpar{KMT04} Table~C.

\subsection{Star Formation in S\,140}

The HII region S\,140 is located at the southwestern edge of the L\,1204
dark cloud, along the Cepheus Bubble \citep{ABK}, at a distance of about 900~pc from the Sun
\citep{Cramp74}. The ionization of the clouds is maintained by HD\,211880, a B0V star
\citep{Blair78}. It is separated from L\,1204  by a nearly edge-on ionization front.
The core of the cloud is totally invisible in optical images while even the earliest
infrared and radio observations have suggested that there is a dense cluster in the
center of the core. \citet{Roua77} detected far-infrared emission from a region
in L\,1204, a few arcmin NE of S\,140. They deduced a dust temperature of about 35~K,
computed the total IR intensity, and estimated a mass of 600~M$_{\sun}$
for the observed area. Further infrared and submillimeter studies of the infrared
source, S\,140~IRS  \citep{Tokun78,Diner79,Beich79,Little80,Hackw82,Thorn83} confirmed
that the heating source of the cloud is a small cluster of embedded stars.

Several observational studies
have been carried out to study the region in different wavelength regimes. These are
mostly focused on the photon-dominated region (PDR) at the edge of L\,1204, and on the
embedded infrared sources located right behind it
\citep[e.g.][]{Hayashi85,Keene85,Lester86,Schwa89b,Haseg91,Golyn91,Smirnov92,Plume94,Wilner94,Zhou94,Schneid95,
Minchin95a,Minchin95c,Stoerz95,Park95,Prei02,Bally02,Poel06,Poel05}.
VLA observations of the 6~cm H$_2$CO line by \citet{Evan87} revealed absorption of the cosmic
background radiation towards a 4\arcmin $\times$ 3\arcmin \ region of the S\,140 molecular cloud
with structures on scales from 20\arcsec \ to 4\arcmin. They attributed these structures to
clumps with masses around 40~M$_{\sun}$ and suggested that the clumps represent the
first stages of the fragmentation of this portion of the cloud (although they did not
rule out the possibility that the absorption maxima are low density holes surrounded
by high-density regions). VLA observation of NH$_3$ by \citet{Zhou93}, however, showed
absence of significant NH$_3$~(1,1) emission at the H$_2$CO absorption peaks,
indicating that the peaks correspond to low density ``holes'' rather than high-density
clumps. The high density molecular gas was studied by  \citet{Unger86}, who mapped
the region using NH$_3$~(1,1) and (2,2), and found that the column density and rotational
temperature peak at the position of the embedded infrared source. The kinetic
temperature is peaked at 40\,K and decreasing smoothly to 20\,K within the neighborhood
of the infrared source. \citet{Zeng91} studied the hyperfine structures of HCN~(1-0) emission
from the high density molecular core.

Several optical, near-, mid- and far-infrared, and radio surveys were carried out looking
for young stars in the region \citep[e.g.][]{Roua77,Beich79,Hayashi87,Evan89,Persi95,OSP,
Bally02,Prei02}. \citet*{Beich79} provided the first catalog of young stellar objects,
consisting of three infrared sources, IRS\,1,~2,~3. Later \citet{Evan89} added two
additional sources (VLA~4 and NW) to the catalog from observations using the VLA at
6 and 2~cm. The positions of the sources can be found in Table~14.
From the spectral indices of IRS\,1--3 they concluded that the radio emission from these
sources originates from optically thin HII regions ionized by Lyman-continuum photons
from single, main sequence stars with spectral type of B1.5-B2. \citet{Evan89} also
carried out near-IR photometry using the NOAO infrared camera at 1.2, 1.65, and
2.2~\micron. They detected all known far-IR sources except IRS\,2 and found additional
11 sources in the near-IR. At least five of these near-IR sources appear to be
discrete sources, suggesting that a deeply embedded young cluster is forming in
the region. Another cluster, containing about 100 near-IR sources associated with
S\,140, was discovered north of the region by a K$^{\prime}$-band imaging survey
by \citet{Hodapp}.

\citet{Joyce86} carried out a near-infrared polarization study and found an extremely high
level of 2.2~$\mu$m polarization towards S\,140~IRS\,1, indicating an outflow directed
nearly along our line of sight. This finding was later confirmed by \citet{Hayashi87}
who observed the HII region in $^{12}$CO and $^{13}$CO. \citet{Minchin93} found that
the blue and redshifted lobes of the CO bipolar outflow have position angles of
160$\deg$ and 340$\deg$, respectively.
 The high-resolution CS map obtained by \citet{Hayashi92} reveals a prominent
V-shaped ridge or a ring around the S\,140 IR cluster encircling the blue and red
lobes of the molecular outflow, with no emission detected in the vicinity of the
IR sources. The observations suggest that the CS ring is a remnant of a nearly
pole-on massive gaseous disk interacting with the high-velocity outflow.

\begin{figure*}[!ht]
\centerline{
\includegraphics[draft=False,width=4.25in]{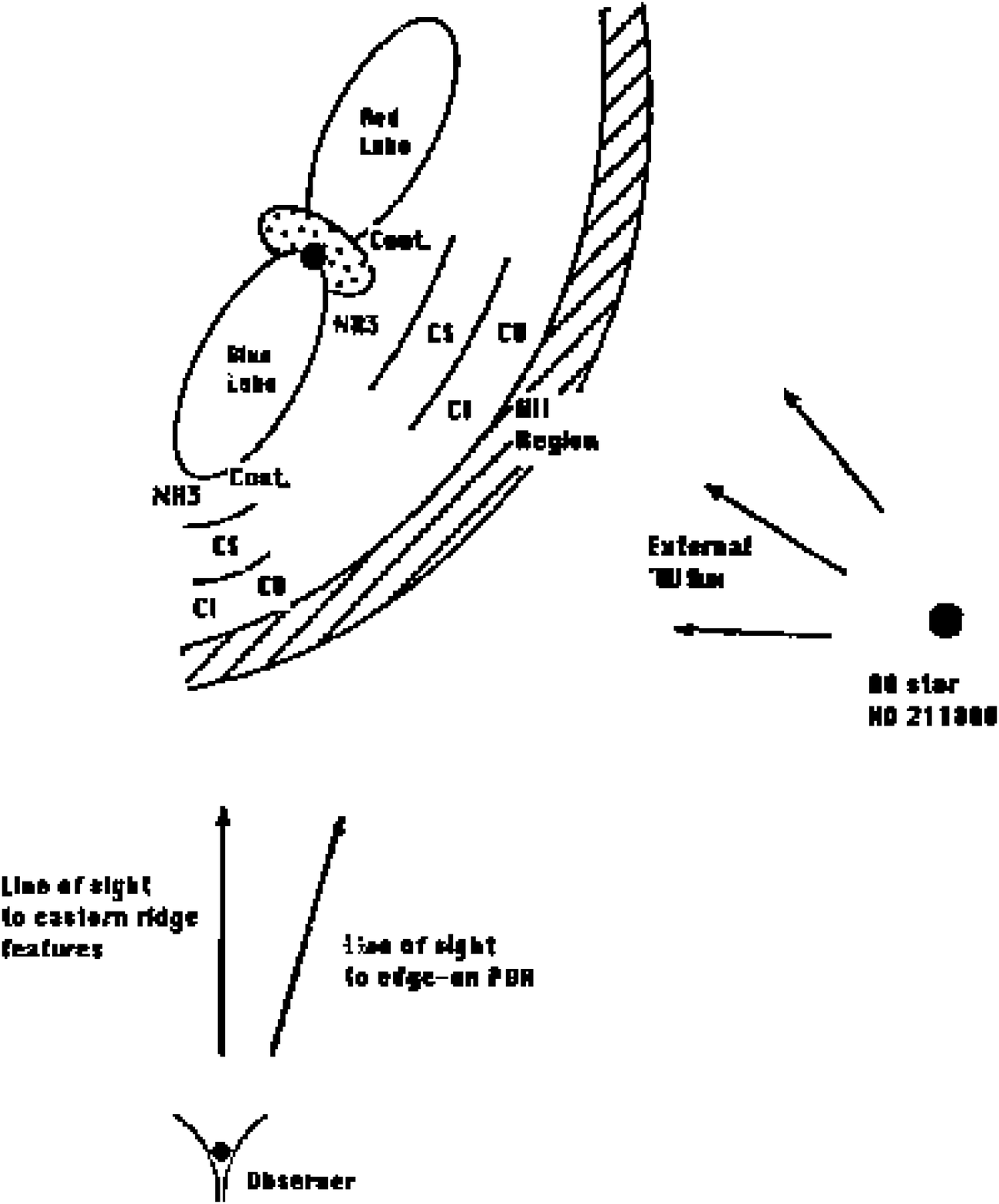}
}
\caption{Schematic representation of the S\,140/L\,1204 region showing the plane
that contains the observer, the external illuminating star HD\,211880, the
HII region/molecular cloud interface and the embedded molecular outflow.
Fig 2. of \citet{Minchin95b}.}
\label{fig:Minch}
\end{figure*}

\begin{figure*}[!htb]
\centering
\includegraphics[draft=False,width=5.25in]{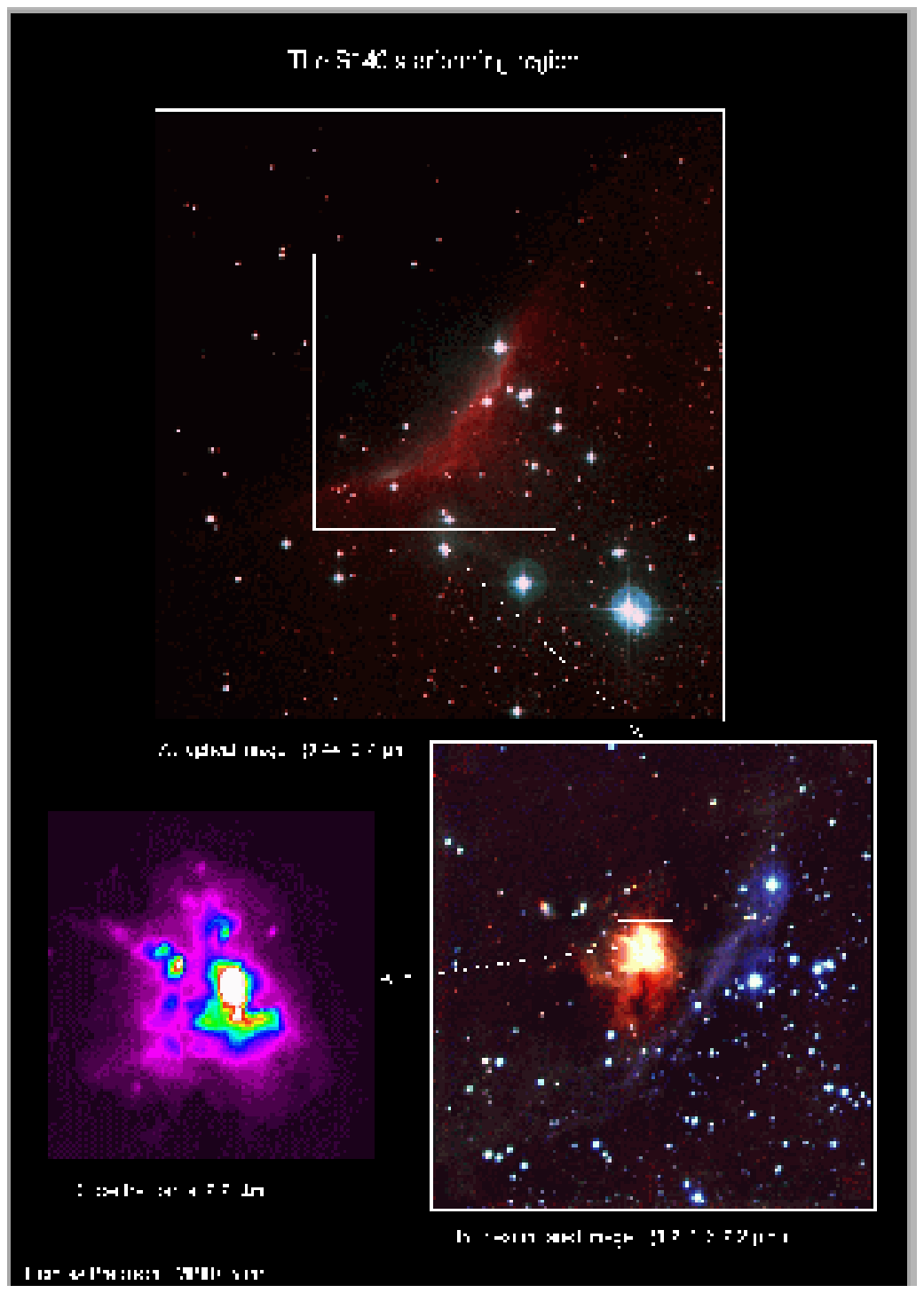}
\caption{Optical and NIR images of S\,140 \citep{Prei02}.}
\label{fig_S140_opt_nir}
\end{figure*}

A self-consistent model of the region, consistent with all the molecular, atomic and submm
continuum data was provided by \citet{Minchin95b} (see Fig.~\ref{fig:Minch}). According
to their model the eastern ridge is the dense, clumpy edge of the blueshifted outflow
lobe that is closest to the observer. This outflow has expanded towards the edge of the
molecular cloud so its blueshifted lobe is bounded by the HII region. Outside this edge is
an externally illuminated PDR. The CI emission emanates from the outer edge of the
cloud, with the CS emission tracing the compressed high density gas between the expanding
outflow and PDR regions. The NH$_3$ and continuum emission emanate from the inner edge
of the outflow lobe, shielded from the external UV field.

Optical and near-infrared images of S\,140, adopted from \citet{Prei02} are displayed in
Fig.~\ref{fig_S140_opt_nir}. According to the catalog of \citet{Porr03} the S\,140 region
contains two young stellar groups. One is S\,140 itself, another one is S\,140\,N identified
by \citet{Hodapp}.

\citet{Schwa89a} found that the IRS\,1 radio source consists of a
core source with a jetlike appendage pointing toward an extended radio source suggesting
ejection of an interstellar bullet of material from IRS\,1.

\citet{Harker97} observed the protostellar system in S\,140 at 2.2,
3.1 and 3.45~\micron.  They developed a simple model of the region
which has been used to derive the physical conditions of the dust and
gas. IRS\,1 is surrounded by a dense dusty disk viewed almost
edge-on. Photons leaking out through the poles of the disk illuminate
the inner edge of a surrounding shell of molecular gas as seen at
locations NW and VLA4.  Their thick disk model can explain both the
observed K$-$[3.45] color and scattered light intensity
distributions. The observed K$-$[3.45] color of the bluest regions
implies a cool radiation field with a color temperature of
850-900K. Most likely, these cool temperatures are the result of
reprocessing of the protostellar radiation field by dust close to the
protostar.

K band (2.0-2.3~\micron) and H$_2$ observations have revealed two bipolar outflows in the
region \citep{Prei02,Weig02}, one of them with an orientation similar
to the CO outflow (160/340$\deg$) and the other one in the 20/200$\deg$ direction.
Both bipolar outflows seem to be centered on IRS\,1.

\begin{figure*}[!htbp]
\centering
\includegraphics[draft=False,width=8cm]{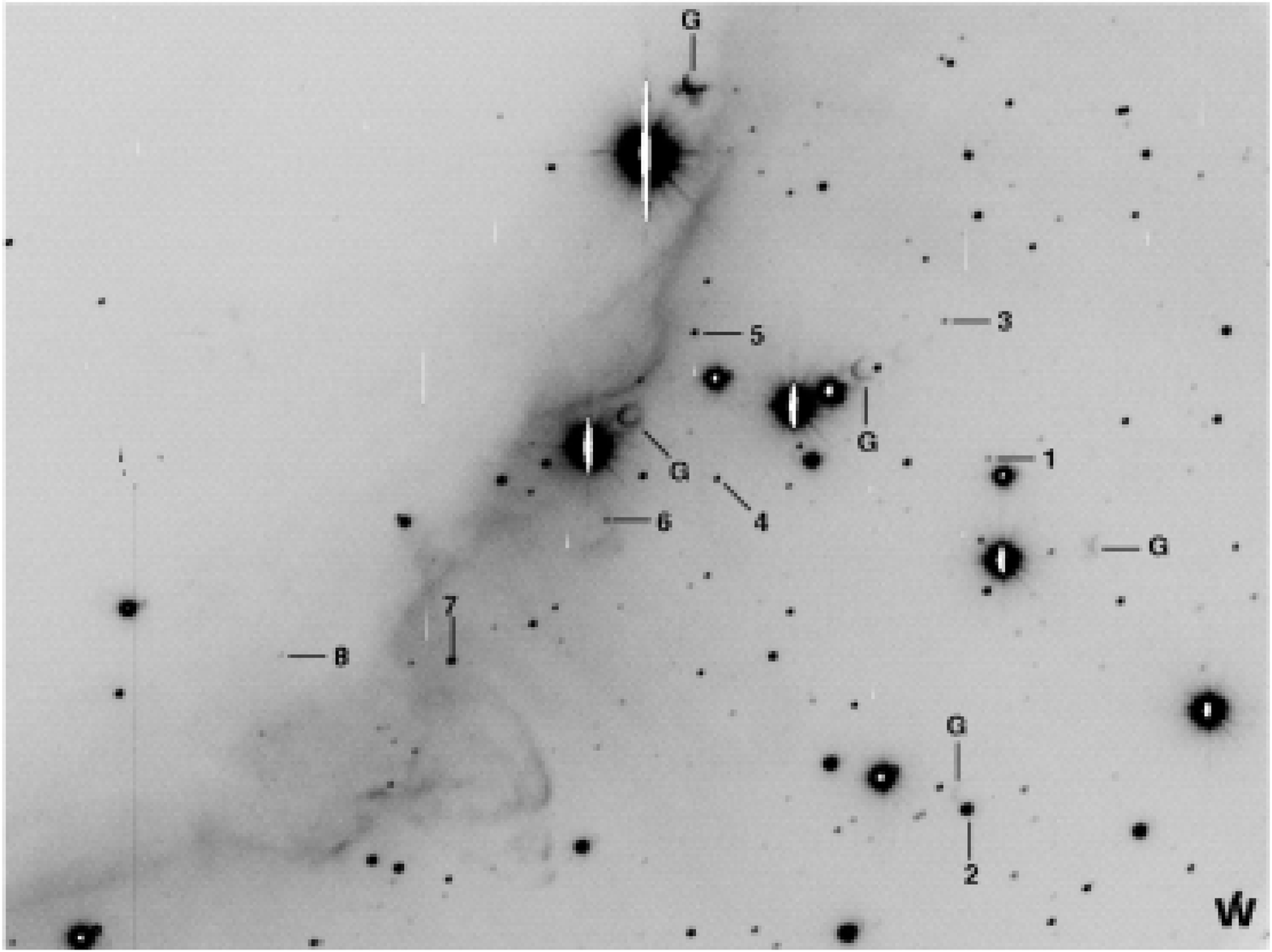}
\caption{Finding chart for H$\alpha$ emission objects in S\,140. G designate ghost images. \citep{OSP}}
\label{fig_S140_ha}
\end{figure*}

\citet{OSP} catalogued 8 stars with visible H$\alpha$ emission (Table~15,
Fig.~\ref{fig_S140_ha}). The emission line stars are mostly concentrated around the
tip of the bright rim, similarly to the distribution found for near-IR clusters in the
vicinity of an IRAS source.

\begin{table}[!htbp]
\stepcounter{table}
\label{Tab_S140_1}
\begin{center}
\smallskip
{\small {Table \thetable.~~~~ Position of far-infrared sources in S\,140}}\\
\smallskip
{\footnotesize
\begin{tabular}{lcc}
\tableline
\noalign{\smallskip}
number & RA (J2000)& Dec (J2000)\\
\noalign{\smallskip}
\tableline
\noalign{\smallskip}
IRS 1 & 22 19 18.4  & +63 18 55  \\
IRS 2 & 22 19 18.2  & +63 19 05 \\
IRS 3 & 22 19 19.6  & +63 18 50 \\
VLA 4 & 22 19 17.5  & +63 18 41  \\
NW    &  22 19 18.8 & +63 18 57 \\
\noalign{\smallskip}
\tableline
\end{tabular}}
\end{center}
\end{table}

\begin{table}[!htbp]
\caption{List of H$\alpha$ emission objects in S\,140 \citep{OSP,Ikeda08} }
\label{Tab_S140_2}
\smallskip
\begin{center}
{\footnotesize
\begin{tabular}{lccc}
\tableline
\noalign{\smallskip}
number & RA (J2000)& Dec(J2000) & EW(H$\alpha$) \\
\noalign{\smallskip}
\tableline
\noalign{\smallskip}
1 &22 18 47.8  & +63 18 18  & $\cdots$  \\
2 &22 18 48.5  & +63 16 40  &  13.6\\
3 &22 18 49.6  & +63 18 56  & $\cdots$  \\
4 &22 18 59.0  & +63 18 12  & 94.8 \\
5 &22 18 59.4  & +63 19 07  & $\cdots$  \\
6 &22 19 03.5  & +63 18 01  & $\cdots$ \\
7 &22 19 09.9  & +63 17 21  &  26.4 \\
8 &22 19 16.9  & +63 17 22  & 163.5 \\
\noalign{\smallskip}
\tableline
\end{tabular}}
\end{center}
\end{table}

\begin{table}[!hbp]
\caption{List of HH objects in S\,140}
\label{Tab_S140_3}
\smallskip
\begin{center}
{\footnotesize
\begin{tabular}{l@{\hskip2mm}c@{\hskip2mm}c@{\hskip2mm}l@{\hskip2mm}c}
\tableline
\noalign{\smallskip}
Name & RA (J2000)& Dec (J2000) & Remark from \citet{Bally02} & Ref. \\
\noalign{\smallskip}
\tableline
\noalign{\smallskip}
HH 615	&22 19 15.6	& +63 17 29	&[S II] jet aimed at HH 616A & 1 \\
HH 616A	&22 19 05.9	& +63 16 43	&Northern tip & 1 \\
HH 616B	&22 19 05.9	& +63 16 26	&Middle tip & 1 \\
HH 616C	&22 19 05.7	& +63 16 19	&Southern tip & 1 \\
HH 616D	&22 19 07.1	& +63 16 40	&Inner shock & 1 \\
HH 616E	&22 19 12.8	& +63 16 43	&[S II] edge, southern rim of HH 616 & 1 \\
HH 616F	&22 19 14.3	& +63 16 28	&[S II] edge, southeastern rim of HH 616 & 1\\
HH 617	&22 19 03.0	& +63 17 53	&Northern bow; tip of northern breakout & 1 \\
HH 623	&22 19 55.0	& +63 19 30	&Faint knot east of S\,140IR & 1 \\
HH 618A	&22 19 53.0	& +63 19 29	&Western part of pair, east of S\,140IR & 1 \\
HH 618B	&22 19 54.9	& +63 19 30	&Eastern part of pair, east of S\,140IR  & 1\\
Filament  & 22 18 52.1	& +63 16:08	&$H\alpha$ filament at P.A. = $300^{\circ }$ & 1 \\
HH 251	&22 19 34.4	& +63 32 57	& $-$ & 2\\
HH 252	&22 19 37.8	& +63 32 38	& $-$ & 2\\
HH 253	&22 19 45.0	& +63 31 45	& $-$ & 2\\
HH 254	&22 19 49.6	& +63 31 14	& $-$ & 2\\
HH 619	&22 19 16.4	& +63 32 49	&Two knots in east-west flow & 1\\
HH 620	&22 19 27.6	& +63 32 50	&Cluster of three knots south of nebular star & 1\\
HH 621	&22 19 21.5	& +63 34 44	&Cluster of knots: HH 251-254 counterflow & 1\\
HH 622	&22 19 50.6	& +63 35 18	&Pair of knots at P.A. = 220$^{\circ }$ from nebular star & 1\\
HH 609	&22 21 28.8	& +63 30 02	&Southwestern [S II] knot in chain of two & 1\\
HH 610	&22 21 33.3	& +63 37 34	&Tiny knot west of reflection nebula & 1\\
HH 611	&22 21 39.5	& +63 36 53	&Compact groups of [S II] knots & 1\\
HH 612	&22 21 54.5	& +63 34 39	&Compact diffuse [S II] knot & 1\\
HH 613	&22 21 58.5	& +63 33 23	&Faint [S II] group & 1 \\
HH 614	&22 22 01.2	& +63 27 56	&Diffuse [S II] complex & 1\\
\noalign{\smallskip}
\tableline
\noalign{\smallskip}
\multicolumn{5}{l}{\parbox{0.8\textwidth}{\footnotesize References: (1) \citet{Bally02}; (2) \citet{Eiro93}.}}
\end{tabular}}
\end{center}
\end{table}

\begin{table}[!htb]
\caption{List of Class II/I/0 objects in S\,140 \citep{Mege04}}
\label{Tab_S140_4}
\smallskip
\begin{center}
{\footnotesize
\begin{tabular}{ccccccc}
\tableline
\noalign{\smallskip}
RA (J2000)& Dec (J2000) & [3.6]& [4.5]& [5.8]&[8.0] & Class \\
\noalign{\smallskip}
\tableline
\noalign{\smallskip}
22 18 21.6 & +63 15 32 &12.38 &12.07 &11.65 &11.09 & Class II\\
22 18 37.2 & +63 13 01 &12.91 &12.49 &12.20 &11.65 & Class II\\
22 18 48.5 & +63 16 40 &9.94 &9.65 &9.22 &8.77 & Class II\\
22 18 47.6 & +63 18 17 &13.03 &12.82 &12.26 &11.38 & Class II\\
22 18 58.8 & +63 18 11 &11.84 &11.32 &10.50 &9.87 & Class II\\
22 19 03.4 & +63 18 00 &12.10 &11.78 &10.99 &10.13 & Class II\\
22 19 24.5 & +63 14 26 &11.81 &11.50 &11.50 &10.92 & Class II\\
22 19 09.7 & +63 17 20 &11.97 &11.62 &11.32 &10.62 & Class II\\
22 19 28.3 & +63 15 07 &12.35 &11.56 &11.43 &10.58 & Class II\\
22 19 25.9 & +63 18 24 &11.70 &10.90 &10.23 &9.50 & Class II\\
22 19 20.4 & +63 19 38 &10.74 &9.98 &9.59 &8.95 & Class II\\
22 19 28.5 & +63 18 49 &11.92 &11.32 &10.51 &9.54 & Class II\\
22 19 27.1 & +63 19 22 &9.80 &9.16 &7.93 &7.06 & Class II\\
22 19 48.7 & +63 16 41 &11.37 &11.22 &11.23 &10.46 & Class II\\
22 19 29.1 & +63 21 01 &13.64 &12.89 &12.29 &11.30 & Class II\\
22 19 38.1 & +63 19 32 &12.85 &12.56 &11.78 &10.83 & Class II\\
22 20 19.2 & +63 16 23 &13.01 &12.65 &12.27 &11.31 & Class II\\
22 20 21.0 & +63 16 14 &13.09 &12.62 &11.81 &10.90 & Class II\\
22 20 07.5 & +63 18 45 &13.32 &12.76 &12.24 &11.18 & Class II\\
22 19 37.0 & +63 25 31 &12.54 &12.36 &11.63 &10.65 & Class II\\
22 20 27.3 & +63 17 07 &11.51 &11.20 &11.02 &10.47 & Class II\\
22 20 27.2 & +63 17 58 &12.31 &11.58 &11.10 &10.34 & Class II\\
22 19 37.9 & +63 17 10 &11.43 &11.18 &10.83 &9.54 & Class II\\
22 19 15.6 & +63 19 33 &11.29 &9.75 &8.90 &7.87 & Class 0/I\\
22 19 25.7 & +63 18 49 &10.61 &9.56 &8.74 &8.12 & Class 0/I\\
22 19 30.9 & +63 18 32 &11.16 &10.13 &9.19 &8.54 & Class 0/I\\
22 19 32.5 & +63 19 24 &9.96 &8.38 &6.24 &4.83 & Class 0/I\\
22 19 39.4 & +63 19 03 &11.94 &10.92 &10.43 &9.70 & Class 0/I\\
22 19 43.5 & +63 20 08 &11.91 &11.18 &10.57 &9.30 & Class 0/I\\
22 19 52.3 & +63 19 01 &14.73 &12.78 &11.86 &11.00 & Class 0/I\\
22 19 48.3 & +63 20 27 &14.24 &12.42 &11.62 &10.86 & Class 0/I\\
22 19 45.5 & +63 21 21 &14.59 &13.01 &12.39 &11.47 & Class 0/I\\
22 20 18.5 & +63 18 57 &12.12 &10.73 &10.07 &8.91 & Class 0/I\\
22 20 19.4 & +63 19 05 &13.90 &12.75 &11.90 &11.05 & Class 0/I\\
22 19 35.1 & +63 20 26 &14.80 &13.40 &12.43 &12.14 & Class 0/I\\
\noalign{\smallskip}
\tableline
\end{tabular}}
\end{center}
\end{table}

\citet{Tafa93} observed the dense gas in the L\,1204/S\,140 molecular complex using
CS(J = 1-0) and NH$_3$. The large-scale CS(J = 1-0) maps show that L\,1204 is formed
by three filamentary clouds, each being fragmented into cores of a few hundred solar
masses and surrounded by low-level emission. The most prominent core is associated
with S\,140,  the star-forming activity, however, is not restricted to the vicinity
of the HII region, but extends throughout the
complex; very red IRAS sources lie close to most of the cores, and molecular outflows
have been detected in half of them. The ammonia observations reveal velocity shifts
of about 0.5-0.8~km/s in the dense gas inside the cores with embedded stars.
These velocity shifts, although small, are systematic and tend to divide the cores
into two velocity regimes with little overlap. Fast rotation of the cores or the
interaction between the bipolar outflows and the dense gas (or a combination of both)
are the most likely causes for these velocity shifts. \citet{Minchin96} studied the
structure of the magnetic field by measuring the 800~\micron \ polarization
at three positions towards S\,140.
Several studies reported detection of water maser emission toward the S\,140~IRS region
\citep[e.g.][]{ Lekht93, Tofani95, Lekht01, Trini03}.

\citet{Bally02} carried out a wide field CCD survey of Herbig--Haro objects in the S\,140
HII region and reported several new Herbig--Haro objects in the vicinity of S\,140
(Table~\ref{Tab_S140_3}).
They found two large bow shocks, HH\,616 and HH\,617. The northern shock, HH\,617, is probably
associated with the molecular hydrogen outflow from IRS\,3, while the source of the larger
velocity southern bow shock, HH\,616, is still unclear. It appears to trace an outflow from
an unknown source south of S\,140.

Recently a survey using InfraRed Array Camera (IRAC) on board the {\it Spitzer Space Telescope}
was carried out by \citet{Mege04}. They used the IRAC color plane to identify 12 Class 0/I and
23 Class II objects (Table~\ref{Tab_S140_4}). The list of \citet{Mege04} contains 5 stars (1, 2, 4, 6, 7
in Table~\ref{Tab_S140_2}) with H$\alpha$ emission from \citet{OSP}.

Using H- and K$_s$-band imaging polarimetry for S\,140 and
spectropolarimetry from 1.26 to 4.18 \micron \ for IRS\,1,
\citet{Yao98} discovered two reflection nebulae, illuminated by IRS\,1
and IRS\,3, which seem to be physically connected. Based on the
location and orientation of the reflection lobes around IRS\,1,
\citet{Yao98} suggest that S\,140 IRS\,1 may drive a quadrupolar
outflow. \citet{Schertl00} studied the structure of the envelope
around the central protostar in IRS\,1 using high resolution
bispectrum speckle interferometry and speckle polarimetry. Their high
resolution images showed bright emission which can be attributed to
light reflected from the inner walls of a cavity in the circumstellar
material around IRS\,1. Given that the orientation of the evacuated
cavity agrees with the direction of the molecular outflow they suggest
that the cavity has been carved out by the strong outflow from IRS\,1.
Recently \citet{Hoare} obtained multiepoch high-resolution radio
continuum maps of IRS\,1 using the full MERLIN array. The observations
revealed a highly elongated source that changes over time and is
perpendicular to the larger scale bipolar molecular outflow. He
explained the phenomenon with an equatorial wind driven by radiation
pressure from the central star and inner disk acting on the gas in the
surface layer of the disk.  \citet{Jiang08} obtained K-band
polarimetric images with the Coronagraphic Imager with Adaptive Optics
(CIAO) mounted on the Subaru telescope. They found that S140 IRS~1
shows well-defined outflow cavity walls and a polarization disk which
matches the direction of previously observed equatorial disk wind
\citep{Hoare}, thus confirming that the polarization disk is actually
the circumstellar disk.  \citet{Prei01} obtained a bispectrum speckle
interferometric K-band image with a resolution of 150~mas and a
seeing-limited molecular hydrogen line emission image of IRS\,3.
Their speckle image resolves IRS\,3 into three point sources, a close
binary with separation 0\farcs63 and a third component 1\farcs3
away. A rough assessment of the system stability suggests that the
IRS\,3 triple system is unstable. The speckle image also reveals
extended diffuse emission of very complex morphology around IRS\,3.

\citet{Trini07} present results of 1.3~cm continuum and H$_2$O maser
emission observations made with the VLA in its A configuration toward
IRS\,1 and also present results of continuum observations at 7~mm and
re-analyse observations at 2, 3.5 and 6~cm (previously published).
IRS\,1A is detected at all wavelengths, showing an elongated
structure. Three water maser spots are detected along the major axis
of the radio source IRS\,1A. They have also detected a new continuum
source at 3.5~cm (IRS\,1C) located some 0\farcs6 northeast of
IRS\,1A. The presence of these two YSOs (IRS\,1A and 1C) could explain
the existence of the two bipolar molecular outflows observed in the
region. In addition, they have also detected three continuum clumps
(IRS\,1B, 1D and 1E) located along the major axis of IRS\,1A, and
they discuss two possible models to explain the nature of IRS\,1A: a
thermal jet and an equatorial wind.

Several papers have studied the physical processes in photon dominated
regions of S\,140 using sub-mm and radio observations \citep[see
e.g.][and references therein]{Li02,Poel05,Poel06, Rodri07}.
\citet{Ashby00} detected H$_2$O in S\,140 using the Submillimeter Wave
Astronomy Satellite. They used Monte Carlo simulation to model the
radiative transport and to interpret the detected 557~GHz line
profiles. Their model required significant bulk flow in order to
explain the relatively single-peaked H$_2$O line.  However, they were
not able to discriminate between infall and outflow.

\subsection{L\,1188}

\begin{figure*}
\centering
\includegraphics[draft=False,width=5cm]{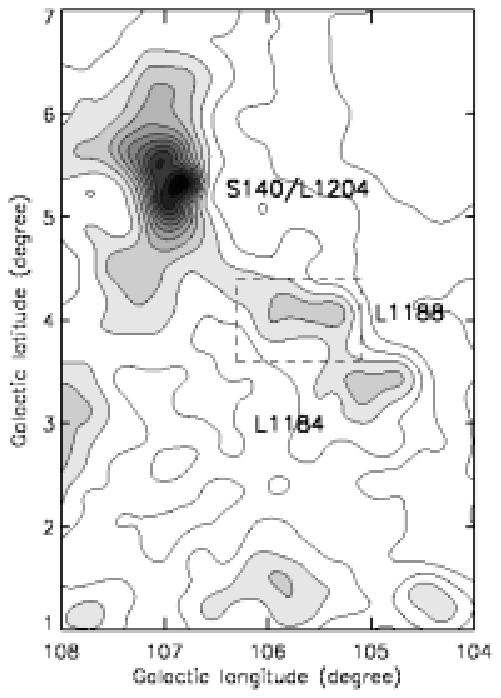}
\includegraphics[draft=False,width=5cm,angle=90]{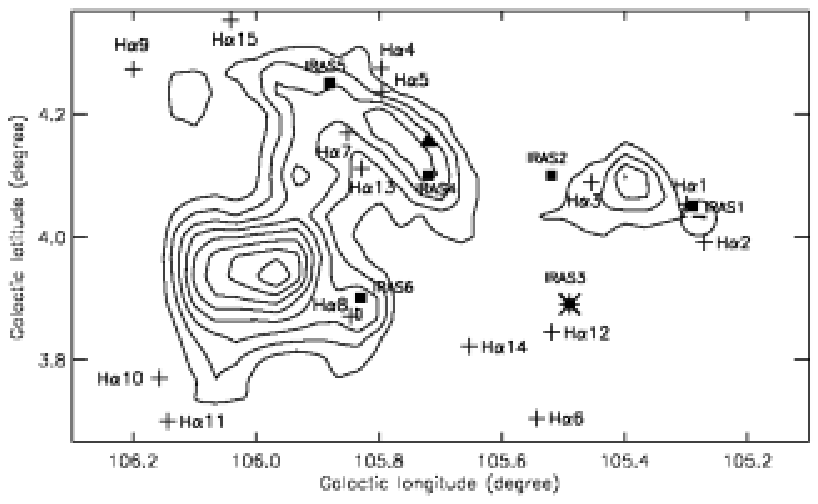}
\caption{Left: 100~\micron \  optical depth image of the L\,1188/L\,1204 region. Right:
Distribution of the molecular gas, IRAS point sources and H$\alpha$ emission stars
in the region of L\,1188  \citep[From][]{ADM95}.}
\label{fig_l1188}
\vspace{8mm}
\includegraphics[draft=False,width=\textwidth]{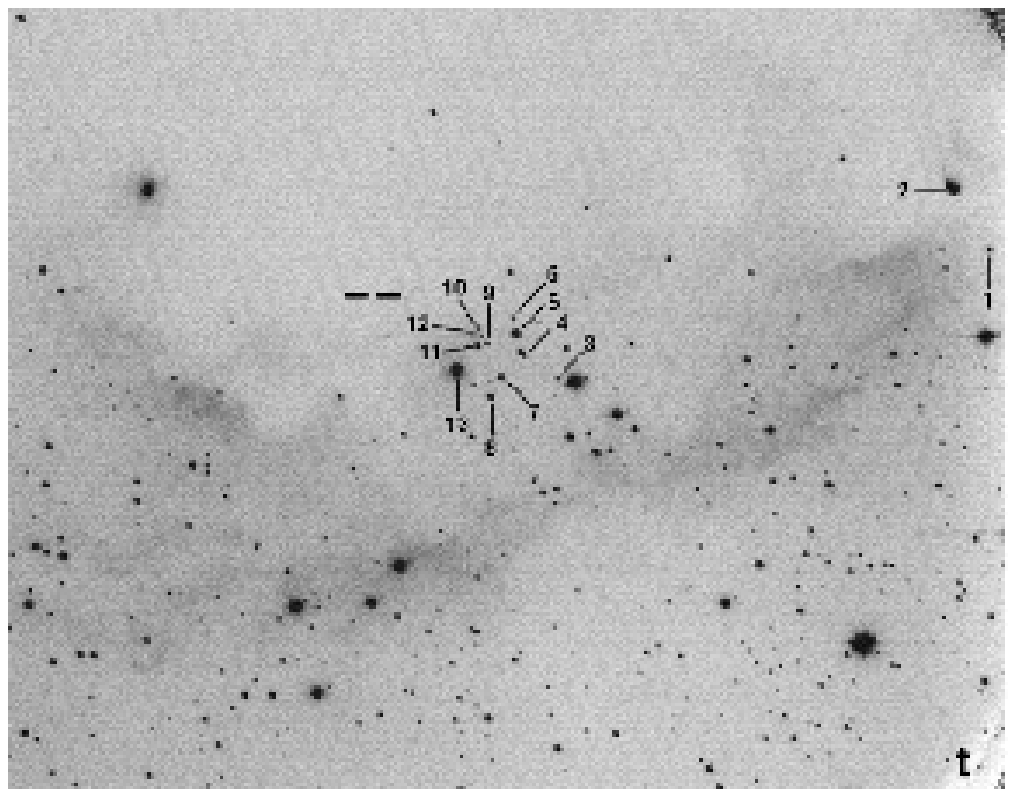}
\caption{Finding chart for H$\alpha$ emission objects in BRC\,44, a bright-rimmed
dark cloud associated with S145 \citep{OSP}.}
\label{Fig_brc44}
\end{figure*}

L\,1188 is one of the molecular clouds along the Cepheus Bubble. \citet{ADM95} mapped
the cloud in $^{13}$CO, and found a molecular mass of $\sim 1800~M_{\sun}$ within a field
of 74\arcmin$\times$44\arcmin. They selected 6 IRAS point sources as candidate YSOs in the
field, and found 15 H$\alpha$ emission stars during a photographic objective prism
survey. Figure~\ref{fig_l1188} shows the 100~\micron \  optical depth image of the
L\,1188/L\,1204 region, suggesting that these clouds are connected to each other,
and the distribution of the candidate YSOs in and around L\,1188. \citet{Konyves}
studied the SEDs of the IRAS sources associated with L\,1188 using 2MASS, MSX,
IRAS, and ISOPHOT data.

\subsection{S\,145}

S\,145 is an extended HII region, located at (l,b)=(107\fdg67, +5\fdg69), in the north-eastern part
of the Cepheus Bubble. Both its distance and velocity suggest its relation to the
bubble \citep{Patel98,KMT04}. S\,145 is associated with a bright rimmed cloud
BRC~44 \citep{SFO}, in which \citet{OSP} found 13 H$\alpha$ emission stars
(Table~\ref{Tab_brc44}, and Fig.~\ref{Fig_brc44}).

\begin{table}[!ht]
\caption{H$\alpha$ emission stars associated with BRC\,44 from \citet{OSP}
(EW-s and remarks revised by \citet{Ikeda08} are shown.)}
\label{Tab_brc44}
\begin{center}
{\footnotesize
\begin{tabular}{lcccl}
\noalign{\smallskip}
\tableline
\noalign{\smallskip}
N & RA(J2000) & Dec(J2000) & EW (\AA) & Remarks \\
\noalign{\smallskip}
\tableline
\noalign{\smallskip}
1   & 22 28 19.01 & 64 13 54.0 &  55.8 &	    \\
2   & 22 28 21.00 & 64 14 13.2 &  ~4.2 & reflection nebula ?	     \\
3   & 22 28 41.77 & 64 13 11.6 & $\cdots$ & contam. from nearby star	\\
4   & 22 28 43.54 & 64 13 19.1 & $\cdots$ & double star 	 \\
5   & 22 28 43.98 & 64 13 26.0 &  50.6 &	    \\
6   & 22 28 44.12 & 64 13 31.2 &  20.1 & very weak cont.	 \\
7   & 22 28 44.74 & 64 13 12.0 & $\cdots$ & contam. from nearby stars	 \\
8   & 22 28 45.31 & 64 13 05.6 &  22.5 & bad pix.	   \\
9   & 22 28 45.48 & 64 13 23.0 & $\cdots$ & contam. from No. 11 star	  \\
10  & 22 28 45.77 & 64 13 24.9 & $\cdots$ & contam. from Nos. 5 and 12 stars	\\
11  & 22 28 45.96 & 64 13 22.1 &  53.5 & contam. from No. 9 star       \\
12  & 22 28 46.24 & 64 13 25.6 & $\cdots$ & contam. from No. 5 star	 \\
13  & 22 28 47.12 & 64 13 14.3 &  ~9.5 &	    \\
\noalign{\smallskip}
\tableline
\noalign{\smallskip}
\end{tabular}}
\end{center}
\end{table}

\subsection{S\,134}

Two star-forming molecular clouds, associated with this HII region, can be
found in the literature. \citet{YDHS98} observed a head-tail structured molecular cloud
and CO outflow associated with IRAS~22103+5828, whereas \citet{Dobashi01} reported
on a CO outflow and molecular cloud associated with IRAS~22134+5834. The latter
source proved to be a high-mass protostar  at a very early evolutionary
stage \citep{Sridharan}. It is associated with H$_2$O maser emission \citep{Cesaroni}.

\section{Star Formation in the Association Cep OB3}

\citet*{BHJ59} made the first detailed photometric
investigation of the association Cep\,OB3. They found 40 early-type members
at 725~pc. \citet{Blaauw64} found evidence for two subgroups, Cep~OB3a and Cep~OB3b,
with ages of 8 and 4~Myr, respectively. The luminous stars of the younger subgroup,
Cep~OB3b, excite the HII region S\,155. \citet{Garmany} suggested an expansion age of
0.72 Myr, based on the relative motion of the two subgroups.
Seventeen of the 40 Cep~OB3 members compiled by \citeauthor{BHJ59} are
contained in the Hipparcos Catalog. However, \citet{deZeeuw}
could not identify Cep~OB3 as a moving group using the Hipparcos data.
\citet{Hoogerw01} have shown that the parent association of the runaway star
$\lambda$~Cep is probably not Cep~OB2, but Cep~OB3.

Several photometric studies \citep*{Crawford,Garrison,JTGE}
refined the \citeauthor{BHJ59}  membership list, and extended it to
fainter stars.  \citet{Moreno93} performed {\it JHK$^{\prime}$LM\/} photometry for the 40 luminous
stars in \citeauthor{BHJ59}'s list. A comprehensive summary of
all previous membership studies was given by \citet{JTGE}, who obtained
ages of 7.5 and 5.5~Myr for the two subgroups.

\citet{SVSG} found an expanding HI shell centered on the young
subgroup in Cep\,OB3 but did not detect significant H\,I
associated with the older subgroup.

\begin{figure*}[!ht]
\centering
\includegraphics[draft=False,width=\textwidth]{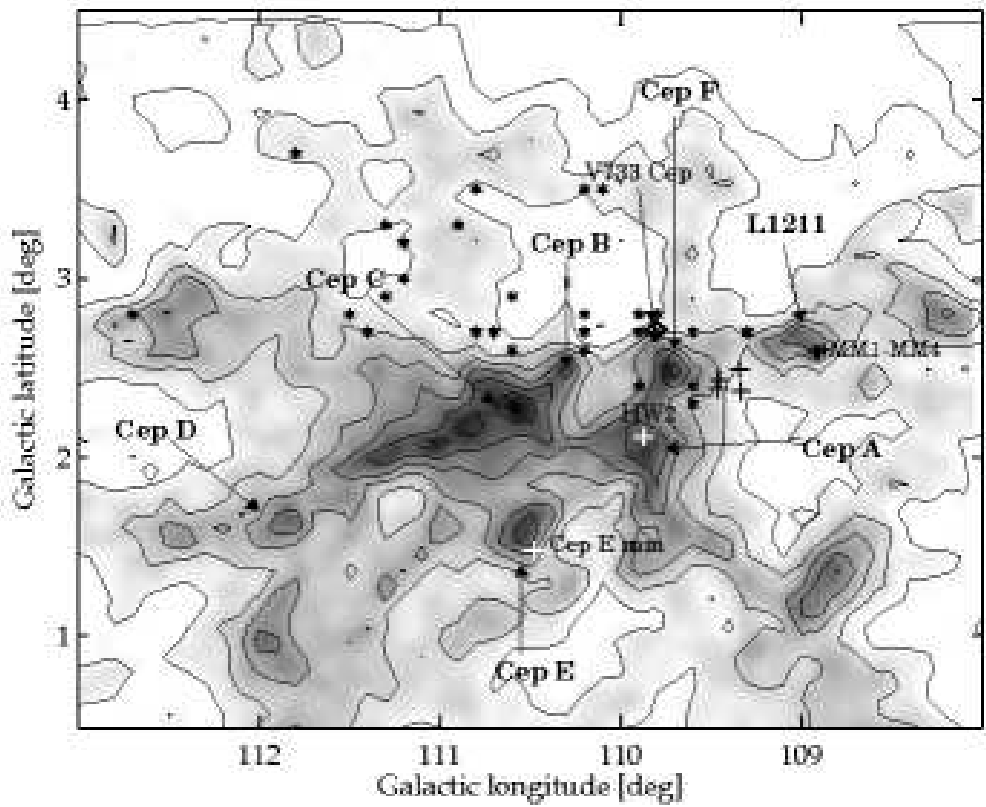}
\caption{Distribution of the visual extinction \citep{DUK} and the young, luminous stars in the
region of Cep~OB3b. The dense clumps Cep A--Cep F, identified in the distribution of CO
by \citet{Sargent77}, the dark cloud Lynds~1211, as well as the most prominent associated
young stars are labeled. The lowest contour of the extinction is at $A_V=1$~mag,
and the increment is 0.7~mag. Star symbols mark the luminous members of Cep~OB3, listed by
\citet{GS92}.}
\label{fig_cepob3_ext}
\end{figure*}

\citet{Sargent77,Sargent79} mapped the vicinity of Cep\,OB3 in the J=1-0 transition
of $^{12}$CO, and found a 20\,pc$\times$60\,pc molecular cloud complex at the
average radial velocity of $-10$~km\,s$^{-1}$, close to the velocity range
of the association members and S\,155. \citeauthor{Sargent77}'s CO observations
revealed several clumps in the Cep\,OB3 molecular cloud. She labeled them as
{\it Cep~A,B,C,D,E,F\/}, and concluded that some of them, especially Cep~A, are sites of
triggered star formation due to the interaction of the expanding HII region S\,155
and the molecular cloud. \citet{EL77} considered Cep~OB3 as one
of the examples of sequential star formation.
 \citet{FTHP} measured the thermal radio emission from
Cep~OB3 and S\,155. Ammonia maps around the IRAS sources associated with the Cep~A--Cep~F
clouds are presented in \citet*{HWW93}.
The distribution of the luminous stars of Cep~OB3 and the associated molecular cloud
as shown up in the extinction map of the region \citep{DUK} is displayed in Fig.~\ref{fig_cepob3_ext}.

\begin{figure*}[!ht]
\centerline{
\includegraphics[draft=False,width=9.cm,angle=90]{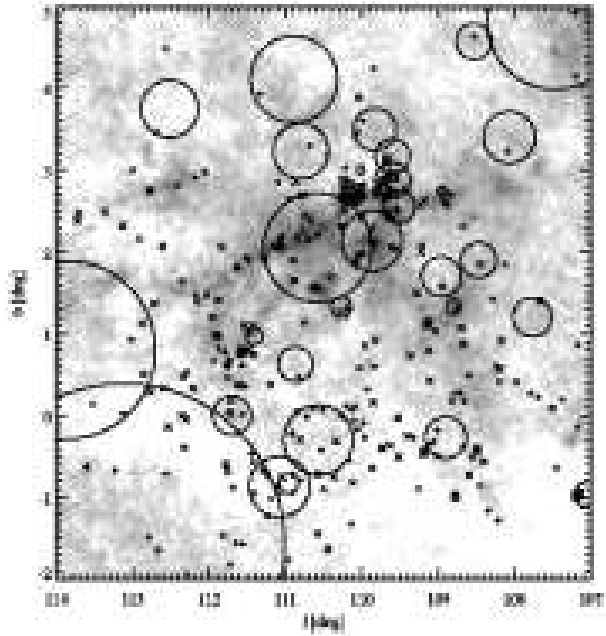}
}
\caption{Pre-main sequence stars and candidates in the Cep~OB3 region
overlaid on the map of visual extinction, obtained from 2MASS data
based on interstellar reddening using the NICER method
\citep{Lombardi}.  Large circles denote the clouds from
Table~\ref{Tab_cloud} associated with young stars. The meaning of
the different symbols are as follows: Filled triangles - T Tauri stars;
Filled squares - Herbig Ae/Be stars; Filled circles - Weak-line T
Tauri stars; Open squares - Photometric candidate and possible PMS
members; X - H$\alpha$ emission stars; + - T Tauri candidates selected
from 2MASS.}
\label{fig_cepob3s}
\end{figure*}

\subsection{Pre-main Sequence Stars and Candidates in Cep~OB3b}

\citet{NF99} discovered over 50 X-ray point sources
in the region of Cep OB3 with ROSAT PSPC and HRI, the majority of which
are probably T~Tauri stars. Using the ratio of high-mass to low-mass stars to constrain
the initial mass function, \citet{NF99} found that it is consistent with that for field stars.
Most of the T~Tauri stars are close to, but outside the molecular cloud.

\citet{PNJD} identified 10 T~Tauri stars and 6 candidates using {\it UBVI\/} photometry
and follow-up multi-fiber spectroscopy. Their optical survey covered an area of
some 1300~arcmin$^2$. The newly discovered pre-main sequence stars have masses in the
range $\sim 0.9 -3.0$\,M$_{\sun}$ and ages from $<$~1\,Myr to nearly 10\,Myr.
Out of the 10 definite TTS, four have a ROSAT X-ray counterpart in \citet{NF99}.

\begin{figure*}
\centerline{
\includegraphics[draft=False,width=5.25in]{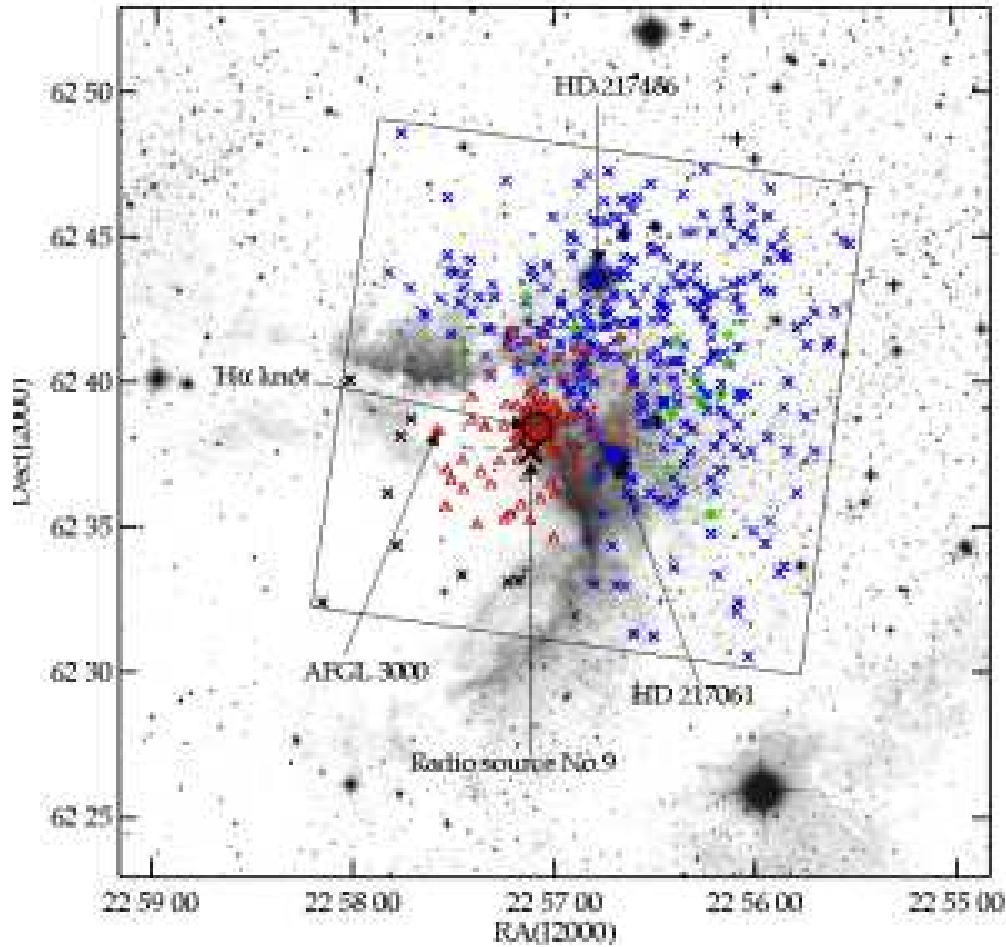}
}
\caption{R-band image covering $0\fdg5\times 0\fdg5$ of the Cep\,B and Cep OB3b
neighborhood from the Digitized Sky Survey (DSS).  North is up, and east is to the left. The Chandra
$17\arcmin \times 17\arcmin$ ACIS-I field  \citep{Getman06} is outlined by the square, and the dashed rectangle shows the
region in which \citet{OSP} searched for H$\alpha$ emission stars. Cep\,B, the hottest component
of the Cepheus molecular cloud, is at the bottom left corner of the  Chandra field. To the north and west lies
Cep~OB3b, the younger of two subgroups of the Cep~OB3 association. The interface between
Cep~B and Cep~OB3 is delineated by the H\,II region S\,155. The most massive and optically bright
stars in the field, HD\,217086 (O7n) and HD\,217061 (B1V), are labeled. Black plusses indicate the T~Tauri stars
identified by \citet{PNJD}. Blue crosses show the X-ray sources which are probably
members of a cluster belonging to Cep~OB3b. Red triangles indicate the X-ray emitting
members of an embedded cluster in the molecular cloud Cep~B, whereas green diamonds
show the X-ray sources whose 2MASS counterparts are indicative of K-band excess, originating from accretion
disks. Small, thick red plusses within the dashed rectangle show the H$\alpha$
emission stars found by \citet{OSP}. Black circle outlines the bright H$\alpha$ knot on the ionization front,
associated with a compact cluster and studied in detail by \citet{Moreno93} and \citet{TOH95}.
A star symbol shows the infrared source AFGL~3000, and a thick black cross is the bright radio continuum
source No.~9 discovered by \citet{FTHP}.}
\label{fig_xray}
\end{figure*}

\citet{Mikami} presented a list and finding charts of H$\alpha$ emission stars
in the region of Cep~OB3. Their objective prism survey covered an area of 36~square
degrees. They found 108 H$\alpha$ emission stars, 68 of which are new. The surface distribution
of the H$\alpha$ emission stars outlines a ring-like area, which almost coincides with that of the
heated dust shown by the IRAS images. The surveyed area is much larger than that occupied by
the stars of Cep~OB3, and extends to the south of the associated molecular cloud.  It includes NGC\,7419,
a cluster at 2\,kpc, and King~10, below the Galactic plane. Further objects, not associated with
Cep~OB3 but included in the surveyed area and lying along the shell-like surface, are S\,157,
NGC\,7654, and S\,158.

A portion of the Cep~OB3b and the molecular cloud Cepheus~B have been
observed with the ACIS detector on board the Chandra X-ray Observatory
\citep{Getman06}. The observations resulted in the discovery of two
rich clusters of pre-main sequence stars. The cluster projected
outside the molecular cloud is part of the association Cep~OB3b. The
X-ray observations detected 321 pre-main sequence members. This is the
best census of the stellar population of the region.  The results
suggest that the X-ray luminosity function, and thus probably the IMF,
of Cepheus~OB3b differs from that of the Orion Nebula Cluster: more
stars of $M < 0.3 M_{\sun}$ can be found in Cepheus~OB3b than in
Orion.

Figure~\ref{fig_cepob3s} shows the distribution of pre-main sequence stars and catalogued
clouds overlaid on the visual extinction in the Cep~OB3 region \citep{DUK}.
Figure~\ref{fig_xray} shows the field of view and the main results of the X-ray observations.

\subsection{Star Formation in the Molecular Cloud associated with Cep~OB3}

\subsubsection{Cepheus A} is a very active high-mass star-forming region within the
molecular cloud associated with Cep~OB3. It shows strings of sources
whose spectra suggest that some are thermal and some nonthermal
\citep{HW84,Hughes85,Hughes88}, several compact HII regions
\citep{Beich79,Rodriguez80a}, OH, H$_2$O, and CH$_3$OH masers
\citep*{BL79,WHH80,Lada81,Cohen84,Mehringer97,Patel07},
and strong infrared emission \citep{Koppen79,Beich79,Evans81} within an area smaller than
1~arcmin. Cep~A has been therefore an exciting target for high-resolution
interferometric observations and has a huge literature.

\begin{figure*}[!ht]
\centerline{
\includegraphics[draft=False,width=12.0cm]{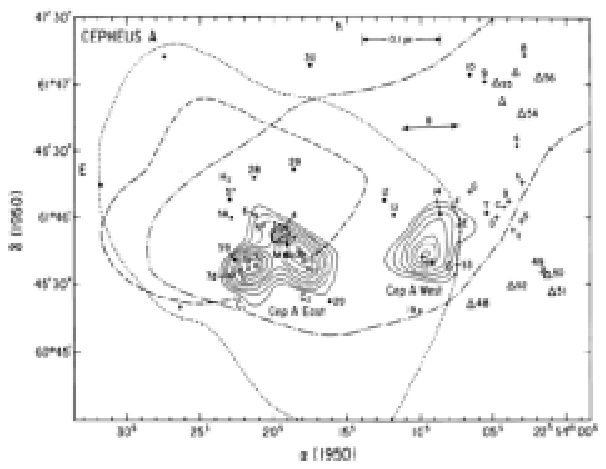}
}
\caption{Structure of Cepheus A \citep{HL85}. Solid lines show the 20~cm continuum contours
\citep{Rodriguez83}, dot-dash and long-dashed lines show the distribution of the redshifted
and blueshifted CO, respectively \citep{Rodriguez80a}, and a short-dashed line shows the
extent of the NH$_3$ emission \citep*{HMR82}. Triangles are reflection nebulae, plus
signs indicate HH objects, and filled circles are visible stars. The hatched circular
area is an extended 20~\micron \ emission area \citep{Beich79}. Small numbered open circles show
the 6~cm continuum sources, detected by \citet{HW84} and shown in more detail in
Fig.~\ref{fig_HW84}.}
\label{fig_cepa}
\end{figure*}

\begin{figure*}[!ht]
\centerline{
\includegraphics[draft=False,width=10.0cm]{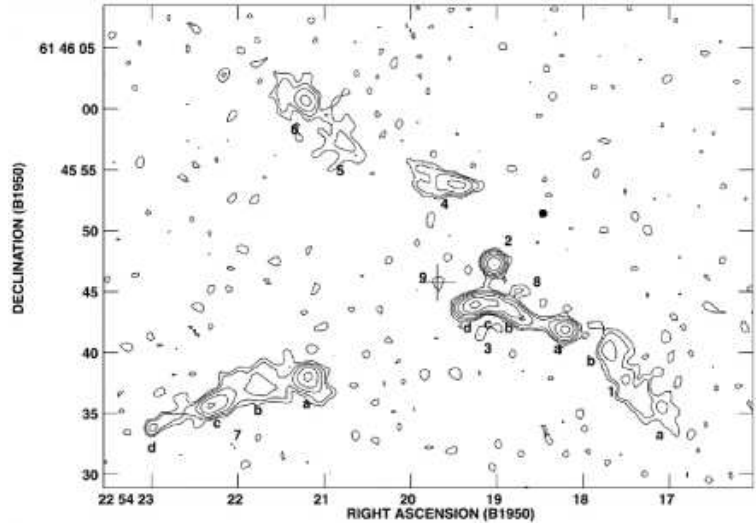}
}
\caption{VLA map at 6~cm of Cep~A East, observed by \citet{HW84}, displaying two chains
of 14 compact sources.}
\label{fig_HW84}
\end{figure*}

A powerful molecular outflow was discovered by \citet{Rodriguez80a} and studied in
further detail by among others \citet{Richardson87}, \citet{Torrelles87}, \citet*{Hayashi88},
\citet{Bally90}, \citet{Torrelles93}, \citet{Narayanan96}, and \citet{Froebrich02}.
Water maser emission has been detected from several centers of activity
\citep{Torrelles96}, and numerous
thermal and nonthermal radio sources \citep{Garay96,Hughes01}.
The HH object HH~168 (original name GGD\,37), consisting of several knots
\citep{HL85}, lies about 2\arcmin \ west of Cep~A. It was studied in detail by
\citet{HL85}, \citet{Lenzen84}, \citet{Hartigan86}, \citet{Lenzen88}, \citet{Garay96},
\citet{Wright96}, and \citet{HMB}. An apparent
counter-flow of HH~168, HH~169 was discovered by \citet{Lenzen88} 2\arcmin \ northeast of Cep A.
The objects are part of a larger, elliptical region containing several fainter
HH objects \citep{CRM}. A comprehensive summary of the literature of HH~168 and 169
can be found in \citet{HHcat}.

\citet{HW82} mapped Cep~A at 21~cm. The map has shown the presence
of two sources, Cep~A West and Cep~A East, separated by $\sim 1\farcs5$. Cep~A West is
associated with optical nebulosities and an optically visible star at its peak contour level.
It is named {\em HW~object\/}  \citep{HL85}, and appears to be an H\,II region on the
near side of the cloud.
\citet{Hughes89} obtained radio maps of Cep~A West, and found it to consist of two
compact sources, \normalsize{W\,1} and \normalsize{W\,2}. The first component is constant in time, while the
second is variable, and there is a third, diffuse component, \normalsize{W\,3}. The HW object
was found to be nonstellar, radiating mainly in H$\alpha$, and suggested to be an
HH object. \citet{Garay96} studied in detail the three sources within Cep~A West,
and found that the energy source, powering the activity observed in Cep~A West, is probably
\normalsize{W\,2}, associated with a low-luminosity embedded pre-main sequence star, whereas
emission of the shocked gas flowing from \normalsize{W\,2} can be observed from the diffuse
component \normalsize{W~3}. \citet{Wright96}, based on observations by ISO SWS, studied the molecular
hydrogen emission from the GGD~37 complex  in Cep~A West.

OH and H$_2$O maser sources are situated near the center of Cep~A
East, which appears younger and more heavily extincted than Cep~A
West. \citet{HW84} performed radio observations of Cep~A East, with
resolutions down to 1\arcsec \ at 21~cm and 6~cm, using both the
Westerbork Synthesis Radio Telescope and VLA. The maps have shown two
strings of 14 compact radio sources, numbered as HW 1a, 1b, 2, 3a--d,
4, 5, 6, 7a--d, (see Fig.~\ref{fig_HW84}) which were interpreted as
HII regions, being produced by about 14 stars, each of which mimics
main-sequence B3 stars; the length of each string is about 0.1 pc.
\citeauthor{Hughes93} \citeyearpar{Hughes88,Hughes93} reported on the
variability and high proper motion of some compact radio sources of
Cep~A East and discovered two new, highly variable compact radio
sources (sources 8 and 9). He suggested that, contrary to the original
interpretation, some of the compact sources are probably not H\,II
regions, but Herbig--Haro objects. \citet{Garay96}, based on
multifrequency, high resolution radio continuum observations,
classified the 16 compact sources into two groups: sources 2, 3a,3c,
3d, 8, and 9 harbor an energy source, whereas sources 1a, 1b, 4, 5, 6,
7a, 7b, 7c, and 7d are excited by an external source of energy.  Of
the stellar sources, HW\,2, 3c, and 3d are probably associated with
high luminosity stars, while the variable sources 3a, 8, and 9 are
probably low-mass pre-main sequence stars. The nature of source 3b
remained uncertain.  \citet{Torrelles98} detected a new continuum
source (Cep A:VLA 1) in an 1.3~cm VLA map.  \citet{Goetz98} present
new infrared images, including near-infrared broadband (K,
L$^{\prime}$, and M$^{\prime}$) and spectral line ([Fe II]
emission line at 1.644~\micron \ and H$_2$ 1-0 S[1] line at
2.122~\micron) observations of Cep~A East. The images show two
regions of shock-excited line emission from separate bipolar flows.
Figure~\ref{fig_cepa}, adopted from \citet{HL85}, shows the schematic
structure of the region of Cep~A, and Fig.~\ref{fig_HW84} shows the
distribution of the radio continuum sources in Cep~A East, discovered
by \citet{HW84}.

Both the compact radio continuum and H$_2$O maser sources in Cep~A exhibit remarkable
variations on various time scales. \citeauthor{Hughes01}
\citeyearpar{Hughes85,Hughes88,Hughes93,Hughes01} reported on the variability of
sources HW\,2, 3c, and 3d, and pointed out that the strong variability results in appreciable
changes in the spectra. Variations of H$_2$O maser emission have been detected by
\citeauthor{Mattila85} \citeyearpar{Mattila85,Mattila88}, \citet{Cohen85}, and \citet{Rowland86}.

\citet{Patel07}, using the  Submillimeter Array (SMA), detected the
321.226 GHz,
$10_{29}-9_{36}$ ortho-H$_2$O maser emission from Cep~A. The 22.235~GHz, $6_{16}-5_{23}$ water
masers were also observed with the Very Large Array 43 days following the SMA observations.
Three of the nine detected submillimeter maser spots are associated with the centimeter
masers spatially as well as kinematically, while there are 36 22~GHz maser spots without
corresponding submillimeter masers. The authors interpret the submillimeter masers in
Cepheus~A to be tracing significantly hotter regions
(600-2000~K) than the centimeter masers.

\citet{Rodriguez94} obtained multifrequency VLA radio continuum
observations of HW\,2, the most luminous radio continuum source of the region. They have
shown HW\,2 to be a powerful thermal radio jet, and suggest that it is responsible
for at least part of the complex outflow and excitation phenomena observed in the region.
HW\,2 proved to be a complex object, consisting of several components
\citep[e.g.][]{Gomez99,Curiel02,Curiel06,Jimenez07,Brogan07}, including a hot core \citep{MP05}.
\citet{Torrelles01} report three epochs of VLBA water maser observations toward HW\,2.
VLBA data show that some of the masers detected previously with the VLA \citep{Torrelles98}
unfold into unexpected and remarkable linear/arcuate ``microstructures,'' revealing, in particular
three filaments (R1, R2, R3) with length sizes $\sim$~3--25~mas (2--18 AU) and unresolved in the
perpendicular direction ($\la~0.1$~AU), an arcuate structure (R4-A) of  $\approx$~20~mas size (15~AU),
and a curved chain of masers (R5) of $\approx$~100~mas size
($\approx$~72~AU). Some of these structures unfold into even smaller linear ``building blocks''
(down to scales of 0.4~AU) shaping the larger structures.

\citet{Jimenez07} present VLA and PdBI subarcsecond images ($0.15\arcsec-0.6\arcsec$)
of the radio continuum emission at 7~mm and of the SO$_2$ $J=19_{2,18}-18_{3,15}$ and
$J=27_{8,20}-28_{7,21}$ lines toward the Cep~A HW\,2 region. The SO$_2$ images reveal
the presence of a hot core internally heated by an intermediate-mass protostar, and
a circumstellar rotating disk around the HW\,2 radio jet of size $600\times100$~AU and mass
~1~M$_{\sun}$. The high-sensitivity radio continuum image at 7~mm shows,
in addition to the ionized jet, an extended  emission to the west (and marginally
to the south) of the HW2 jet, filling the southwest cavity of the HW\,2 disk.

\citet{Torrelles07} report SMA 335~GHz continuum observations with angular resolution of
$\sim$\,0\farcs3, together with VLA ammonia observations with $\sim$\,1\arcsec \ resolution
toward Cep~A HW\,2. The observations have shown a  flattened disk structure of the dust emission
of $\sim$\,0\farcs6 size (450 AU), peaking on HW\,2.
In addition, two ammonia cores were observed, one associated with a
hot core previously reported and an elongated core with a double peak separated by
$\sim$\,1\farcs3, with signs of heating at the inner edges of the gas facing HW\,2. The
double-peaked ammonia structure, as well as the double-peaked CH$_3$CN structure
reported previously (and proposed to be two independent hot cores), surround both the
dust emission as well as the double-peaked SO$_2$ disk structure found by
\citet{Jimenez07}.

\begin{figure*}[!ht]
\centerline{
\includegraphics[draft=False,width=8.5cm]{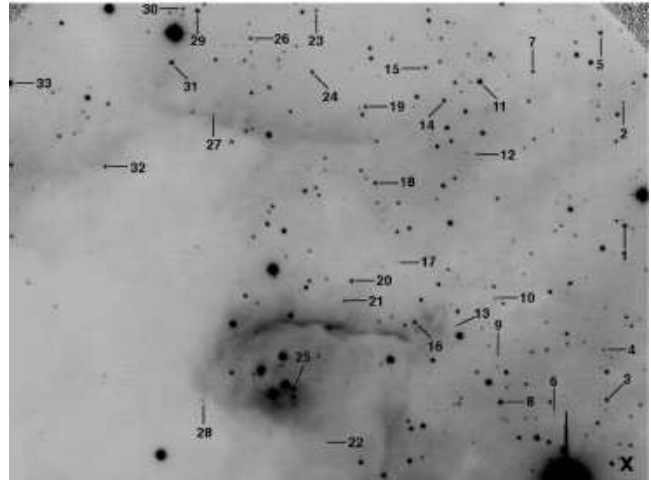}
}
\caption{H$\alpha$ emission stars in Cep~B found by \citet{OSP}. }
\label{fig_cepb}
\end{figure*}

\citet{Pravdo05} report the discovery of X-rays from both components of Cepheus~A, East and
West, with the XMM-Newton observatory. They detected prominent X-ray emission from the complex
of compact radio sources and call this source HWX.
Its hard X-ray spectrum and complex spatial distribution may arise from one or more
protostars associated with the radio complex, the outflows, or a combination of the two.
They also detected 102 X-ray sources, many presumed to be pre-main sequence stars on the
basis of the reddening of their optical and IR counterparts.

\citet{Sonnentrucker} report the first fully sampled maps of the distribution of interstellar
CO$_2$ ices, H$_2$O ices and total hydrogen nuclei, as inferred from the 9.7\,\micron \
silicate feature, toward Cepheus~A East with the IRS instrument on board the Spitzer Space Telescope.
They find that the column density distributions for these solid state features all peak at, and
are distributed around, the location of HW2. A correlation between
the column density distributions of CO$_2$ and water ice with that of total
hydrogen indicates that the solid state features mostly arise from
the same molecular clumps along the probed sight lines.

\citet{Comito07} employed the Plateau de Bure Interferometer to acquire (sub-)arcsecond-resolution
imaging of high-density and shock tracers, such as methyl cyanide (CH$_3$CN) and silicon
monoxide (SiO), towards the HW2 position. They find that on the 1~arcsec ($\sim$~725 AU) scale,
the flattened distribution of molecular gas around HW2 appears to be due
to the projected superposition, on the plane of the sky, of at least three
protostellar objects, of which at least one is powering a molecular outflow at a
small angle with respect to the line of sight. The presence of a protostellar disk
around HW2 is not ruled out, but such structure is likely to be detected on a smaller
spatial scale, or using different molecular tracers.

\subsubsection{Cepheus B}  is located at the edge of the HII region S\,155.
\citet{FTHP} and \citet{PT81} suggested that a younger subgroup of
Cep\,OB3 originated from the Cep~B/S\,155 complex. Several features of the Cep~B/S\,155
interface indicate triggered star formation in Cep~B, for instance a bright H$\alpha$ nebula
located near the ionization front, referred to as the H$\alpha$ knot by \citet{Moreno93} and
\citet{TOH95}, a compact radio continuum source (source \#9) detected by \citet{FTHP},
and the bright infrared source AFGL\,3000.

\citet{Moreno93} studied the  S\,155/Cep~B interface with H$\alpha$
and \linebreak $BV(RI)_C$
imaging, and identified  a cluster of pre-main sequence stars in the H$\alpha$ knot.
\citet{TOH95} performed radio and near infrared observations of the H$\alpha$ knot.
The unresolved radio source \#9  lies on top of the diffuse emission.
Far infrared and high resolution CO observations indicate that an embedded
B1--B0.5 star is the source of heat for the molecular hot spot and
the source of ionization of \#9.  More than 100 low luminosity stars
have been found in an area of about $3\arcmin\times2$\arcmin, and most of them lie above and to
the right of the main sequence. Many of them are associated with reflection nebulosities.
\citet{TOH95} concluded that they are pre-main sequence stars. They identified new Herbig~Ae/Be stars
among the cluster members. \citet{OSP} found 33 H$\alpha$ emission stars in Cep~B.
The list of these candidate pre-main sequence stars is given in Table~\ref{Tab_cepb_ha},
and the finding chart, adopted from \citet{OSP} is displayed in Fig.~\ref{fig_cepb}.
\citet{Getman06} identified 64 members of the cluster embedded in Cep~B, based on
deep X-ray observations with the Chandra Observatory (see Fig.~\ref{fig_xray}).
\citet{Mookerjea06} studied the emission from the photon dominated regions in Cepheus B,
based on $15\arcmin \times 15\arcmin$ fully sampled maps of [C\,I] at 492~GHz and
$^{12}$CO (4-3) observed at 1\arcmin \ resolution. They estimated the column
densities of neutral carbon in Cepheus B and studied
the factors which determine the abundance of neutral carbon relative to CO.

\begin{table}[!ht]
\caption{H$\alpha$ emission stars associated with bright rimmed cloud Cep~B \citep{OSP}}
\label{Tab_cepb_ha}
\begin{center}
{\footnotesize
\begin{tabular}{lccrl}
\noalign{\smallskip}
\tableline
\noalign{\smallskip}
N & RA(J2000) & Dec(J2000) & EW$^{*}$ & Remarks$^{*}$ \\
\tableline
\noalign{\smallskip}
1   & 22 56 37.97 & 62 39 51.1 &  69.7 &	    \\[-1pt]
2   & 22 56 38.12 & 62 40 58.7 & 104.7 & very weak cont.	 \\[-1pt]
3   & 22 56 39.33 & 62 38 15.5 &  80.1 &	    \\[-1pt]
4   & 22 56 39.58 & 62 38 43.1 &  17.2 & very weak cont.	 \\[-1pt]
5   & 22 56 39.93 & 62 41 37.1 &  14.6 & M-star ?	   \\[-1pt]
6   & 22 56 43.54 & 62 38 07.5 & $\cdots$ & invisible cont.	     \\[-1pt]
7   & 22 56 45.33 & 62 41 15.8 &   8.0 & M-star ?	   \\[-1pt]
8   & 22 56 47.79 & 62 38 14.0 &  22.4 &	    \\[-1pt]
9   & 22 56 48.02 & 62 38 40.2 & 125.6 & very weak cont.	 \\[-1pt]
10  & 22 56 48.23 & 62 39 11.1 &  62.3 & very weak cont.	 \\[-1pt]
11  & 22 56 49.54 & 62 41 10.0 &  21.4 &	    \\[-1pt]
12  & 22 56 49.77 & 62 40 30.1 &  81.3 & very weak cont.	 \\[-1pt]
13  & 22 56 51.41 & 62 38 55.8 & $\cdots$ & invisible cont.	     \\[-1pt]
14  & 22 56 52.40 & 62 40 59.6 &  63.6 & contam. from nearby star	     \\[-1pt]
15  & 22 56 53.87 & 62 41 17.7 &  19.0 &	    \\[-1pt]
16  & 22 56 54.65 & 62 38 57.8 &  16.8 & weak cont., contam. from bright rim	 \\[-1pt]
17  & 22 56 56.11 & 62 39 30.8 & $\cdots$ & invisible cont.	     \\[-1pt]
18  & 22 56 57.83 & 62 40 14.0 &  59.8 &	    \\[-1pt]
19  & 22 56 58.65 & 62 40 56.0 & $\cdots$ & H$\alpha$ ? 	 \\[-1pt]
20  & 22 56 59.68 & 62 39 20.2 &  76.4 &	    \\[-1pt]
21  & 22 57 00.22 & 62 39 09.4 &  24.9 & weak cont.	     \\[-1pt]
22  & 22 57 01.88 & 62 37 52.1 & $\cdots$ & invisible cont.	     \\[-1pt]
23  & 22 57 02.63 & 62 41 48.7 & $\cdots$ & contam. from nearby star	\\[-1pt]
24  & 22 57 02.93 & 62 41 14.9 &   4.9 &	    \\[-1pt]
25  & 22 57 04.31 & 62 38 21.1 & $\cdots$ & contam. from neighboring stars	    \\[-1pt]
26  & 22 57 07.86 & 62 41 33.2 &  23.1 & weak cont.	     \\[-1pt]
27  & 22 57 10.82 & 62 40 51.0 & $\cdots$ & invisible cont., contam. from bright rim	 \\[-1pt]
28  & 22 57 11.48 & 62 38 14.1 &  39.6 & very weak cont.	 \\[-1pt]
29  & 22 57 12.10 & 62 41 48.1 & $\cdots$ & contam. from No. 30 star	  \\[-1pt]
30  & 22 57 13.26 & 62 41 49.3 & $\cdots$ & contam. from No. 29 star	  \\[-1pt]
31  & 22 57 14.11 & 62 41 19.8 & $\cdots$ & double star, both show H$\alpha$ emission	   \\[-1pt]
32  & 22 57 19.38 & 62 40 22.5 & $\cdots$ & H$\alpha$ ?, contam. from bright. rim      \\[-1pt]
33  & 22 57 27.04 & 62 41 07.9 &   6.4 &	    \\[-1pt]
34N & 22 57 04.93 & 62 38 23.2 &  14.2 &	    \\[-1pt]
35N & 22 57 05.91 & 62 38 18.4 &  10.1 &	    \\[-1pt]
36N & 22 56 36.14 & 62 36 45.9 & $\cdots$ & invisible cont.	     \\[-1pt]
37N & 22 56 35.29 & 62 39 07.8 &   8.1 &	    \\
\noalign{\smallskip}
\tableline
\noalign{\smallskip}
\end{tabular}}
\end{center}
{\footnotesize $^{*}$Column revised by \citet{Ikeda08}}
\end{table}

\subsubsection{Cepheus C} The mass of this clump, estimated from the
formaldehyde observations obtained by \citet{Few83} is $\sim$~3600~N$_{\sun}$,
which ranks Cep~C as the most massive clump of the Cep~OB3 molecular cloud.
The region contains a cluster of  infrared sources
\citep{Hodapp} and is associated with water maser
emission \citep{WW86} and an outflow \citep{Fukui}.
The Cep~C cluster, first identified in a near-IR survey by \citet{Hodapp},
was included in the Young Stellar Cluster survey performed by the {\it Spitzer Space Telescope\/}
\citep{Mege04}.
In addition to the near-IR cluster, the IRAC data show Class~I and II sources
distributed over a 3~pc diameter region. The molecular gas traced
by the C$^{18}$O is visible in the  IRAC images as filamentary dark clouds
obscuring a diffuse nebulosity  extending across the entire mosaic.
Two Class~I objects appear outside the C$^{18}$O emission; $^{13}$CO emission
is found toward both of these sources.

\subsubsection{Cepheus E} is the second most massive and dense clump ($M \sim$2100~M$_{\sun}$)
of the Cep~OB3 molecular cloud  according to the H$_2$CO map  \citep{Few83}.
An outflow was identified in Cepheus~E based on millimeter CO observations
\citep{Sargent77,Fukui}, followed by near-infrared and higher spatial
resolution CO studies \citep{Hodapp,Eisloffel96,LH197,NC98}.

The outflow is quite compact, and driven by the source
IRAS~23011+6126, also known as Cep~E-mm. The outflow is deeply
embedded in a clump of density 10$^{5}$~cm$^{-3}$ and nearly invisible
at optical wavelengths, with the exception of its southern lobe, which
is breaking through the molecular cloud and is seen as HH~377
\citep{DRB97,NC97,NG01,Ayala00}.  \citet*{Lefloch} have shown that
IRAS~23011+6126 is a Class~0 protostar. The properties of the outflow
have been thoroughly analyzed by \citet{Eisloffel96}, \citet{MM01} and
\citet{Smith03}.  H$_2$ and [FeII] images obtained by
\citet{Eisloffel96} have shown two, almost perpendicular outflows
emanating from Cep~E, suggesting that the driving source is a Class~0
binary.  Submillimeter and near-infrared line and continuum
observations by \citet{LH197} led to a similar conclusion.  With the
assumption that the morphology of the jet results from precession,
\citet{Terquem} inferred an orbital separation of 4--20~AU and disk
radius of 1--10~AU for the binary.

Hot molecular bullets were detected in the outflow by
\citet*{Hatchell}. A comparative study of the Cep E-mm source, in the
context of other well known Class~0/I sources, was carried out by
\citet{Froebrich03}.

Submillimeter observations by \citet{Chini} and far-infrared
photometry by \citet{Froebrich03} resulted in $L_\mathrm{submm} /
L_\mathrm{bol} = 0.017\pm0.001$, an envelope mass $M_\mathrm{env} =
7.0~M_{\sun}$, an estimated age of $3\times10^{4}$~yr, and an H$_2$
luminosity of 0.07~L$_{\sun}$, which confirm that Cep~E-mm belongs to
the Class~0 objects. At a distance\footnote{Throughout the
literature of Cep~E, the distance of 730\,pc is used.  This value is
not an independent estimate for this cloud, but rounded from the
725\,pc derived by \citet{BHJ59} and \citet{Crawford} for Cep~OB3
\citep[see][]{LH197}.} of 730~pc,  Cep~E-mm is one of the brightest Class~0
protostars known and likely to become an intermediate-mass
(3~M$_{\sun}$) star \citep{MM01,Froebrich03}.

The Cep~E outflow and its protostellar source have been observed using
the three instruments aboard the Spitzer Space Telescope \citep{NC04}.
The new observations have shown that the morphology of the outflow in the
mid-infrared is remarkably similar to that of the near-infrared observations.
The Cep~E-mm source or IRAS~23011+6126 was detected in all four IRAC channels.
The IRAC and MIPS integrated fluxes of the Cep~E-mm source are consistent with
the Class~0 envelope models.

\subsubsection{Cepheus F} (L\,1216) contains
V733~Cep (Persson's star), the only known
bona fide FUor in the star forming regions of Cepheus, located at the coordinates
22:53:33.3, +62:32:23 (J\,2000). The brightening  of this star
was discovered by \citet{Persson} by comparing the old and
new Palomar Sky Survey plates. \citet{Reipurth07} have shown that the optical
spectrum of Persson's star exhibits all the features characteristic
of \normalsize{FU Ori} type stars. They also identified a molecular outflow associated with
the star. At an assumed distance of 800\,pc the observed apparent magnitude
$R \sim 17.3$ mag, together with the extinction $A_V \sim 8$ mag,
estimated from the strength of the water vapor features in the infrared
spectrum, corresponds to a luminosity of about 135\,$L_{\sun}$.
The star erupted sometime between 1953 and 1984.

\citet{Reipurth07} identified several nebulous near-infrared sources in L\,1216
around IRAS 22151+6215
\citep[Table~20, adopted from][]{Reipurth07}.
To the south of the aggregate of infrared sources containing Persson's star there is
an extended far-infrared source, Cep\,F(FIR) \citep{SDN83}, with a luminosity
of about 500\,L$_{\sun}$. Several IRAS sources can be found around this object
\citep[see Table~\ref{Tab_L1216_2}, adopted from][]{Reipurth07}, which
most probably form an embedded cluster containing a Herbig Ae/Be star.
A compact HII region without an obvious
IRAS counterpart, Cep\,F(HII), was discovered by \citet*{HTF81} to the south of
Cep\,F(FIR).

\begin{table}
\begin{center}
\parbox{0.5\textwidth}{\small Table 20.~~~Near-infrared
  sources in Cep\,F around IRAS 22151+6215 (Reipurth et al. 2007)}\\
{\footnotesize
\begin{tabular}{lccc}
\noalign{\smallskip}
\tableline
\noalign{\smallskip}
ID & RA(2000) & Dec(2000) & K$^{\prime}$ \\
\noalign{\smallskip}
\tableline
\noalign{\smallskip}
IRS\,1 & 22 53 40.7 & 62 32 02 & 16.3 \\
IRS\,2 & 22 53 41.1 & 62 31 56 & 13.7 \\
IRS\,3 & 22 53 40.9 & 62 31 49 & 18.1 \\
IRS\,4 & 22 53 41.0 & 62 31 48 & 17.4 \\
IRS\,5 & 22 53 41.2 & 62 31 48 & 15.6 \\
IRS\,6 & 22 53 43.3 & 62 31 46 & 14.2 \\
\noalign{\smallskip}
\tableline
\end{tabular}}
\end{center}
\label{Tab_L1216_1}
\vspace{-8mm}
\end{table}
\stepcounter{table}

\begin{table}
\caption{IRAS sources around Cep\,F(FIR) (Reipurth et al. 2007)}
\label{Tab_L1216_2}
\begin{center}
{\footnotesize
\begin{tabular}{lcccccrc}
\noalign{\smallskip}
\tableline
\noalign{\smallskip}
IRAS & RA(2000) & Dec(2000) & 12\micron \  & 25\micron \  & 60\micron \  & 100\micron \  & $L_{IRAS}$ \\
\noalign{\smallskip}
\tableline
\noalign{\smallskip}
22507+6208 & 22 50 47.9 & 62 08 16 & 0.54: & 1.00 & ~12.99: & 53.98~     & 18  \\
22152+6201 & 22 51 14.1 & 62 01 23 & 1.61  & 4.86 & ~13.11  & $<$104.62~ & 23 \\
22518+6208 & 22 51 53.5 & 62 08 02 & 2.07 & 1.92: & ~34.53 & 210.21: & 56  \\
22521+6205 & 22 52 08.0 & 62 05 51 & 0.62 & 1.84 & $<$21.48~ & 257.23: & 49 \\
\noalign{\smallskip}
\tableline
\end{tabular}}
\end{center}
\end{table}

\subsubsection{L\,1211} is a class~5 dark cloud (Lynds 1962) about 1\deg \ west of
Cepheus~A (see Fig.~\ref{fig_cepob3_ext}). The mass of this cloud, derived from
$^{13}$CO measurements, is 1900~M$_{\sun}$ \citep{YDMOF}.
Its angular proximity to the group of Cepheus A-F clouds and its similar LSR velocity
suggest that it is related to the group, and therefore
lies at a similar distance from the Sun \citep[725~pc, see][]{BHJ59,Crawford,Sargent77}.
\citet{Fukui} has reported a bipolar outflow around the embedded source IRAS~22453+6146.
\citet{HWW93} have mapped the ammonia emission around this source, finding a dense
molecular core, and \citet{Hodapp} has imaged the region in the K$^\prime$ band,
finding a small cluster of sources associated with diffuse emission.

\begin{figure*}
\centerline{
\includegraphics[draft=False,width=5.25in]{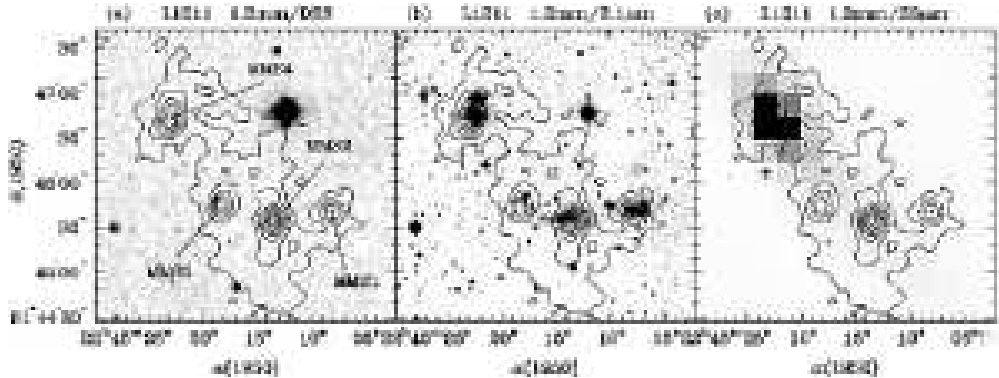}
}
\caption{Distribution of 1.2 mm emission (contours) in L\,1211 superimposed on images
of emission at different wavelengths. (a)  Optical DSS (red) image.
(b)  K$^{\prime}$ (2.1~\micron) image from Hodapp (1994).
(c)  25~\micron \  IRAS HIRES image.
Contours are at intervals of 25 mJy per  beam starting at 25 mJy per beam.
Adopted from \citet{Tafalla}.}
\label{fig_L1211}
\end{figure*}

\citet{Tafalla} conducted millimeter continuum and line observations
of a dense core in L\,1211. They have found a small cluster of at least 4 millimeter
sources with no optical counterparts, but each associated with near infrared diffuse emission.
The strongest mm source has no NIR point-like counterpart, and constitutes a good candidate
for a Class~0 object. The other mm objects appear associated with NIR sources and most
likely belong to Class~I, as also suggested by the spectral energy distributions derived
from combining mm data with IRAS {\it HIRES\/} fluxes. As evidenced by mm~line data, the mm
sources are embedded in an elongated, turbulent core of about 150\,M$_{\sun}$ of mass
and 0.6\,pc length. Two of the millimeter sources power bipolar
molecular outflows, another signature  of their extreme youth. These outflows are referred
to as the L1211-MMS1 and L1211-MMS4 outflows.
L\,1211 is included in the far-infrared (ISOPHOT) photometric studies of embedded objects
performed by \citet{Froebrich03}.
Table~22 lists the coordinates, millimeter fluxes and estimated masses of
the mm-sources in L\,1211, and Fig.~\ref{fig_L1211} shows their appearance at different
wavelengths  \citep[adopted from][]{Tafalla}.
Table~\ref{Tab_HH_CepOB3} lists the Herbig--Haro objects in the
Cepheus~OB3 molecular cloud.

\begin{table}[!ht]
\begin{center}
\parbox{0.5\textwidth}{\small Table 22. \hspace{8mm} L\,1211 millimeter sources}
{\footnotesize
\begin{tabular}{lcccc}
\noalign{\smallskip}
\tableline
\noalign{\smallskip}
Source & RA(2000) & Dec(2000) & Int. flux & Mass$^*$ \\
 &&& (mJy) & (M$_{\sun}$) \\
\noalign{\smallskip}
\tableline
\noalign{\smallskip}
MMS 1 & 22 46 54.5 & 62 01 31 & ~45 & 0.3 \\
MMS 2 & 22 47 07.6 & 62 01 26 & 215 & 1.3 \\
MMS 3 & 22 47 12.4 & 62 01 37 & ~85 & 0.5 \\
MMS 4 & 22 47 17.2 & 62 02 34 & 135 & 0.8 \\
\noalign{\smallskip}
\tableline
\noalign{\smallskip}
\end{tabular}}
\smallskip
{\parbox{0.53\textwidth}{\footnotesize $^*$ Assuming optically thin
    dust at 30~K with an opacity of 0.01 cm$^{2}$ g$^{-1}$}}
\end{center}
\label{Tab_L1211}
\end{table}
\stepcounter{table}

\begin{table}[!h]
\caption{Herbig--Haro objects in the Cepheus OB3 molecular clouds.}
\label{Tab_HH_CepOB3}
\begin{center}
{\footnotesize
\begin{tabular}{ l c c l l r l }
\tableline
\noalign{\smallskip}
Name & RA(2000) & Dec(2000) & Source & Cloud & d & Reference \\
\noalign{\smallskip}
\tableline
\noalign{\smallskip}
$[$H89$]$ W3  & 22 56 08.8 & +62 01 44 &                 & Cep A   &  700 & 2 \\
HH 168        & 22 56 18.0 & +62 01 47 & HW 2            & Cep A   &  700 & 1,5 \\
HH 169        & 22 56 34.8 & +62 02 36 & HW 2            & Cep A   &  700 & 3 \\
HH 174        & 22 56 58.5 & +62 01 42 & HW 2            & Cep A   &  700 & 4 \\
HH 377        & 23 03 00.0 & +61 42 00 & IRAS 23011+6126 & Cep E   &  700 & 6 \\
\noalign{\smallskip}
\tableline
\end{tabular}
}
\end{center}
\smallskip
{\footnotesize
References: 1 -- \citet{GGD}; 2 -- \citet{Hughes89}; 3 -- \citet{CRM};  4 --  \citet{Bally99};
5 -- \citet{HMB}; 6 -- \citet{DRB97}.}
\end{table}

\section{Star Formation in Cepheus OB4}

\subsection{Structure of Cep OB4}

\begin{figure*}[!ht]
\centerline{
\includegraphics[draft=False,width=11cm]{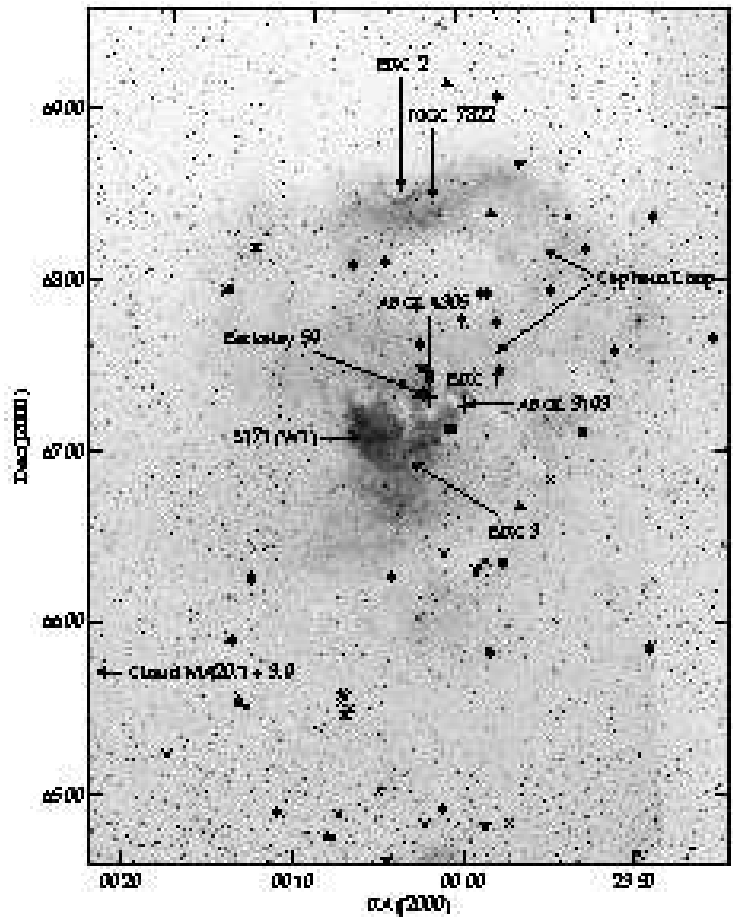}
}
\caption{A large scale red photograph of the Cep~OB4 region.
The infrared sources and the radio continuum loop are indicated on
the red DSS image. The HII region W1 is also known as S171. Star symbols show the
luminous stars of Cep~OB4 \citep{RGA83}, and crosses mark the H$\alpha$ emission
stars \citep{MacConnell}.}
\label{fig_cepob41}
\end{figure*}

Cep~OB4 was discovered by \citet{BW59}, who noticed the presence of 16
early-type stars in a small region around (l,b)=(118\fdg4,+4\fdg7),
including the cluster Berkeley~59. Cep~OB4 is related to a dense,
irregular dark cloud containing several emission regions, including
the dense H\,II region S\,171 (W\,1) in the central part, and
NGC\,7822 to the north of S\,171 \citep*{Lozinskaya}, see
Fig.~\ref{fig_cepob41}. We note that in the original catalogs both
W\,1 \citep[RA(J2000)=00 02 52; Dec(J2000)=+67 14,][]{Westerhout} and
S\,171 \citep[RA(J2000)=00 04 40.3; Dec(J2000)=+67 09,][]{Sharpless}
are associated with NGC\,7822, situated about one degree north of the
HII region, according to its catalog coordinates (RA(J2000)=00 03.6;
Dec(J2000)=+68 37). The {\it Simbad\/} data server also associates
these objects with each other.  A detailed description of the
association and related objects was given by \citet{MacConnell}. He
identified 42 members earlier than B8 at 845~pc.
\citeauthor{MacConnell}'s UBV photometric study of the luminous
members of Cep~OB4 revealed a correlation between the luminosity and
reddening of the stars: the O and early B stars were found only within
the cloud, whereas later B type stars are found only outside the cloud
due to the incompleteness of their detection. Based on the absence of
supergiants, an earliest spectral type of O7\,V, and the gravitational
contraction time of a B8 star, \citeauthor{MacConnell} estimated an
age between 0.6 and 6~Myr.

\citet{Lozinskaya}  studied the morphology and kinematics of the H\,II regions
associated with Cep~OB4 based on monochromatic images of the $[$OIII$]$, $[$NII$]$, $[$SII$]$
and H$\alpha$ lines, and found two expanding shells: one
shell, of radius $\sim$0\fdg7, connects NGC\,7822 and S\,171. Most Cep\,OB4 members are
located inside this shell; their energy input into the interstellar medium
can account for its observed size and expansion velocity of 10 km\,s$^{-1}$. The
other shell, of radius $\sim$1\fdg5, is centered on S\,171 and has an expansion velocity
of $\sim$30--40~km\,s$^{-1}$; it may be the result of a supernova explosion or of the stellar
wind of a massive star that so far has escaped detection.

\citet*{Olano} found that the space distribution and kinematics of the interstellar matter
in the region of Cep~OB4 suggest the presence of a big expanding shell, centered on
(l,b)~$\sim$~(122\deg,+10\deg). Assuming a distance of 800~pc for the center
they derived a radius of some 100~pc, expansion velocity of
4~km\,s$^{-1}$, and HI mass of $9.9 \times 10^{4}$~M$_{\sun}$ for the Cepheus~OB4~Shell,
whose approximate position is plotted in Fig.~\ref{cepheus:Fig1}.

Only 19 of the 42 classical members of Cep~OB4 are listed in the Hipparcos
Catalog \citep{deZeeuw}. This may be caused by a combination of crowding effects
and the large extinction toward Cep~OB4, $A_V > 3$~mag \citep{MacConnell}. Based on their
proper motions and parallaxes, three \citeauthor{MacConnell} stars (HIP 117724,
118192, 118194) are not associated with Cep~OB4. \Citet{deZeeuw} found that the Hipparcos
parallaxes of the other classical members are consistent with a distance of 800--1000~pc.

\citet*{RGA83} mapped the Cep~OB4 region in the 6\,cm transition of H$_2$CO, and detected
neutral gas at the velocities $-$13, $-7$, and $-1$~km\,s$^{-1}$. They established
that Cep~OB4 consisted of two kinematically distinct components, W1 west and W1 east,
and a loop-shaped, optically thin, thermal shell, the Cep Loop.
They modeled the observed morphology and kinematics as follows.
The association Cep~OB4 consists of two subgroups with differing ages and kinematic
properties. The older, dispersed, subgroup extends over an area about 4 degrees
(60\,pc) in diameter, clustered towards the Cep Loop. The younger subgroup, the
young cluster Berkeley~59, extends over an area of 15\arcmin \ (about 4\,pc) in diameter,
located along the southern edge of the Cep Loop.
The average velocity of the OB stars of the older subgroup is $-6$~km\,s$^{-1}$.
Thus \citet{RGA83} propose that the gas component at $-7$~km\,s$^{-1}$ represents
the undisturbed gas associated with the star forming region. Beginning with a
cloud complex having a velocity of $-7$~km\,s$^{-1}$, star formation occurred near the
center of what is now the Cep Loop. The Cep Loop was subsequently  formed by this
first generation of OB stars. Expansion of the Cep Loop into a cloud to the north
resulted in collisional excitation of the HII region NGC\,7822. In the south,
expansion of the Cep Loop resulted in fragmentation of the remainder of the original
dark complex producing clouds at $-13$ and $-1$~km\,s$^{-1}$. Berkeley~59 was formed
in this environment. Ionization of the clouds surrounding Berkeley~59 has resulted
in ionized gas at each velocity component. Ionization is now occurring most
actively in a $-1$~km\,s$^{-1}$ cloud west of Be~59.
\citet{Okada03} studied the photodissociation region associated with S\,171 using mid- to
far-infrared spectroscopy using the ISO SWS, LWS, and PHT-S instruments.
\citet{Gahm06} investigated the structure and velocity of an elephant trunk associated with
NGC~7822.
Figure~\ref{fig_cepob41}, based on Fig.~1 of \citet{RGA83}, shows a large scale red photograph of
the Cep~OB4 region. The infrared sources and the radio continuum loop are indicated.
Star symbols show the luminous stars of Cep~OB4, and crosses mark the H$\alpha$ emission
stars.

\subsection{Low and Intermediate Mass Star Formation in Cep~OB4}

\citet{MacConnell} found 24 H$\alpha$ emission-line (named as
MacC\,H1--H24) objects within the dark cloud in the region of Cep~OB4,
some of which may be T~Tauri stars.  \citet{CK76} obtained optical
spectrophotometry and infrared photometry of some MacC\,H stars,
determined their spectral types and luminosities, as well as masses
and ages using \citeauthor{Iben}'s \citeyearpar{Iben} pre-main
sequence evolutionary tracks. They identified four new nebulous stars
in the same field (MC\,1--MC\,4), three of which have shown H$\alpha$
emission.  Table~\ref{Tab_MacC} lists the H$\alpha$ emission stars
found by \citeauthor{MacConnell}, and nebulous H$\alpha$ emission
stars reported by \citet{CK76}, supplemented by the spectral types
determined by \citet{CK76}.

\begin{figure*}
\centerline{
\includegraphics[draft=False,width=6.5cm]{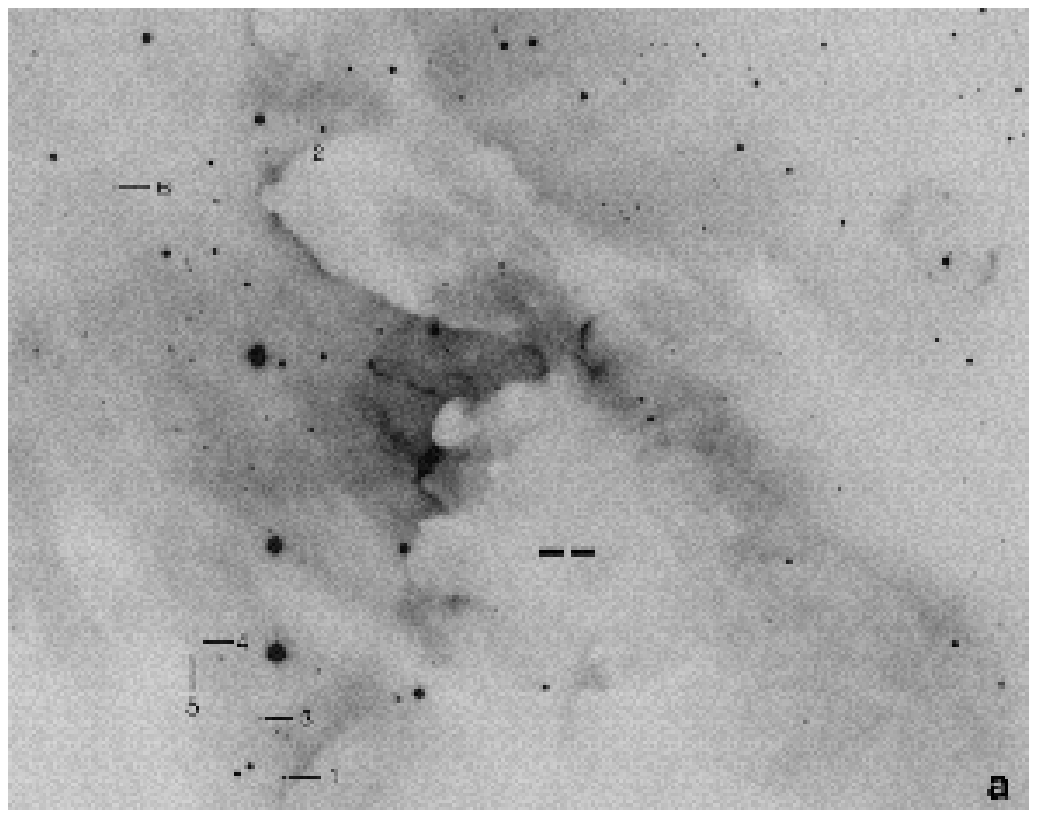}
\includegraphics[draft=False,width=6.5cm]{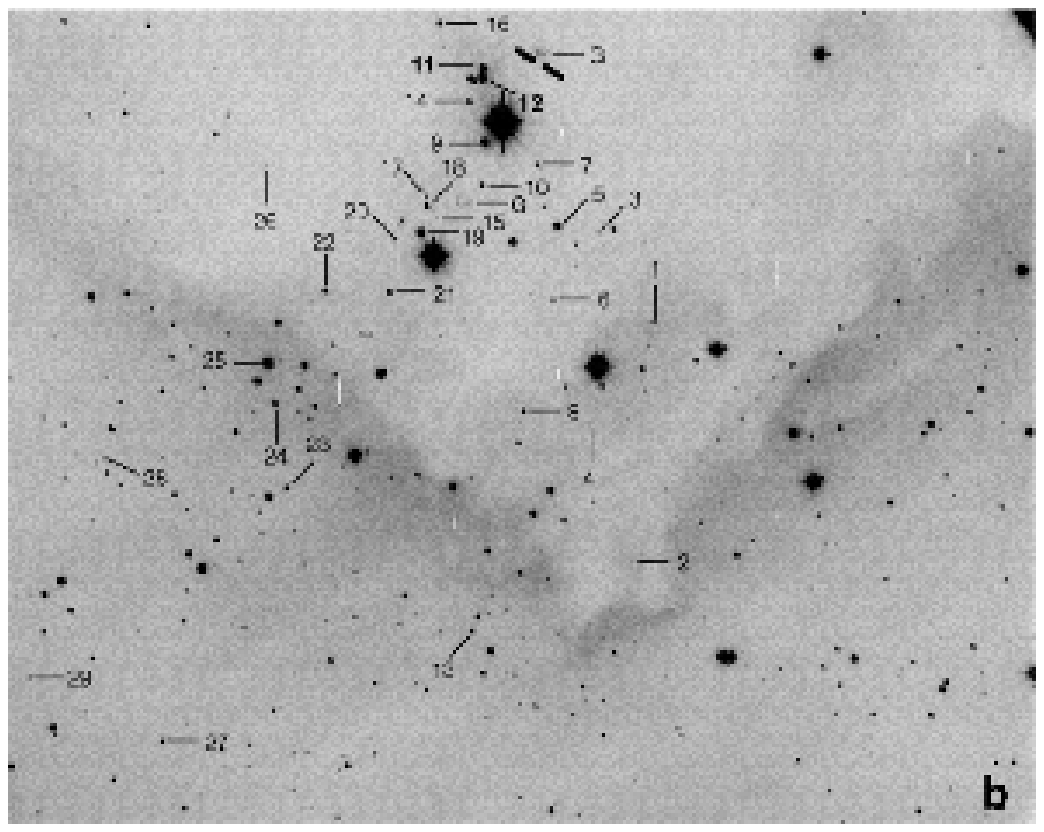}
}
\caption{H$\alpha$ emission stars discovered by \citet{OSP} in BRC\,1 (left) and
 BRC\,2 (right) associated with S\,171. The position of the IRAS source associated with the
 cloud is shown by a pair of thick tick marks.}
\label{fig_brc_s171}
\end{figure*}

\begin{table}[!ht]
\caption{Data for the H$\alpha$ emission stars found by MacConnell (1968) and
Cohen \& Kuhi (1976) in the region of Cep\,OB4}
\label{Tab_MacC}
\begin{center}
{\footnotesize
\begin{tabular}{l@{\hskip2mm}l@{\hskip2mm}c@{\hskip2mm}c@{\hskip2mm}r@{\hskip2mm}r@{\hskip2mm}c@{\hskip2mm}c@{\hskip2mm}c}
\tableline
\noalign{\smallskip}
Name &	Other name & RA(2000) & Dec(2000) & V~~ & B$-$V & U$-$B & Type & Remarks \\
\noalign{\smallskip}
\tableline
\noalign{\smallskip}
H1 & HBC\,742 &    23 52 33.0 & 68 25 55 & 14.97 & 1.47 & +0.27  & B8e$\alpha$ &  \\[-1pt]
H2 &        &    23 41 45.0 & 66 39 36 & 12.47 & 1.1 & $-$0.09  &   & 1  \\[-1pt]
H3 & HBC\,319, Blanco 1 &    23 54 26.6 & 66 54 17 & 14.50 & 1.09 & $-$0.15  & K2 & 2,20  \\[-1pt]
H4 & HBC\,321n, Blanco 3 &    23 58 41.4 & 66 26 11 & 14.67 & 1.96: &    & A9e$\alpha$ & 3  \\[-1pt]
H5 & HBC\,322, Blanco 4 &    23 59 20.2 & 66 23 10 & 15.80 & 1.75: &     & K5 & 4,20  \\[-1pt]
H6 &           &    23 59 12.0 & 66 22 16 & 16.79 &       &     & pT &  \\[-1pt]
H7 & Blanco 5  &    ~0 00 57.3 & 66 28 53 & 16.68 & 0.97  &     & pT & 5  \\[-1pt]
H8 & Blanco 11 &    ~0 15 21.6 & 65 45 32 & 17.33 & 0.60: &     & pT & 6  \\[-1pt]
H9 & Blanco 10 &    ~0 13 29.1 & 65 35 59 & 14.96 & 1.28: & +0.72: & K4 & 7,20 \\[-1pt]
H10 & Blanco 9 &    ~0 12 54.4 & 65 34 09 & 14.72 & 1.51: & +0.52: & K4 & 8,20 \\[-1pt]
H11 & Blanco 7 &    ~0 07 06.1 & 65 40 15 & 16.83 & 0.92: &     & pT &  \\[-1pt]
H12 &       &    ~0 07 03.1 & 65 38 37 & 16.23 & 1.12: &     & F: & 9,20  \\[-1pt]
H13 & Blanco 8    &     ~0 07 18.4 & 65 36 42 & 16.77 & 0.87: &     & pT & 10 \\[-1pt]
H14 &       &    23 41 24.8 & 65 40 40 & 15.74 &     &     &   &  \\[-1pt]
H15 & GG~179 &   ~0 17 35.0 & 65 16 08 & 12.12 & 1.04 & $-$0.10 &   & 11 \\[-1pt]
H16 & Sh~118 &   ~0 07 20.2 & 64 57 21 & 13.79 & 1.17 & $-$0.36 &   & 12  \\[-1pt]
H17 &        &   ~0 04 52.2 & 65 05 49 & 13.49 & 0.83 & $-$0.12 & & 13 \\[-1pt]
H18 & HBC\,323, Blanco 6 &   ~0 02 13.0 & 64 54 22 & 14.21 & 1.29: & +0.02: & K7  & 14,20 \\[-1pt]
H19 & HBC\,320, Blanco 2 &   23 57 34.3 & 64 54 21 & 14.21 & 1.29: & +0.02: & K3 & 15,20  \\[-1pt]
H20 & GG 162    &   23 50 02.3 & 64 41 41 & 12.23 &     &    &   &   \\[-1pt]
H21 &          &   ~0 06 40.6 & 65 34 52 & 11.55 & 0.49 & +0.32 &   & 16  \\[-1pt]
H22 & MWC~1085 & 23 52 12.4 & 67 10 07 & 9.96 & 0.53 & $-$0.22  & B3e & 17 \\[-1pt]
H23 & AS~517   &   23 57 33.9 & 66 25 54 & 10.37 & 0.70 & +0.01  & B5e & 18  \\[-1pt]
H24 & AS~2     &    ~0 12 58.9 & 66 19 19 & 10.68 & 0.77 & +0.25  & B5e  & 19 \\[-1pt]
MC1 &  & ~0 06 57.9 & 65 37 21  & 14.6 &  &  & A5 & 20 \\[-1pt]
MC2 &  & ~0 35 57.5 & 66 19 15  & 14.1  &  & & A2 & 20 \\[-1pt]
MC3 &  & ~0 16 35.0 & 65 43 20  & 16.8 & &  & K5 & 20 \\[-1pt]
MC4 &  & ~0 16 42.0 & 65 44 20  & 14.4 & & & K4 & 20 \\[-1pt]
sH15 &  &  ~0 13 23.9 & 65 35 20 & 13.6  &  &  & K1 & 20 \\
\noalign{\smallskip}
\tableline
\noalign{\smallskip}
\end{tabular}}
\end{center}
\smallskip
{\footnotesize
Remarks:
(1) Faint, blue continuum; could be Be.
(2) Probable Ca II infrared emission; suspected var. in B.
(3) LkHa 259; probably not T Tauri type.
(4) Found independently by Herbig (unpublished); probable Ca II infrared emission; var. in B and V.
(5) Found independently by Herbig (unpublished).
(6) Suspected var. in B (no filter) and V.
(8) Found independently by Herbig (unpublished); probable var. in V.
(7) Found independently by Herbig (unpublished); probable var. in V.
(8) Found independently by Herbig (unpublished); probable var. in V.
(9) Near-red nebulosity; probable Ca II infrared emission; var. in V.
(10) Var. in V.
(11) Faint, blue continuum; could be Be.
(12) Known planetary nebula, Sh~118.
(13) Faint, blue continuum; could be Be.
(14) Var. in U and V.
(15) Var. in U and V.
(16) New Be star; spectral type about B8, very broad Balmer lines, particularly H$\zeta$ and H$\eta$.
(17) Known Be star; No. 30 of MacConnell's Table 3.
(18) Known Be star; No. 33 of MacConnell's Table 3.
(19) Known Be star; No. 32 of MacConnell's Table 3.
(20) Spectral type from Cohen \& Kuhi (1976)}
\end{table}

\citet{Sharma07} obtained slitless spectroscopy and JH photometry for Berkeley~59.
They identified 9 H$\alpha$ emission stars, whose location in the J/J$-$H color-magnitude
diagram indicates that they may be pre-main sequence stars. The age of the cluster
was estimated from the turn-off and turn-on points and is found to lie between
about 1 and 4 million years. \citet{Pandey08} present $UBVI_\mathrm{C}$  CCD photometry of Be~59 with
the aim to study the star formation scenario in the cluster.  Using
slitless spectroscopy, they have identified
48 H$\alpha$ emission stars in the region of Be~59. The ages of
these YSOs range between $<$1 and $\sim$~2~Myr, whereas the mean age of the massive
stars in the cluster region is found to be $\sim$2~Myr. They found evidence for
second-generation star formation outside the boundary of the cluster, which
may be triggered by massive stars in the cluster.
The radial extent
of the cluster is found to be $\sim$10~arcmin (2.9~pc). The interstellar extinction
in the cluster region varies between E(B$-$V) = 1.4 to 1.8 mag. The ratio
of total-to-selective extinction in the cluster region is estimated as
$3.7\pm0.3$. The distance of the cluster is found to be $1.00\pm0.05$~kpc.

Spectroscopic observations of the eclipsing binary system BD+66\deg1673 by \citet{Majaess08}
revealed it to be an O5\,V((f))n star and the probable
ionizing star of the Be~59/Cep OB4 complex.

\citet{YF92} discovered two dense molecular clumps near Be~59 and in the central region
of S\,171. They have shown that the clumps indicate the interaction between the HII region
and the neighboring molecular cloud.

\citet{SFO} identified three bright rimmed clouds, BRC\,1, BRC\,2, and BRC\,3, associated
with S\,171. We note that BRC\,2 is actually associated with NGC 7822 (see Fig.~\ref{fig_cepob41}).
\citet{OSP} found H$\alpha$ emission stars in BRC\,1 and BRC\,2. Each cloud is included
in the SCUBA survey of bright rimmed clouds by \citet{Morgan08}.
Table~\ref{Tab_brc_s171}
lists the coordinates and H$\alpha$ equivalent widths  \citep[revised by][]{Ikeda08} of these stars, and
Fig.~\ref{fig_brc_s171} shows the finding charts.

\begin{table}[!ht]
\caption{H$\alpha$ emission stars associated with bright rimmed clouds of S\,171 and NGC\,7822 \citep{OSP,Ikeda08}.}
\label{Tab_brc_s171}
\begin{center}
{\footnotesize
\begin{tabular}{lcccl}
\tableline
\noalign{\smallskip}
N & RA(J2000) & Dec(J2000) & EW(H$\alpha$) & Remarks \\
\tableline
\noalign{\smallskip}
\multicolumn{5}{c}{BRC 1} \\
1   & 23 59 42.50 & 67 22 27.8 & $\cdots$ & invisible cont., contam. from nearby star      \\[-1pt]
2   & 23 59 43.72 & 67 25 50.0 &  ~26.6 & weak cont.	     \\[-1pt]
3   & 23 59 43.99 & 67 22 46.8 & $\cdots$ & invisible cont.	     \\[-1pt]
4   & 23 59 47.61 & 67 23 11.2 & $\cdots$ & H$\alpha$ ?, invisible cont.	\\[-1pt]
5   & 23 59 47.98 & 67 23 06.9 & $\cdots$ & H$\alpha$ ?, weak cont.	   \\[-1pt]
6   & 23 59 52.75 & 67 25 37.5 &  ~32.6 & weak cont.	     \\
\multicolumn{5}{c}{BRC 2} \\
1   & 00 03 51.03 & 68 33 15.8 & $\cdots$ & H$\alpha$ ? 	 \\[-1pt]
2   & 00 03 52.32 & 68 31 58.8 & $\cdots$ & invisible cont.	     \\[-1pt]
3   & 00 03 54.52 & 68 33 44.6 & $\cdots$ & contam. from nearby stars	\\[-1pt]
4   & 00 03 54.98 & 68 32 42.8 & 145.7 & very weak cont.	 \\[-1pt]
5   & 00 03 57.12 & 68 33 46.7 &  ~15.6 &	    \\[-1pt]
6   & 00 03 57.35 & 68 33 23.0 &  ~99.6 &	    \\[-1pt]
7   & 00 03 58.33 & 68 34 06.5 &  ~13.1 &	    \\[-1pt]
8   & 00 03 59.12 & 68 32 47.3 &  ~18.2 &	    \\[-1pt]
9   & 00 04 01.70 & 68 34 13.8 &   ~~2.8 &	    \\[-1pt]
10  & 00 04 01.81 & 68 34 00.0 &  ~14.0 &	    \\[-1pt]
11  & 00 04 01.80 & 68 34 37.4 &  ~~5.5 &	    \\[-1pt]
12  & 00 04 01.88 & 68 34 34.5 &  ~21.2 &	    \\[-1pt]
13  & 00 04 02.32 & 68 31 36.2 & $\cdots$ & H$\alpha$ ? 	 \\[-1pt]
14  & 00 04 02.65 & 68 34 26.6 &  ~24.9 &	    \\[-1pt]
15  & 00 04 04.69 & 68 33 49.3 &  ~25.8 & weak cont.	     \\[-1pt]
16  & 00 04 04.59 & 68 34 52.2 &  ~23.0 &	    \\[-1pt]
17  & 00 04 05.28 & 68 33 56.0 & 136.6 & very weak cont.	 \\[-1pt]
18  & 00 04 05.26 & 68 33 53.1 &  ~50.4 &	    \\[-1pt]
19  & 00 04 05.66 & 68 33 44.3 &  ~94.5 &	    \\[-1pt]
20  & 00 04 07.33 & 68 33 42.3 & $\cdots$ & invisible cont.	     \\[-1pt]
21  & 00 04 07.60 & 68 33 25.1 &  ~12.8 &	    \\[-1pt]
22  & 00 04 11.66 & 68 33 25.4 &  ~46.7 &	    \\[-1pt]
23  & 00 04 13.97 & 68 32 21.8 &  ~85.7 & weak cont.	     \\[-1pt]
24  & 00 04 14.71 & 68 32 49.1 &  ~42.1 &	    \\[-1pt]
25  & 00 04 15.18 & 68 33 02.0 &  ~16.1 &	    \\[-1pt]
26  & 00 04 15.41 & 68 34 05.5 &  ~26.6 & very weak cont.	 \\[-1pt]
27  & 00 04 21.66 & 68 30 59.6 & $\cdots$ & H$\alpha$ ? 	 \\[-1pt]
28  & 00 04 25.36 & 68 32 31.0 &  ~12.3 & weak cont.	     \\[-1pt]
29  & 00 04 29.99 & 68 31 19.9 & $\cdots$ & H$\alpha$ ? 	 \\[-1pt]
30N & 00 03 59.88 & 68 33 41.7 &  ~~1.4 &	    \\
\noalign{\smallskip}
\tableline
\noalign{\smallskip}
\end{tabular}}
\end{center}
\end{table}

The bright rimmed cloud BRC\,2 contains a compact cluster of H$\alpha$
emission stars. The {\em S\,171 cluster\/} was observed by the IRAC
on board the {\it Spitzer Space Telescope\/} \citep{Mege04}.
The cluster of young stars is situated near the edge of the cloud with a
dense group of five Class~I sources at the northern apex of the cluster.
This morphology suggests that star formation is triggered by a
photoevaporation-driven shock wave propagating into the cloud, as
first proposed for this region by \citet{SFO}.
In addition to the stars in the cluster, {\it Spitzer\/} detected six Class~II and
two Class~I objects spread throughout the molecular cloud. The presence
of these stars suggests that a distributed mode of star formation is
also occurring in the cloud.

A new generation of low mass stars has been born in the molecular clouds in the neighborhood of
the young luminous stars.
\citet{Yang90} reported on the discovery of a molecular cloud in Cep~OB4. The cloud
M\,120.1+3.0, appearing dense and filamentary, is composed of two parts. Each of the
two parts has a size of 6\,pc$\times$1\,pc, and a mass of 800\,M$_{\sun}$.
The cloud is associated with 12 low luminosity ($L < 20\,L_{\sun}$) IRAS sources,
and the locations of the sources show remarkable coincidence with the distribution
of the dense molecular gas. Two molecular outflows have been discovered towards two IRAS
sources, IRAS 00213+6530 and IRAS 00259+6510. The results indicate that low-mass
star formation took place recently in the cloud.

\section{Cepheus OB6}

The nearby association Cep~OB6 first appeared in the literature in
1999. \Citet{deZeeuw} identified this moving group of 27 stars in the
Hipparcos data base. The stars show a modest concentration at (l,b)$
\approx (104\fdg0,-0\fdg5$). The final sample contains 20 stars, 6 B,
7 A, 1 F, 2 G and 3 K type in the area $110\deg < l < 110\deg$, and
$-2\deg < b < +2\deg$. The brightest member is the K1Ib supergiant
$\zeta$~Cep (HIP~109492). The color--magnitude diagram is very narrow,
and strengthens the evidence that these stars form a moving group,
that is, an old OB association. The earliest spectral type is B5III,
suggesting an age of some 50 million years.  The mean distance of the
association is $270\pm12$~pc. The supergiant $\delta$~Cephei, the
archetype of Cepheid variables, also belongs to Cep~OB6.
\citet{Makarov07} during his study of the Galactic orbits of nearby
stars found that a few members of the AB~Dor moving group were in
conjunction with the coeval Cepheus~OB6 association 38~Myr ago.  He
proposed that the AB~Dor nucleus formed 38~Myr ago during a close
passage of, or encounter with, the Cepheus~OB6 cloud, which may have
triggered formation of the latter association as well. No younger
subgroup of Cep~OB~6 has been identified.

\acknowledgements This work was supported by the Hungarian OTKA grant
\linebreak T049082.  We are grateful to Jeong-Eun Lee for sending us
the results
on L\,1251\,B before publication, to Mikl\'os R\'acz for his help with
some figures, to L\'aszl\'o Szabados for a careful reading of the
manuscript, and to Tom Megeath for the data in Table~\ref{Tab_S140_4}.
We thank Giovanni Benintende, Richard Gilbert, John Bally, Robert
Gendler, and Davide De Martin for the use of figures 6, 8, 10, 12,
and 16, respectively.  Bo Reipurth's referee report led to an enormous
improvement of this chapter.  We used the {\it Simbad\/} and {\it
ADS\/} data bases throughout this work.

\end{document}